\documentclass[aps,prd,twocolumn,showpacs,superscriptaddress,floatfix]{revtex4-2}  
\usepackage{graphicx}  
\usepackage{dcolumn}   
\usepackage{longtable}
\usepackage{bm}        
\usepackage{amssymb}   
\usepackage{booktabs}
\usepackage[caption=false]{subfig}

\usepackage[caption=false]{subfig}
\usepackage{xcolor}
\usepackage{subfig}
\usepackage{amsmath}
\usepackage{comment}
\usepackage{subfiles}
\usepackage{placeins}
\usepackage{array}
\usepackage[normalem]{ulem}
\usepackage{xcolor}
\usepackage{float}
\usepackage{todonotes}
\usepackage{dblfloatfix}
\usepackage{multirow}
\usepackage{xspace}
\newcounter{todocounter}

\newcommand{\tkialpha}{$\delta\alpha_{T}$\xspace}
\newcommand{\tkicoplanar}{$\theta_{coplanarity}$\xspace}
\newcommand{\tkiptmu}{P$_{\mu T}$\xspace}

\newcommand{\water}{H$_2$0}
\newcommand{\chisq}{\ensuremath{\chi^2}\xspace}
\newcommand{\ndf}{\ensuremath{N_{dof}}\xspace}

\newcommand{\tkiptproton}{\ensuremath{P^P_T}\xspace}
\newcommand{\tkidelta}{\ensuremath{\delta P_{T}}\xspace}
\newcommand{\tkidptx}{\ensuremath{\delta P_{T}}$_{x}$\xspace}
\newcommand{\tkidpty}{\ensuremath{\delta P_{T}}$_{y}$\xspace}

\newcommand{\tkipl}{\ensuremath{\delta P_{L}}\xspace}
\newcommand{\tkipn}{\ensuremath{{P}_{n}}\xspace}
\newcommand{\tkiphi}{\ensuremath{\phi_{T}}\xspace}

\newcolumntype{L}[1]{>{\raggedright\hspace{0.5cm}\arraybackslash}p{#1}}
\newcolumntype{C}[1]{>{\centering\hspace{0.5cm}\arraybackslash}p{#1}}
\newcolumntype{R}[1]{>{\raggedleft\hspace{0.5cm}\arraybackslash}p{#1}}
\newcolumntype{P}{>{\hspace{0.3em}}c<{\hspace{0.3em}}}
\newcolumntype{Q}{l<{\hspace{0.3em}}}

\begin{document}

\title{Measurement of the {\it A} dependence of the $\nu_{\mu}$ charged-current quasielastic-like cross section as a function of muon and proton kinematics at $<$E$_{\nu}>\sim$ 6~GeV}
\newcommand{\minerva}{MINERvA}

\newcommand{\Rutgers}{Rutgers, The State University of New Jersey, Piscataway, New Jersey 08854, USA}
\newcommand{\Hampton}{Hampton University, Dept. of Physics, Hampton, VA 23668, USA}
\newcommand{\Dortmund}{Institute of Physics, Dortmund University, 44221, Germany }
\newcommand{\Otterbein}{Department of Physics, Otterbein University, 1 South Grove Street, Westerville, OH, 43081 USA}
\newcommand{\JMU}{James Madison University, Harrisonburg, Virginia 22807, USA}
\newcommand{\Florida}{University of Florida, Department of Physics, Gainesville, FL 32611}
\newcommand{\UCIrvine}{Department of Physics and Astronomy, University of California, Irvine, Irvine, California 92697-4575, USA}
\newcommand{\CBPF}{Centro Brasileiro de Pesquisas F\'{i}sicas, Rua Dr. Xavier Sigaud 150, Urca, Rio de Janeiro, Rio de Janeiro, 22290-180, Brazil}
\newcommand{\PUCP}{Secci\'{o}n F\'{i}sica, Departamento de Ciencias, Pontificia Universidad Cat\'{o}lica del Per\'{u}, Apartado 1761, Lima, Per\'{u}}
\newcommand{\INRM}{Institute for Nuclear Research of the Russian Academy of Sciences, 117312 Moscow, Russia}
\newcommand{\Jlab}{Jefferson Lab, 12000 Jefferson Avenue, Newport News, VA 23606, USA}
\newcommand{\Pittsburgh}{Department of Physics and Astronomy, University of Pittsburgh, Pittsburgh, Pennsylvania 15260, USA}
\newcommand{\Guanajuato}{Campus Le\'{o}n y Campus Guanajuato, Universidad de Guanajuato, Lascurain de Retana No. 5, Colonia Centro, Guanajuato 36000, Guanajuato M\'{e}xico.}
\newcommand{\Athens}{Department of Physics, University of Athens, GR-15771 Athens, Greece}
\newcommand{\Tufts}{Physics Department, Tufts University, Medford, Massachusetts 02155, USA}
\newcommand{\WM}{Department of Physics, William \& Mary, Williamsburg, Virginia 23187, USA}
\newcommand{\FNAL}{Fermi National Accelerator Laboratory, Batavia, Illinois 60510, USA}
\newcommand{\Purdue}{Department of Chemistry and Physics, Purdue University Calumet, Hammond, Indiana 46323, USA}
\newcommand{\MCLA}{Massachusetts College of Liberal Arts, 375 Church Street, North Adams, MA 01247}
\newcommand{\UMD}{Department of Physics, University of Minnesota -- Duluth, Duluth, Minnesota 55812, USA}
\newcommand{\Northwestern}{Northwestern University, Evanston, Illinois 60208}
\newcommand{\UNI}{Facultad de Ciencias F\'{i}sicas, Universidad Nacional Mayor de San Marcos, CP 15081, Lima, Per\'{u}}
\newcommand{\Rochester}{Department of Physics and Astronomy, University of Rochester, Rochester, New York 14627 USA}
\newcommand{\Austin}{Department of Physics, University of Texas, 1 University Station, Austin, Texas 78712, USA}
\newcommand{\USM}{Departamento de F\'{i}sica, Universidad T\'{e}cnica Federico Santa Mar\'{i}a, Avenida Espa\~{n}a 1680 Casilla 110-V, Valpara\'{i}so, Chile}
\newcommand{\Geneva}{University of Geneva, 1211 Geneva 4, Switzerland}
\newcommand{\Chicago}{Enrico Fermi Institute, University of Chicago, Chicago, IL 60637 USA}
\newcommand{\hired}{}
\newcommand{\OregonState}{Department of Physics, Oregon State University, Corvallis, Oregon 97331, USA}
\newcommand{\oxford}{Oxford University, Department of Physics, Oxford, OX1 3PJ United Kingdom}
\newcommand{\umiss}{University of Mississippi, Oxford, Mississippi 38677, USA}
\newcommand{\upenn}{Department of Physics and Astronomy, University of Pennsylvania, Philadelphia, PA 19104}
\newcommand{\AMU}{Department of Physics, Aligarh Muslim University, Aligarh, Uttar Pradesh 202002, India}
\newcommand{\wroclaw}{University of Wroclaw, plac Uniwersytecki 1, 50-137 Wroa\l{}aw, Poland}
\newcommand{\Mohali}{Department of Physical Sciences, IISER Mohali, Knowledge City, SAS Nagar, Mohali - 140306, Punjab, India}
\newcommand{\CINVESTAV}{Departamento de Fisica Col. San Pedro Zacatenco, 07360 Mexico, DF, Av. Instituto PolitÃ©cnico Nacional, Mexico}
\newcommand{\york}{York University, Department of Physics and Astronomy, Toronto, Ontario, M3J 1P3 Canada}
\newcommand{\ND}{Department of Physics and Astronomy, University of Notre Dame, Notre Dame, Indiana 46556, USA}
\newcommand{\ICL}{The Blackett Laboratory,  Imperial College London,  London SW7 2BW, United Kingdom}
\newcommand{\warwick}{Department of Physics, University of Warwick, Coventry, CV4 7AL, UK}
\newcommand{\qmul}{G O Jones Building, Queen Mary University of London, 327 Mile End Road, London E1 4NS, UK}

\newcommand{\kleykampThanks}{now at Department of Physics and Astronomy, University of Mississippi, Oxford, MS 38677}
\newcommand{\mascencioThanks}{Now at Iowa State University, Ames, IA 50011, USA}
\newcommand{\ricfregianThanks}{now at Department of Physics and Astronomy, University of California at Davis, Davis, CA 95616, USA}
\newcommand{\finerThanks}{Now at Los Alamos National Laboratory, Los Alamos, New Mexico 87545, USA}
\newcommand{\adrianThanks}{Now at Department of Physics, Drexel University, Philadelphia, Pennsylvania 19104, USA}
\newcommand{\bamThanks}{Now at University of Minnesota, Minneapolis, Minnesota 55455, USA}
\newcommand{\byaeggyThanks}{Now at Department of Physics, University of Cincinnati,  Cincinnati, Ohio 45221, USA}
\newcommand{\lazazuetareyesThanks}{now at Syracuse University, Syracuse, NY 13244, USA}


\author{J.~Kleykamp}\thanks{\kleykampThanks}  \affiliation{\Rochester}
\author{S.~Akhter}                        \affiliation{\AMU}
\author{Z.~~Ahmad~Dar}                    \affiliation{\WM}  \affiliation{\AMU}
\author{N.S.~Alex}                        \affiliation{\Rochester}
\author{V.~Ansari}                        \affiliation{\AMU}
\author{M.~V.~Ascencio}\thanks{\mascencioThanks}  \affiliation{\PUCP}
\author{M.~Sajjad~Athar}                  \affiliation{\AMU}
\author{M.~Betancourt}                    \affiliation{\FNAL}
\author{J.~L.~Bonilla}                    \affiliation{\Guanajuato}
\author{A.~Bravar}                        \affiliation{\Geneva}
\author{G.~Caceres}\thanks{\ricfregianThanks}  \affiliation{\CBPF}
\author{G.A.~D\'{i}az~}                   \affiliation{\FNAL}  \affiliation{\Rochester}
\author{H.~da~Motta}                      \affiliation{\CBPF}
\author{J.~Felix}                         \affiliation{\Guanajuato}
\author{L.~Fields}                        \affiliation{\ND}
\author{R.~Fine}\thanks{\finerThanks}     \affiliation{\Rochester}
\author{A.M.~Gago}                        \affiliation{\PUCP}
\author{H.~Gallagher}                     \affiliation{\Tufts}
\author{P.K.Gaur}                         \affiliation{\AMU}
\author{R.~Gran}                          \affiliation{\UMD}
\author{E.Granados}                       \affiliation{\Guanajuato}  \affiliation{\Guanajuato}
\author{D.A.~Harris}                      \affiliation{\york}  \affiliation{\FNAL}
\author{A.L.~Hart}                        \affiliation{\qmul}
\author{A.~Klustov\'{a}}                  \affiliation{\ICL}
\author{M.~Kordosky}                      \affiliation{\WM}
\author{D.~Last}                           \affiliation{\Rochester} \affiliation{\upenn}
\author{A.~Lozano}\thanks{\adrianThanks}  \affiliation{\CBPF}
\author{X.-G.~Lu}                         \affiliation{\warwick}  \affiliation{\oxford}
\author{S.~Manly}                         \affiliation{\Rochester}
\author{W.A.~Mann}                        \affiliation{\Tufts}
\author{C.~Mauger}                        \affiliation{\upenn}
\author{K.S.~McFarland}                   \affiliation{\Rochester}
\author{M.~Mehmood}                       \affiliation{\york}
\author{B.~Messerly}\thanks{\bamThanks}   \affiliation{\Pittsburgh}
\author{O.~Moreno}                        \affiliation{\WM}  \affiliation{\Guanajuato}
\author{J.G.~Morf\'{i}n}                  \affiliation{\FNAL}
\author{J.K.~Nelson}                      \affiliation{\WM}
\author{C.~Nguyen}                        \affiliation{\Florida}
\author{A.~Olivier}                       \affiliation{\ND}  \affiliation{\Rochester}
\author{V.~Paolone}                       \affiliation{\Pittsburgh}
\author{G.N.~Perdue}                      \affiliation{\FNAL}  \affiliation{\Rochester}
\author{C.~Pernas}                        \affiliation{\WM}
\author{K.-J.~Plows}                      \affiliation{\oxford}
\author{M.A.~Ram\'{i}rez}                 \affiliation{\upenn}  \affiliation{\Guanajuato}
\author{R.D.~Ransome}                     \affiliation{\Rutgers}
\author{N.~Roy}                           \affiliation{\york}
\author{D.~Ruterbories}                   \affiliation{\Rochester}
\author{H.~Schellman}                     \affiliation{\OregonState}
\author{C.~J.~Solano~Salinas}             \affiliation{\UNI}
\author{D.~S.~Correia}                    \affiliation{\CBPF}
\author{M.~Sultana}                       \affiliation{\Rochester}
\author{V.S.~Syrotenko}                   \affiliation{\Tufts}
\author{E.~Valencia}                      \affiliation{\Guanajuato}  \affiliation{\WM}
\author{N.H.~Vaughan}                     \affiliation{\OregonState}
\author{A.V.~Waldron}                     \affiliation{\qmul}  \affiliation{\ICL}
\author{B.~Yaeggy}\thanks{\byaeggyThanks}  \affiliation{\USM}
\author{L.~Zazueta}\thanks{\lazazuetareyesThanks}  \affiliation{\WM}

\collaboration{The MINERvA Collaboration}
\noaffiliation

\begin{abstract} 
The first simultaneous measurements of the $\nu_{\mu}$ quasielastic-like cross section on C, CH, H$_2$0, Fe, and Pb targets as a function of kinematic imbalance variables in the plane transverse to the incoming neutrino direction are presented.  These variables combine the muon and proton information to provide a new way to disentangle the effects of the nucleus in quasielastic-like processes.  The data were obtained using a wide-band $\nu_{\mu}$ beam with $<$E$_{\nu}>\sim$ 6~GeV.  Cross-section ratios of the different target materials to CH are also shown.  These measurements are used to explore the nature of the cross-section $A$-scaling, as well as initial and final state interaction effects.  Comparisons are made to predictions from a number of commonly used neutrino Monte Carlo event generators.  The range of predictions of the different models tends to cover the data but the degree and consistency of the agreement suffers in regions, and on higher $A$ targets, where the final state interactions are expected to be more pronounced.

\end{abstract}

\maketitle


\section{Introduction}

In neutrino charged current quasielastic scattering (CCQE), $\nu_\mu + n \to \mu^- + p$, the interaction occurs on a nucleon neutron within a nucleus. Often an estimate of the neutrino energy is made under the assumption that the neutron is free, or bound with no Fermi smearing, and with no final state nuclear effects.
Under this assumption, the energy can be estimated simply from the final state muon and proton energies, or from the final state muon momentum and angle with respect to the neutrino direction.  
 However,  initial and final state nuclear effects can occur which can add additional smearing to 
 the neutrino energy reconstruction.  Such interactions are not true CCQE but are often indistinguishable from CCQE interactions experimentally.  In fact, any interaction without pions in the final state is characterized as 
 quasielastic-like (CCQE-like). Measurements of nuclear effects in CCQE-like scattering provide important tests of nuclear models, which are needed to be able to predict the effect that the nucleus has on neutrino energy reconstruction.

This work presents the simultaneous measurement of the $\nu_{\mu}$ quasielastic-like (CCQE-like) cross section across different nuclear target materials.  These results are based on events with both a reconstructed muon and a reconstructed proton in the final state. The sample used is a subset of the one used in a more inclusive muon-only analysis \cite{jk_1trk_prl}, and includes the addition of the reconstructed proton requirement. 
The added proton information allows the reconstruction of kinematic information in the plane transverse to the incoming neutrino direction. 
 Momentum imbalance in this plane gives hints to the effects that keep this from being a simple elastic scattering process.  Transverse kinematic imbalance (TKI) variables
 exhibit enhanced sensitivity to nuclear effects \cite{Lu:2015tcr}.  The results here are presented differentially as a function of TKI variables, and in terms of the individual muon and proton momenta and angles with respect to the neutrino beam direction. 

\begin{figure}[tb]
\centering
\includegraphics[width=0.48\textwidth]{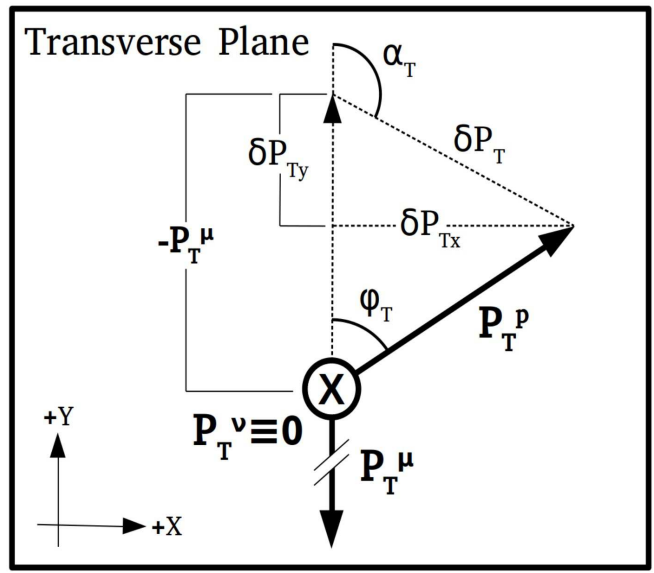}
\caption{Schematic of transverse kinematic imbalance variables used by this analysis.   The variables represent directions and magnitudes of momentum vectors that lie in the plane transverse to the beam direction.
}
\label{fig:TKI}
\end{figure}

Figure~\ref{fig:TKI} 
illustrates the definitions of most of the TKI variables used in this work. These variables are based on transverse momentum conservation and are easy to interpret in the limit of CCQE interactions without nuclear effects, i.e., no Fermi motion for the target nucleon and no final state interactions or initial state nuclear effects. The transverse momenta with respect to the beam direction of the proton and muon tracks are P$_{p T}$ and \tkiptmu , respectively. The missing transverse momentum in the plane transverse to the beam axis is \tkidelta .   The components of \tkidelta \ perpendicular and parallel to the muon direction are \tkidptx \ and \tkidpty , respectively. The angle between the negative transverse muon momentum and \tkidelta \ is \tkialpha . 
The extent to which the proton and muon transverse momenta are not colinear is expressed as \tkiphi . An additional variable of interest not shown in Fig.~\ref{fig:TKI}, but included in the analysis, is the initial state neutron momentum, \tkipn, which can be inferred from assuming conservation of momentum, given the final state muon and proton momenta and angles.

 Deviations from free-nucleon behavior probe the nature of the nuclear effects.  For example, for scattering off a free nucleon at rest in the lab frame,
the final state muon and proton transverse momenta would sum to zero, so \tkidelta\ would be zero. 
These variables have been discussed extensively in recent years and used by the T2K, MINERvA, and MicroBooNE experiments to probe nuclear effects in neutrino interactions~\cite{Lu:2015tcr,Furmanski:2016wqo,Abe:2018pwo,Dolan:2018sbb,MINERvA:2018hba,Dolan:2018zye,Lu:2019nmf,Harewood:2019rzy,Cai:2019jzk,Cai:2019hpx,Coplowe:2020yea,MicroBooNE:2023wzy}.
This work expands on the earlier MINERvA results using these variables by systematically surveying  CCQE-like interactions on targets of C, CH, H$_2$0, Fe, and Pb.  The results are also compared to predictions from several widely used Monte Carlo generators.

The large number of variables and wide range of target nucleus sizes explored in this work comprise a significant and systematic survey sensitive to {\it A}-dependent nuclear effects.  Due to the sheer volume of results, the analysis is described in detail for a single variable, $\delta P_{T}$, which represents the transverse momentum imbalance (or the residual transverse momentum) of the muon-proton system.  For all other variables, the analysis is similar and only the results are presented.  
The results for each variable will be presented in such a way that comparison of behavior across several target nuclei is possible.


\section{Experiment}

The  Main Injector Experiment for $\nu$-A (MINERvA)~\cite{minervaweb, MINERvA:2013zvz} 
took data in the Neutrino Main Injector (NuMI) neutrino beam at Fermilab~\cite{numibeam}. The NuMI beam is formed by colliding 120~GeV protons from the Main Injector on a graphite target.  The charged pions and kaons resulting from those collisions are focused by two magnetic horns towards a 675~m decay region, and the neutrinos in the beam come primarily from the decays of these mesons.  The polarity of the two horn currents selects the charge sign of the focused mesons which, in turn, determines whether the beam is enriched in  neutrinos or antineutrinos (predominately muon flavor). The work here makes use of data recorded between 2012 and 2019, with the horn polarity optimized for neutrinos (with $\sim$3.8\%  contamination of antineutrinos).

MINERvA accumulated data sets corresponding to 5.4$\times$10$^{20}$ protons-on-target (POT) in the low energy (LE) beam, with $<$E$_\nu$$>$$\sim$3~GeV, and 24.5$\times$10$^{20}$ POT in the medium energy (ME) beam with $<$E$_\nu$$>$$\sim$6~GeV.  
The analysis described here is based on a sample corresponding to 10.61$\times$10$^{20}$ POT in the neutrino enhanced ME configuration.

The MINERvA detector, described in detail in Ref.~\cite{MINERvA:2013zvz}, consists of 208 hexagonal planes each consisting of 127 polystyrene scintillator strips.   The strips run perpendicular to the neutrino beam with three different angular orientations that vary across adjacent planes enabling three-dimensional track reconstruction. The strips are triangular in cross section with a height of $\sim$1.7~cm and a width of $\sim$3.3~cm.  In the upstream part of the detector (closer to the decay region), there is a $\sim$1.25~m-long region with layers of passive nuclear targets interspersed with scintillator planes.  This portion of the detector, the so-called nuclear target region, contains passive targets of carbon, iron, lead, and water.  Data from interactions on these targets are used extensively in this work. The nuclear targets are described in more detail in Refs.~\cite{MINERvA:2013zvz} and~\cite{jk_1trk_prl}.  Downstream of the nuclear target region (along the direction of the beam) is a region of 124 contiguous scintillator planes, called the tracker region of the detector.  Further downstream is a region of the detector with 20 scintillator planes and layers of 2~mm thick Pb placed between every two planes.  This region acts as an electromagnetic calorimeter.  Finally, there is a region where 20 planes of scintillator are interleaved with layers of 2.54~cm thick steel, functioning as a hadronic calorimeter.  The overall length of the detector 
is $\sim$5~m.  The magnetized MINOS near detector \cite{MINOS:2008hdf}, is used as a muon range stack and spectrometer, and its most upstream steel plane lies 2~m downstream of MINERvA.

\section{Simulation}
\label{sec:simulation}

A GEANT4-based simulation~\cite{GEANT4:2002zbu, *Allison:2006ve} of the NuMI beamline is used to calculate the expected NuMI (anti)neutrino flux at the MINERvA detector as described in~\cite{MINERvA:2016iqn, *fluxaddendum}.  The simulation of the hadron production from the target is reweighted to agree with external measurements of hadron production in proton beams~\cite{MIPP:2014shj, *NA49:2006oyk}. In addition, the flux prediction is constrained by an {\it in situ} measurement of neutrino scattering off atomic electrons\cite{MINERvA:2019mkf,MINERvA:2021mpk}.

Neutrino interactions are simulated for use in the analysis and for comparison purposes by several different neutrino event generators and variations in the generator options.  The comparison of these simulations to data will be discussed later in this paper.  The simulated interactions used in the analysis come from the GENIE version 2.12.6 neutrino event generator~\cite{GEANT4:2002zbu}.  The nucleons inside the nucleus are simulated using a relativistic Fermi gas model with the addition of a Bodek-Ritchie high-momentum tail to account for nucleon-nucleon short range correlations~\cite{Bodek:1981wr}.  The maximum momentum for the Fermi motion is k$_{F}$=0.221 GeV/c. The binding energy and mass density of the nucleus are constrained by electron scattering measurements.  For nuclei smaller than calcium, the mass density is modeled using a Gaussian density parametrization.  For heavier nuclei, the 2-parameter Wood-Saxon density function is used~\cite{DeVries:1987atn}.  Quasielastic interactions are simulated using the
Llewellyn-Smith formalism~\cite{LlewellynSmith:1971uhs} with the vector form factors modeled using the BBBA05 model~\cite{Bradford:2006yz}. The axial vector form factor uses the dipole form with an axial mass of M$_{A}$ = 0.99 GeV/c$^{2}$.  Scattering off of correlated pairs of nucleons (so-called \lq\lq 2p2h") are simulated using the IFIC Valencia model with events simulated only with a three-momentum transfer less than 1.2 GeV/c~\cite{Gran:2013kda, Nieves:2011pp}.  Resonance production is simulated using the Rein-Sehgal model with an axial mass of M$_{A}^{RES}$ = 1.12 GeV/c$^{2}$~\cite{Rein:1980wg}. Higher invariant mass interactions are simulated using a leading order model for deep inelastic scattering (DIS) with the Bodek-Yang prescription for the modification at low Q$^{2}$~\cite{Bodek:2004pc}.  Hadronic showers are created using the AGKY hadronization model~\cite{Yang:2009zx}.
The INTRANUKE-hA package within GENIE is used to simulate intranuclear hadron transport~\cite{Dytman:2007zz, *Dytman:2011zz}.  

The base neutrino interaction model described above is modified in order to improve agreement with previously analyzed MINERvA data. The cross section is changed as a function of energy and three momentum transfer based on the random phase approximation (RPA) part
of the Valencia model~\cite{Nieves:2004wx}\cite{Gran1705} appropriate for a Fermi gas~\cite{Martini:2009uj,*Nieves:2017lij}; the screening effect for carbon is used for heavier nuclei as an approximation. 
The cross section for the 2p2h events simulated by the Valencia Model is increased in specific regions of energy and three momentum transfer space based on empirical fits to MINERvA data in the LE beam configuration~\cite{MINERvA:2015ydy}. Also, based on MINERvA measurements~\cite{MINERvA:2014ogb, *MINERvA:2014ani} and a reanalysis of pion production data in bubble chamber data~\cite{Wilkinson:2014yfa}, the non-resonant pion production is reduced significantly.  A known bug for the elastic hA FSI events is fixed by reweighting those events to be no-FSI events  for the C, O, Fe, and Pb nuclei~\cite{Harewood:2019rzy}. This modified version of the interaction model, "MINERvA tune v1.0.1"  is referred to as the MINERvA tune in the remainder of this manuscript.

The response of the MINERvA detector is simulated using GEANT4 version 4.9.3.p6 with the QGSP{\_}BERT physics list~\cite{Kaidalov:1982xg}.  The performance of the optical elements and the electronics are included in the simulation. The water target configuration, i.e., empty versus filled, is also modeled as a function of time.  Through-going muons are used to set the absolute energy scale of minimum ionizing energy depositions by requiring the average and RMS of energy deposits match between data and simulation as a function of time. Data taken with a small scale version of MINERvA exposed to a charged particle test beam was used to set the absolute scale of hadron energy response~\cite{MINERvA:2015yej}.  Overlapping beam-related activity in the detector, i.e., hits from particles produced in other interactions within or outside the detector, is simulated by overlaying hits in both MINERvA and MINOS from data corresponding to random beam spills appropriate to the time periods in the simulation so as to reflect changes in beam intensity.

The longitudinal and transverse distribution of the different nuclear target materials relative to the beam varies somewhat.  This means that the flux seen by each target species is slightly different.  The presented cross section ratios are corrected for this effect, such that the fluxes used in the numerator and denominator of the ratio are the same and equal to the flux incident on the nuclear target in the numerator.

\section{Analysis and results for \tkidelta }
\label{sec:analysis}
This analysis  
selects for CCQE-like interaction candidates on CH, C, H$_2$O, Fe, and Pb targets in NUMI's ME neutrino-enriched beam. The interactions are considered signal if they have a muon with an angle with respect to the beam of $<$17$^{\circ}$ and a momentum within the range 2~GeV/c~$<$~p$_{\mu}$~$<$~20~GeV/c, and a proton with an angle $<$70$^{\circ}$ and a momentum in the range 500~MeV/c~$<$~p$_{p}$~$<$~1100~MeV/c. Additionally, the interaction must not have mesons, baryons heavier than neutrons, 
or photons above 10~MeV. The muon constraints are due to the requirement that their momentum is reconstructed accurately in MINOS. The proton angular constraint is added because MINERvA is unable to reconstruct high angle protons because of the orientation of  the scintillator strips. 
The protons need a 
minimum momentum to penetrate several scintillator planes and form enough energy deposits for a track to be reconstructed.
High momentum protons are more likely to either leave the sensitive volume of the detector or scatter inelastically, leading to poor momentum reconstruction. 
For events with more than one  
proton matching the constraints, the highest momentum 
matching proton 
is used.  The presence of a heavy baryon is indicative of a non-CCQE-like interaction, as are the higher energy photons that mostly come from neutral pion or eta decays.  The photons with an energy less than 10~MeV are accepted since they can come from nuclear de-excitations.

Events selected for this analysis have a negatively charged muon candidate reconstructed in MINOS and at least one proton candidate.  
To reject charged pions, the event must have no  
electron candidates from pion decays (\lq\lq Michel electrons")
near the vertex or any reconstructed track endpoint.  No more than one isolated cluster of energy is allowed in selected events in order to reduce the neutral pion background.  The muon and the proton, or the highest momentum proton candidate if there is more than one candidate, are the tracks used in the TKI variable calculations.  The calculations are not changed based on the number of protons in the final state since secondary protons may not be reconstructed.
The proton energy is determined via range.  The algorithm assumes protons that exit the detector or that interact inelastically are not well reconstructed.  The energy of deposited hits near the end of the track is used to flag protons that may have interacted inelastically or exited the detector, as these processes reduce the  end-of-track Bragg peak that is characteristic of stopping protons. Protons exhibiting a hit pattern consistent with the expected Bragg peak are accepted as well-reconstructed tracks. The Bragg peak of charged pions differs from that of protons, and this is used as an additional veto of pions.

\begin{figure}  
	\centering
\includegraphics[width=0.48\linewidth]{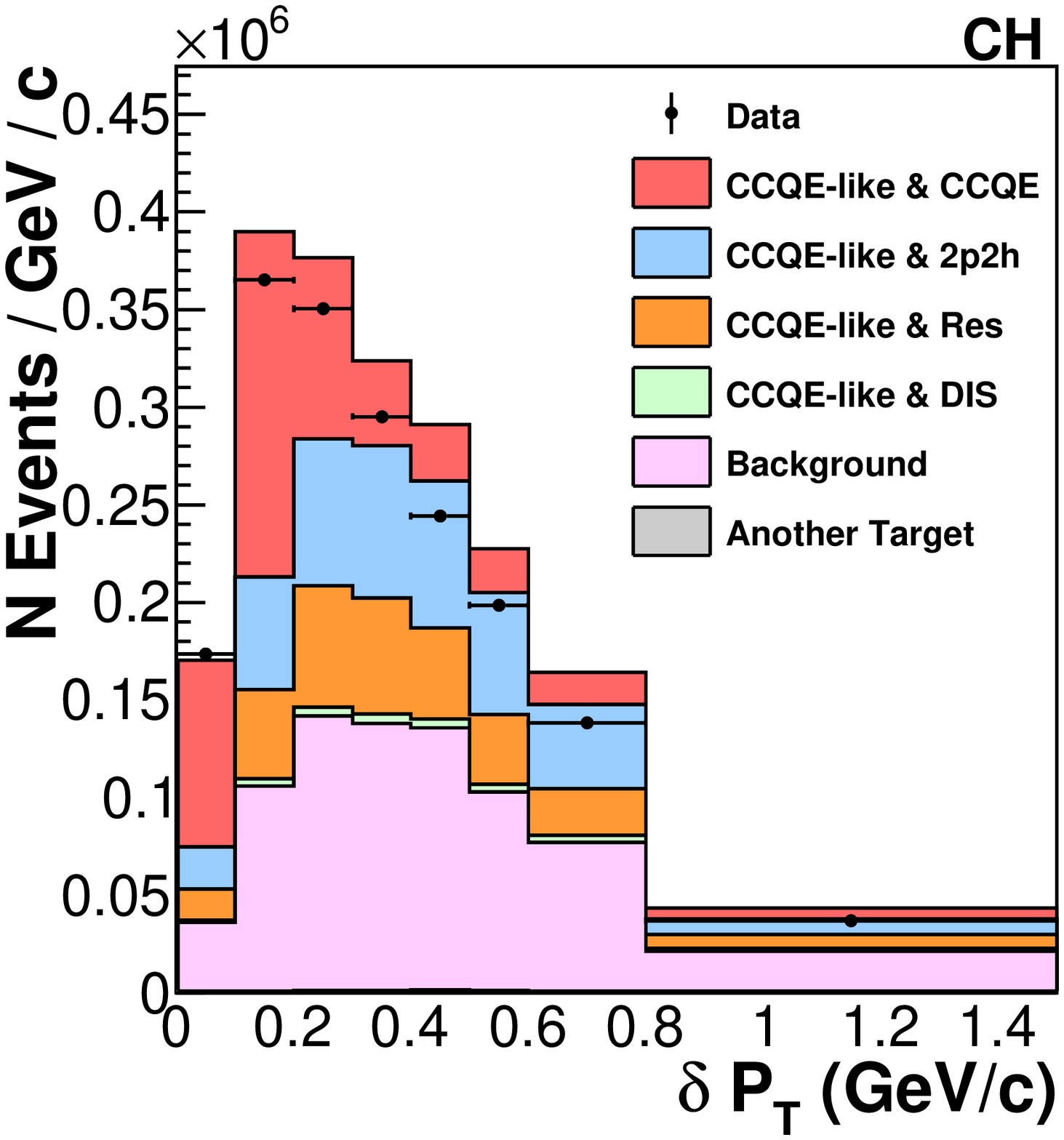}
\includegraphics[width=0.48\linewidth]{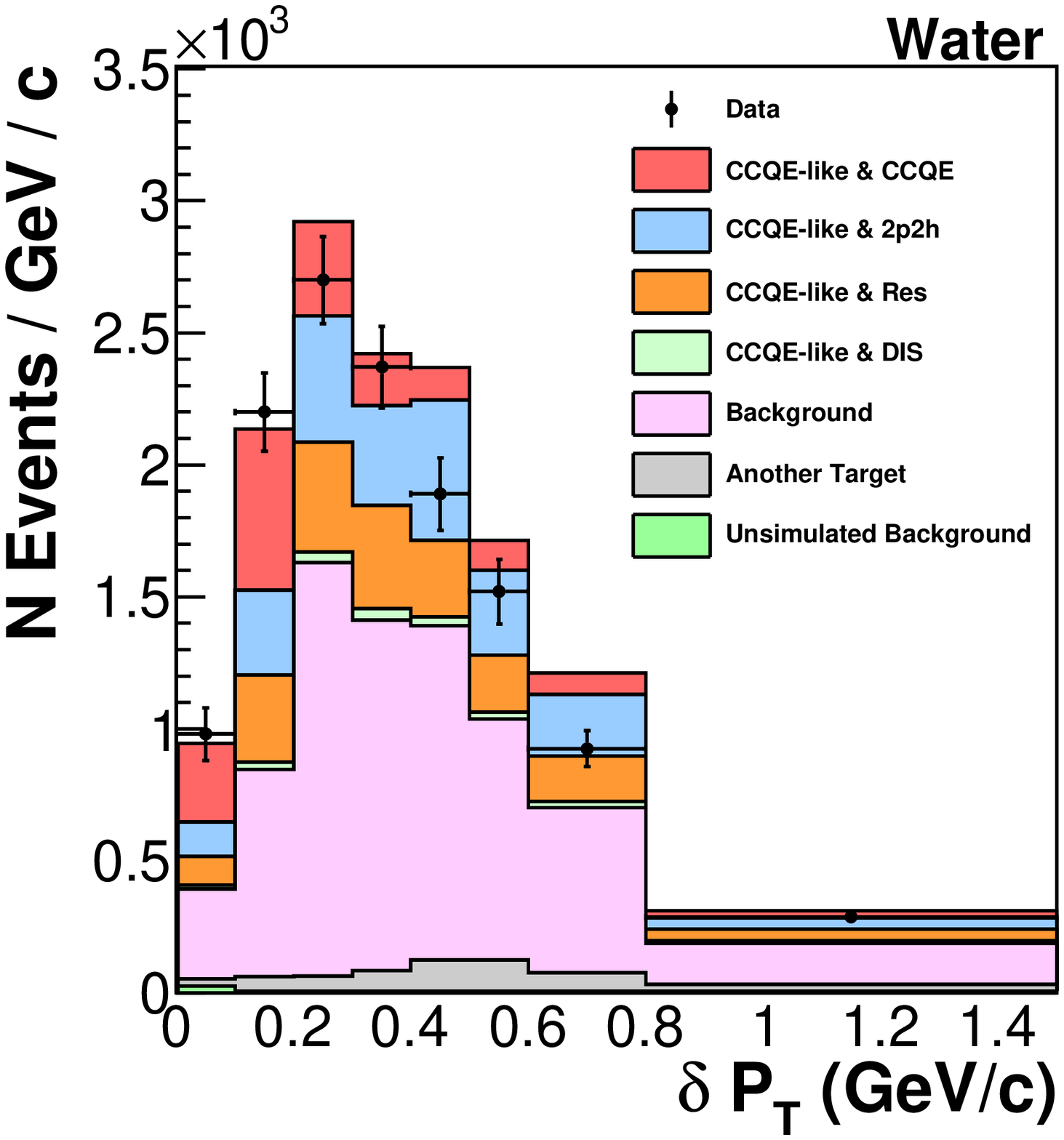}
   	\includegraphics[width=0.48\linewidth]
{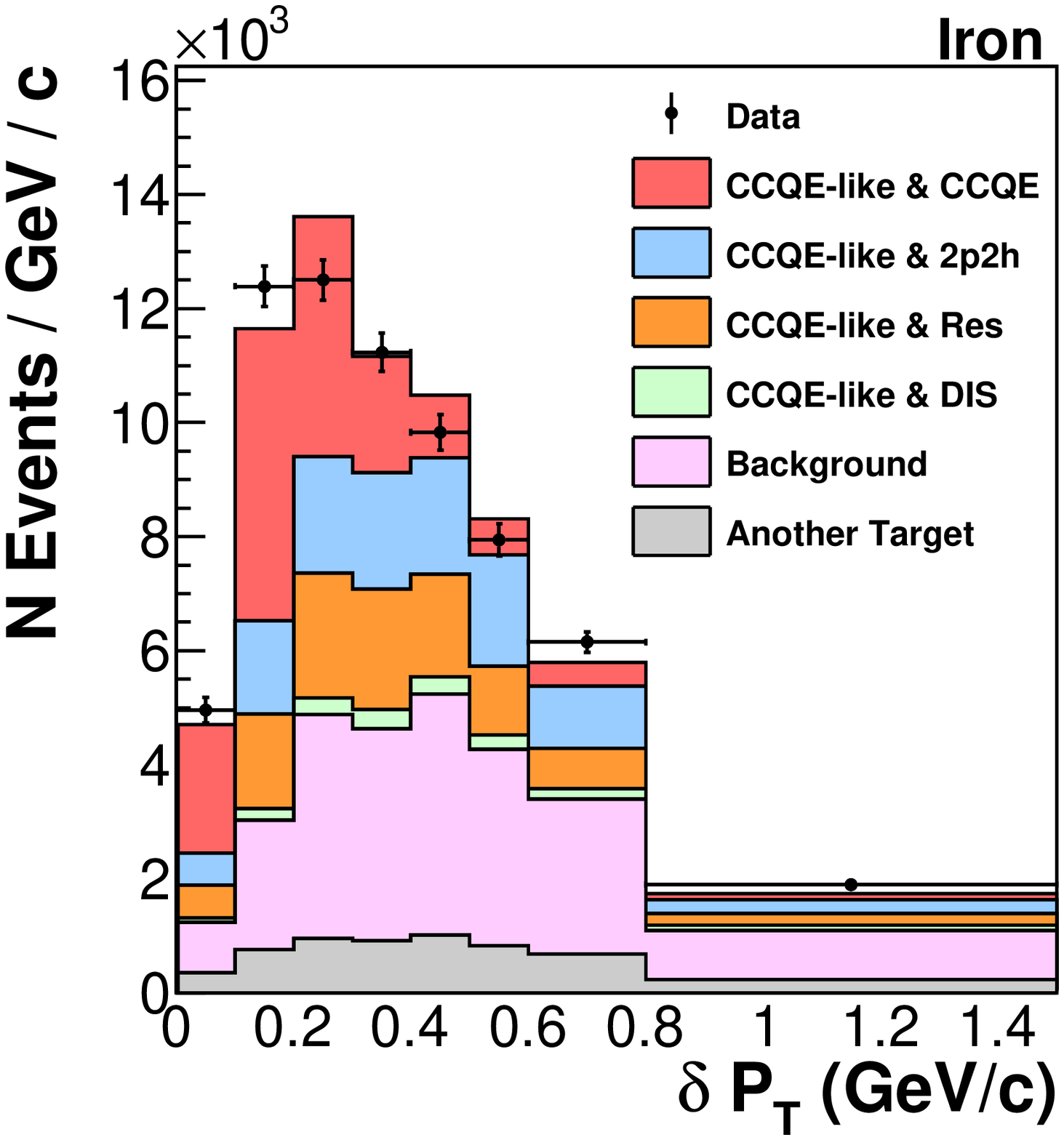}
        \includegraphics[width=0.48\linewidth]
{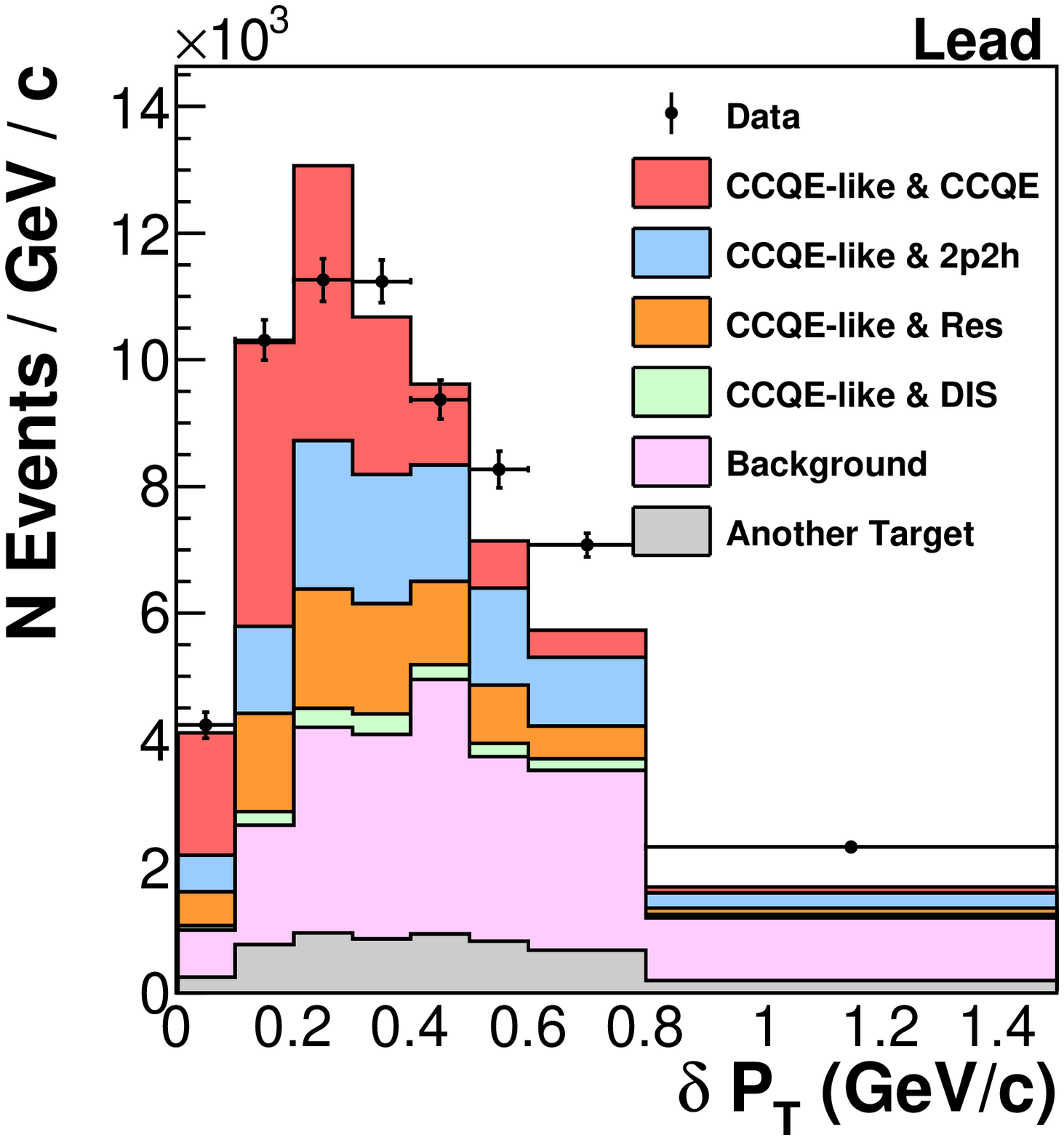}
	\caption{Event rates as a function of \tkidelta \ for CH, H$_{2}$O, Fe, and Pb targets. As explained in Sec.\ref{water-target-explanation}, the water target includes a measure of its `unsimulated background' estimated from the event rate while empty. }
	\label{fig:tkidelta_evntrate_sig}
\end{figure}

Figure~\ref{fig:tkidelta_evntrate_sig} shows the selected event distributions in the data in each target as a function of \tkidelta. Given the signal definition above, 2p2h interactions can also be identified as quasielastic-like since those may also produce at least one final state proton.  Similarly, a $\Delta$-resonance production followed by $\Delta$ decay and pion absorption in the parent nucleus can also produce a final state proton and no pion.  
 These plots show that all three processes are predicted to contribute  quasielastic-like events.  
The data taken on the carbon target are similar to those taken on scintillator and are not shown in select places in the paper, including here, 
since they primarily serve as a cross-check.  
The data are shown as black points along with  statistical error bars.  The corresponding event distributions from the simulation are also shown overlaid with the data on each plot.  The simulated data is broken down in terms of event origin.  Backgrounds on each appropriate target are shown in pink.  These are typically events containing one or more charged or neutral pions escaping the nucleus but which were undetected.  Background events coming from interactions on a different target, generally the  scintillator surrounding one of the passive nuclear targets, are shown in light grey.  This background is called the plastic background.  

The event rate distributions in variable $X$ are converted to differential cross sections using the following equation:
\begin{equation}
{d\sigma \over dX}_i = {\sum_{j} U_{ij}(N^{measured}_{j} - N^{background}_{j}) \over \epsilon_i T \Phi_i \Delta X_i}
\label{equation:cross_section}
\end{equation}
where $\sigma$ is the cross section, $N$ is the number of events, $\epsilon_i$ is the efficiency of reconstructing an event in bin i, $\Phi_i$ is the integrated total neutrino flux in bin i, $T$ is the number of targets for the given target type, and $\Delta X_i$ is the bin width.  $U_{ij}$ is the unfolding matrix that takes the observed distributions and removes the reconstruction effects insofar as possible to recover the best estimate of the true distribution. 

\subsubsection{Sideband background tuning}

The analysis backgrounds are tuned using multiple procedures.  The single neutral pion background is constrained using a previous measurement of single neutral pion production by MINERvA \cite{pi0_paper}.  The amount of plastic background in each nuclear target sample is constrained using the interactions on the scintillator near the nuclear target~\cite{jk_1trk_prl}.  
For neutrino interactions on the correct target, the backgrounds in this analysis are dominated by events with produced charged pions.  The model dependence in backgrounds is reduced through tuning the background level in the simulation to agree with the data in sideband samples that are similar to, but strictly not, signal events.  The sidebands are selected to be rich in background events relative to the signal region.

\begin{figure}  
	\centering
	\includegraphics[width=0.48\linewidth]%
{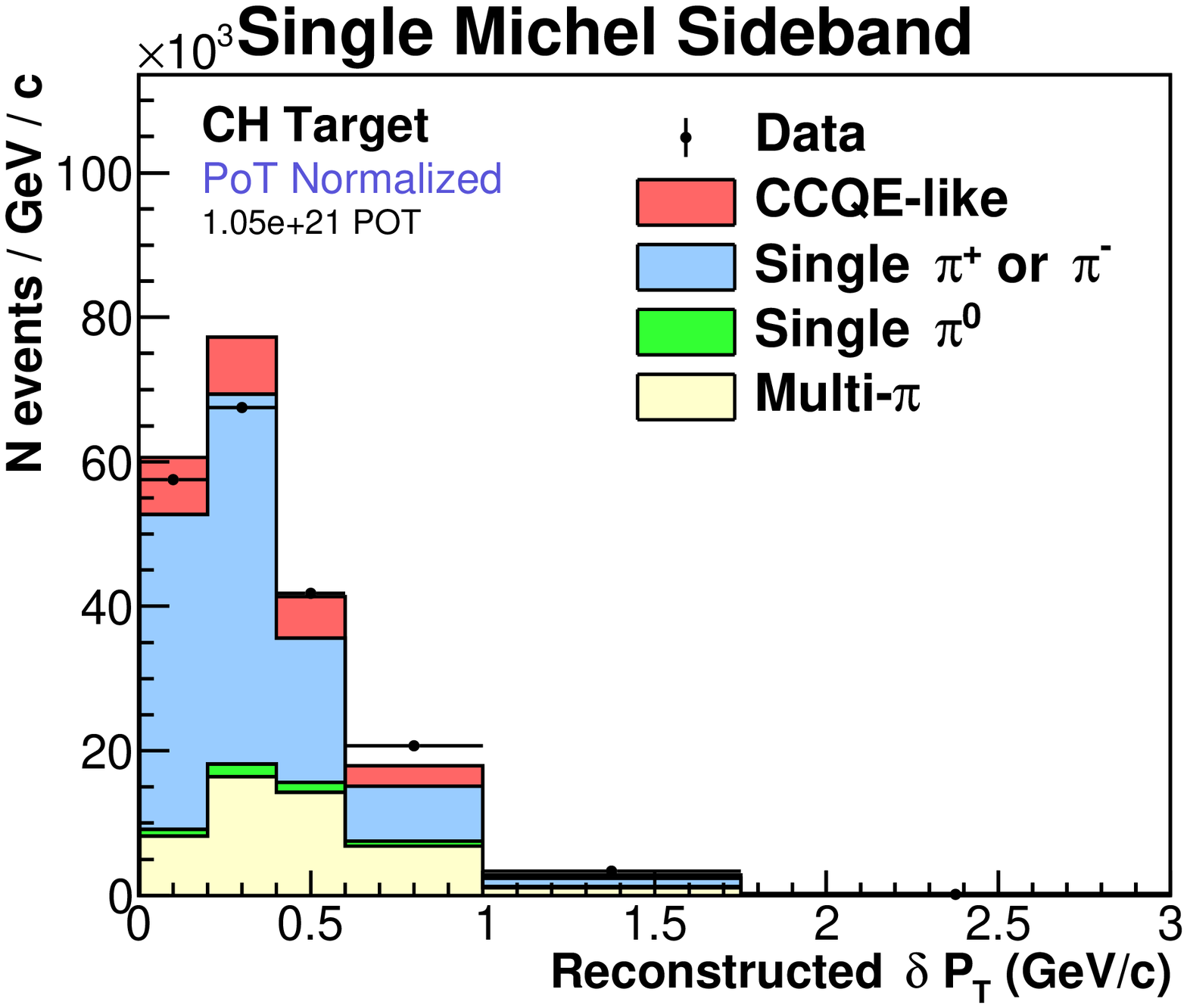}
 	\includegraphics[width=0.48\linewidth]
{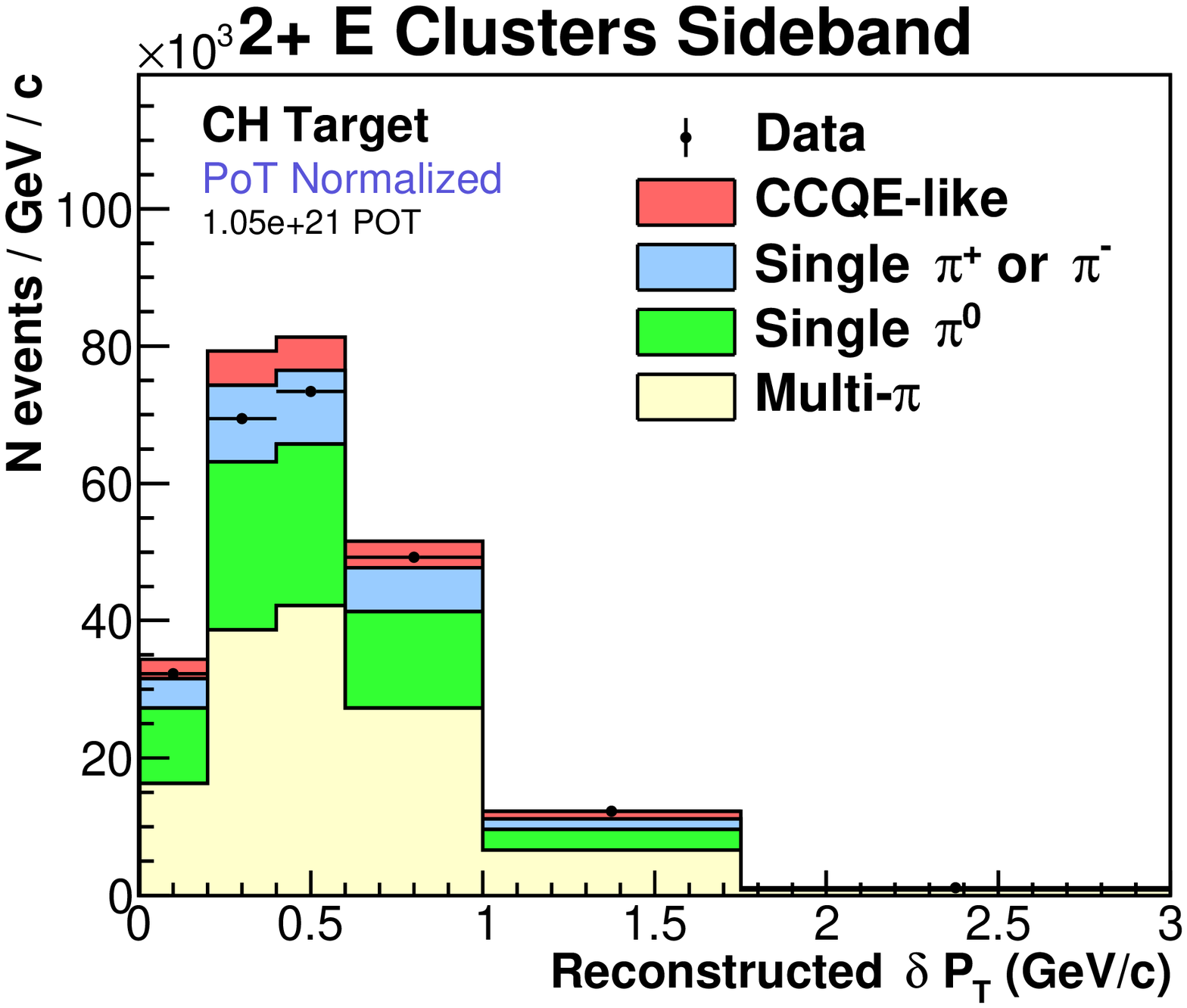}
	\includegraphics[width=0.48\linewidth]
{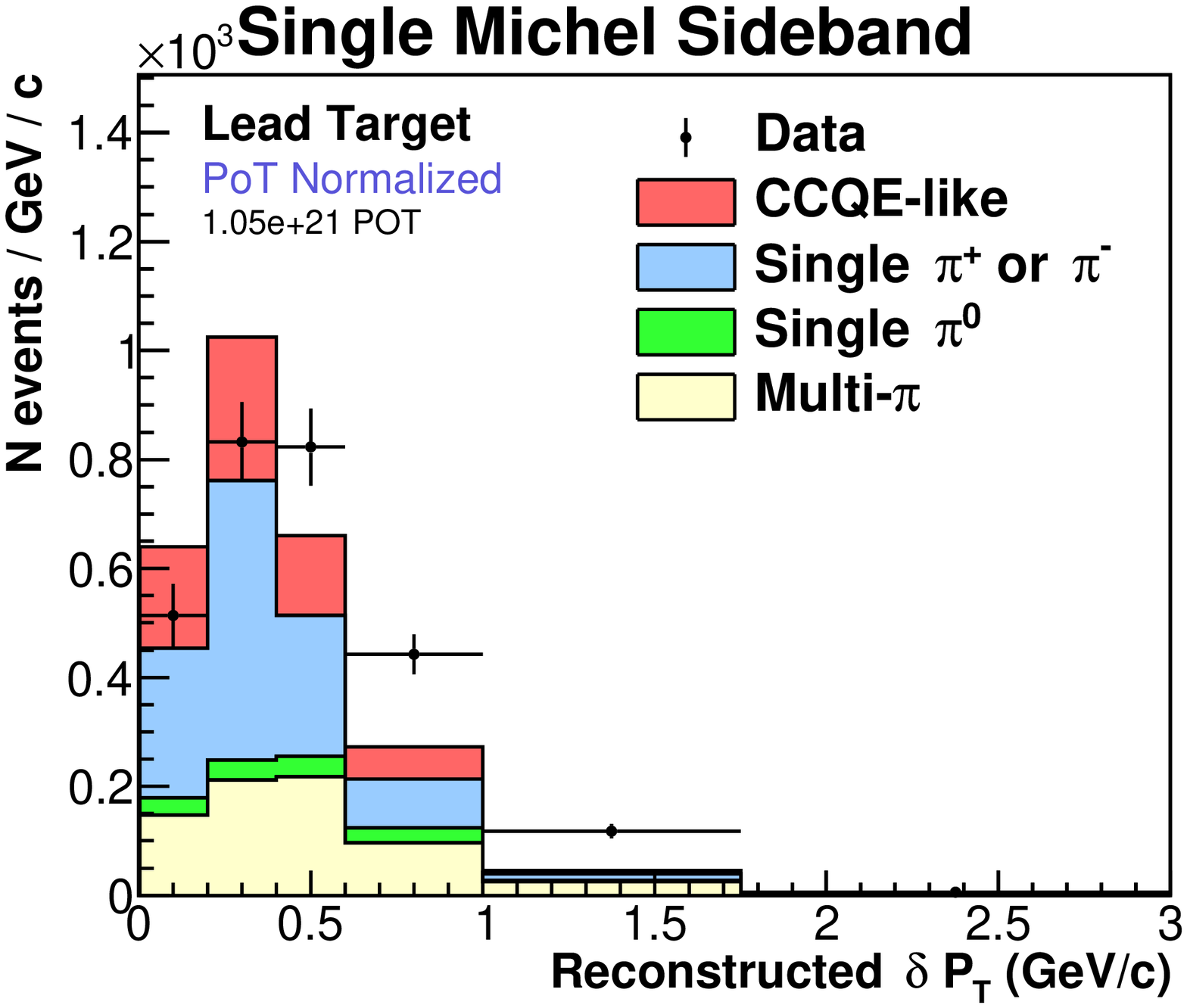}
 	\includegraphics[width=0.48\linewidth]
{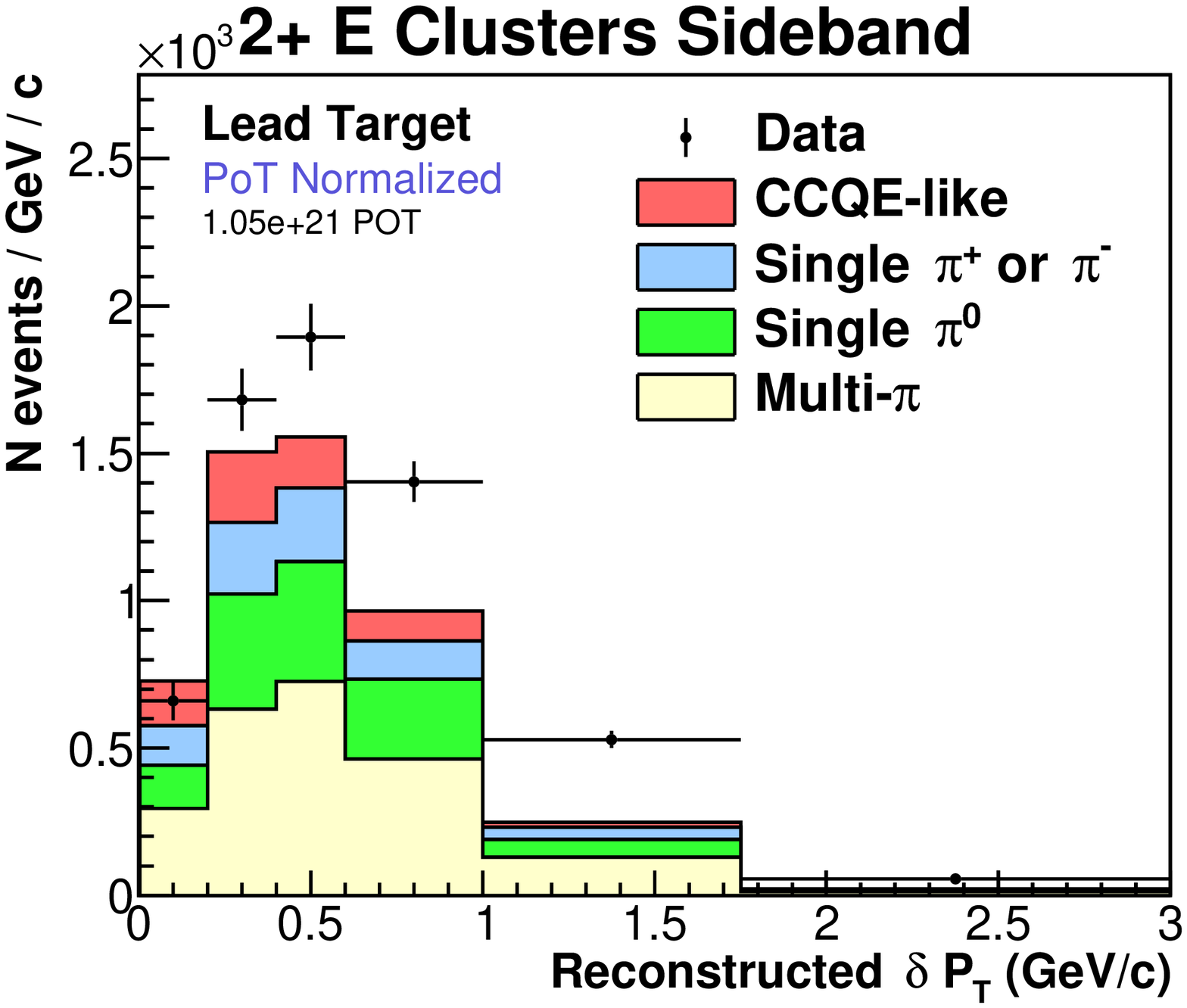}
	\caption{Event rates as a function of \tkidelta\ for the two sideband samples for CH and Pb targets before background tuning.}
\label{fig:tkidelta_sdbn_evntrate_untuned}
\end{figure}

Two sidebands are used in this analysis.  One sideband is formed by 
requiring a Michel electron and keeping the other requirements identical to those of the signal sample. 
This sideband is enriched with charged pions.  The second sideband  
is formed by requiring
at least two extra isolated clusters of energy away from the interaction vertex.  This sample is enriched in neutral pions.  Figure~\ref{fig:tkidelta_sdbn_evntrate_untuned} shows the two sidebands as a function of \tkidelta , before tuning the backgrounds, for events interacting on scintillator and events interacting on lead.  In these sidebands, \tkidelta \ is defined using the muon and leading proton just like it is in the signal sample.  Note that here, as well as in 
Figure~\ref{fig:tkidelta_evntrate_sig}, the agreement of the simulation with the data is significantly worse for the Pb target.  There is a significant signal component in the sideband.  Therefore, the signal region is included in the fit and the signal is allowed to float in the fitting procedure~\cite{jk_1trk_prl}. The fitting is regularized to find smooth tuning parameters as a function of each kinematic variable. 
The same plots after the backgrounds are tuned are shown in the Supplement. 

For the sideband plots and selected other plots in this paper, results for interactions on the scintillator and lead targets are chosen as illustrative examples at the two extremes of target size, i.e., atomic mass A. The plots for the other targets are qualitatively similar to these.  

The sideband tuning and background subtraction are done separately per target material and per each different variable.
As a check on the tuning procedure, the final total cross section calculated for each variable in the context of a single target nucleus type was determined and checked for consistency with the other variables. The event rate in the signal region after tuning for each target is shown in the Supplement.

\subsubsection{Background subtraction}

Table~\ref{tab:statistics} provides the relative numbers of events, purity, and the atomic number and number of neutrons for each nuclear target. 
The efficiency ranges from 5\% to 8\% for events in the passive targets, according to the simulation, while the efficiency in the tracker is 28\%.  These numbers are substantially lower than what was seen for the analogous single track analysis where there is no requirement for a final state proton~\cite{jk_1trk_prl}. The sample purities, estimated from the simulation, range from approximately 50\% to 60\%.  
\begin{table}[h]
\begin{tabular}{ | l | c | c | c | c ||}
\hline 
Target & Z & N & Events & Purity \\
\hline\hline
Tracker (CH) & 7 & 6 & 218,000 & 60\% \\
\hline
Carbon & 6 & 6 & 2,255 & 54\% \\
\hline
Water & 10 & 8 & 1,563 & 47\% \\
\hline
Iron & 26 & 30 & 8,577 & 47\% \\
\hline
Lead & 82 & 124 & 8,660 & 55\% \\ 
\hline 
\end{tabular}
\label{tab:statistics} 
\caption{Event statistics after all cuts, sample purity and regions of Z and N probed by this measurement.}
\end{table}

\subsubsection{Unfolding}

The background-subtracted signal distributions are corrected for detector smearing, i.e., unfolded, using the D'Agostini prescription~\cite{DAgostini:1994fjx,DAgostini:2010hil} supported by a series of reweighted (warped) model validation studies.  The binning and the number of unfolding iterations were optimized for each variable.  The unfolded \tkidelta \ event rate distributions are shown in the Supplement.

\subsubsection{Efficiency}

After unfolding, the signal distributions were corrected for the signal event reconstruction efficiency.  The simulation was used to estimate the efficiency.  The efficiency was defined to be the number of true signal events passing the reconstruction cuts divided by the number of true signal events with true muon and proton tracks satisfying the selection criteria.  

The efficiency is a  function of target position and type and muon and proton track angles and momenta.  
Efficiencies as a function of \tkidelta\ for the scintillator region and for lead are shown in the Supplement.

\subsubsection{Flux}

The neutrino flux at MINERvA varies at the few percent level as a function of the longitudinal position along the detector.  It also varies radially around the detector since the detector symmetry axis and the beam axis are not perfectly aligned. In addition, the passive targets themselves are not symmetric with respect to either the beamline or the detector axis. 
Consequently, the flux varies for each individual nuclear target, in particular for neutrino energies around 8~GeV at the falling edge of the focusing peak, as shown in Fig.~\ref{fig:fluxes}. 
Finally, the water target is filled some of the time and empty some of the time. The data from when the water target was empty is used to subtract any unsimulated background events from data taken with the full water target for the cross section extracted on water. 

In order to provide the most useful cross-section ratio to CH, the same flux should be used for both the numerator and the denominator. Therefore, the CH cross section is extracted with different fluxes for use in each ratio as appropriate depending on the target. This is done by splitting the CH target into twelve equally-spaced angular regions.  Then an appropriately weighted linear combination of these sections is used to combine the event rates and create a CH cross section with a flux equal to each targets' flux~\cite{jk_1trk_prl}.  The fit for the section weights is regularized to retain CH statistics.

\subsubsection{Number of targets}
\label{water-target-explanation}

The number of targets for each target type was determined from the mass of the target and its known composition.  The mass of each nuclear target was calculated using its known density and measured dimensions. The uncertainties on the fiducial pure carbon, iron and lead masses are estimated to be less than 1\%. The water target shape and dimensions were measured with lasers after data taking was completed. The water target has a fill meter which  
indicates how much water is contained. The laser measurements were used to create a model to relate fill meter readings to actual water mass. 
The mass of the water target 
was determined using the model of the target itself combined with a measurement of how much water was used to fill the target.
The uncertainty on the water mass is estimated to be 1\%. 

Vertex positions are reconstructed inside the passive nuclear targets based on the point of closest approach of the reconstructed muon and proton tracks. Since a small adjustment can change a reconstructed interaction on iron to an interaction on lead, the fiducial volume of each target is defined with a buffer of 20~mm between targets in the same plane. A data-driven fit is used to accurately position the targets in the simulation. The simulation is then used to maximize the selection efficiency and purity selecting interactions on a particular nuclear target, and minimize the selection of interactions from the surrounding scintillator.

\subsubsection{Cross section}

\begin{figure}  
	\centering
 \includegraphics[width=.95\linewidth]{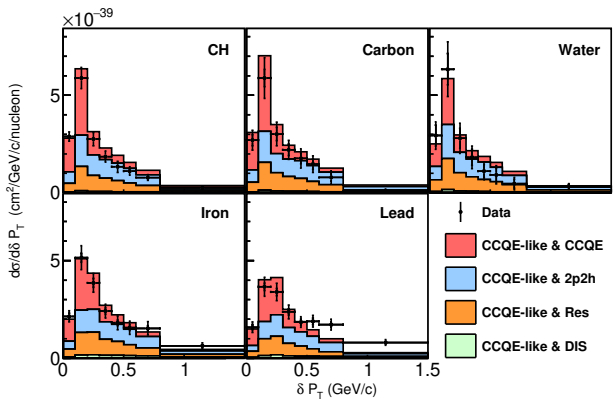}
	\caption{The differential cross section per nucleon as a function of \tkidelta \ for CH, C, H$_{2}$O, Fe, and Pb targets.  The predicted contributions as a function of the QE-like interaction process are also shown for each target material.}
	\label{fig:dpt_xsec}
\end{figure}

The differential cross sections were determined using Eq.~\ref{equation:cross_section}. 
The measured differential cross sections (per nucleon) for each target as a function of \tkidelta \ are shown in Fig.~\ref{fig:dpt_xsec}.  The data is shown as black points with horizontal bars showing the bin and vertical bars showing the estimated 1$\sigma$ uncertainty.  The inner (outer) horizontal hashmarks represent the statistical (statistical+systematic) 1$\sigma$ uncertainty.  The histogram overlayed on the data is the prediction from the central value Monte Carlo simulation (MINERvA Tune) broken down by the original event morphology.  Note that the 2p2h process, which can produce two nucleons in the final state preferentially shows up at high \tkidelta\ because the second nucleon's momentum is not included and therefore contributes to the missing momentum.

The cross-section results in \tkidelta , presented in Fig.~\ref{fig:dpt_xsec}, show a shift in the distribution toward higher \tkidelta \ for the higher A targets (Fe and Pb) relative to the lower A targets.  This 
is expected given the larger chance for Final State Interactions (FSI) that would occur with the larger nucleus.  The simulation describes the data and this evolution reasonably well except for the case of Pb, where the shift to higher \tkidelta\ is even higher in the data than in the prediction.

\subsubsection{Systematic uncertainties}

\begin{figure}
	\centering
 \includegraphics[width=\linewidth]{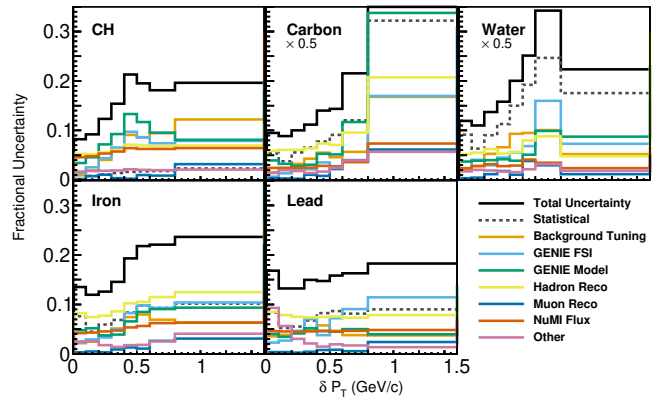}
	\caption{Contributions to the uncertainties in the differential cross sections as a function of \tkidelta .}
\label{figure:xsec_dpt_uncert}
\end{figure}

The systematic uncertainties on the cross-section measurement were determined using a multi-universe technique where the cross section is re-extracted after varying each source of uncertainty.  The correlations between different bins and nuclear targets are taken into account.  Uncertainties in the flux, neutrino interaction model, and detector effects were considered.  Summary distributions of the uncertainty in the measured cross section in each target as a function of \tkidelta \ are shown in Fig.~\ref{figure:xsec_dpt_uncert}.  The dashed line in each figure is the statistical uncertainty.  Uncertainties in the hadron production from the target and focusing dominate the flux uncertainty.  It is constrained by external hadron production data and neutrino-electron scattering measurements made during the same run period\cite{MINERvA:2019mkf,MINERvA:2021mpk}.  The neutrino interaction uncertainties are dominated by modeling of the backgrounds and  FSI.  The proton reconstruction uncertainty, due to detector effects, dominates the cross-section uncertainty.  The main driver is the uncertainty in the conversion of the scintillator light signal to the absolute energy for the proton hits in the Bragg peak. 

\begin{figure}
    \centering
\includegraphics[width=\linewidth]{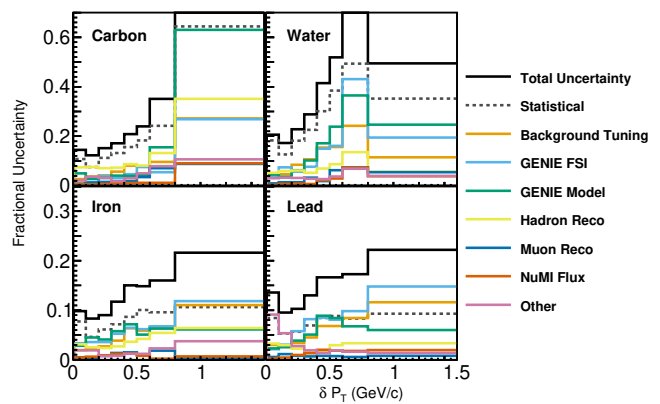}
    \caption{Contributions to the uncertainties in the cross section ratios to CH as a function of \tkidelta.}
\label{fig:xsec_dpt_uncert_rat}
\end{figure}

Given the large correlations in the systematic uncertainties between different targets, there is some reduction in the total uncertainty in measurements of the cross-section ratio between targets.  The systematic uncertainties on the cross section ratios as a function of \tkidelta\ are shown in Fig.~\ref{fig:xsec_dpt_uncert_rat}.  In particular, the flux uncertainty and the hadron and muon reconstruction uncertainties largely cancel in the ratio.  The cross-section ratio to that of scintillator for each target are shown in the next section for \tkidelta \ and in Sec.~\ref{sec:appendixA} for the other variables.  The cross section for each target is scaled by the number of neutrons in the target to make comparisons more straightforward.  In the absence of nuclear effects the cross sections would simply scale linearly as a function of the number of neutrons and therefore the ratios presented should simply be unity.  Although the ratio between water and carbon is measured to be consistent with unity there are clear deviations from unity for the iron and lead targets.  

\subsection{Comparisons to models}

The neutrino interaction model used in the simulation compared to the data in Fig.~\ref{fig:dpt_xsec} is the MINERvA tune version of the GENIE neutrino generator, as described in Sec.~\ref{sec:simulation}. 
It is interesting to compare the measured results to newer implementations of GENIE and a few other neutrino event generators commonly used. 

Even though the GENIE hA configuration is used in both the GENIE2 MINERvA tune and two of the GENIE3 options, it evolved significantly between the GENIE versions.   
The hA configuration uses a single mean free path based step to decide first if there was a reinteraction and then uses two tables as a function of kinetic energy to choose which among several reaction fates should occur.  These fates include single nucleon knockout (with and without charge exchange), absorption on multiple nuclei, and pion production.   These tables were tuned to reproduce models and data from nucleon- and pion-nucleus scattering, (similar to the hN tuning) but the one-step feature simplifies how systematics are applied.
The MINERvA tune GENIE2 applies the elastic scattering bugfix described in Sec.~\ref{sec:simulation} in a similar way to the change made in GENIE3.   Those events produce visible distortions in TKI variables, but are replaced as no-scattering events using a weight.  GENIE3 also introduces changes that decrease the probability of no interaction, the effect is stronger for heavier nuclei.  Finally, the GENIE3 hA interactions modify the fractions of pion fates to be $A$ dependent.  In particular, the absorption component that turns one-pion events into the zero pion signal process becomes a higher fraction of the fates for higher $A$.

These comparisons  as a function of $A$ are useful in understanding the predictive power of generators, and can give some insight in the modeling process in general, particularly when their use extends to large $A$. Outside of the central value Monte Carlo (MINERvA tune), these comparisons are done in the context of the NUISANCE framework~\cite{Stowell:2016jfr}.

Table~\ref{tab:generators} lists the models shown in comparison to the data, delineated by the names used in the plots. For the four GENIE3 generator comprehensive model configurations, the untuned versions are used G18\_xxx\_00\_000 and hereafter the short form is used e.g. G18\_10a. We have analyzed the changes between GENIE2 and the GENIE3 configurations in detail and in the table summarize the features relevant for these cross sections cross section comparisons.

{\footnotesize 
\begin{longtable*}{p{3.5cm}|p{13cm}} \hline\hline
Name & Description \\ \hline
   MINERvA tune   &  This is the generator described in Sec.~\ref{sec:simulation} that is used for the central value Monte Carlo in this analysis and shown in Fig.~\ref{fig:dpt_xsec}. \\ \hline
   GENIE v3 G18\textunderscore 01a      & For this analysis, the hA FSI option is updated in several ways as described in the text, only one of which is part of the MINERvA tune.  Also relevant for this analysis, it uses the so-called empirical 2p2h instead of the MINERvA tuned version of the Valencia model.  
  \\ \hline
   GENIE v3 G18\textunderscore 01b  &  This is the same as G18\textunderscore 01a  except the FSI option is hN.  Hadrons from the reaction are stepped through the nucleus and the possibility of interacting is tested at each step.  This often leads to multiple reactions and more variation in the outgoing proton number and angle in hN compared to hA.\\ \hline
  GENIE v3 G18\textunderscore 10a   & 
   For the QE model, this version uses the Valencia local Fermi gas with RPA for the initial state~\cite{Nieves:2004wx}, plus the Valencia model for the 2p2h interactions~\cite{Nieves:2011pp,Gran:2013kda,Schwehr:2016pvn}.  Both are ingredients in the MINERvA tune where the 2p2h process is further enhanced and the QE process in GENIE3 produces an intrinsically different proton energy distribution.  For resonances, the form factors are significantly changed from G18\_01 and the MINERvA tune, including the portion that has no pion in the final state and is a signal process in this analysis.   The lepton mass and pion pole effects~\cite{Kuzmin:2003ji}\cite{Berger:2007rq} are not relevant at MINERvA energies, and GENIE’s tune~\cite{GENIE:2021zuu} is not used.   
  \\ \hline  
  GENIE v3 G18\textunderscore 10b &  This is the same as G18\textunderscore 10a except the FSI option is hN. \\ \hline
 NuWro LFG  & This generator is NuWro~\cite{Juszczak:2009qa} version 19.02 with a local Fermi gas as the initial state.  \\ \hline
 NuWro SF  &  This generator is NuWro~\cite{Juszczak:2009qa} version 19.02 with a  spectral function as the initial state. \\ \hline
GiBUU T0  & This generator is GiBUU~\cite{Buss:2011mx} release 2019, patch 8 (Sept 8, 2020).  In this version the scaling parameter $T$ that sets the size of the enhancement of 2p2h is set to zero.  \\ \hline
GiBUU T1 & This generator is similar to the generator above except in this version the enhancement parameter $T$ that sets the size of the 2p2h contribution is set to one. This means it has twice the 2p2h contribution than that in GiBUU $T0$ above.  \\ \hline
 NEUT LFG & This is the neutrino generator NEUT~\cite{neut} version 5.4.1.  This version uses a local Fermi gas for the initial state, a 2p2h hadron tensor technique based on the Nieves model, coherent $\pi$ production based on the Berger-Sehgal model, and a single $\pi$ production model by Kabirnezhad~\cite{minoo_pi}. \\ 
 \hline \hline
 \caption{Generators used for comparisons with the data. Additional content on the physics content of generators is available in~\cite{Betancourt:2018bpu} and~\cite{Avanzini:2021qlx}.}
 \label{tab:generators}
 \end{longtable*}
}

\begin{figure}
    \centering
    \includegraphics[width=\linewidth]{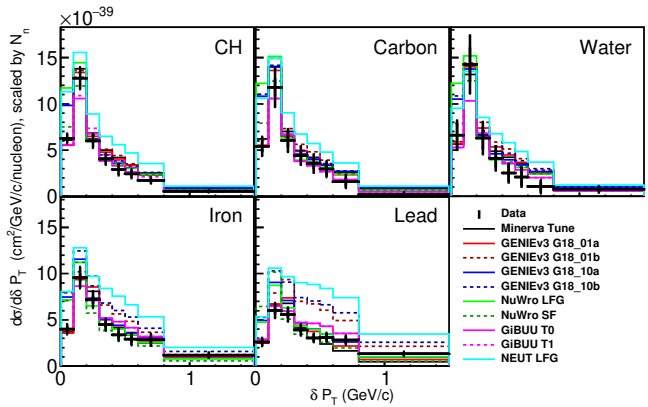}
    \caption[The \tkidelta \ cross section comparison for multiple targets. ]{Comparisons of the cross section per number of neutrons in the target ($N_n$) as a function of \tkidelta \ for different targets and for different neutrino event generators.  The ratios of the data and the other generators to the default GENIE prediction can be found in Supplement Sec.~\ref{sec:appendixA}.
    }
    \label{Figure:Gencompare_dpt}
\end{figure}

Figure~\ref{Figure:Gencompare_dpt} shows the comparison of the measured cross section in the data with different nuclear targets to the ensemble of models above as a function of \tkidelta .  In the top row, the cross-section data on CH, C, and H$_{2}$O targets are shown along with an overlay of each of the models.  
The bottom two plots show the same for the Fe and Pb targets, respectively.
 
Qualitatively, though there is significant spread and differing trends, the models broadly seem to describe the data.  NEUT broadly overestimates the cross section in all the targets.  The disagreement between NEUT and the iron and lead data sets in particular could be due to the absence of nuclear de-excitation in the model and the assumption that the non-H nucleus is isoscalar\cite{Hayatoprivate}. The hN versions of GENIE show behavior similar to NEUT in shape and normalization for the larger targets.  NuWro and GENIE v3 G18\textunderscore 10 tend to overpredict the data at low \tkidelta .  The \chisq\ between the cross sections and the models can be found in Table~\ref{tab:dpt}.  The ratio between the measured cross section and the MINERvA tune, along with the ratio between the various models and the MINERvA tune can be found in Fig.~\ref{fig:models_dpt_rat}.

\begin{figure}
    \centering
        \includegraphics[width=\linewidth]{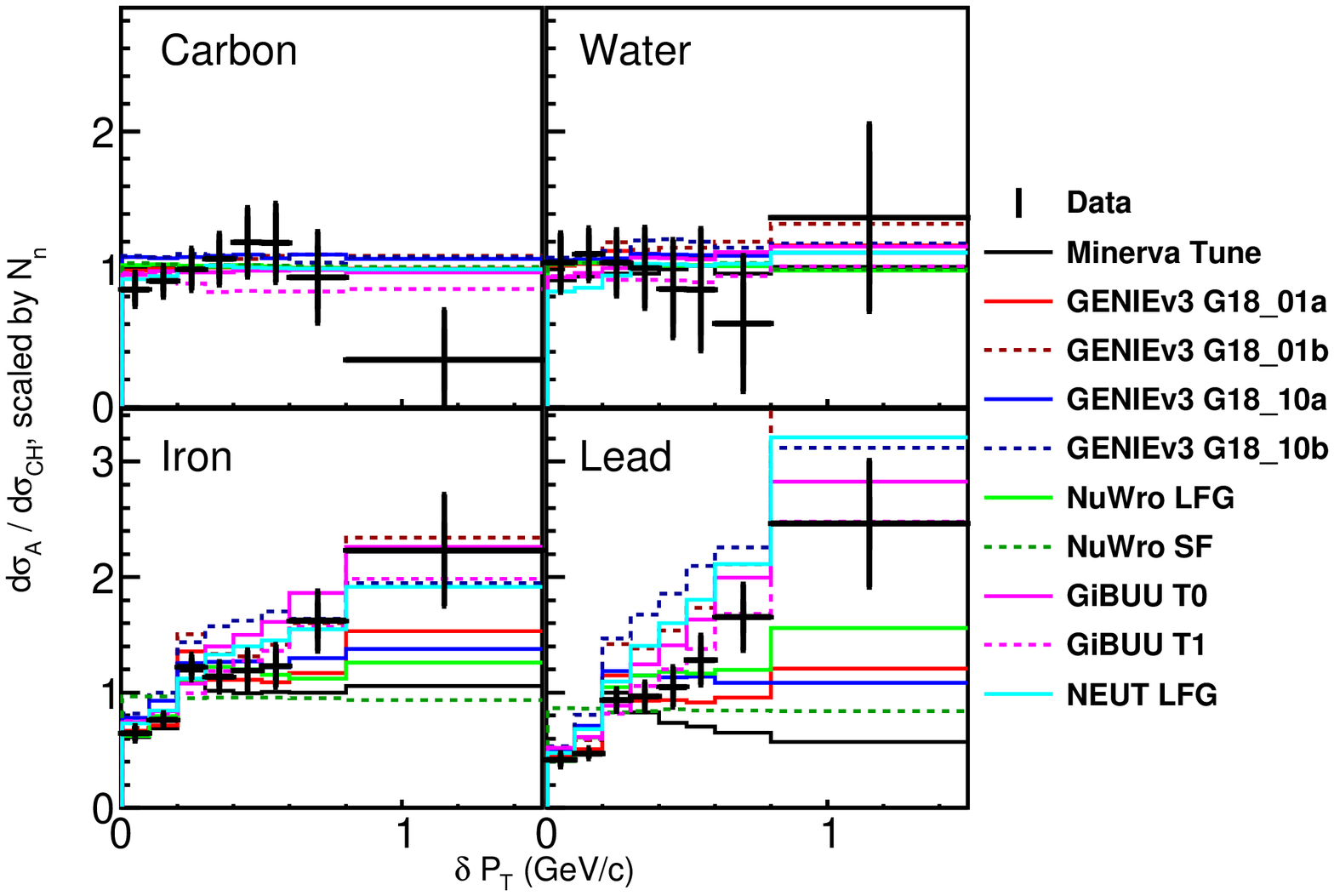}
    \caption{\tkidelta\ cross-section ratio generator comparison for multiple targets.  Note that changes to the FSI model (GENIE ``a'' to ``b'')
     in GENIE change the cross section ratio at high \tkidelta\  much more than changes to the 2p2h model (GiBUU T0 to GiBUU T1) or changes to the initial state (GENIE 1 to GENIE 10) or (NuWro LFG to NuWro SF).   }
    \label{Figure:Gencompare_dpt_ratio}
\end{figure}

The cross-section ratio as a function of \tkidelta \ for each nuclear target to scintillator is shown in Fig.~\ref{Figure:Gencompare_dpt_ratio}.  For each target the same cross-section ratio according to the different models is shown.  

The models describe the data in the cross-section ratio plots for the smaller $A$ targets.  For the Fe and Pb targets the model spread is large at high \tkidelta  , where FSI is expected to have the largest effect.  In these regions NuWro and the hA versions of GENIE underpredict the data significantly while NEUT, GiBUU, and the hN versions of GENIE follow the data trend at high \tkidelta \ fairly well.  The \chisq\ between the cross section ratios and the models can be found in Table~\ref{tab:dpt}.

\section{Results and Discussion for other observables as a function of $A$}
\label{sec:results}
\subsection{Cross Sections as a function of Acoplanarity}

\begin{figure}[h]  
	\centering
 \includegraphics[width=\linewidth]{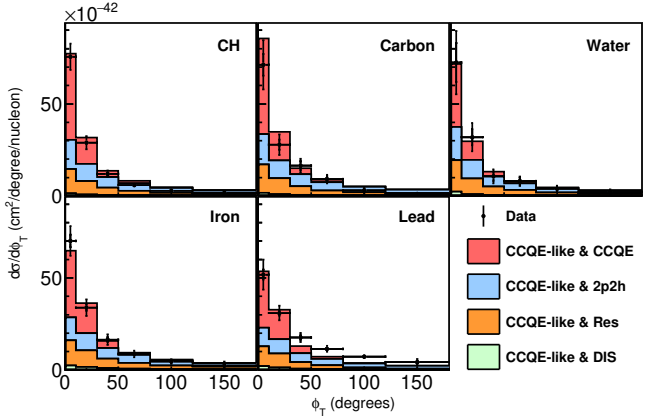}
	\caption{The differential cross section as a function of $\phi_{T}$ for CH, C, H$_{2}$O, Fe, and Pb targets, along with the predictions for the different quasielastic-like signal processes.}
	\label{figure:xsec_coplan}
\end{figure}

The differential cross section as a function of $\phi_{T}$ for interactions on the CH, C, H$_{2}$O, Fe, and Pb targets, respectively, are shown in Fig.~\ref{figure:xsec_coplan}. 
The simulations (MINERvA tune) are represented by the histograms and further distinguished by their respective interaction processes.
A similar format is used in the cross-section plots for each of the other variables in this section.  The uncertainties broken down by source for both the absolute cross sections and the cross section ratios as a function of $\phi_T$ can be found in Fig.~\ref{fig:xsec_err_coplan}.

The variable $\phi_{T}$ (see Fig.~\ref{fig:TKI}) is a measure of the extent to which the proton momentum is not back-to-back with the muon momentum in the transverse plane.  Such a deviation from back-to-back might be expected to happen with the effects of FSI as the proton transits the nucleus.  From the way that $\phi_{T}$ is defined, a larger value means more of a deviation from back-to-back and the greater the FSI effects.  The MINERvA tune shown uses a GENIE2 hA FSI model which does not change the angular distribution of the outgoing protons from quasielastic interactions for larger nuclei, but a full cascade like hN would broaden the angular distributions.   The prevalence of resonance feed-in to the signal in this model is also not strongly modified with A.   
The most noticeable difference between the data and the simulation is in the case of the Pb target where there is significantly more cross section at larger $\phi_{T}$, indicating more FSI broadening in the data.

\begin{figure}[h]
    \centering
 \includegraphics[width=\linewidth]
 {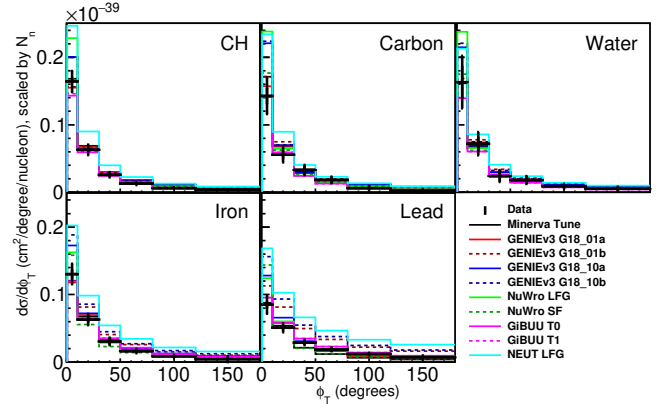}
    \caption[$\phi_{T}$ cross-section comparison for multiple targets ]{Cross section measurements and predictions as a function of $\phi_T$ for different targets and for a selection of different generator and model choices.
    }
    \label{fig:models_coplan}
\end{figure}

The differential cross section as a function of $\phi_{T}$ for the different targets as compared to a range of models is given in Fig.~\ref{fig:models_coplan}. As seen earlier, NEUT appears to have a higher cross section than seen in the data.  The hN versions of GENIE are very similar to NEUT in the larger targets and predict more events at higher $\phi_{T}$ where the FSI effect is largest.  The \chisq\ between the cross sections and the models can be found in Table~\ref{tab:phi}.


\begin{figure}
    \centering
            \includegraphics[width=\linewidth] {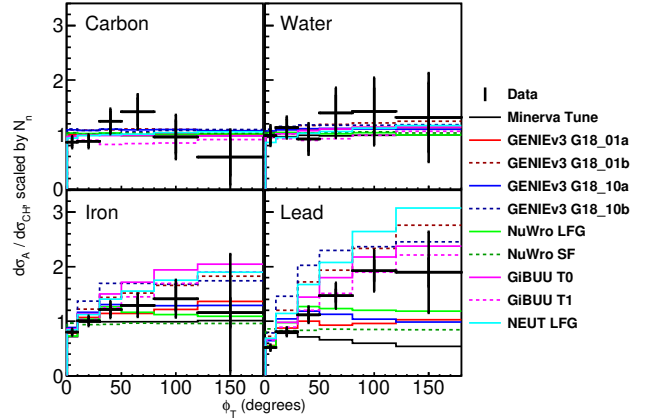}
    \caption{$\phi_{T}$ cross-section ratio comparison for multiple targets. Note that changes to the FSI model (GENIE ``a'' to ``b'')
     in GENIE change the cross section ratio at high $\phi_T$  much more than changes to the 2p2h model (GiBUU T0 to GiBUU T1) or changes to the initial state (GENIE 1 to GENIE 10) or (NuWro LFG to NuWro SF). }
        \label{Figure:Gencompare_phi_ratio}
\end{figure}

The ratio of the differential cross section as a function of the $\phi_{T}$ for each nuclear target (C, H$_{2}$O, Fe, Pb) to that for scintillator is shown in Fig.~\ref{Figure:Gencompare_phi_ratio}. The 
models describe the data fairly well for the smaller targets.  The model variation is greatest for the larger targets at high $\phi_{T}$.  The MINERvA tune, NuWro, and the hA versions of GENIE underpredict the ratio in the data taken on Pb at higher $\phi_{T}$.  The \chisq\ between the cross section ratios and the models can be found in Table~\ref{tab:phi}.  The ratio between the measured cross section and the MINERvA tune, along with the ratio between the various models and the MINERvA tune can be found in Fig.~\ref{fig:models_coplan_rat}.

\subsection{Cross Sections as a function of \tkialpha } 

\begin{figure}[h]  
	\centering
 \includegraphics[width=\linewidth]{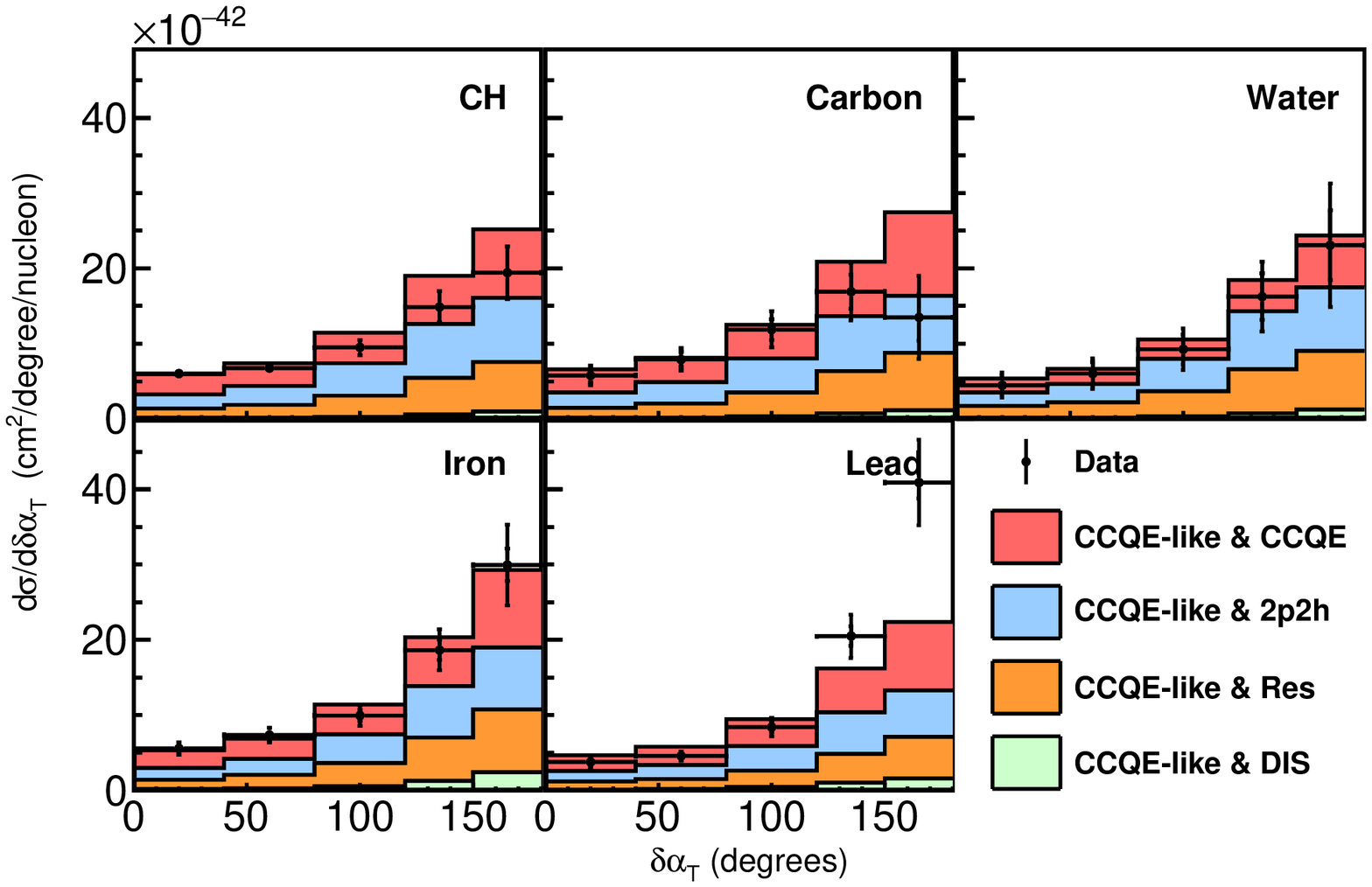}
	\caption{The differential cross section as a function of \tkialpha \ for CH, C, H$_{2}$O, Fe, and Pb targets, along with the predictions for the different quasielastic-like signal processes.}
	\label{figure:xsec_alph}
\end{figure}

Figure~\ref{figure:xsec_alph} shows the differential cross section as a function of \tkialpha \ for interactions on the CH, C, H$_{2}$O, Fe, and Pb targets, respectively. 
\tkialpha \ is the angle that measures the direction of the transverse momentum imbalance between the incoming neutrino and the sum of the lepton and hadron momenta.  It is sensitive to the intranuclear momentum transfer that comes from nucleon correlations and FSI.  MINERvA has measured \tkialpha \ previously for interactions on scintillator in the LE run~\cite{MINERvA:2018hba}.  The uncertainties broken down by source for both the absolute cross sections and the cross section ratios as a function of \tkialpha\ can be found in Fig.~\ref{fig:xsec_err_alpha}.

Here the largest discrepancies between the data and the simulation occur at large \tkialpha \ where the variable is most sensitive to FSI effects.  A larger \tkialpha \ indicates the proton losing momentum relative to the case without nuclear effects.  The simulation significantly overpredicts the cross section at high \tkialpha \ for the 
carbon and scintillator targets and underpredicts what is seen at high \tkialpha \ for lead.

\begin{figure}
    \centering
        \includegraphics[width=\linewidth]{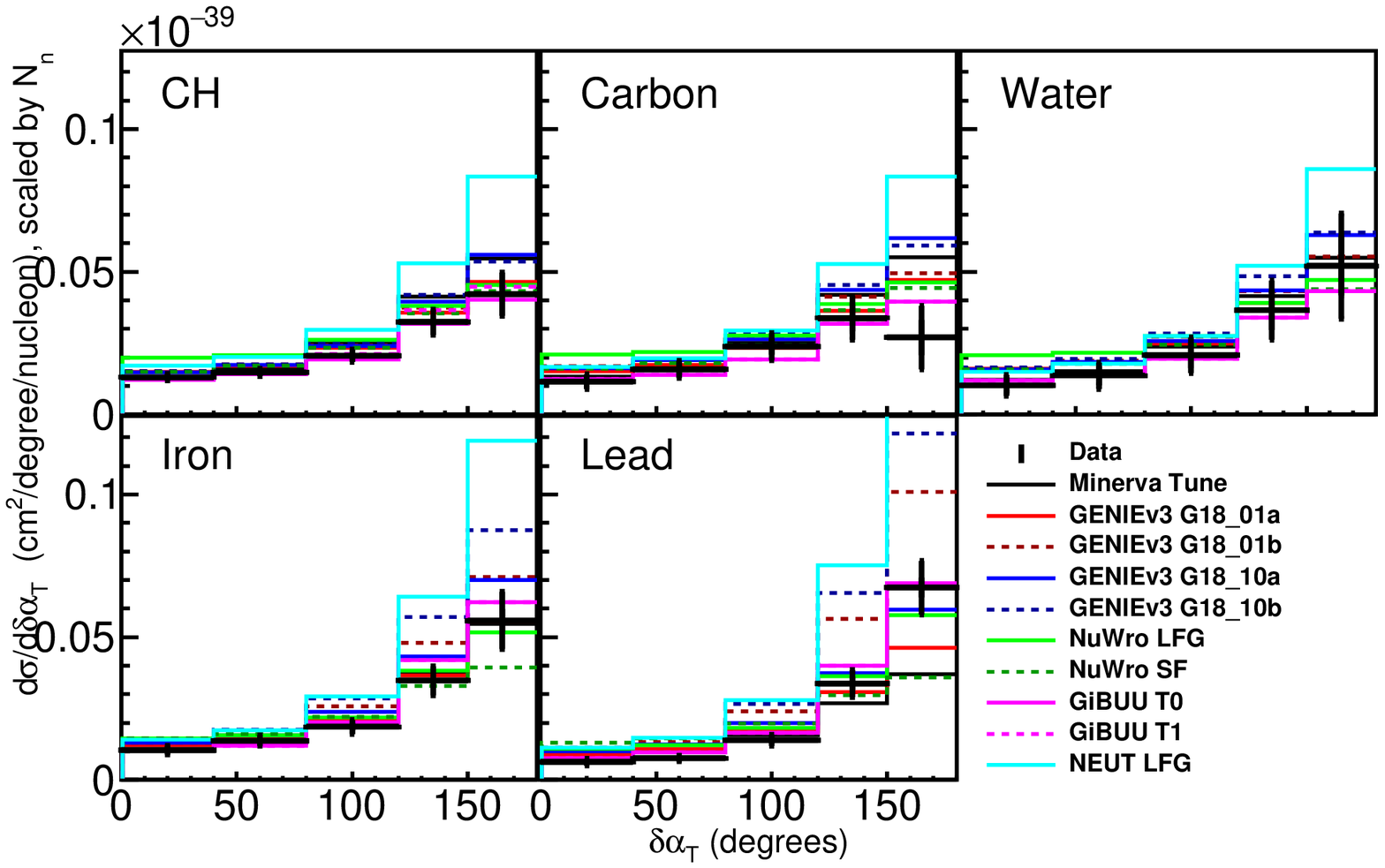}
    \caption[\tkialpha \ cross-section comparison for multiple targets ]{Cross section measurements and predictions as a function of \tkialpha\ for different targets and for a selection of different generator and model choices.}
    \label{fig:models_alphat}
\end{figure}

The differential cross section as a function of \tkialpha \ for the different targets is shown in Fig.~\ref{fig:models_alphat}.  Also shown for comparison are results for a range of models. The observed comparisons are similar to what was seen for $\phi_{T}$ and \tkidelta :  NEUT predicts more events than seen in the data, particularly in regions where the FSI is expected to be large.  The hN versions of GENIE give predictions approaching NEUT in those high FSI regions at large \tkialpha \ for the larger targets.  The other generators do fairly well in simulating what is seen in the data although the MINERvA tune, NuWro SF and GENIE v3 G18\textunderscore 01a underpredict the data in Pb where the FSI is expected to be most pronounced.  The \chisq\ between the cross sections and the models can be found in Table~\ref{tab:dalphat}.

\begin{figure}
    \centering
        \includegraphics[width=\linewidth]{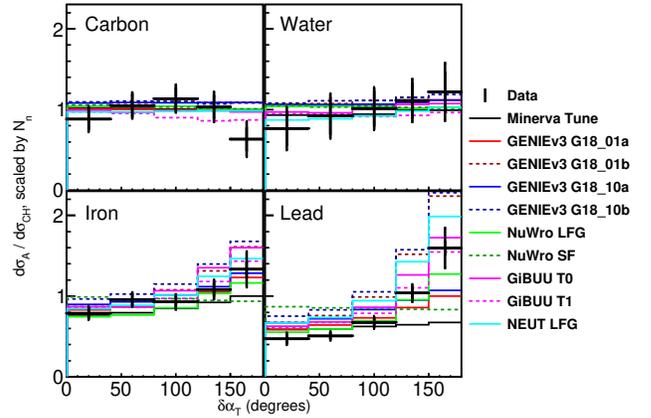}
    \caption{Cross-section ratios as a function of \tkialpha\ for different generators and multiple targets.  Note that changes to the FSI model (GENIE ``a'' to ``b'')
     in GENIE change the cross section ratio at both intermediate and high \tkialpha\  more than changes to the 2p2h model (GiBUU T0 to GiBUU T1) or changes to the initial state (GENIE 1 to GENIE 10) or (NuWro LFG to NuWro SF). }
        \label{Figure:Gencompare_alpha_ratio}
\end{figure}

The ratio of the differential cross section as a function of the \tkialpha \ for each nuclear target (C, H$_{2}$O, Fe, Pb) to that for scintillator is shown in Fig.~\ref{Figure:Gencompare_alpha_ratio}. The ratios show reasonable agreement between the data and the MINERvA tune except at high \tkialpha \ in Fe and Pb, where the model spread grows.  NEUT and the hN versions of GENIE predict a higher ratio where at large \tkialpha , where the FSI is greatest.  In the same region the hA versions of GENIE and NuWro predict a smaller ratio than seen in the data.  The \chisq\ between the cross section ratios and the models can be found in Table~\ref{tab:dalphat}.  

\subsection{Cross Sections as a function of the Transverse Momentum Imbalance perpendicular to the Muon Transverse Momentum }

\begin{figure}[h]  
	\centering
  \includegraphics[width=\linewidth]{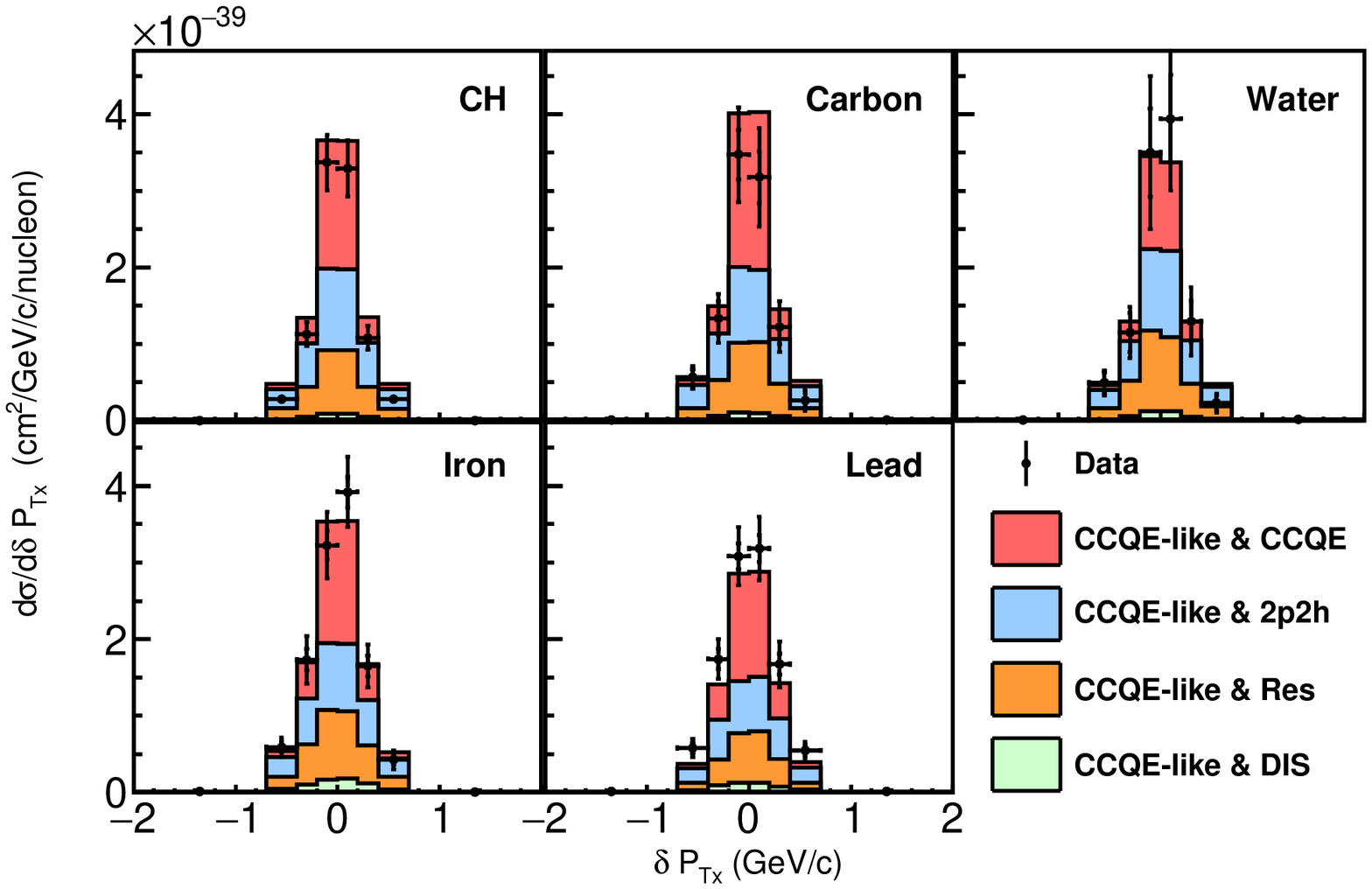}
	\caption{The differential cross section as a function of \tkidptx \ for CH, C, H$_{2}$O, Fe, and Pb targets, along with the predictions for the different quasielastic-like signal processes.}
	\label{figure:xsec_dptx}
\end{figure}

The differential cross section as a function of \tkidptx \ for interactions on the CH, C, H$_{2}$O, Fe, and Pb targets, respectively, are shown in Fig.~\ref{figure:xsec_dptx}. This variable is the component of \tkidelta \ that is perpendicular to the muon direction.  The width is sensitive to nuclear effects and it should be symmetric around zero.  The uncertainties broken down by source for both the absolute cross sections and the cross section ratios as a function of \tkidptx\ can be found in Fig.~\ref{fig:xsec_err_dptx}.

These results show no evidence of the left-right  asymmetry in \tkidptx \ for interactions on scintillator that was observed earlier with marginal significance in the LE data~\cite{MINERvA:2019ope}. Also, the asymmetry is not in evidence in the data taken on any of the other nuclear targets.  Improvements in the background tuning and subtraction procedure allowing for asymmetric backgrounds is thought to be the probable reason behind this change.

The width of the QE part of this variable is expected to arise from Fermi smearing.  FSI and 2p2h and resonant processes give broader contributions to the width of this variable.  Note that the data is somewhat narrower than the MINERvA tune expectation for interactions on scintillator and carbon.  This is similar to what was seen in the LE data~\cite{MINERvA:2019ope}.  For the interactions on Pb, the data is broader than the simulation.

\begin{figure}
    \centering
        \includegraphics[width=\linewidth]
         {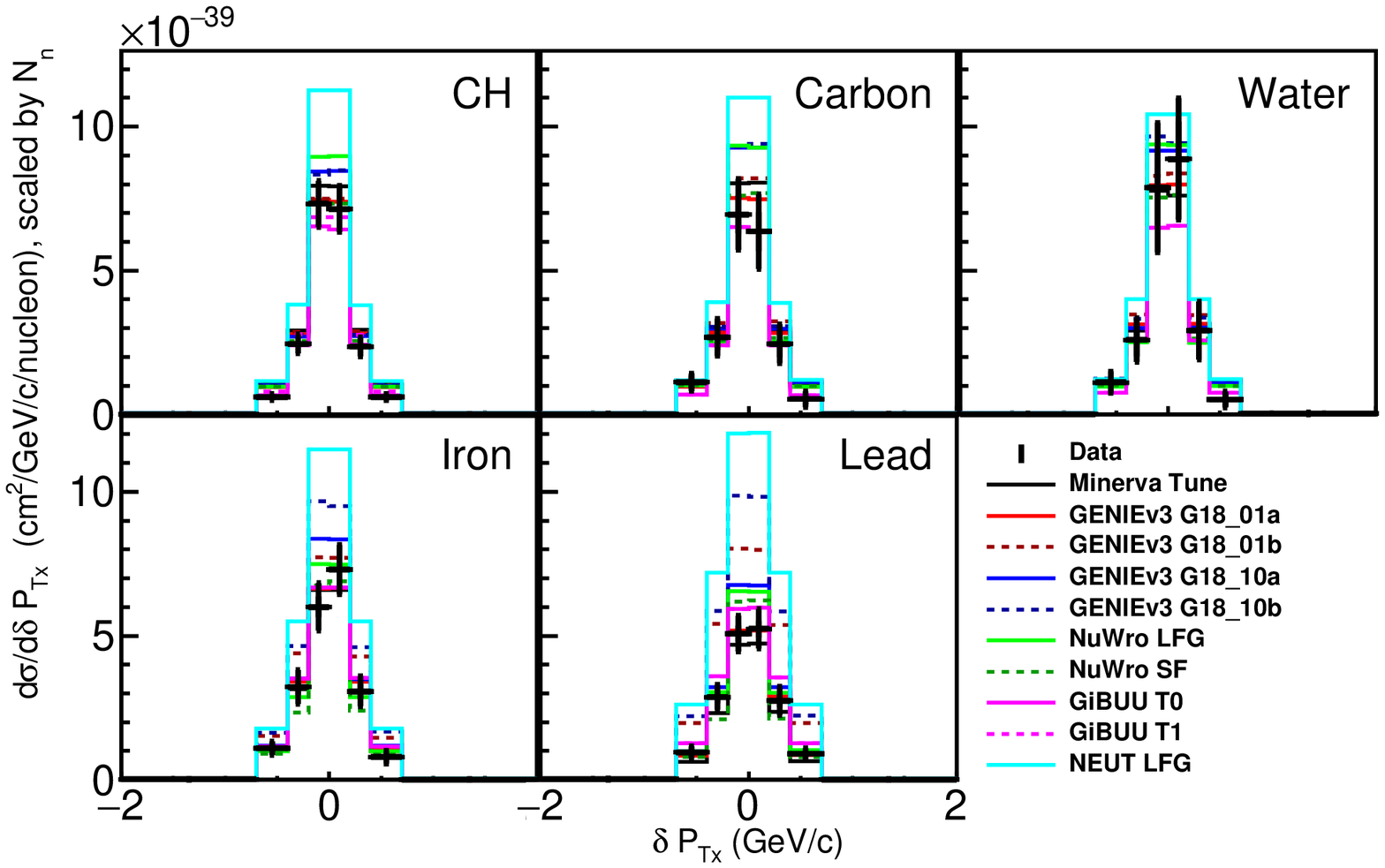}
    \caption{Cross section measurements and predictions as a function of \tkidptx\ for different targets and for a selection of different generator and model choices.}
    \label{fig:models_dptx}
\end{figure}

The differential cross section as a function of \tkidptx \ for the different targets as compared to a range of models is given in Fig.~\ref{fig:models_dptx}. Again the data are covered by the range of models.  NEUT and the hN versions of GENIE tend to overpredict the cross section as compared to the data, particularly for the larger targets. The other models do a fairly good job describing the data.  The \chisq\ between the cross sections and the models can be found in Table~\ref{tab:dptx}.

\begin{figure}
    \centering
        \includegraphics[width=\linewidth]{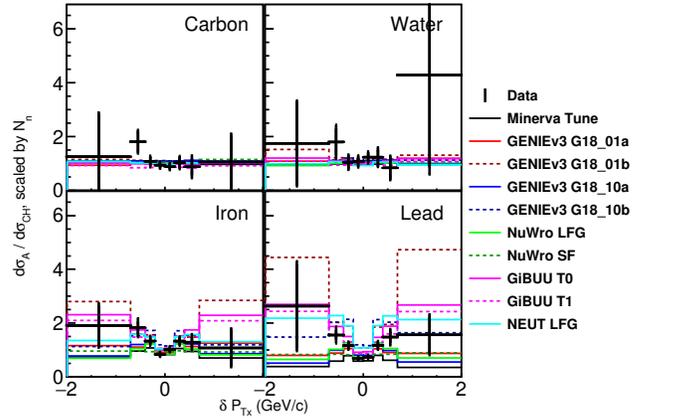}
    \caption{\tkidptx\ cross-section ratio comparison for multiple targets.  Note that changes to the FSI model (GENIE ``a'' to ``b'')
     in GENIE change the cross section ratio at extreme positive and negative values of \tkidptx\ more than changes to the 2p2h model (GiBUU T0 to GiBUU T1) or changes to the initial state (GENIE 1 to GENIE 10) or (NuWro LFG to NuWro SF).}
        \label{Fig:ratios_dptx}
\end{figure}

The ratio of the differential cross section as a function of  \tkidptx \ for each nuclear target (C, H$_{2}$O, Fe, Pb) to that for scintillator is shown in 
Fig.~\ref{Fig:ratios_dptx}. The models reproduce the data fairly well in these ratios except in the tails of Fe and Pb where the FSI is expected to play a large role.  
The \chisq\ between the cross section ratios and the models can be found in Table~\ref{tab:dptx}.

\subsection{Cross Sections as a function of the Transverse Momentum Imbalance parallel to the Muon Transverse Momentum }

\begin{figure}[h]  
	\centering
  \includegraphics[width=\linewidth]{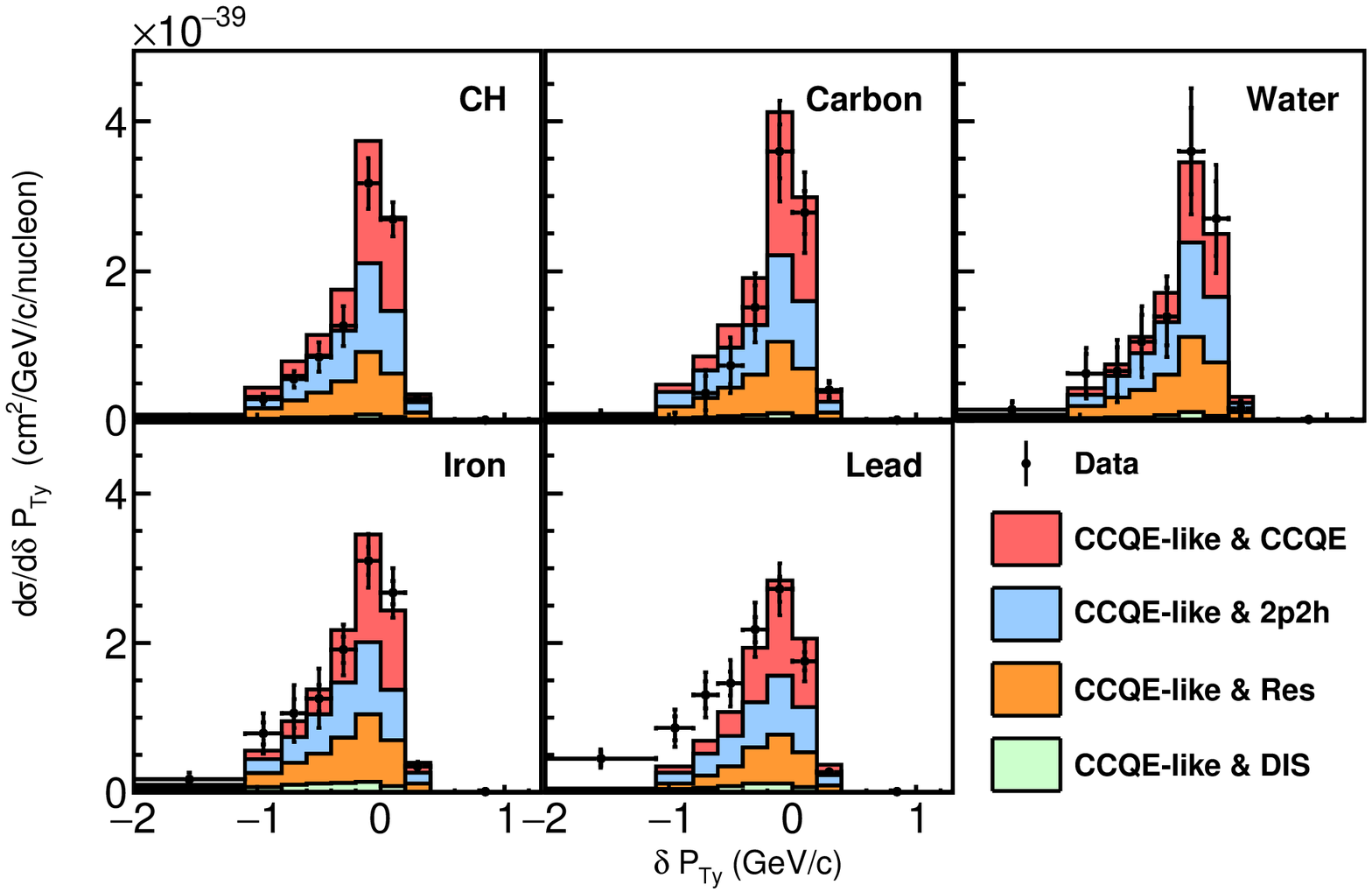}
	\caption{The differential cross section as a function of \tkidpty \ for CH, C, H$_{2}$O, Fe, and Pb targets,along with the predictions for the different quasielastic-like signal processes.}
	\label{figure:xsec_dpty}
\end{figure}

Results for the differential cross section as a function of \tkidpty \ for interactions on the CH, C, H$_{2}$O, Fe, and Pb targets, respectively, are given in Fig.~\ref{figure:xsec_dpty}. 
This variable reflects changes to the outgoing proton energy after the interaction.  The negative tail indicates that the proton often loses energy traversing the nucleus.  The plots show that the tail increases in cross section relative to the peak as the nuclear size increases.  The MINERvA tune considerably underpredicts the cross section in the tail for the Pb target.  The uncertainties broken down by source for both the absolute cross sections and the cross section ratios as a function of \tkidpty\ can be found in Fig.~\ref{fig:xsec_err_dpty}.

\begin{figure}
    \centering
        \includegraphics[width=\linewidth]{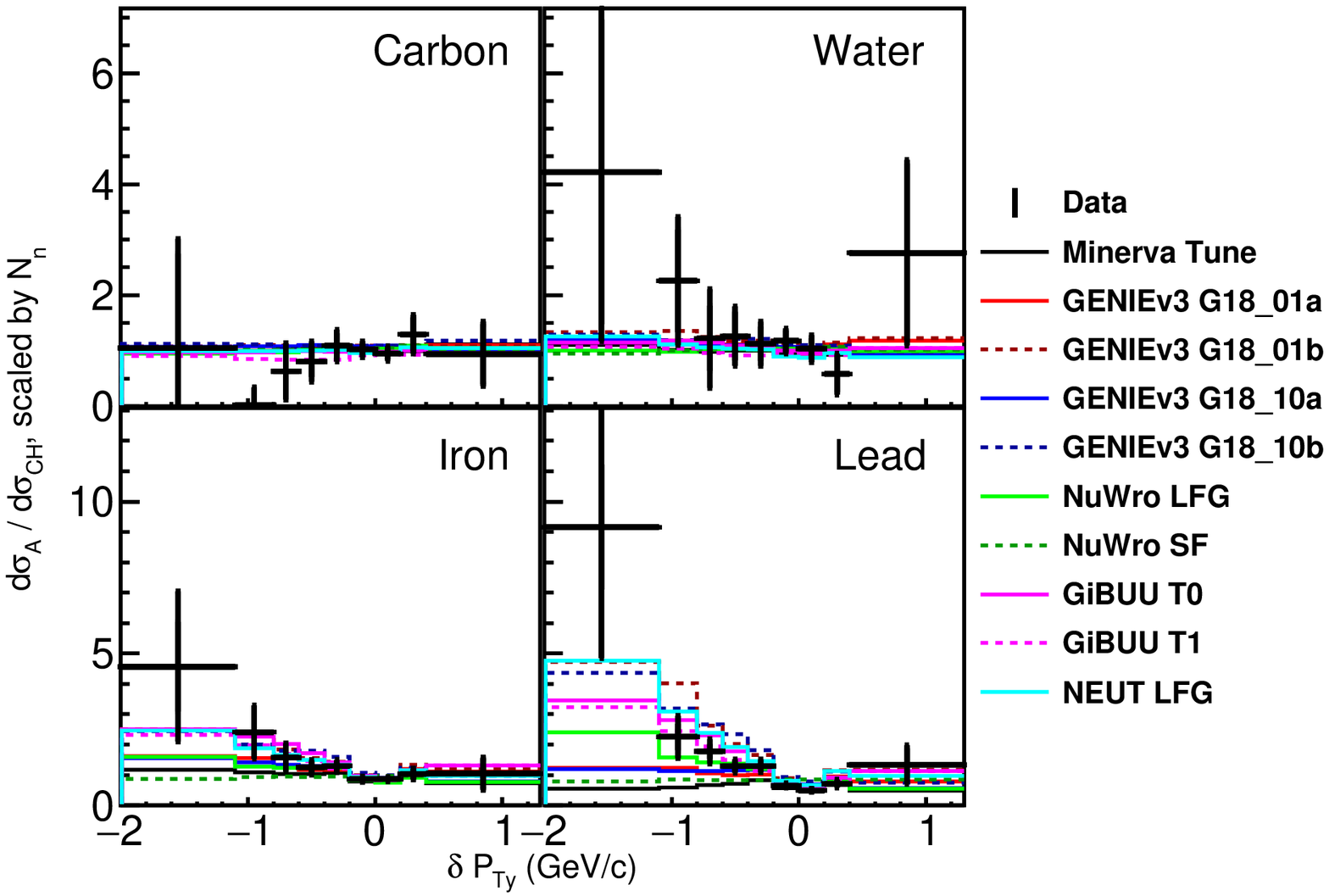}
     \caption{Cross section measurements and predictions as a function of \tkidpty\ for different targets and for a selection of different generator and model choices.}
    \label{fig:models_dpty}
\end{figure}

The differential cross section as a function of \tkidpty \ for the different targets as compared to a range of models is given in Fig.~\ref{fig:models_dpty}. NEUT significantly overpredicts the cross section.  That overprediction is most pronounced for the larger targets, particularly in the tail where FSI is most important.  The hN versions of GENIE approach the prediction of NEUT in Pb.  The MINERvA tune and NuWro SF underpredict the cross section in the tail of \tkidpty \ for the Pb target. The \chisq\ between the cross sections and the models can be found in Table~\ref{tab:dpty}.

\begin{figure}
    \centering
        \includegraphics[width=\linewidth]{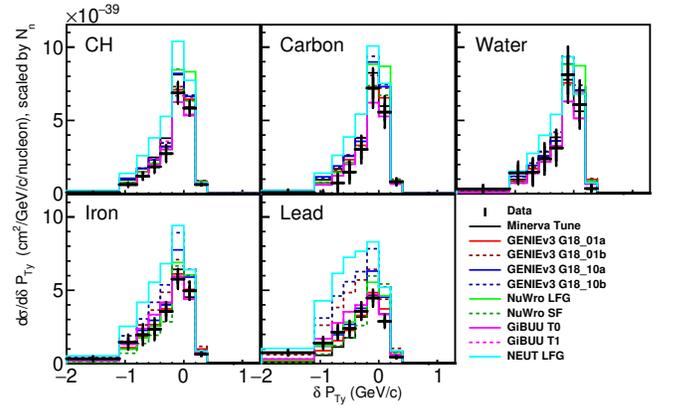}
    \caption{\tkidpty\ cross-section ratio comparison for multiple targets.  Note that changes to the FSI model (GENIE ``a'' to ``b'')
     in GENIE change the cross section ratio at low values of \tkidpty\  more than changes to the 2p2h model (GiBUU T0 to GiBUU T1) or changes to the initial state (GENIE 1 to GENIE 10) or (NuWro LFG to NuWro SF).}
        \label{fig:ratiomodels_dpty}
\end{figure}

Figure~\ref{fig:ratiomodels_dpty} shows the differential cross-section ratio as a function of  \tkidpty \ for each nuclear target (C, H$_{2}$O, Fe, Pb) to that for scintillator. The models in general describe the data fairly well except in the tails of the larger targets where the model spread becomes pronounced.
The \chisq\ between the cross section ratios and the models can be found in Table~\ref{tab:dpty}.

\subsection{Cross Sections as a function of the momentum imbalance along the neutrino direction} 

\begin{figure}[h]  
	\centering
  \includegraphics[width=\linewidth]{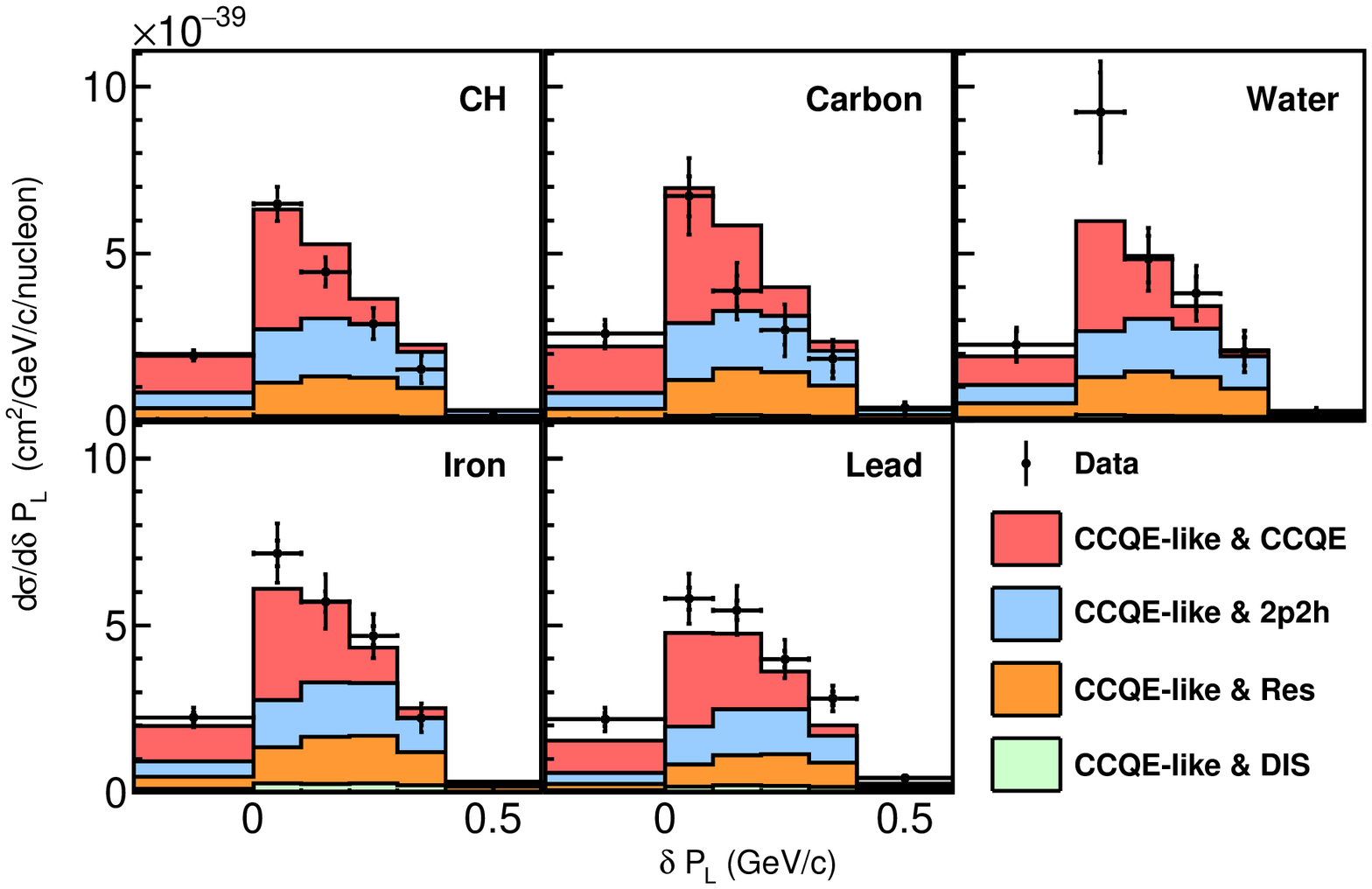}
	\caption{The differential cross section as a function of \tkipl\ for CH, C, H$_{2}$O, Fe, and Pb targets, along with the predictions for the different quasielastic-like signal processes.}
	\label{figure:xsec_pl}
\end{figure}

Figure~\ref{figure:xsec_pl} gives the differential cross section as a function of \tkipl \ for interactions on the CH, C, H$_{2}$O, Fe, and Pb targets, respectively. 
\tkipl \ represents the longitudinal component of the momentum imbalance between the initial neutrino momentum and the sum of the final state lepton and hadron momenta.  This imbalance results from nuclear effects.  This variable and its extraction are discussed in a previous MINERvA paper~\cite{MINERvA:2018hba}. \tkipl\ can be calculated with the following equations,
\begin{equation}
    \tkipl = \frac{1}{2}R - \frac{m_{A'}^2 + \tkidelta^2}{2R}
    \label{eq:pl}
\end{equation}
\begin{equation}
    R \equiv m_{A} + p^\mu_L + p^p_L - E^\mu - E^p
\end{equation}
where $m_{A}$ and $m_{A'}$ are the masses of the atomic nucleus before and after interaction, and $p_L$ and $E$ are the momenta and energy of the muon and proton. Qualitatively, the MINERvA tune reproduces the data fairly well except for the case of interactions on Pb where the data exceeds the prediction. The uncertainties broken down by source for both the absolute cross sections and the cross section ratios as a function of \tkipl\ can be found in Fig.~\ref{fig:xsec_err_pl}.

\begin{figure}
    \centering
            \includegraphics[width=\linewidth]{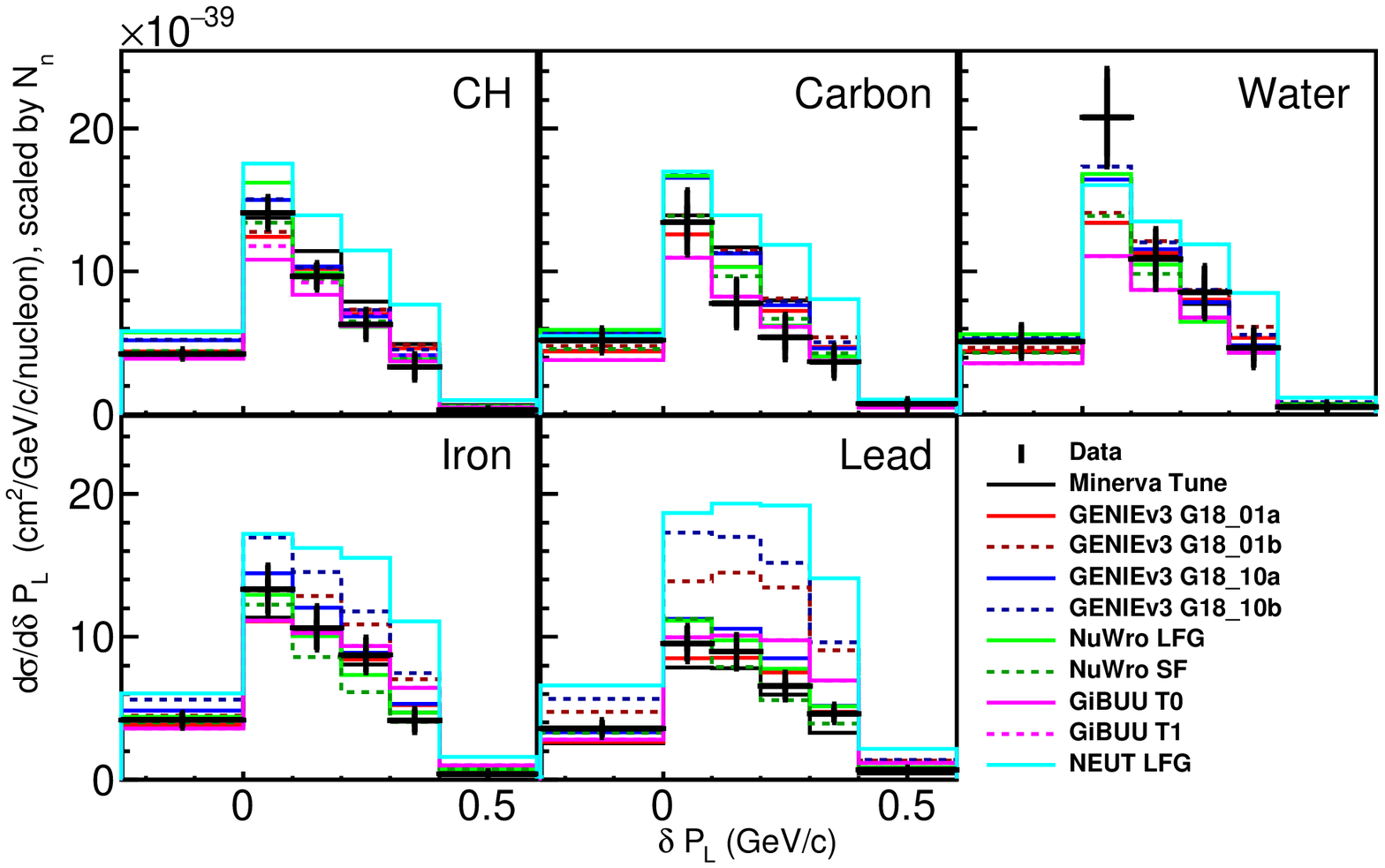}
    \caption{Cross section measurements and predictions as a function of \tkipl\ for different targets and for a selection of different generator and model choices. }
    \label{fig:models_pl}
\end{figure}

The results for a range of models are compared to the differential cross section as a function of \tkipl \ for the different targets in Fig.~\ref{fig:models_pl}.   Relative to the data,  NEUT has a larger cross section and the hN GENIE models and NEUT predict stronger nuclear effects than seen in the data for the larger targets.
The \chisq\ between the cross sections and the models can be found in Table~\ref{tab:pl}.


\begin{figure}
    \centering
       \includegraphics[width=\linewidth]{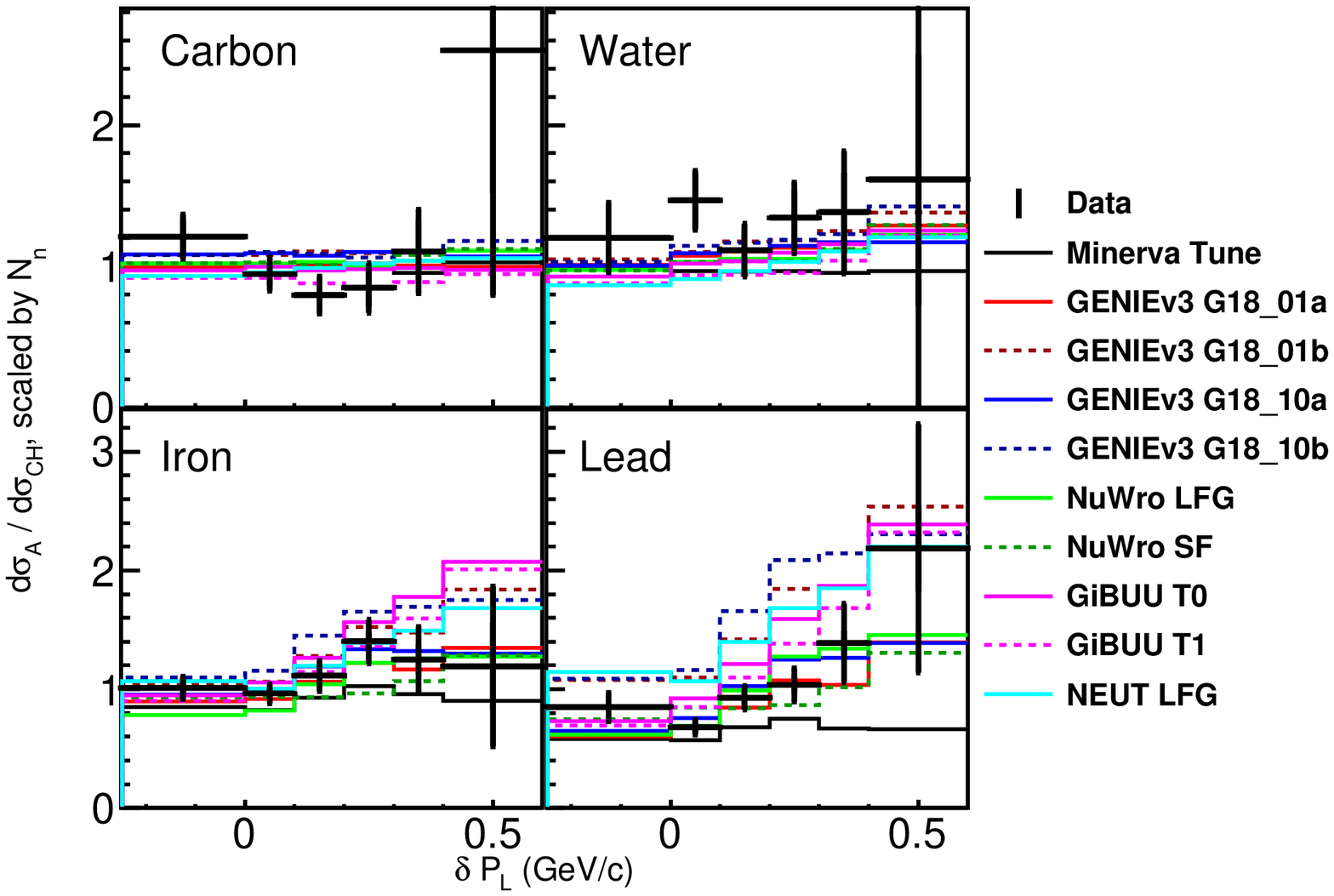}
    \caption{\tkipl\ cross-section ratio generator comparison for multiple targets.  Changes to the FSI model (GENIE ``a'' to ``b'') in GENIE change the cross section ratio throughout \tkipl\, and change the ratio more than changes to the 2p2h model (GiBUU T0 to GiBUU T1) or changes to the initial state (GENIE 1 to GENIE 10) or (NuWro LFG to NuWro SF).}
        \label{Figure:Gencompare_pl_ratio}
\end{figure}

A plot of the ratio of the differential cross section as a function of  \tkipl \ for each nuclear target (C, H$_{2}$O, Fe, Pb) to that for scintillator is given in 
Fig.~\ref{Figure:Gencompare_pl_ratio}.  The data is fairly well described by the models for the smaller targets.  However,the model spread at high \tkipl \ for the Fe and Pb targets is large.
The \chisq\ between the cross section ratios and the models can be found in Table~\ref{tab:pl}.

\subsection{Cross Sections as a function of the struck neutron momentum} 
\begin{figure}[h] 
	\centering
\includegraphics[width=\linewidth]{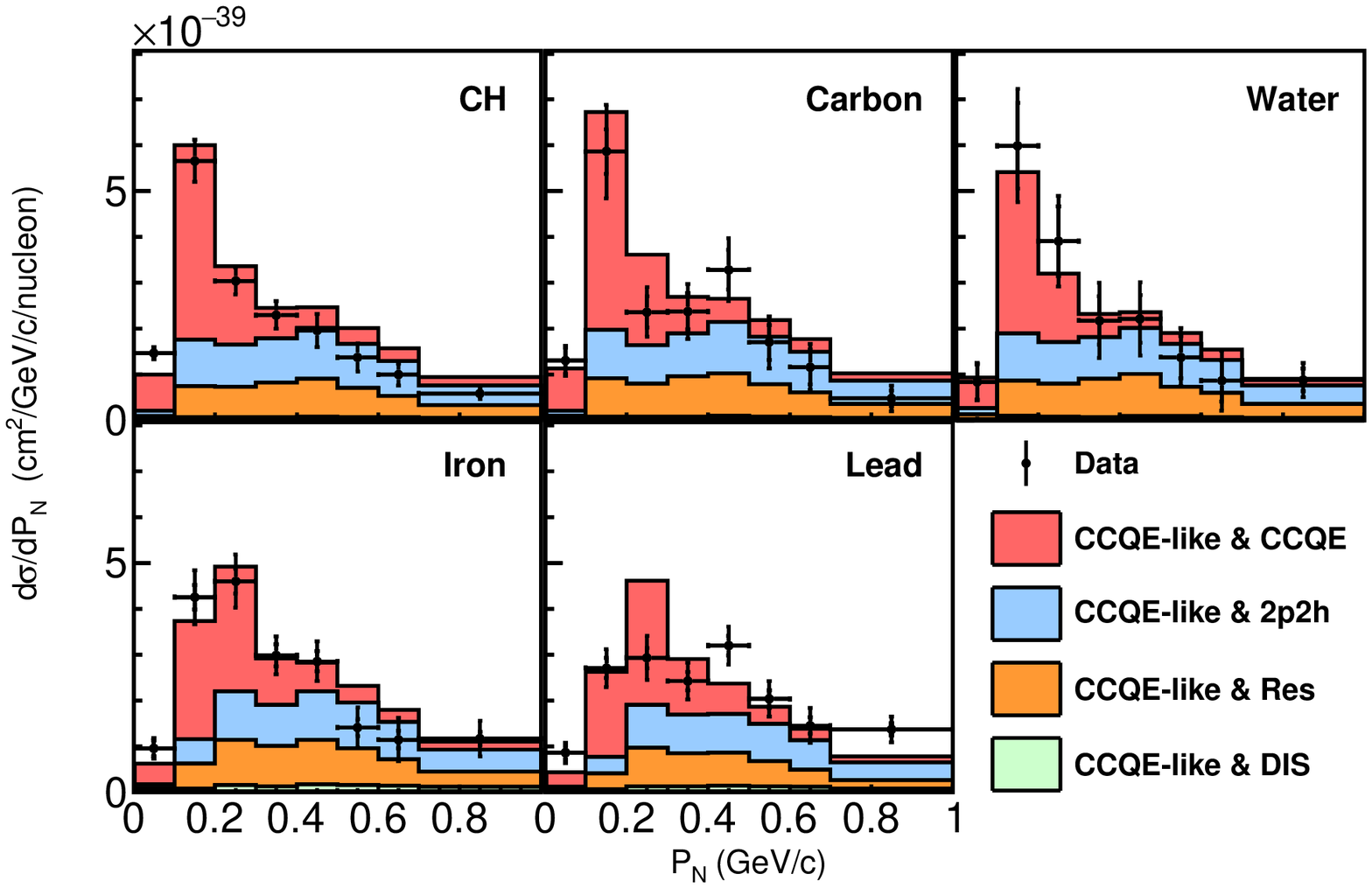}
	\caption{The differential cross section as a function of \tkipn\ for CH, C, H$_{2}$O, Fe, and Pb targets, along with the predictions for the different quasielastic-like signal processes.}
	\label{figure:xsec_pn}
\end{figure}

The momentum imbalance between the initial neutrino momentum and the sum of the final state lepton and hadron momenta.  In the limit of no intranuclear momentum transfer, it represents the momentum of the struck neutron (\tkipn).  This observable can be extracted using the following equation:
\begin{equation}
    \tkipn = \sqrt{\tkidelta^2 + \tkipl^2}
    \label{eq:pn}
\end{equation}
where \tkidelta\ is the net transverse momentum of the muon and proton system and \tkipl\ is defined in Equation \ref{eq:pl}. MINERvA has measured it before on scintillator in the LE run~\cite{MINERvA:2018hba}.  The peak at low \tkipn \ comes from events with little FSI and has a position and width reflecting the Fermi momentum of the struck neutron.  The tail at larger \tkipn \ largely comes from FSI processes that decelerate the proton and pion absorption. A smaller (larger) tail implies less (more) FSI. The differential cross section as a function of \tkipn \ for interactions on the CH, C, H$_{2}$O, Fe, and Pb targets, respectively, are shown in Fig.~\ref{figure:xsec_pn}. 

The MINERvA tune qualitatively describes the data fairly well for the smaller nuclei.   It also follows the data with the reduction of the no-FSI peak and enhancement of the FSI-induced tail with increasing $A$.  The uncertainties broken down by source for both the absolute cross sections and the cross section ratios as a function of \tkipl\ can be found in Fig.~\ref{fig:xsec_err_pl}.

\begin{figure}
    \centering 
\includegraphics[width=\linewidth]{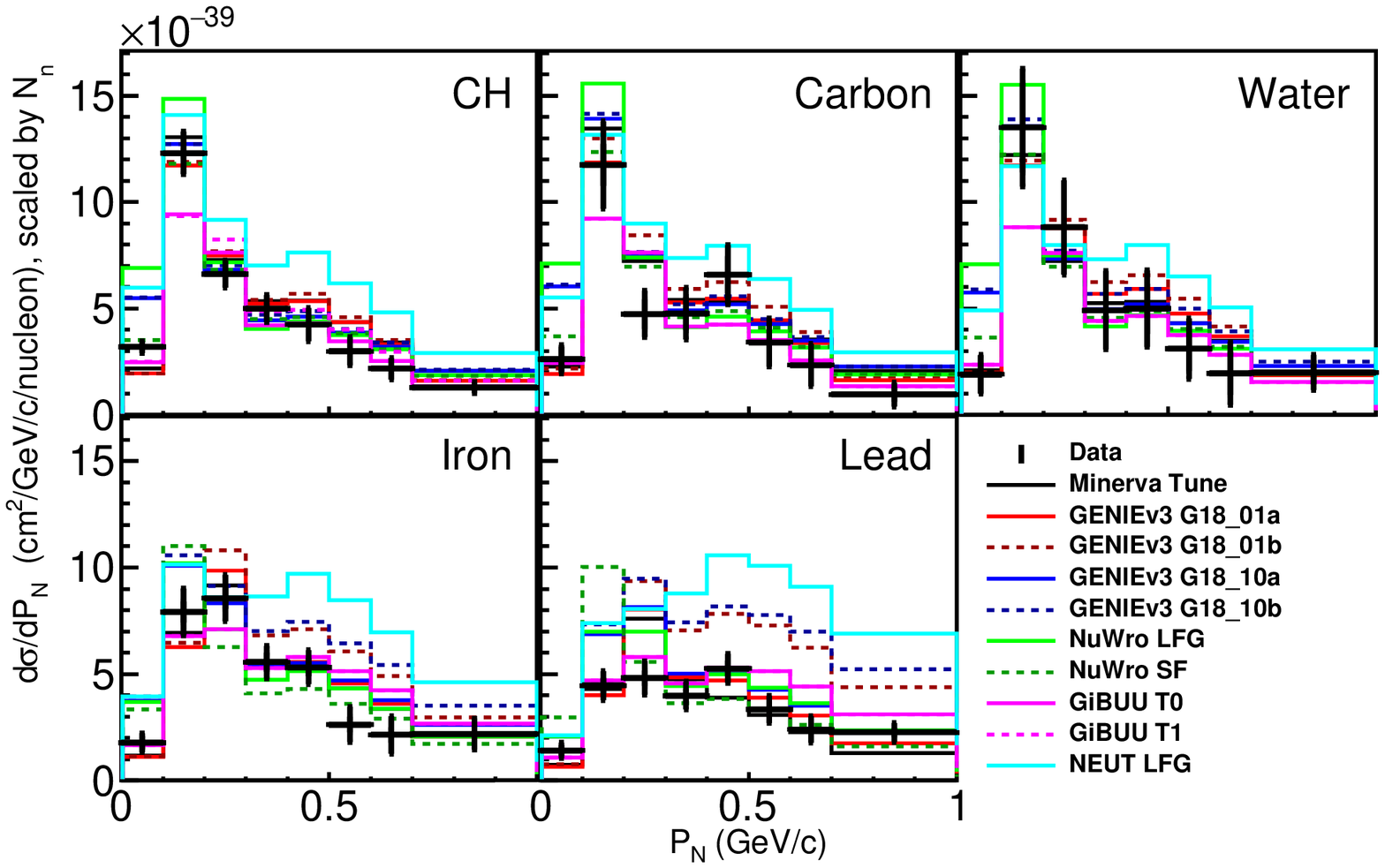}
    \caption{Cross section measurements and predictions as a function of \tkipn\ for different targets and for a selection of different generator and model choices.}
    \label{fig:models_pn}
\end{figure}

\begin{figure}
    \centering
                \includegraphics[width=\linewidth]{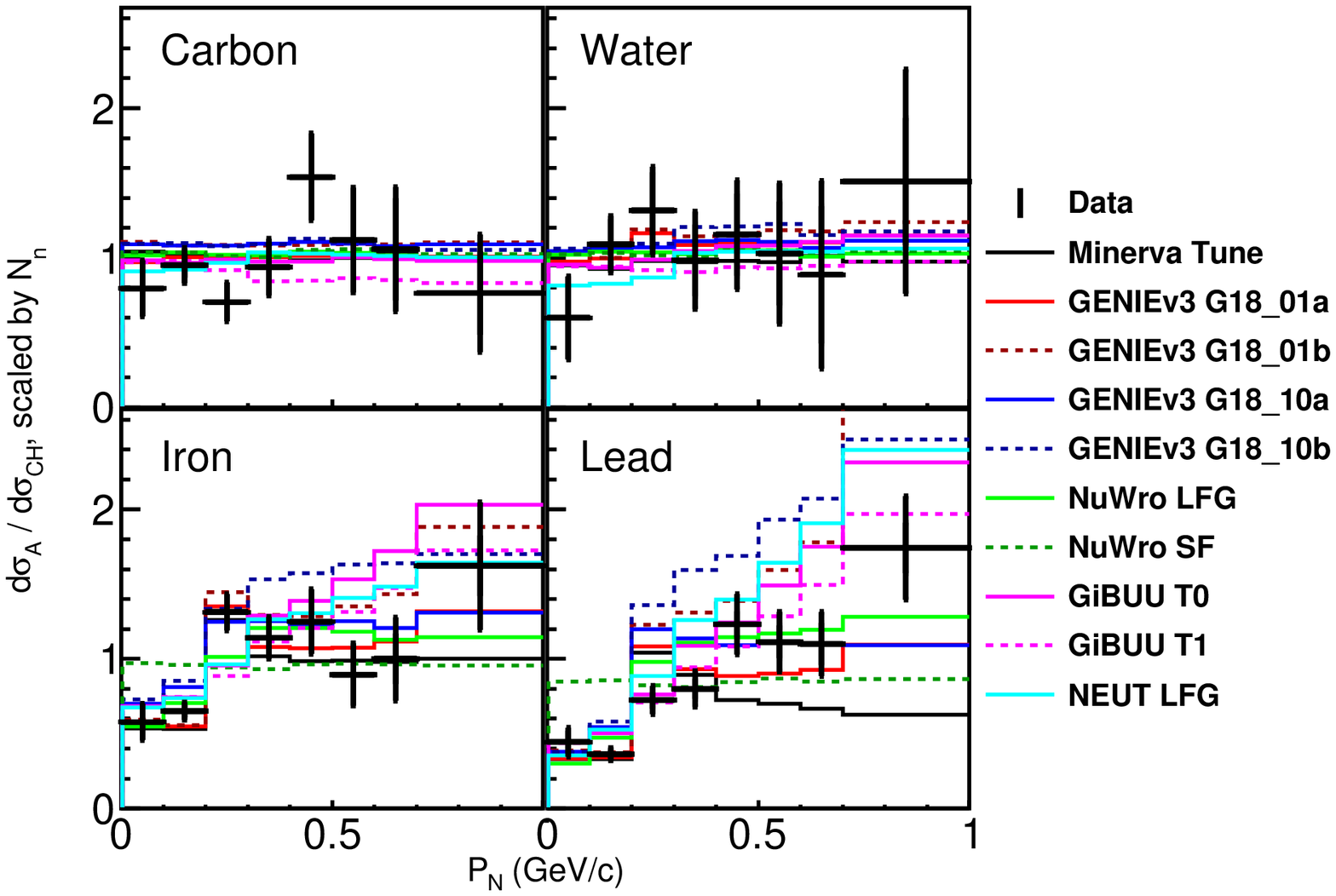}

    \caption{\tkipn\ cross-section ratio generator comparison for multiple targets.  Changes to the FSI model (GENIE ``a'' to ``b'') in GENIE change the cross section ratio most dramatically at high \tkipn\, and those changes are  larger than changes to the 2p2h model (GiBUU T0 to GiBUU T1) or changes to the initial state (GENIE 1 to GENIE 10) or (NuWro LFG to NuWro SF).}
        \label{Figure:Gencompare_pn_ratio}
\end{figure}

The differential cross section as a function of \tkipn \ for the different targets as compared to a range of models is given in Fig.~\ref{fig:models_pn}. In this comparison, NEUT and the hN versions of GENIE predict substantially more cross section in the decelerating FSI in the tail than seen in the data, particularly for the larger nuclear targets.  NuWro tends to have more events in the low-side tail indicating more acceleration of the proton than seen in the data.  GiBUU and the hA verstions of GENIE do a fairly good job describing the data in this variable. The \chisq\ between the cross sections and the models can be found in Table~\ref{tab:neutronmomentum}.
The ratio of the differential cross section as a function of  \tkipn \ for each nuclear target (C, H$_{2}$O, Fe, Pb) to that for scintillator is shown in 
Fig.~\ref{Figure:Gencompare_pn_ratio}. The models agree with the data well for the smaller targets.  For the Fe and Pb targets, the model spread is large.  NuWro SF has a much larger contribution to the ratio at low \tkipn \ than the data or any of the other models.  At higher \tkipn \ in the larger targets, NEUT and the GENIE hN  models tend to be higher than the data and the GENIE hA models and NuWro SF are lower than the data.  For interactions on Pb, the cross section ratio data disagree with the MINERvA Tune especially above \tkipn\ above 0.5~GeV.  This could be evidence for additional decelerating FSI in the tail and FSI processes that accelerate the proton on the low side of the \tkipn \ peak.  The \chisq\ between the cross sections and the models can be found in Table~\ref{tab:neutronmomentum}.

\subsection{Cross Sections as a function of the proton momentum}

\begin{figure}[h] 
	\centering
  \includegraphics[width=\linewidth]{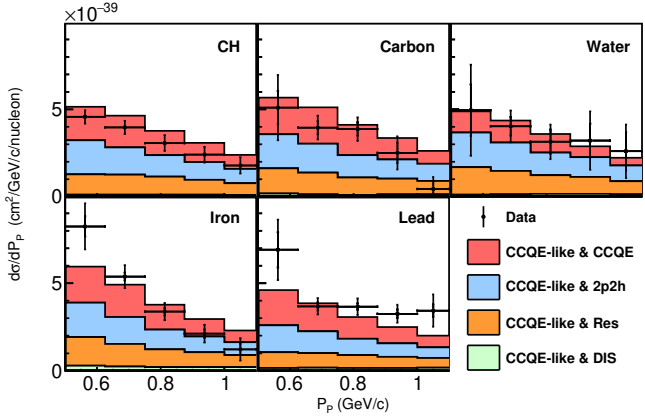}
	\caption{The differential cross section as a function of P$_p$ for CH, C, H$_{2}$O, Fe, and Pb targets, along with the predictions for the different quasielastic-like signal processes.}
	\label{figure:xsec_proton_p}
\end{figure}

The differential cross section in terms of P$_{p}$ for interactions on the various nuclear targets are shown in Fig.~\ref{figure:xsec_proton_p}. P$_{p}$ is the magnitude of the proton momentum in GeV/c.  This might be sensitive to nuclear effects since the proton passes through some nuclear matter before emerging.  The uncertainties broken down by source for both the absolute cross sections and the cross section ratios as a function of P$_p$ can be found in Fig.~\ref{fig:xsec_err_proton_p}.

The MINERvA tune simulation models the data fairly well for each target except Fe and Pb.  For the Fe target, there is an indication of a shape difference, where the data has a slightly softer momentum distribution.  This is different from what is seen in Pb where there is a markedly harder distribution at high momentum than the model.

\begin{figure}
    \centering
\includegraphics[width=\linewidth]{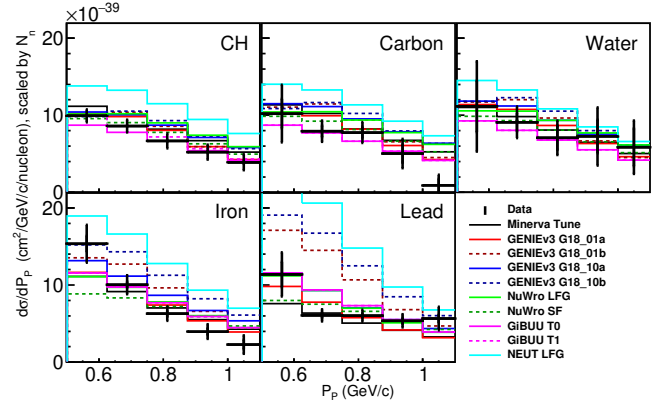}
    \caption{Cross section measurements and predictions as a function of P$_p$ for different targets and for a selection of different generator and model choices.}
    \label{fig:models_proton_p}
\end{figure}

The differential cross sections in P$_{p}$ for the different targets are compared to a range of models in Fig.~\ref{fig:models_proton_p}. NEUT exhibits a higher cross section than the data, in general, with a shape that agrees with the data for the smaller targets.  For the Pb target, NEUT and the hN GENIE models have a considerably softer momentum shape than the data, indicating too much FSI as seen in the other variables.  The other models do a fairly good job describing the data.
The \chisq\ between the cross sections and the models can be found in Table~\ref{tab:protonmomentum}.

\begin{figure}
    \centering
\includegraphics[width=\linewidth]{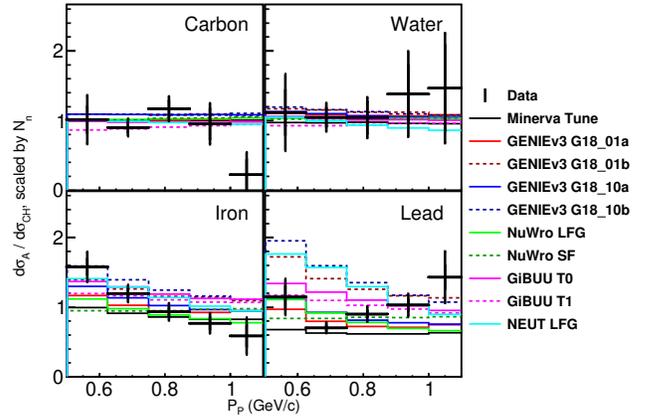}
    \caption{P$_{p}$ cross-section ratio generator comparison for multiple targets.  Changes to the FSI model (GENIE ``a'' to ``b'') in GENIE change the cross section ratio most at low proton momenta, and those changes are larger than changes to the 2p2h model (GiBUU T0 to GiBUU T1) or changes to the initial state (GENIE 1 to GENIE 10) or (NuWro LFG to NuWro SF).}
        \label{Figure:Gencompare_proton_p_ratio}
\end{figure}

The ratio of the differential cross section for each target relative to that for scintillator as a function of  P$_{p}$  is shown in 
Fig.~\ref{Figure:Gencompare_proton_p_ratio}. The ratio is higher for NEUT than the data for all of the targets and the other models.  For the Pb target the hN versions of GENIE approach NEUT and exhibit a higher ratio at low momentum that the data, as does NEUT.  This may indicate that the FSI strength for NEUT and the hN versions of GENIE becomes excessive for Pb as compared to what happens for the smaller targets.
The \chisq\ between the cross section ratios and the models can be found in Table~\ref{tab:protonmomentum}.

\subsection{Cross Sections as a function of the proton transverse momentum}

\begin{figure}[h]  
	\centering
   \includegraphics[width=\linewidth]{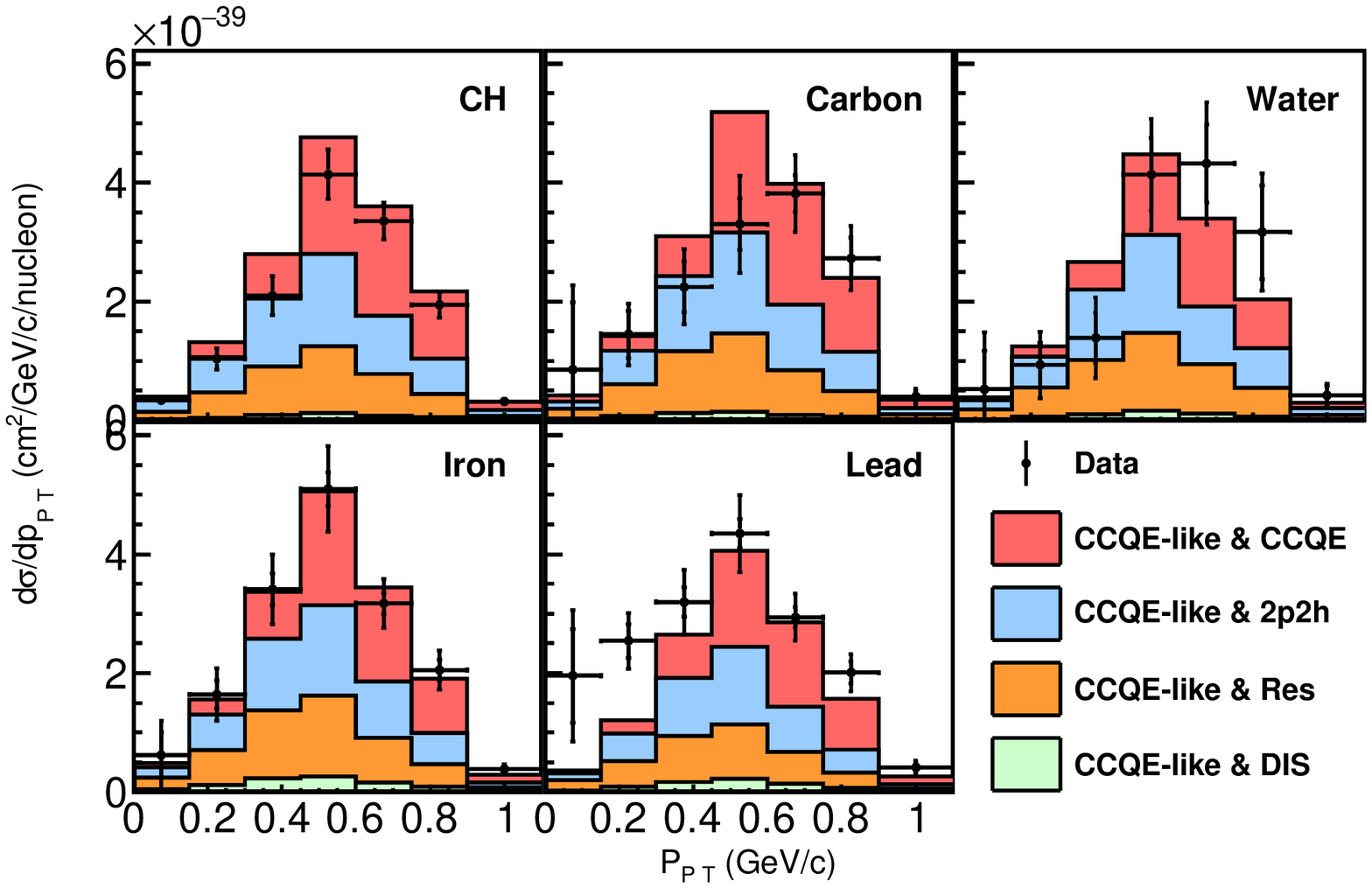}
	\caption{The differential cross section as a function of P$_{p T}$ for CH, C, H$_{2}$O, Fe, and Pb targets, along with the predictions for the different quasielastic-like signal processes.}
	\label{figure:xsec_proton_pt}
\end{figure}

The differential cross section as a function of P$_{p T}$ for interactions on the CH, C, H$_{2}$O, Fe, and Pb targets, respectively, are shown in Fig.~\ref{figure:xsec_proton_pt}. P$_{p T}$ is the magnitude of the  momentum of the proton transverse to the beam direction. Nuclear effects might be expect to smear this quantity out.   The uncertainties broken down by source for both the absolute cross sections and the cross section ratios as a function of P$_{p T}$ can be found in Fig.~\ref{fig:xsec_err_proton_pt}.

The data appear to be shifted to slightly higher P$_{p T}$ relative to the MINERvA tune simulation for the smaller nuclei. Fe is in good agreement.  For Pb, the distribution seems a little flatter than  the simulation, with more cross section at low P$_{p T}$ than expected.

\begin{figure}
    \centering
\includegraphics[width=\linewidth]{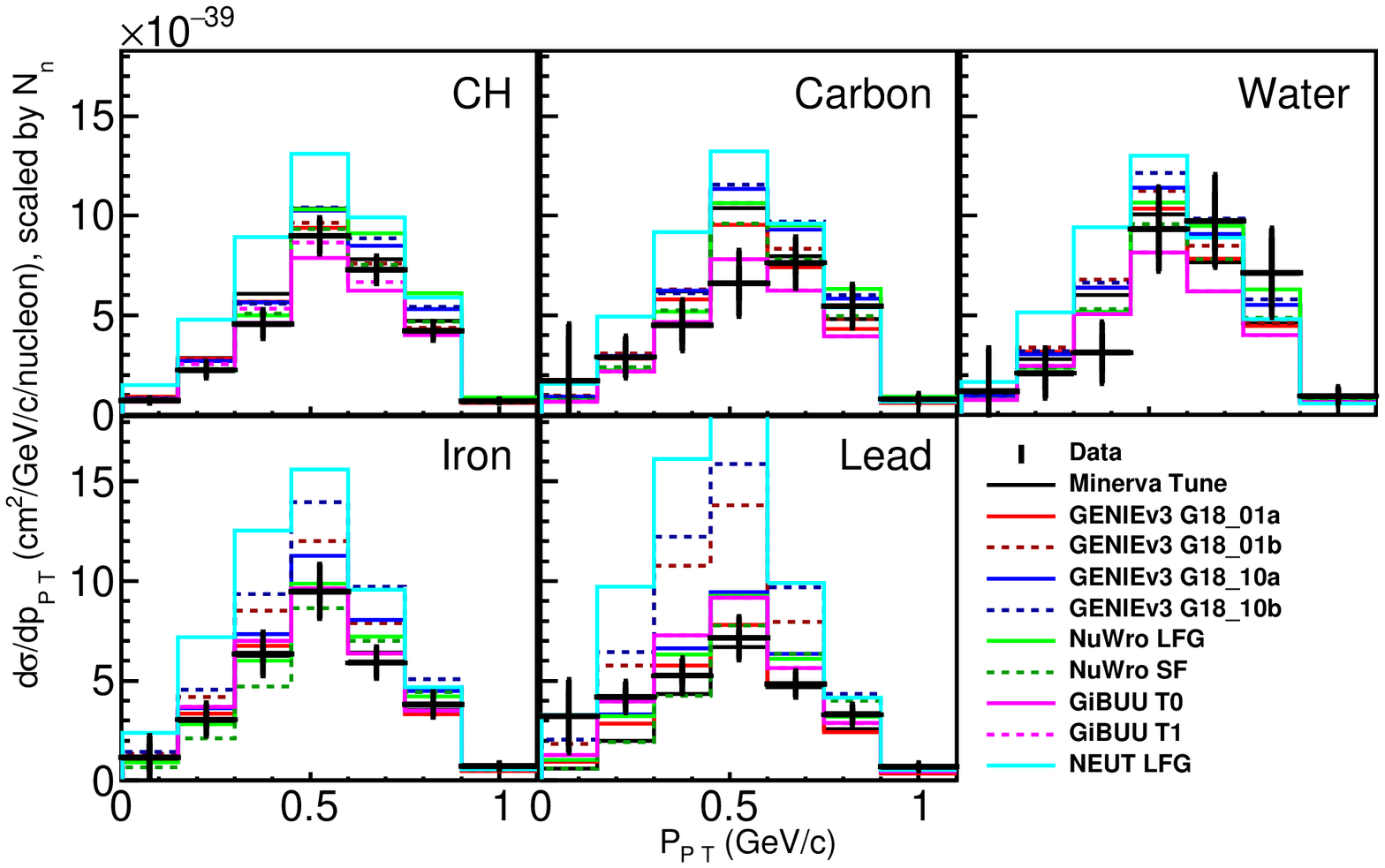}
    \caption{Cross section measurements and predictions as a function of P$_{p T}$ for different targets and for a selection of different generator and model choices.}
    \label{fig:models_proton_pt}
\end{figure}

The differential cross section as a function of P$_{p T}$ for the different targets as compared to a range of models is given in Fig.~\ref{fig:models_proton_pt}. The models all seem to peak at a slightly smaller P$_{p T}$ than is seen in the data for the smaller targets.  NEUT has a higher cross section than the data and this is particularly pronounced at low P$_{p T}$.  The hN versions of GENIE do this as well for the larger targets. Otherwise the models do a fairly good job describing the data.  The \chisq\ between the cross sections and the models can be found in Table~\ref{tab:protonpt}.

\begin{figure}
    \centering
            \includegraphics[width=\linewidth]{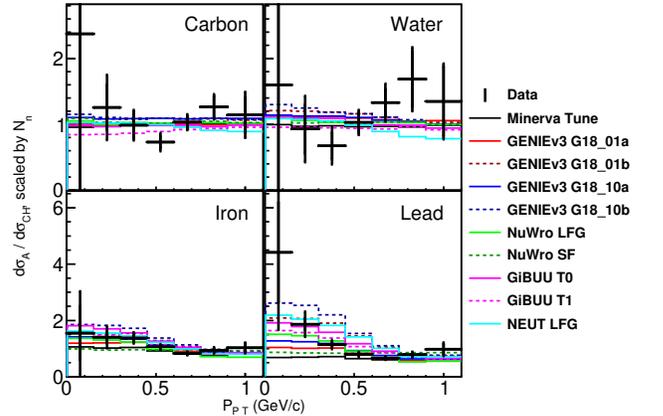}
    \caption{Cross-section ratio as a function of proton transverse momentum, compared to several generators for multiple targets.  Changes to the FSI model (GENIE ``a'' to ``b'') in GENIE change the cross section ratio most at low proton angles, and those changes are larger than changes to the 2p2h model (GiBUU T0 to GiBUU T1) or changes to the initial state (GENIE 1 to GENIE 10) or (NuWro LFG to NuWro SF).}
        \label{Figure:Gencompare_proton_pt_ratio}
\end{figure}

The ratio of the differential cross section as a function of  P$_{p T}$ for each nuclear target (C, H$_{2}$O, Fe, Pb) to that for scintillator is shown in 
Fig.~\ref{Figure:Gencompare_proton_pt_ratio}. The model spread becomes pronounced at low P$_{p T}$ for the Pb target.  At intermediate P$_{p T}$ in Pb, NEUT and the hN versions of GENIE have a higher ratio than the data. At lower P$_{p T}$ they agree with the data and most of the other models are lower, though the errors on the data are large.  GiBUU describes the data fairly well across targets and range in P$_{p T}$.  The \chisq\ between the cross section ratios and the models can be found in Table~\ref{tab:protonpt}.

\subsection{Cross Sections as a function of the proton angle}

\begin{figure}[h]  
	\centering
   \includegraphics[width=\linewidth]{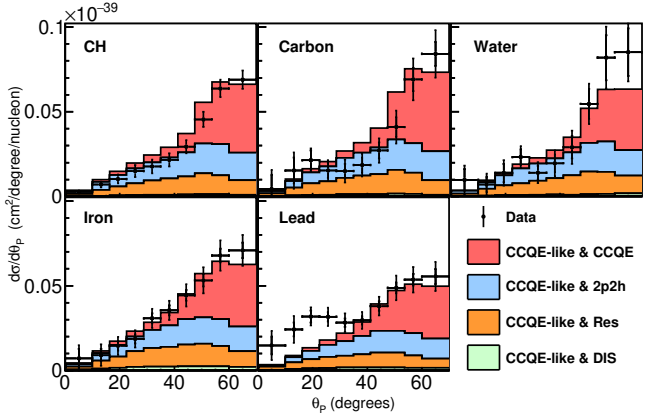}
	\caption{The differential cross section as a function of $\theta_p$ for CH, C, H$_{2}$O, Fe, and Pb targets, along with the predictions for the different quasielastic-like signal processes.}
	\label{figure:xsec_proton_theta}
\end{figure}

Figure~\ref{figure:xsec_proton_theta} shows the observed differential cross section as a function of $\theta_{p}$ for interactions on each of the nuclear targets, respectively.
$\theta_{p}$ is the opening angle of the proton direction relative to the beam direction.  The uncertainties broken down by source for both the absolute cross sections and the cross section ratios as a function of $\theta_p$ can be found in Fig.~\ref{fig:xsec_err_proton_theta}.

The data is fairly well modeled by the MINERvA tune for each target except Pb.  Note the inflection around 40 degrees where the CCQE process tends to kick in.  For the data taken on Pb, relative to the other targets, it seems that a different process is significant below 40 degrees that is not well modeled by the MINERvA tune.  

\begin{figure}
    \centering
 \includegraphics[width=\linewidth]{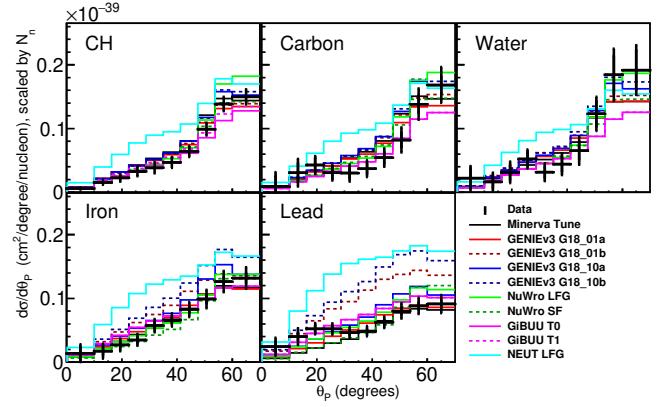}
    \caption{Cross section measurements and predictions as a function of $\theta_p$ for different targets and for a selection of different generator and model choices.}
    \label{fig:models_proton_theta}
\end{figure}

A range of models are compared to the data in terms of the differential cross section as a function of $\theta_{p}$ for each of the nuclear targets in Fig.~\ref{fig:models_proton_theta}.
  NEUT tends to have significantly more cross section in the intermediate $\theta_{p}$ region dominated by 2p2h resonant events.  For the larger targets, NEUT and the hN versions of GENIE exhibit significantly higher cross sections than the data at all but the smallest angle. Besides NuWro SF, the other models tend to be closer to the data relative to the MINERvA tune in the small angle region making it seem less like a process is missing in the models.  The \chisq\ between the cross sections and the models can be found in Table~\ref{tab:protontheta}.


\begin{figure}
    \centering
 \includegraphics[width=\linewidth]{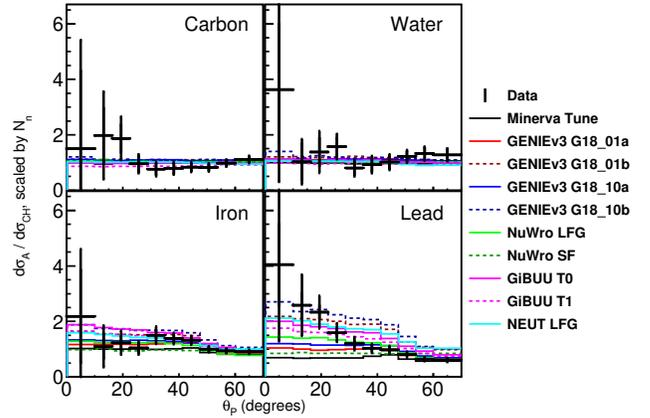}
    \caption{Cross-section ratio as a function of proton angle, for several generators and multiple targets.  Changes to the FSI model (GENIE ``a'' to ``b'') in GENIE change the cross section ratio most at low proton angles, and those changes are larger than changes to the 2p2h model (GiBUU T0 to GiBUU T1) or changes to the initial state (GENIE 1 to GENIE 10) or (NuWro LFG to NuWro SF).}
\label{Figure:Gencompare_proton_theta_ratio}
\end{figure}

The ratio of the differential cross section as a function of  $\theta_{p}$ for each nuclear target (C, H$_{2}$O, Fe, Pb) to that for scintillator is shown in 
Fig.~\ref{Figure:Gencompare_proton_theta_ratio}. For the ratios as a function of $\theta_{p}$, there is qualitative agreement between the data and the MINERvA tune except for the distributions for the Pb target.  For Pb, NEUT, GiBUU, and the hN versions of GENIE agree with the data ratio at low angle but exhibit larger ratios than the data at large angles.  NuWro and the hA versions of GENIE agree well with the data ratio at larger angles and tend to underpredict the data at lower angles.
The \chisq\ between the cross section ratios and the models can be found in Table~\ref{tab:protontheta}.

\subsection{Cross Sections as a function of muon momentum}

\begin{figure}[h]  
	\centering
  \includegraphics[width=\linewidth]{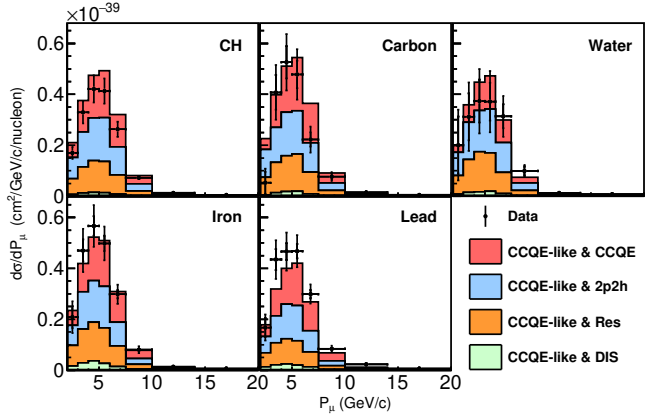}
	\caption{The differential cross section as a function of P$_\mu$ for CH, C, H$_{2}$O, Fe, and Pb targets, along with the predictions for the different quasielastic-like signal processes.}
	\label{figure:xsec_muon_p}
\end{figure}

The differential cross section as a function of P$_\mu$ for interactions on the CH, C, H$_{2}$O, Fe, and Pb targets, respectively, are shown in Fig.~\ref{figure:xsec_muon_p}. P$_\mu$ is the muon momentum in GeV/c.  The uncertainties broken down by source for both the absolute cross sections and the cross section ratios as a function of P$_\mu$ can be found in Fig.~\ref{fig:xsec_err_muon_p}.
Qualitatively, the MINERvA tune does a fairly good job describing the data.  The main area of disagreement is in the interactions on Pb where the data has a bit higher cross section in the peak region of P$_\mu$.

\begin{figure}
    \centering
        \includegraphics[width=\linewidth]{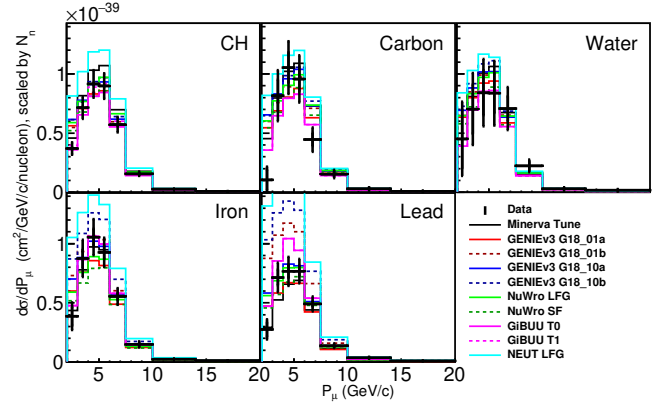}
    \caption{Cross section measurements and predictions as a function of P$_\mu$ for different targets and for a selection of different generator and model choices.}
    \label{fig:models_muon_p}
\end{figure}

The differential cross section as a function of P$_\mu$ for the different targets as compared to a range of models is given in Fig.~\ref{fig:models_muon_p}. The qualitative features of the data are exhibited by the  models.  NEUT gives a higher cross section than is seen in the data, particularly for the larger targets.  The hN versions of GENIE also show this behavior in in Pb.
The \chisq\ between the cross sections and the models can be found in Table~\ref{tab:muonmomentum}.

\begin{figure}
    \centering            \includegraphics[width=\linewidth]{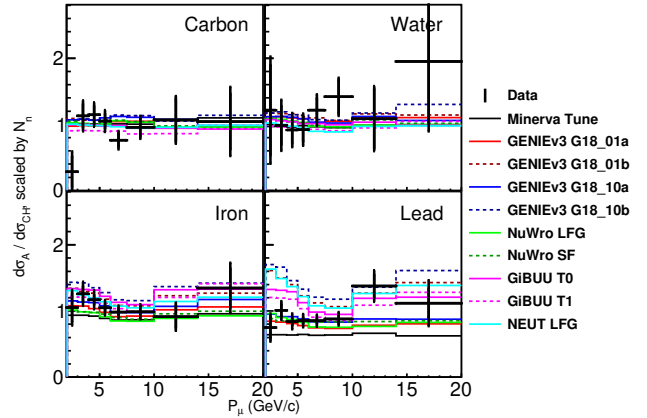}
    \caption{$P_\mu$ cross-section ratio generator comparison for multiple targets.  Changes to the FSI model (GENIE ``a'' to ``b'') in GENIE change the cross section ratio more than changes to the 2p2h model (GiBUU T0 to GiBUU T1) or changes to the initial state (GENIE 1 to GENIE 10) or (NuWro LFG to NuWro SF).  The changes to the ratio are significant across all muon momenta, unlike in other kinematic variables.}
        \label{Figure:Gencompare_pmu_ratio}
\end{figure}

The ratio of the differential cross section as a function of  P$_\mu$ for each nuclear target (C, H$_{2}$O, Fe, Pb) to that for scintillator is shown in 
Fig.~\ref{Figure:Gencompare_pmu_ratio}. Generally, the models cover the data reasonably well for the three smaller targets. For the Pb target, the ratio for the MINERvA tune is significantly lower than the data and the predictions from the other models.  Also on Pb, at lower P$_\mu$ the hN versions of GENIE and NEUT and GiBUU all exhibit higher ratios than seen in the data.
The \chisq\ between the cross section ratios and the models can be found in Table~\ref{tab:muonmomentum}.

\subsection{Cross Sections as a function of Transverse Muon Momentum}

\begin{figure}[h]
	\centering
  \includegraphics[width=\linewidth]{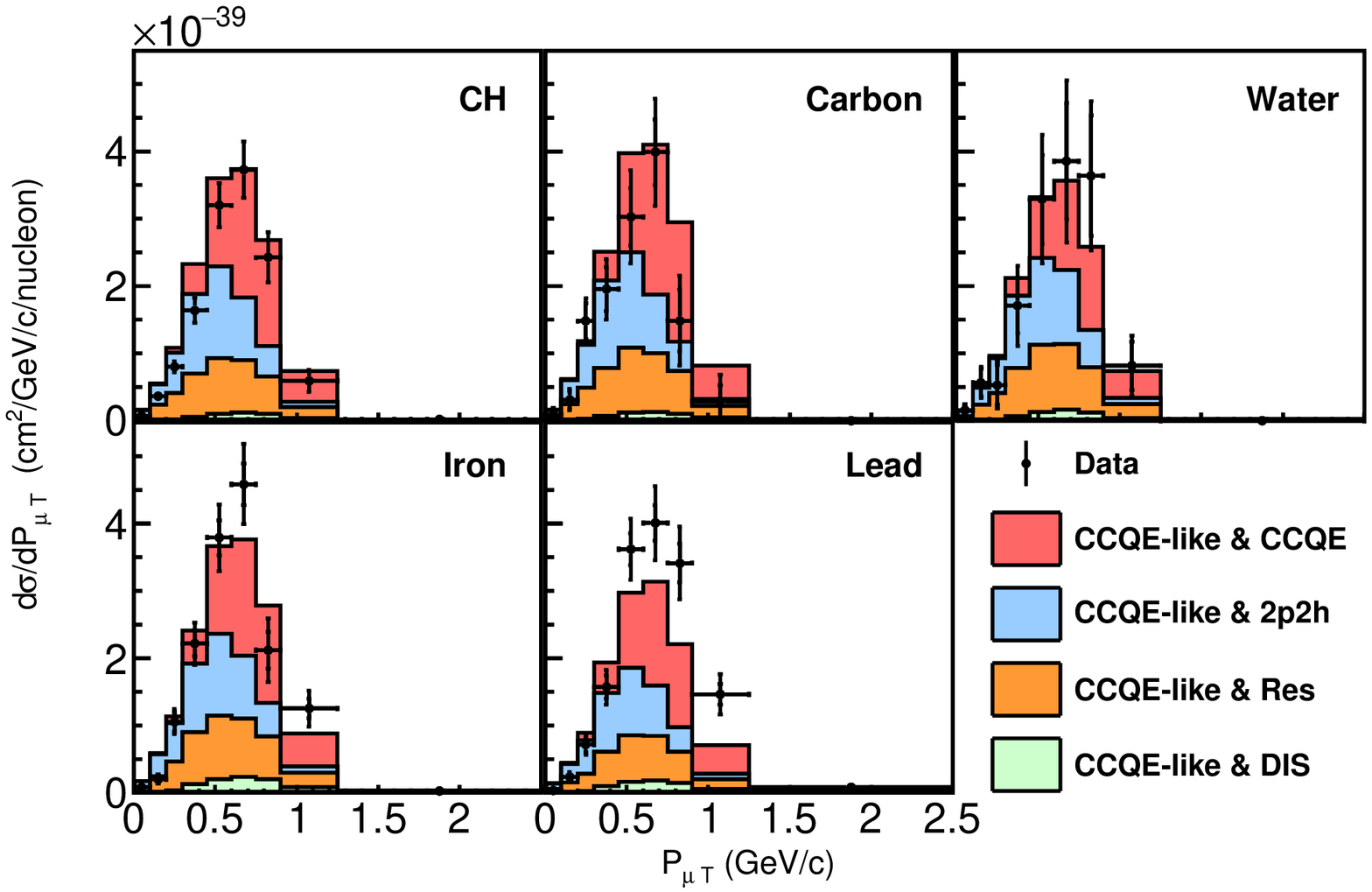}
	\caption{The differential cross section as a function of \tkiptmu\ for CH, C, H$_{2}$O, Fe, and Pb targets, along with the predictions for the different quasielastic-like signal processes.}
	\label{fig:ptmu_xsec}
\end{figure}

Measurements of the differential cross section as a function of \tkiptmu \ for interactions on the nuclear targets are given in Fig.~\ref{fig:ptmu_xsec}.
 \tkiptmu \ is the magnitude of the muon momentum transverse to the beam direction in GeV/c.  The uncertainties broken down by source for both the absolute cross sections and the cross section ratios as a function of \tkiptmu\ can be found in Fig.~\ref{fig:xsec_err_muon_pt}.  

Qualitatively, the MINERvA tune does a fairly good job describing the data for the smaller nuclear targets.  For the larger targets the data seem to have a slightly harder \tkiptmu \ spectrum.  This is pronounced for the data on the Pb target.

\begin{figure}
    \centering            \includegraphics[width=\linewidth]{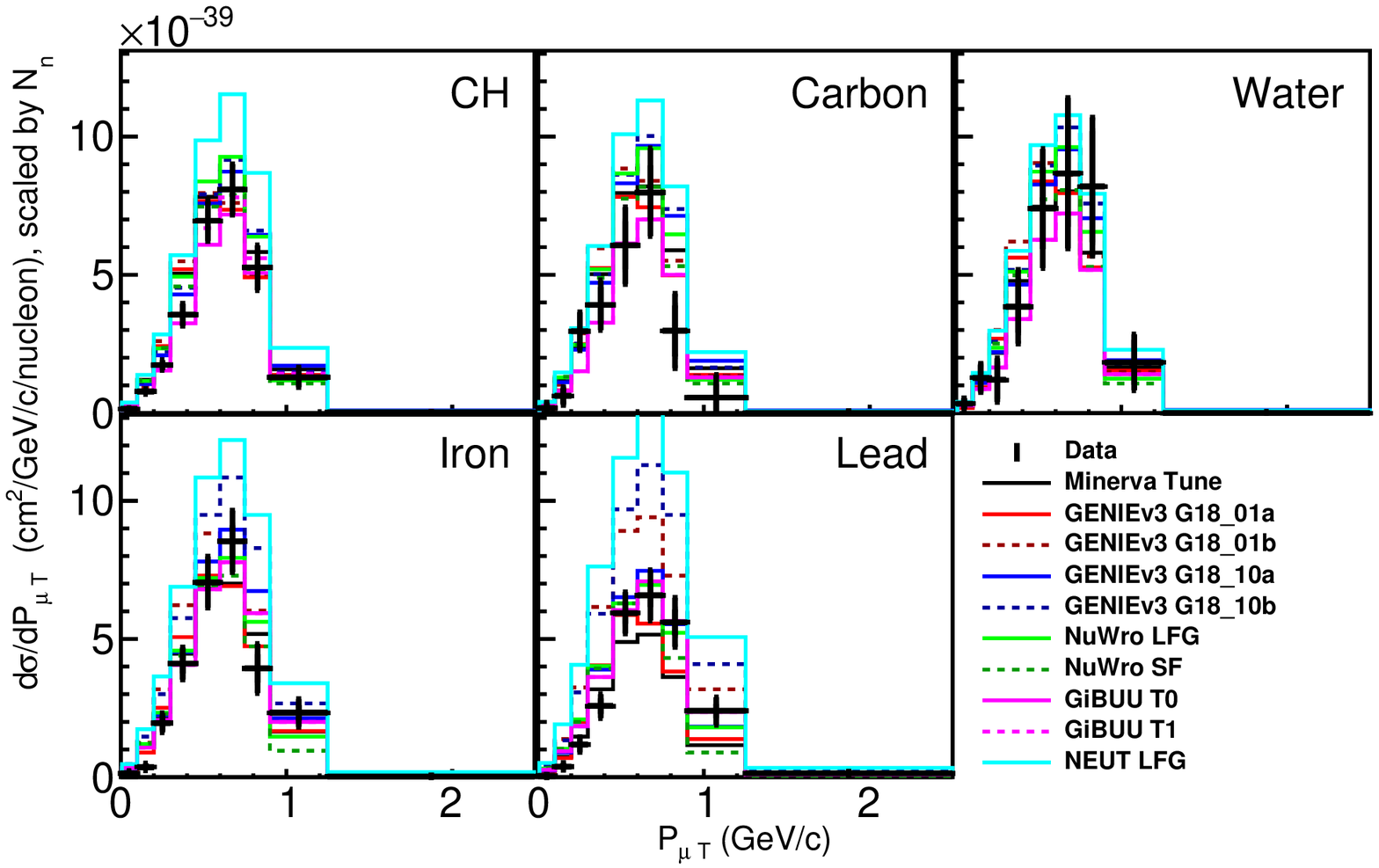}
    \caption{\tkiptmu \ cross-section comparison for multiple targets }{Cross section measurements and predictions as a function of \tkiptmu\ for different targets and for a selection of different generator and model choices.}
    \label{Figure:Gencompare_ptmu}
\end{figure}

Figure~\ref{Figure:Gencompare_ptmu} shows comparisons of a range of models to 
the measured differential cross section as a function of \tkiptmu \ for the different targets. Qualitatively, 
For the smaller targets, the models cover the data fairly well, albeit with a spread.  For the larger nuclear targets, particularly for Pb, NEUT and the hN versions of GENIE exhibit higher cross sections than seen in the data.  Also on Pb, the data exhibit a slightly harder spectrum in \tkiptmu \ than seen in the models.
The \chisq\ between the cross sections and the models can be found in Table~\ref{tab:muonpt}.

\begin{figure}
    \centering
            \includegraphics[width=\linewidth]{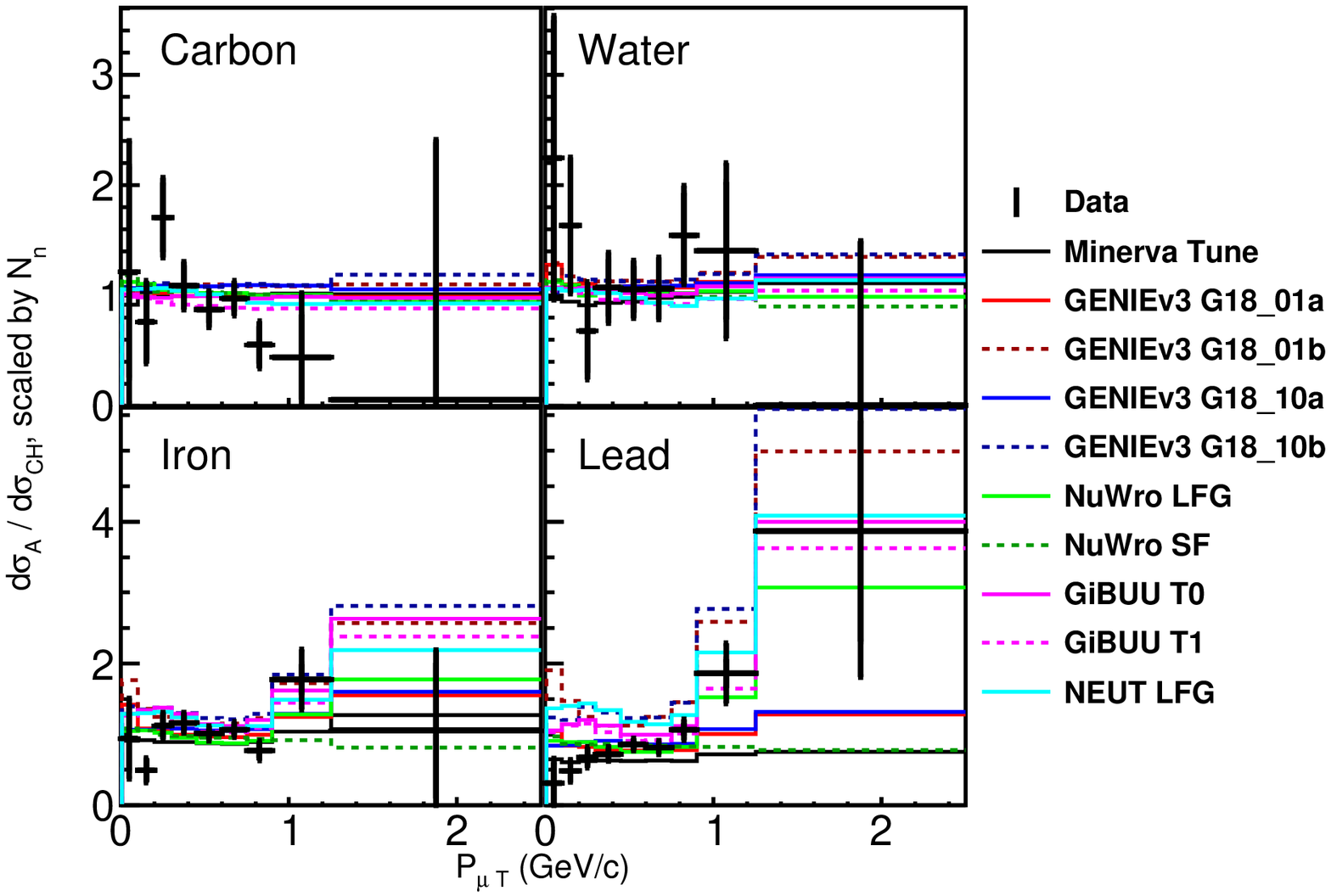}
    \caption{\tkiptmu \ cross-section ratio generator comparison for multiple targets.   Changes to the FSI model (GENIE ``a'' to ``b'') in GENIE change the cross section ratio more than changes to the 2p2h model (GiBUU T0 to GiBUU T1) or changes to the initial state (GENIE 1 to GENIE 10) or (NuWro LFG to NuWro SF).  The changes to the ratio are most significant at high transverse muon momentum.}
        \label{Figure:Gencompare_ptmu_ratio}
\end{figure}

Figure~\ref{Figure:Gencompare_ptmu_ratio} shows the ratio of the 
of the differential cross section as a function of  \tkiptmu \ for each nuclear target to that for scintillator.  The harder \tkiptmu \ spectrum for Pb relative to scintillator is clearly shown as the ratio for the data grows as a function of \tkiptmu and that feature is not so much present for the models.
The \chisq\ between the cross section ratios and the models can be found in Table~\ref{tab:muonpt}.

\subsection{Cross Sections as a function of the neutrino-muon opening angle}

\begin{figure}[h] 
	\centering
  \includegraphics[width=\linewidth]{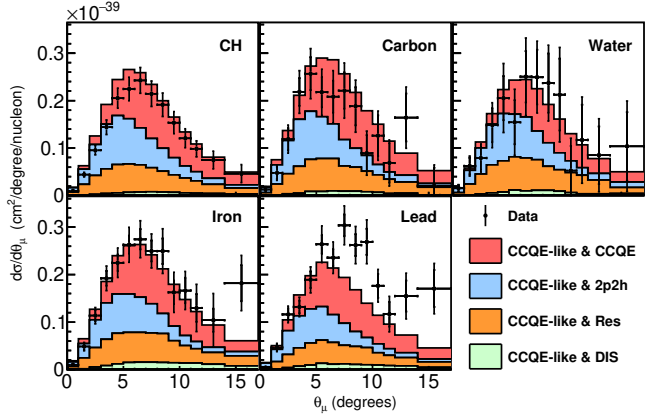}
	\caption{The differential cross section as a function of $\theta_\mu$ for CH, C, H$_{2}$O, Fe, and Pb targets, along with the predictions for the different quasielastic-like signal processes.}
	\label{figure:xsec_muon_theta}
\end{figure}

The differential cross section as a function of $\theta_{\mu}$ for interactions on the CH, C, H$_{2}$O, Fe, and Pb targets, respectively, are shown in Fig.~\ref{figure:xsec_muon_theta}. $\theta_{\mu}$ is the angle of the muon with respect to the neutrino beam direction.  The uncertainties broken down by source for both the absolute cross sections and the cross section ratios as a function of $\theta_\mu$ can be found in Fig.~\ref{fig:xsec_err_muon_theta}.  

In scintillator, the MINERvA tune has more cross section at lower $\theta_{\mu}$ than is seen in the data.  This trend is not present for the data taken on C, as might be expected for consistency with scintillator, although the uncertainties are larger for this data.
There is significantly more cross section in the tail at higher $\theta_{\mu}$ in the data than in the MINERvA tune for data taken on the Pb target.

\begin{figure}[h]
    \centering
    \includegraphics[width=.8\linewidth]{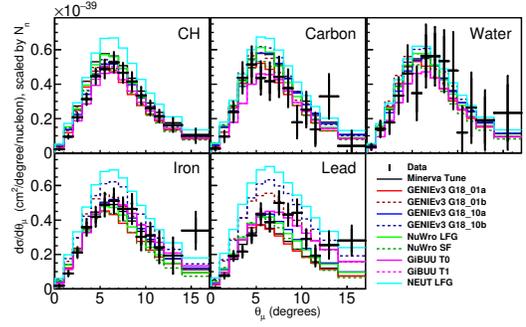}
    \caption{Cross section measurements and predictions as a function of $\theta_\mu$ for different targets and for a selection of different generator and model choices.}
    \label{fig:models_muon_theta}
\end{figure}

The differential cross section as a function of $\theta_{\mu}$ for the different targets as compared to a range of models is given in Fig.~\ref{fig:models_muon_theta}. 
Most of the models investigated have an expectation shifted to slightly smaller angle than that seen in the data.  NEUT exhibits a higher cross section than the data in general.  This is also seen for the hN versions of GENIE for the Pb target.  The \chisq\ between the cross sections and the models can be found in Table~\ref{tab:muontheta}.

\begin{figure}
    \centering
                \includegraphics[width=\linewidth]{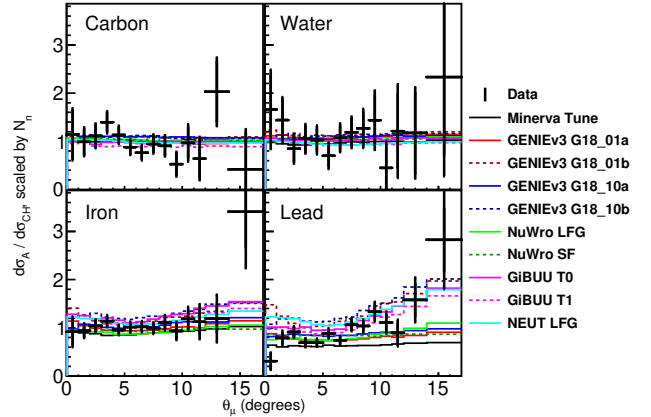}
    \caption{$\theta_\mu$ cross-section ratio generator comparison for multiple targets.  Changes to the FSI model (GENIE ``a'' to ``b'') in GENIE change the cross section ratio more than changes to the 2p2h model (GiBUU T0 to GiBUU T1) or changes to the initial state (GENIE 1 to GENIE 10) or (NuWro LFG to NuWro SF).  The changes to the ratio are most significant at high muon angle.}
        \label{Figure:Gencompare_mutheta_ratio}
\end{figure}

The ratio of the differential cross section as a function of  $\theta_{\mu}$ for each nuclear target (C, H$_{2}$O, Fe, Pb) to that for scintillator is shown in 
Fig.~\ref{Figure:Gencompare_mutheta_ratio}.  
The angle seems well modeled and the scaling with the number of neutrons works well up through the Fe data.  For the data taken on Pb, the data ratio tends to peak at a larger angle than what is seen in the models.
The \chisq\ between the cross section ratios and the models can be found in Table~\ref{tab:muontheta}.

\section{Conclusions}

In this paper, MINERvA presents new measurements of the $\nu_{\mu}$ CCQE-like cross section on C, CH, H$_2$0, Fe, and Pb targets at $E_\nu\sim 6~GeV$.  These data were taken simultaneously in the same detector and same flux and are highly suitable for relative comparison and examining variations in behavior with $A$. 
 The cross sections are presented as a function of assorted transverse kinematic imbalance variables.  These variables are sensitive to both leptonic and hadronic kinematics and are useful for seeing direct evidence of nuclear effects on the final state.  In addition to the numerous differential cross-section results, the ratios  of the different targets to CH per neutron  are shown.  In the ratio, measurements of the total uncertainties are reduced because of the correlation between the dominant systematic uncertainties.  Comparisons are made to  predictions from the GENIE, NuWro, NEUT, and GiBUU neutrino interaction generators.  Tables~\ref{tab:dalphat} to~\ref{tab:protontheta}  provide the $\chi^{2}$ of each comparison.  The number of degrees of freedom is given in the caption for each table.  

The effect of varying the final state interaction model and initial nucleon momentum distributions produce larger effects as Z increase, and the effects from FSI tend to be larger than the effects of changing the initial nucleon momentum distributions at these energies.  While these modifications will change the overall cross section somewhat uniformly as a function of muon momentum, they do make changes in very specific regions in each TKI variable.  The MINERvA data seem to favor more FSI than is currently in its base model, although some modification of the initial state nucleon momenta would also be suggested by these data. 

The behavior of the data on targets up to the size of Fe are reasonably reproduced by the broad range of models.  For interactions on Pb, the model spread is large compared to the data, particularly in regions of the variable where nuclear effects are expected to play an important role.  NEUT tends to over-predict the cross section, particularly for the larger targets.  The hN versions of GENIE approach the NEUT behavior in this respect, particularly for data taken on the Pb target.  In regions where the nuclear effects are important for the larger targets, the hA versions of GENIE and NuWro tend to under-predict the data.  With a few exceptions, GiBUU seems to most consistently model the data reasonably well across targets and variables.

Qualitatively, for most variables the  predicted cross-section ratios 
agree well with the measurements for smaller $A$ targets.  For data taken on Fe and to a greater degree, that on Pb, the ratio of the cross sections are larger than anticipated by the hA versions of GENIE and NuWro in regions where nuclear effects are important and often over-predicted somewhat by NEUT and the hN versions of GENIE.  

These data show clearly that nuclear effects are important and their influence increases with  atomic mass {\it A}.  Simple neutron-number-weighted scaling of the cross section gives a reasonable expectation of the results seen for smaller nuclei.  For data taken on the larger nuclei, FSI effects tend to lead to more complex behavior than simple neutron scaling.  Often that behavior is covered in the spread of models examined here. 

These results have implications for neutrino oscillation experiments because neutrino interaction modeling feeds into experimental oscillation results and systematic uncertainties in various ways.  The modeling is used to project near detector constraints on the data to the far detector analysis and to assess systematic sensitivities.  That may or may not involve different nuclear targets and may be more or less sensitive to final state effects.  The results presented here illustrate the successes and failures of commonly used models in reproducing what is seen for the simplest final state that can correlate the leptonic and hadronic sides of the interaction.  Of some relevance for T2K, the cross-section ratios of data taken on H$_{2}$O to that on scintillator are fairly well described by the models.  For argon-based experiments, such as DUNE, the good news is that the behavior seen on the larger {\it A} targets is described by the broad range seen in the models.  However, the nuclear effects are clearly important and the details in the modeling matter.  These measurements can be used to help constrain model predictions with different nuclear targets.

\begin{acknowledgements}
This document was prepared by members of the MINERvA collaboration using the resources of the Fermi National Accelerator Laboratory (Fermilab), a U.S. Department of Energy, Office of Science, Office of High Energy Physics HEP User Facility. Fermilab is managed by Fermi Forward Discovery Group, LLC, acting under Contract No. 89243024CSC000002.  These resources included support for the MINERvA construction project, and support
for construction also
was granted by the United States National Science Foundation under
Award No. PHY-0619727 and by the University of Rochester. Support for
participating scientists was provided by NSF and DOE (USA); by CAPES
and CNPq (Brazil); by CoNaCyT (Mexico); by ANID PIA / APOYO AFB180002, CONICYT PIA ACT1413, and Fondecyt 3170845 and 11130133 (Chile); 
by CONCYTEC (Consejo Nacional de Ciencia, Tecnolog\'ia e Innovaci\'on Tecnol\'ogica), DGI-PUCP (Direcci\'on de Gesti\'on de la Investigaci\'on  - Pontificia Universidad Cat\'olica del Peru), and VRI-UNI (Vice-Rectorate for Research of National University of Engineering) (Peru); NCN Opus Grant No. 2016/21/B/ST2/01092 (Poland); by Science and Technology Facilities Council (UK); by EU Horizon 2020 Marie Skłodowska-Curie Action; by a Cottrell Postdoctoral Fellowship from the Research Corporation for Scientific Advancement; by an Imperial College London President's PhD Scholarship.  We thank the MINOS Collaboration for use of its near detector data. Finally, we thank the staff of
Fermilab for support of the beam line, the detector, and computing infrastructure.
\end{acknowledgements}

\pagebreak

\clearpage
\FloatBarrier
\section{Supplemental Material}
\renewcommand\thefigure{Supp.\arabic{figure}}
\setcounter{figure}{0}
\renewcommand\thetable{Supp.\Roman{table}}
\setcounter{table}{0}
\label{sec:supplemental}

\subsection{Details of Cross Section Extraction Procedure} 
\label{Sec:supplementalanalysis}

\begin{figure}  
	\centering
	\includegraphics[width=0.48\linewidth]
{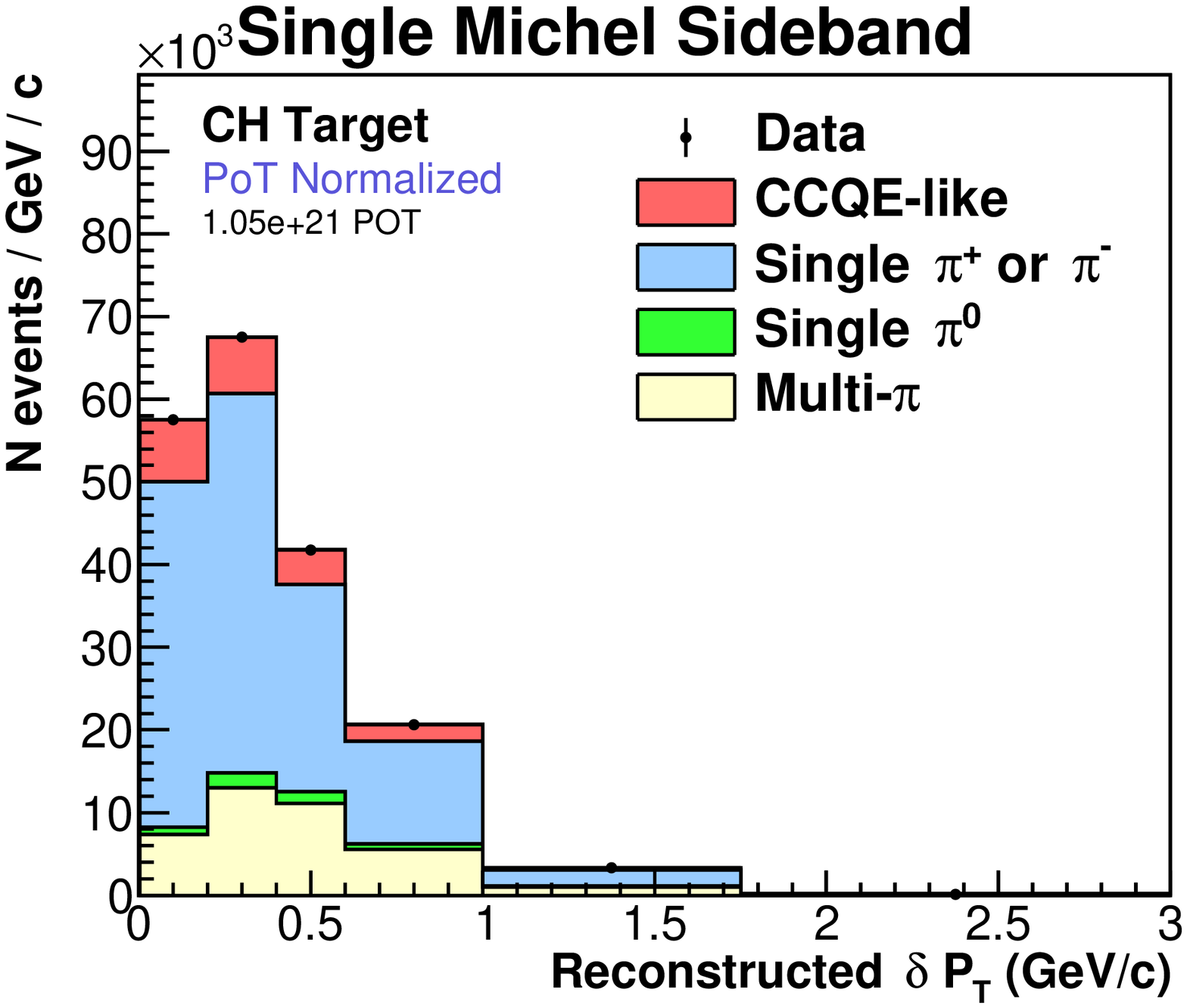}
 	\includegraphics[width=0.48\linewidth]
{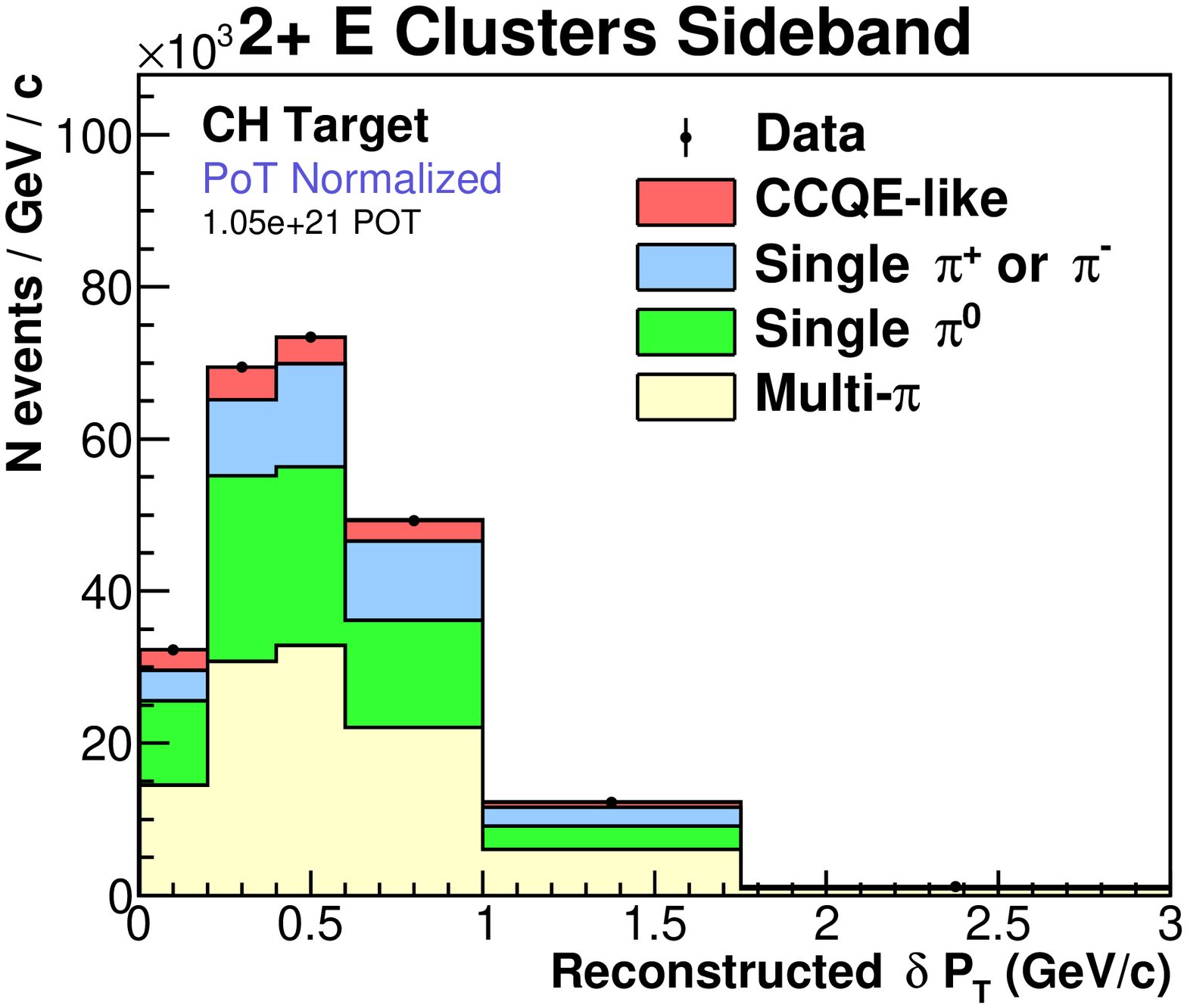}
	\includegraphics[width=0.48\linewidth]
{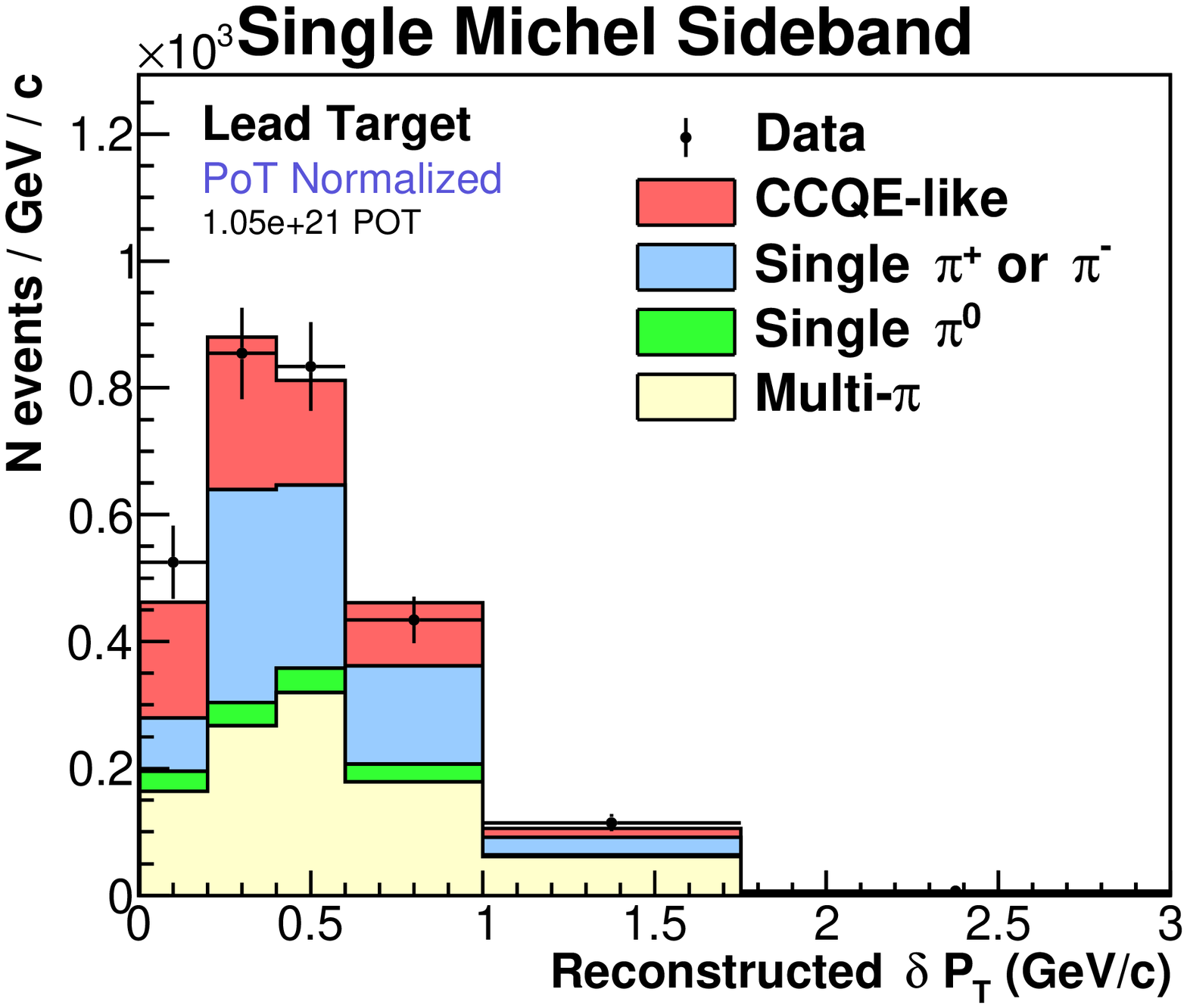}
 	\includegraphics[width=0.48\linewidth]
{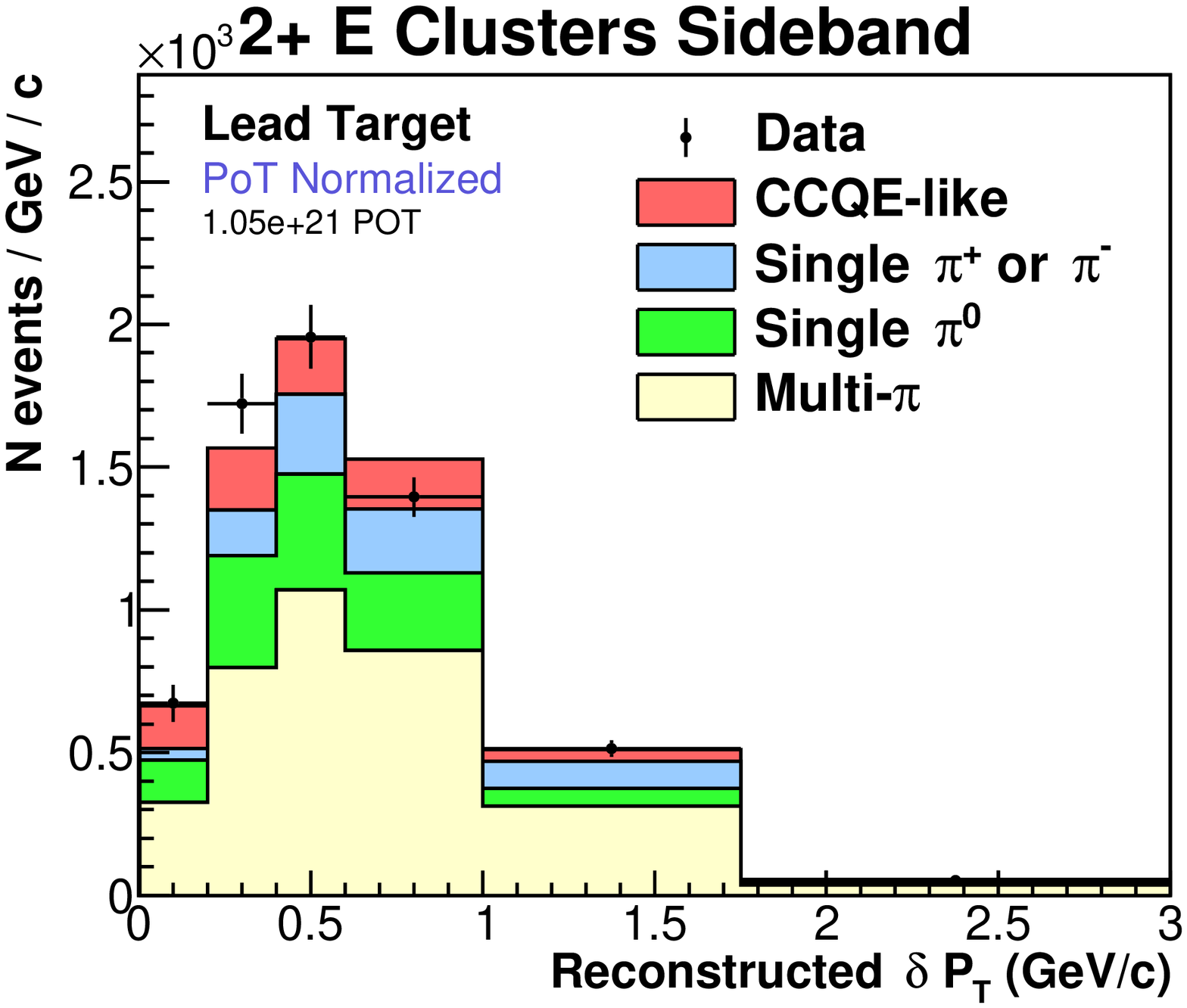}
	\caption{Event rates as a function of \tkidelta\ for the two sideband samples for scintillator (CH) and lead targets after background tuning. }
	\label{fig:tkidelta_sdbn_evntrate_tuned}
\end{figure}

This section provides more detailed description of the cross section extraction procedure for \tkidelta, starting with the sideband tuning procedure, background subtraction, unfolding, and finally efficiency correction. Figure~\ref{fig:tkidelta_sdbn_evntrate_tuned}  shows the sideband sample with a Michel electron after tuning the backgrounds for events interacting on scintillator and events interacting on lead as a function of \tkidelta.  Note that the background fractions vary significantly between scintillator and lead, underscoring the importance of measuring backgrounds for each target nucleus separately wherever possible.  

\begin{figure}[h!]
	\centering
	\includegraphics[width=0.49\linewidth]%
{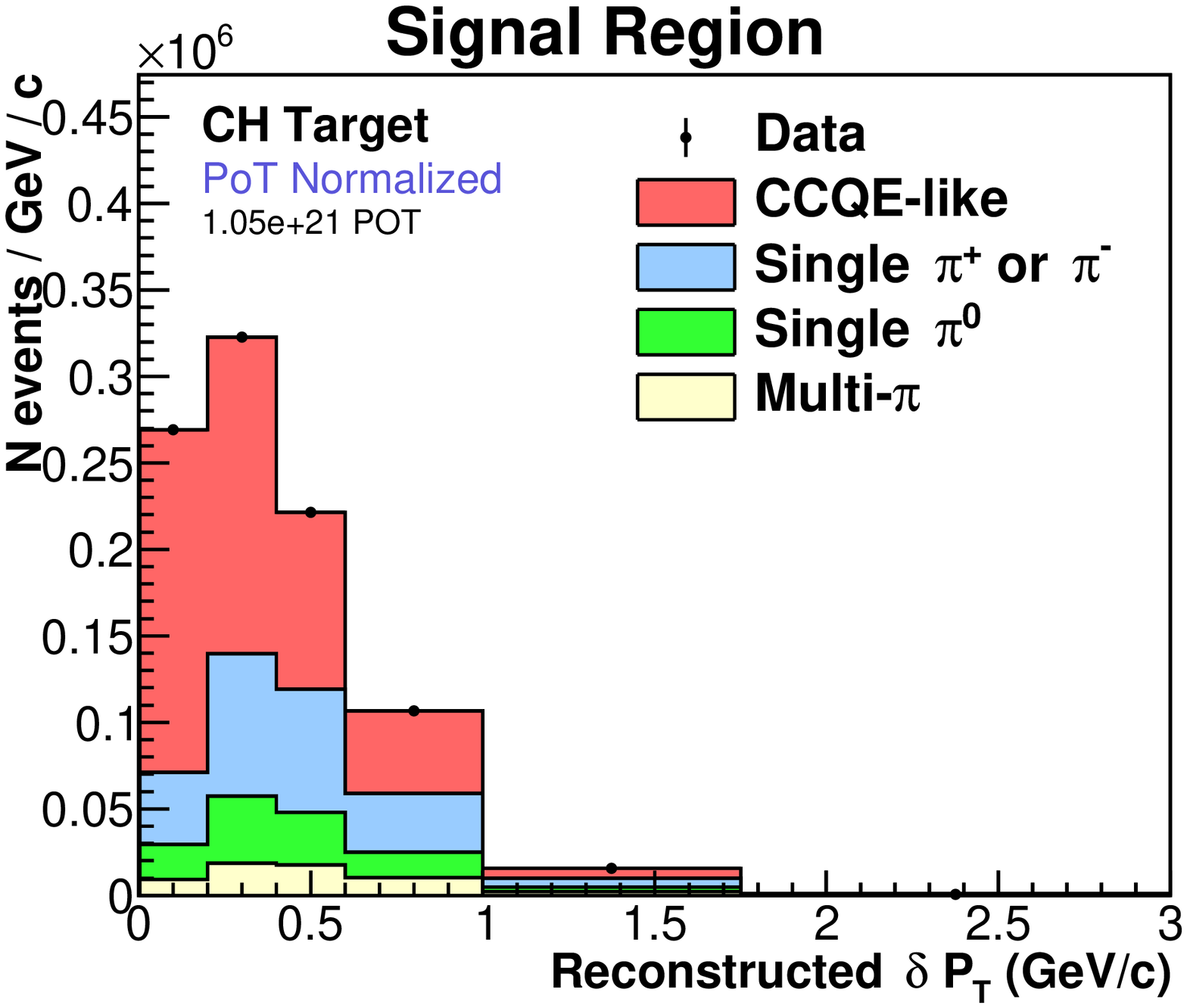}
\includegraphics[width=0.49\linewidth]%
{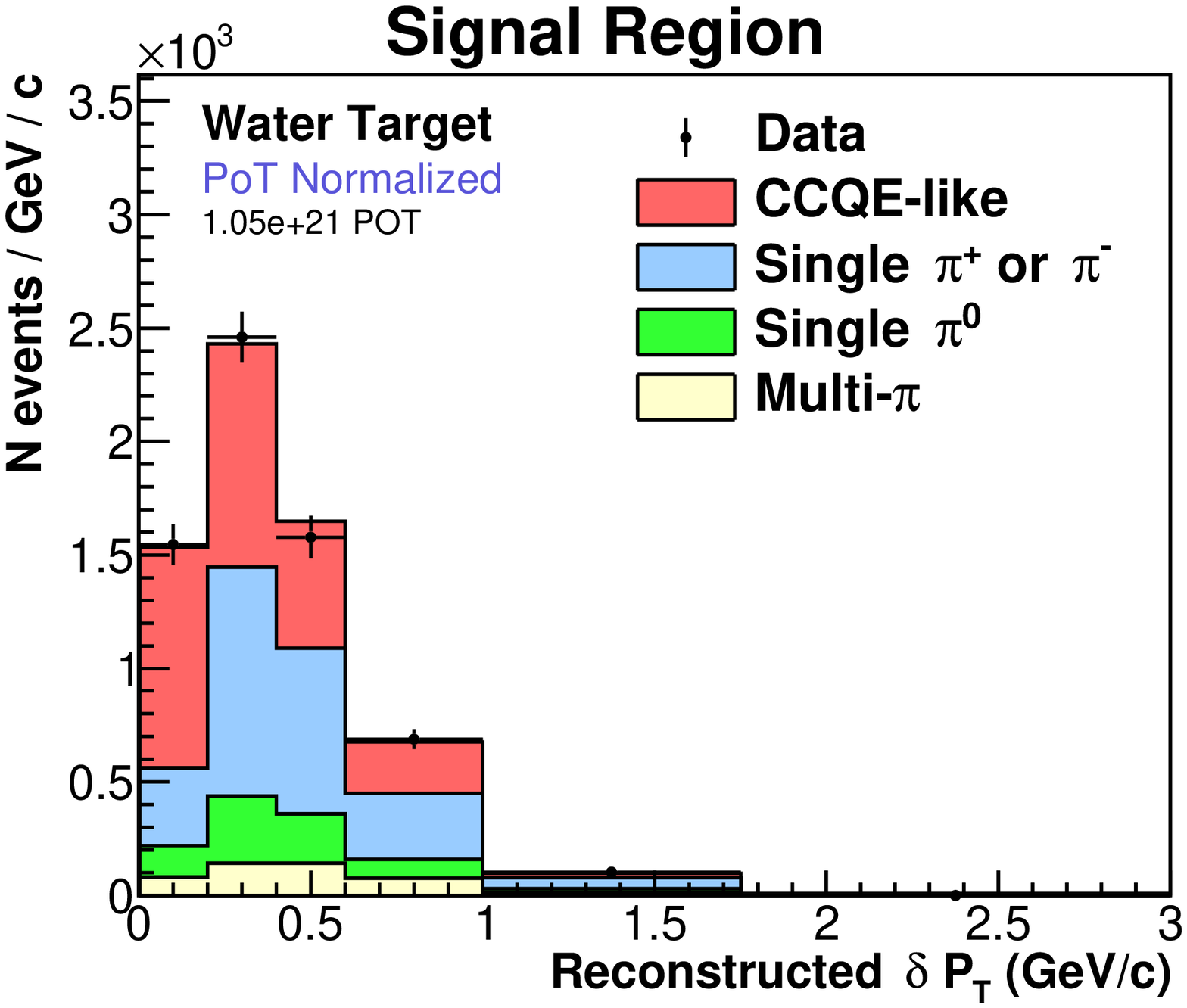}
\includegraphics[width=0.49\linewidth]%
{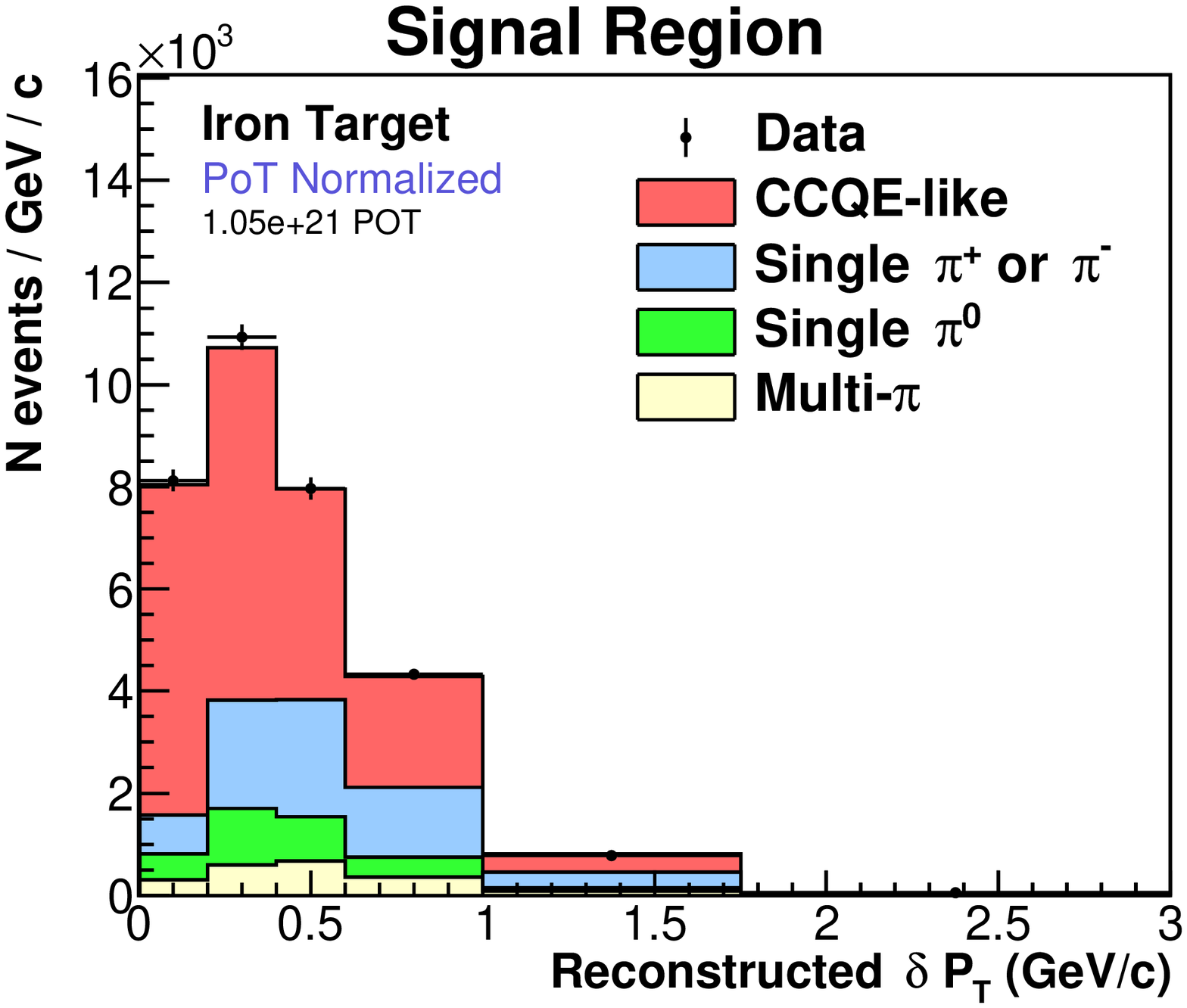}
\includegraphics[width=0.49\linewidth]%
{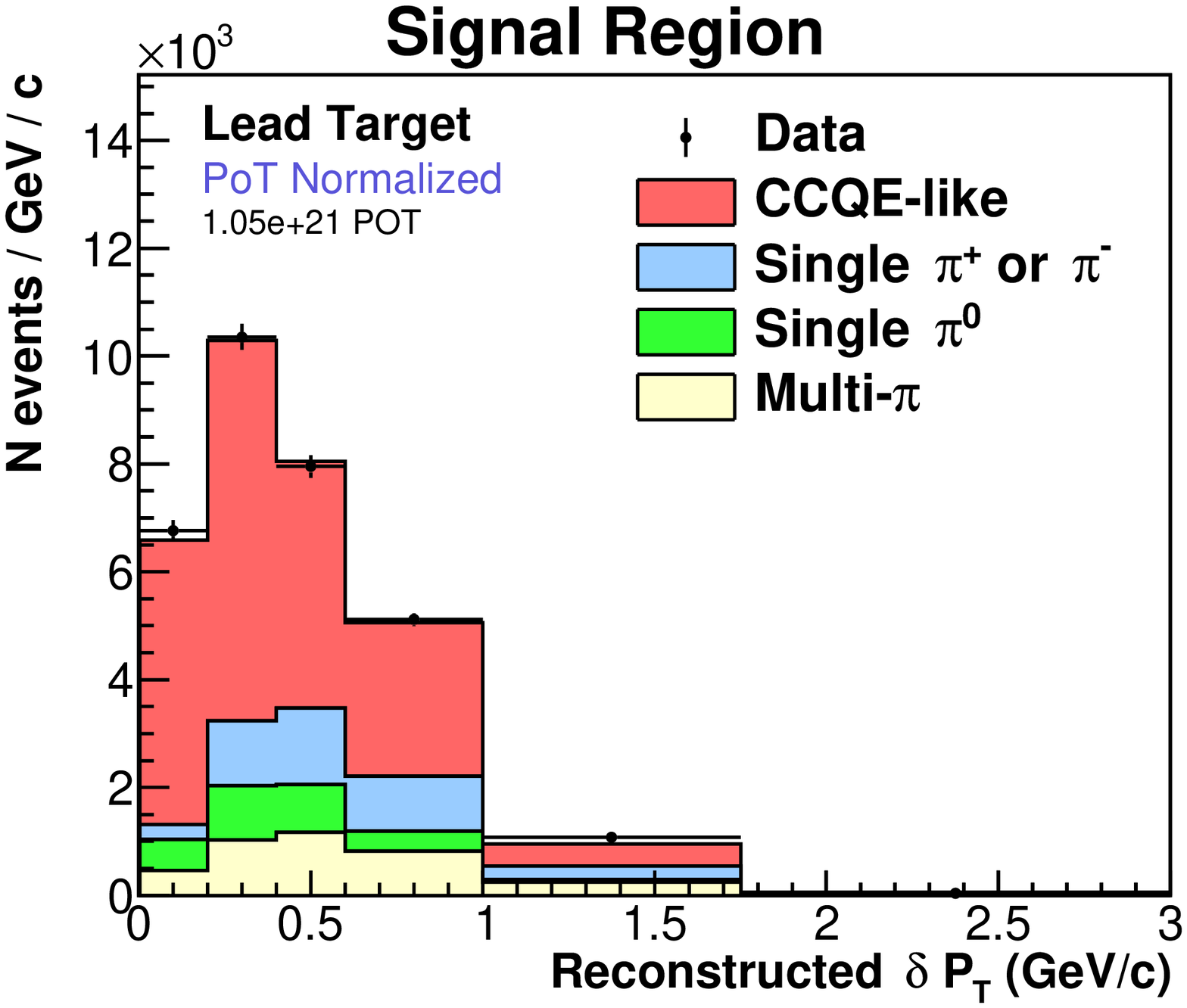}
	\caption{Signal sample event rates as a function of \tkidelta \ for CH, C, H$_{2}$O, Fe, and Pb targets after background tuning but before background subtraction.}
	\label{fig:tkidelta_evntrate_tuned}
\end{figure}

Figure~\ref{fig:tkidelta_evntrate_tuned} shows the event rate in the signal region after tuning for each target except carbon, before background subtraction.  The predicted contributions from each background after tuning are also shown. 


\begin{figure}[h!] 
	\centering
\includegraphics[width=0.49\linewidth]
{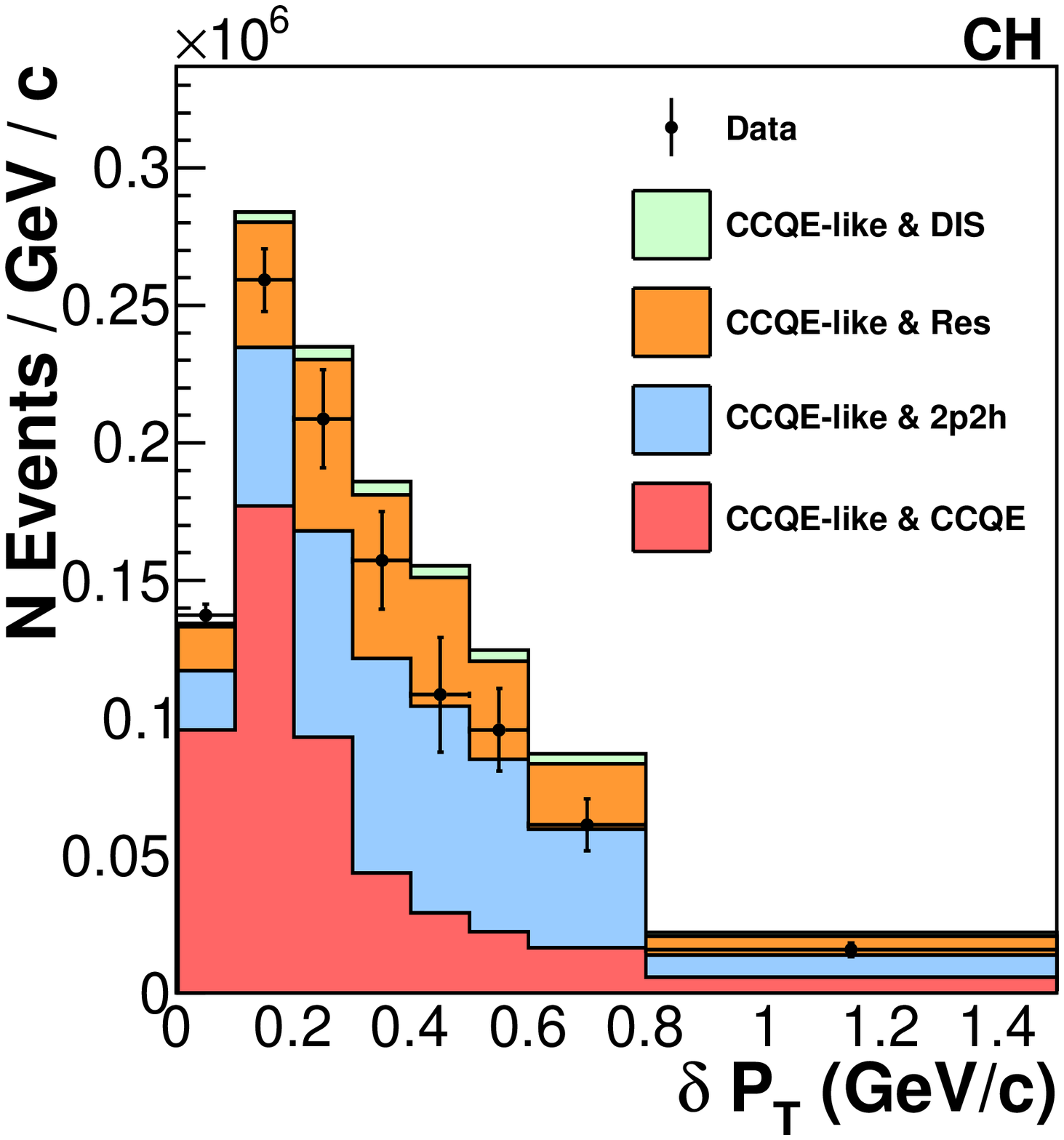}	\includegraphics[width=0.49\linewidth]
{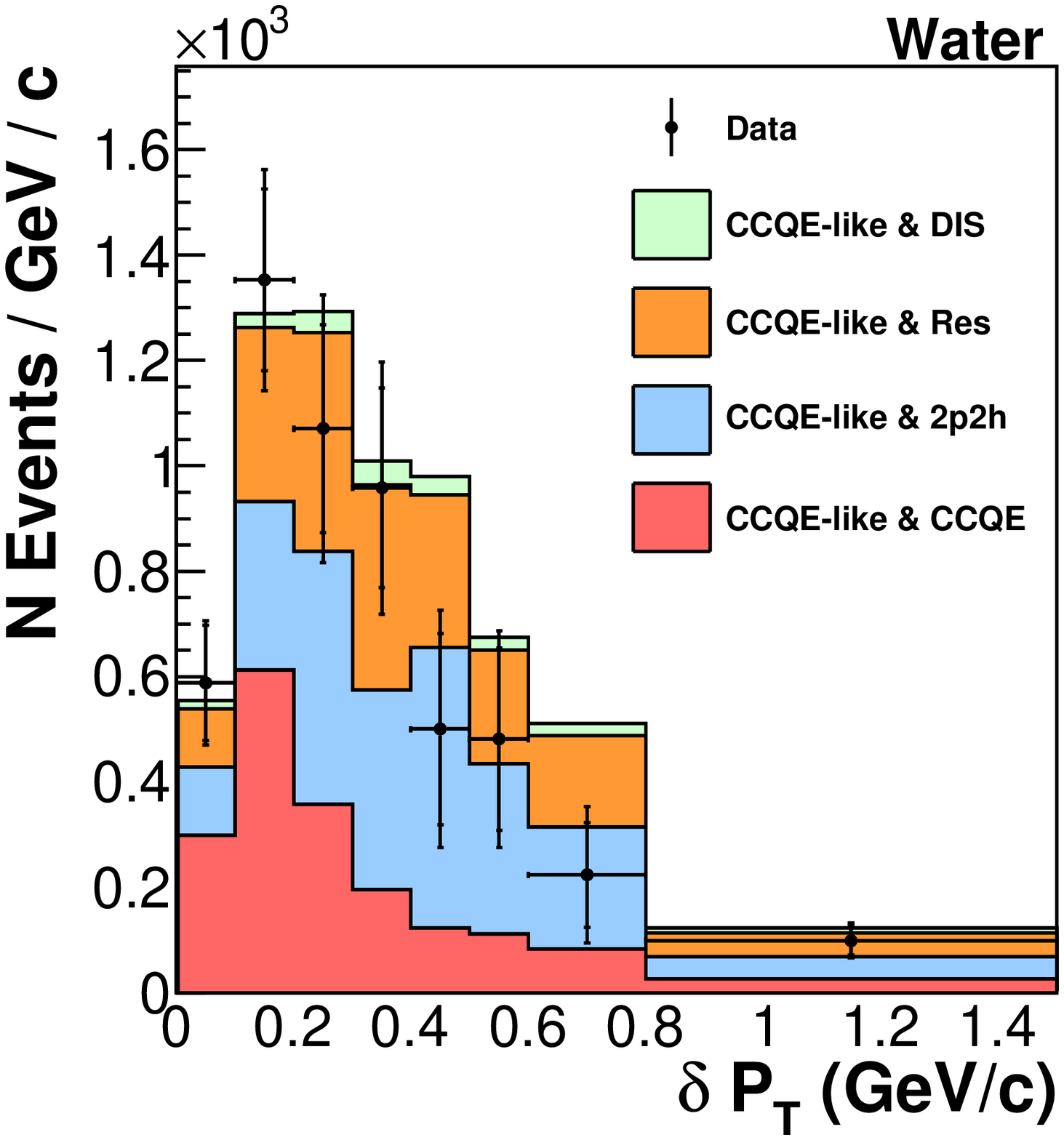}
\includegraphics[width=0.49\linewidth]
{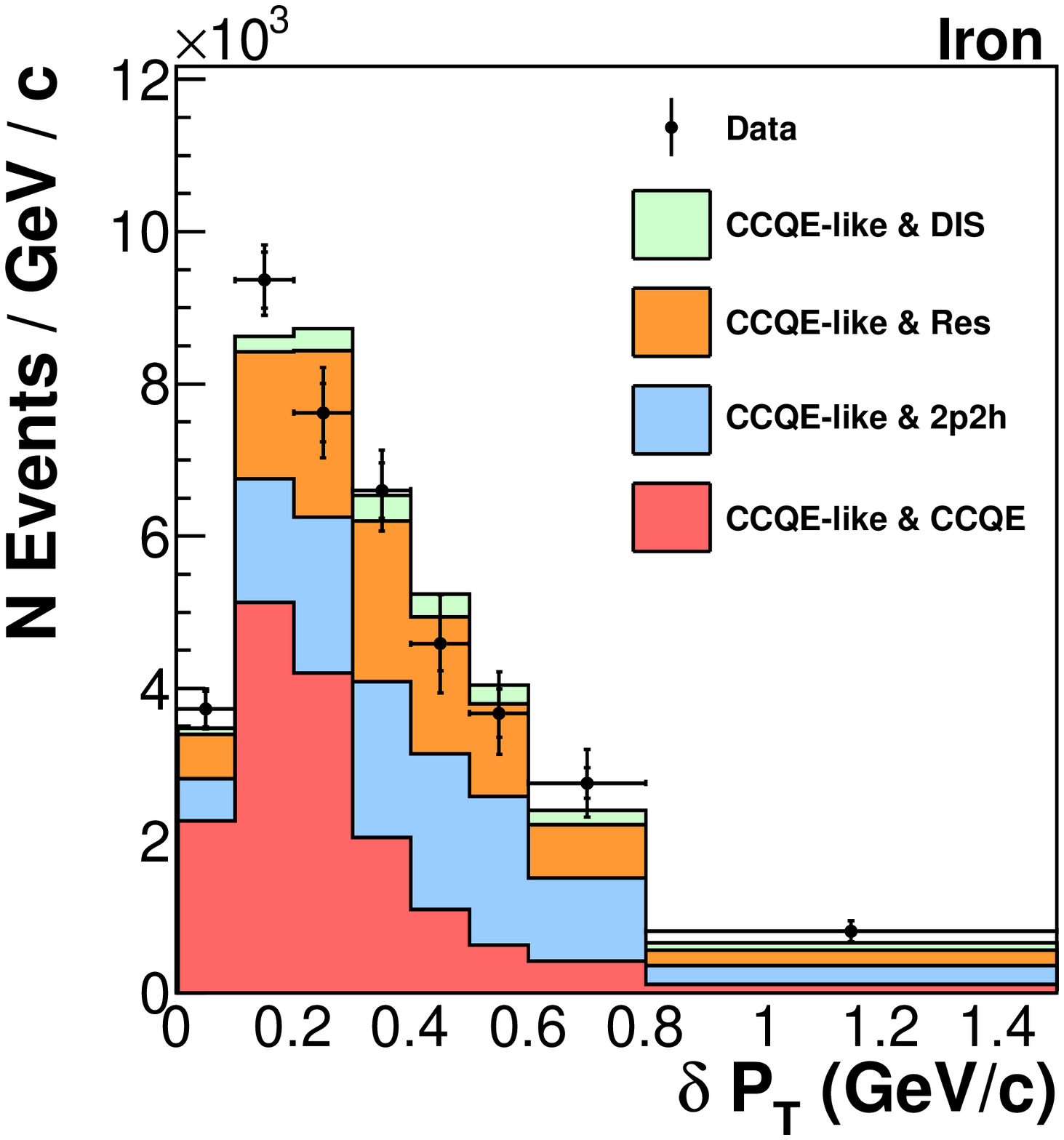}
\includegraphics[width=0.49\linewidth]
{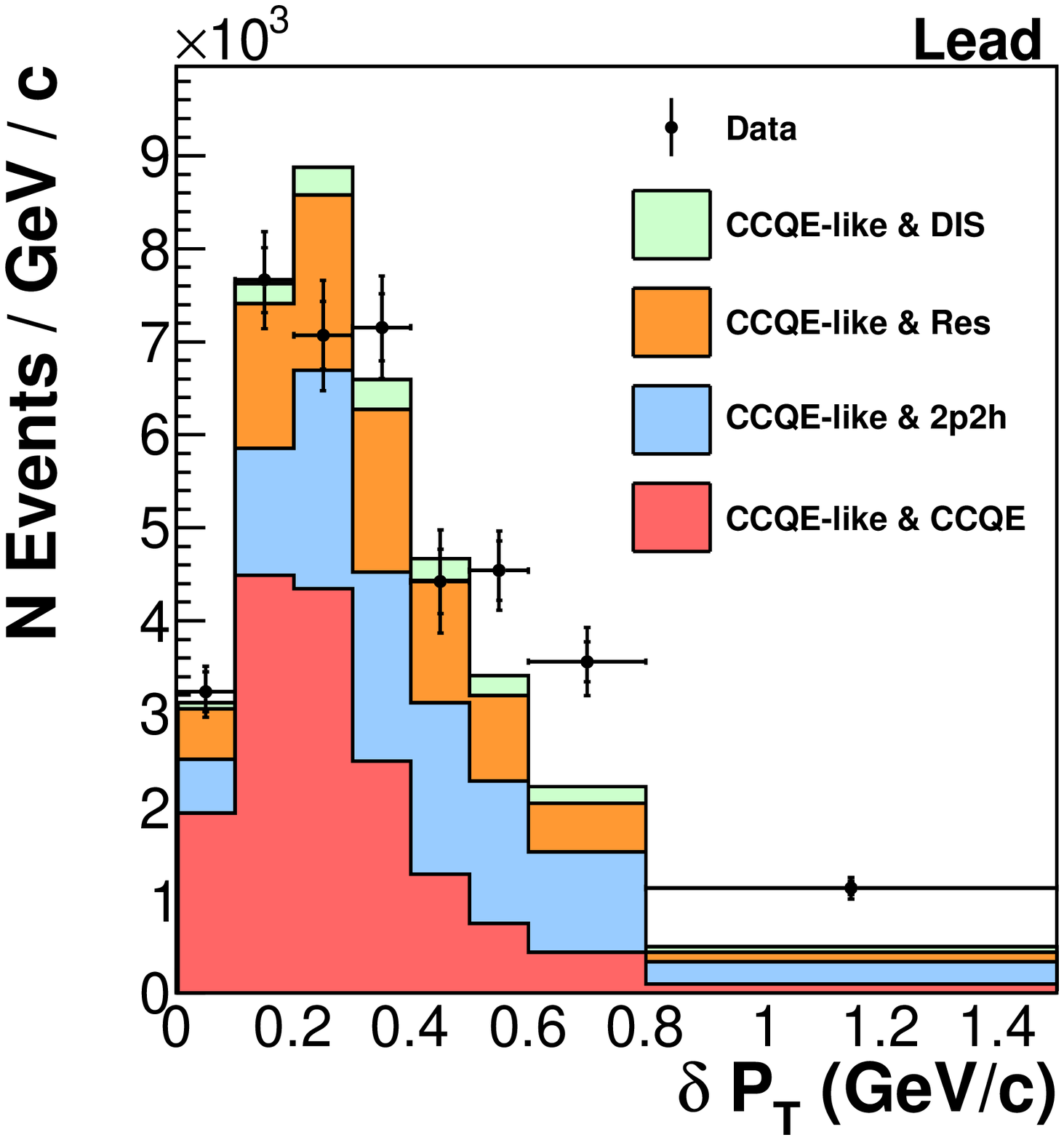}
	\caption{Signal sample event rates as a function of \tkidelta \ for CH, C, H$_{2}$O, Fe, and Pb targets after the tuned background has been subtracted.}
\label{fig:tkidelta_evntrate_backgroundsubtr}
\end{figure}

Figure~\ref{fig:tkidelta_evntrate_backgroundsubtr} shows the signal for each target as a function of \tkiptmu\ after the backgrounds are subtracted.  In this procedure, the background removed is that predicted in the simulation after the background tuning.  The figure shows the various signal contributions to the quasielastic-like interaction.  Note that the quasielastic contribution to quasielastic-like events are predicted to be the most important at low \tkidelta\ but are only a small fraction of the event rate at high \tkidelta .  

\begin{figure}  
	\centering
\includegraphics[width=0.49\linewidth] {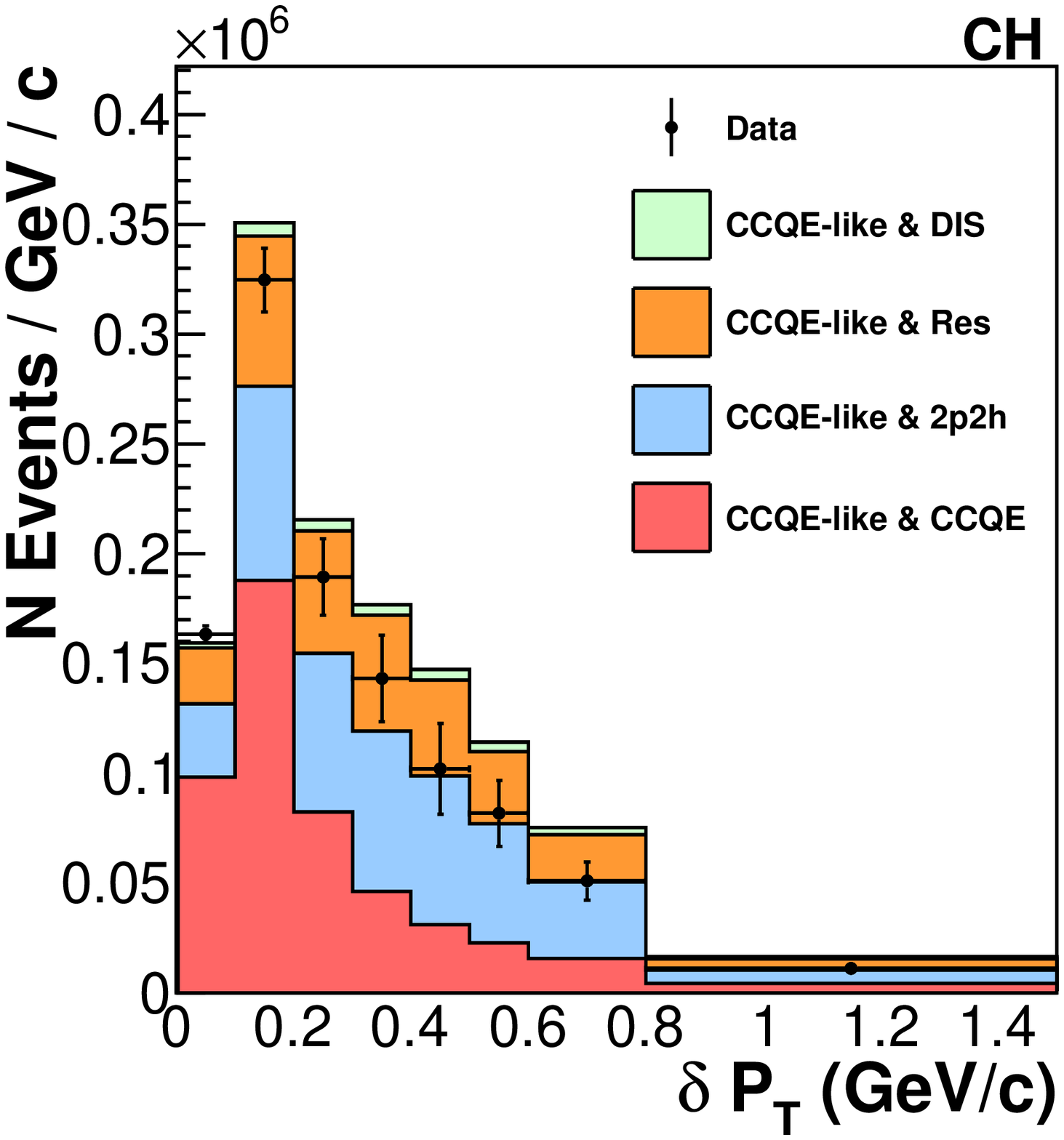}
  \includegraphics[width=0.49\linewidth]
{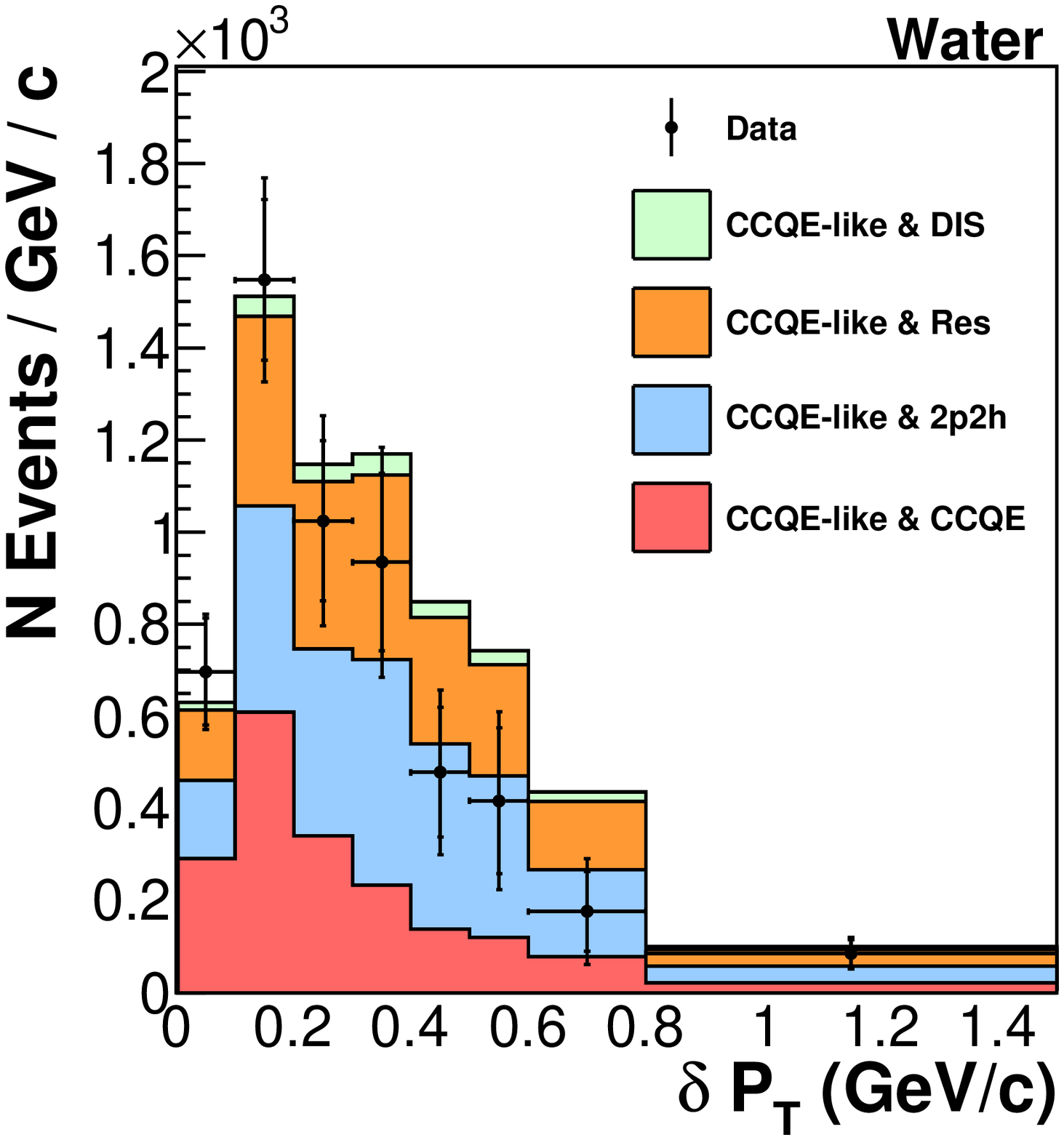}
   	\includegraphics[width=0.49\linewidth]
{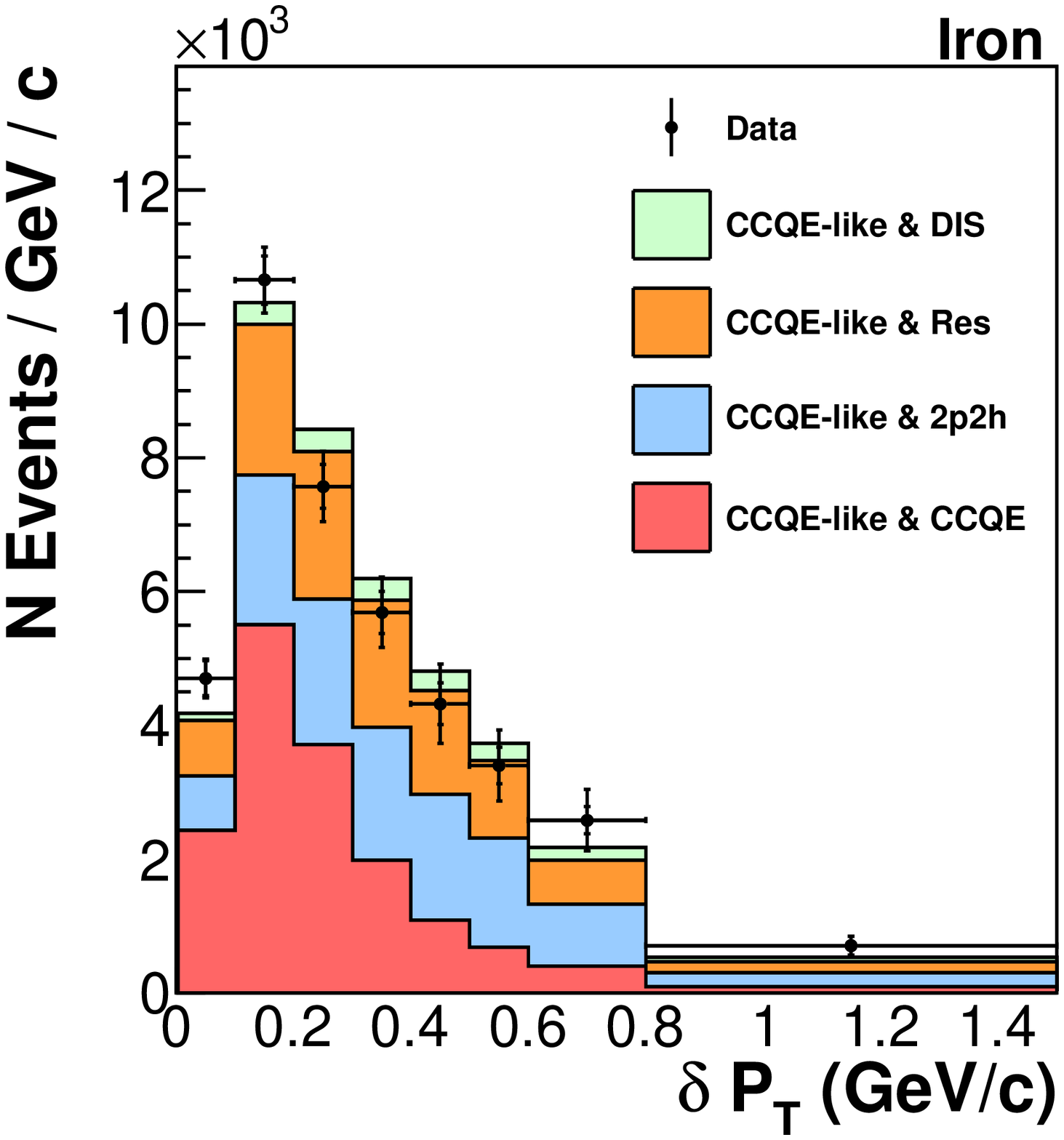}
    \includegraphics[width=0.49\linewidth]
{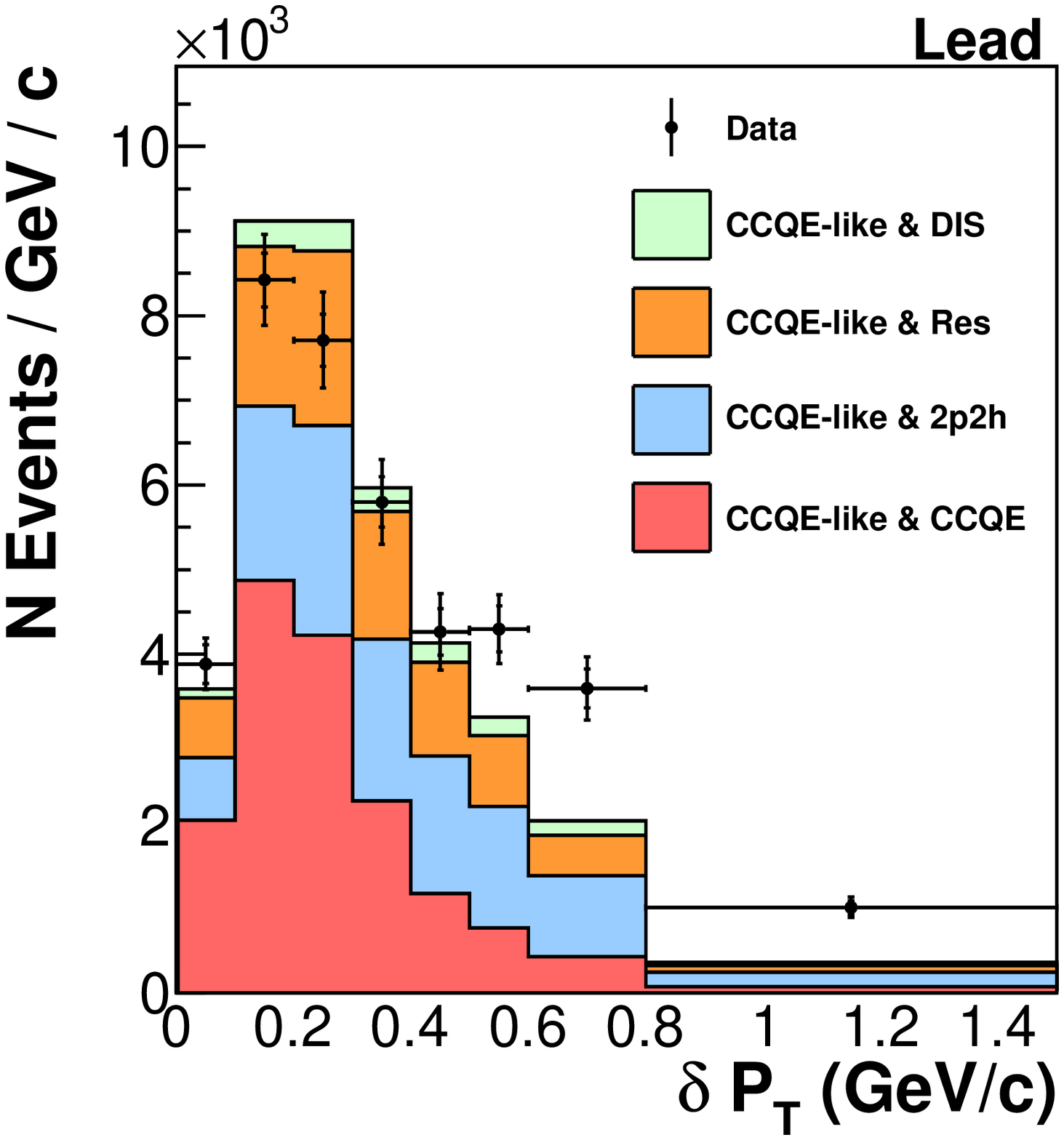}
	\caption{Signal sample event rates as a function of \tkidelta \ for CH, C, H$_{2}$O, Fe, and Pb targets after unfolding. }
	\label{fig:tkidelta_evntrate_unfolded}
\end{figure}

Once the backgrounds have been subtracted, the remaining event distributions can be unfolded as described in section~\ref{sec:analysis}.  The unfolded \tkidelta \ event rate distributions for water, iron, lead and scintillator are shown in Fig.~\ref{fig:tkidelta_evntrate_unfolded}.  

\begin{figure}[h]
	\centering
	\includegraphics[width=.48\linewidth]
 {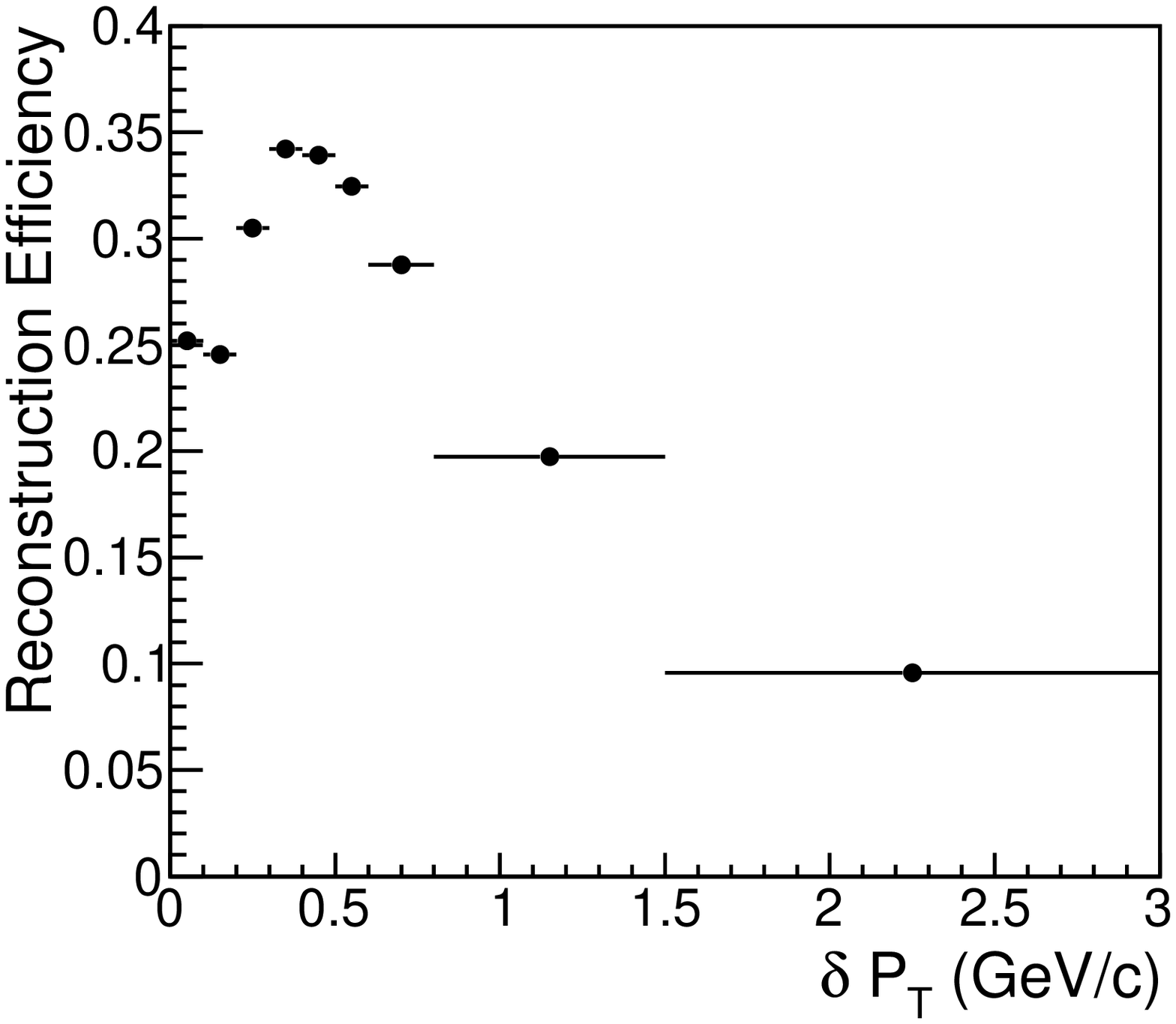}
 	\includegraphics[width=.48\linewidth]
   {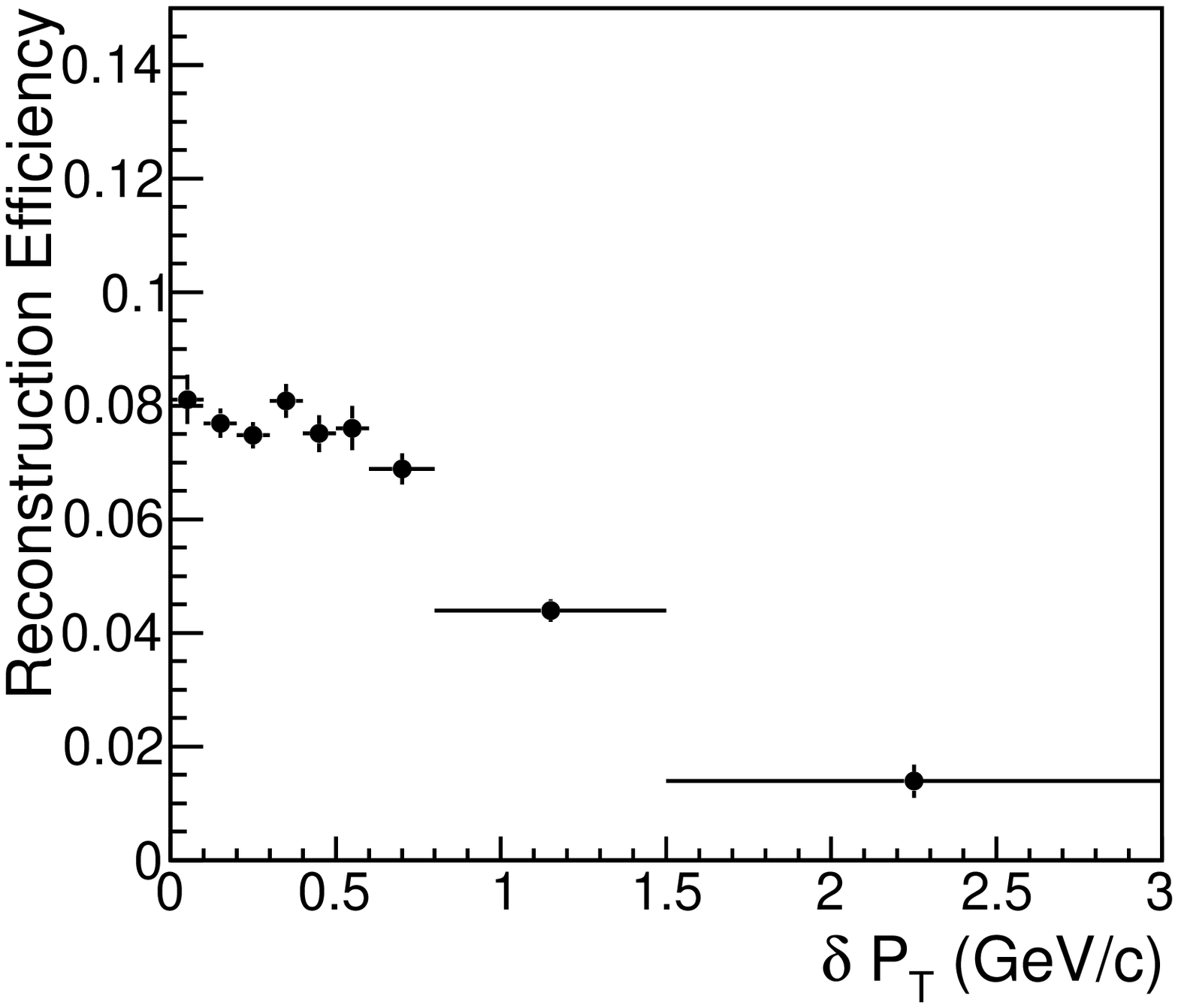}
	\caption{The signal efficiency as  a function of \tkidelta\ for events from interactions on CH on the left and Pb on the right.}
	\label{fig:tkidelta_efficiency}
\end{figure}

After the distributions have been unfolded, then the analysis divides by the efficiency as a function of \tkidelta\ for each target separately in the detector. 
 Figure~\ref{fig:tkidelta_efficiency} shows the average efficiency as function of \tkidelta \ for scintillator on the left and for lead on the right. 

\begin{figure}[tb]
\centering
\includegraphics[width=\linewidth]{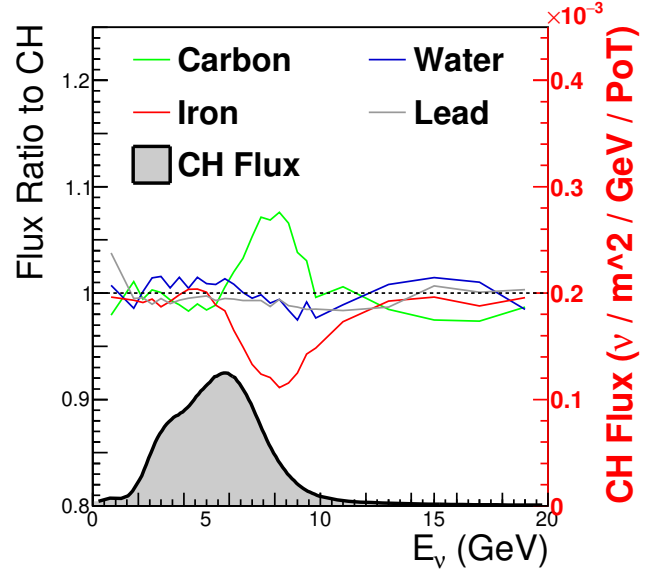}
\caption{The ratio of the flux in each target material to that in the scintillator target.}
\label{fig:fluxes}
\end{figure}

As described in Sec.\ref{sec:analysis}, the flux varies somewhat across the face of the detector.  The variations for each individual target are accounted for in the analysis, and the average variation of the flux for each nuclear target type is shown in Fig.~\ref{fig:fluxes}.  Because the lead targets are distributed in locations that are appoximately symmetric around the center of the detector, the net effect for lead is small.  Similarly, the water target itself is also symmetric around the center of the detector so there is no correction needed for the water to scintillator comparisons.  

\clearpage
\subsection{Supplemental:  Cross-section Ratios to GENIE}
\label{sec:appendixA}
Ratios between the data and the default tuned GENIE simulation are shown in the plots in this section.  The ratios between the various model choices in different generators and the default tuned GENIE simulation are also shown in this section.  The descriptions of the different model choices for each generator are described in Tab.~\ref{tab:generators}.  In many kinematic regions the measurement uncertainties are considerably smaller than the variations between different models, especially for the heavier nuclear targets and for scintillator where the statistics are the highest.  
\begin{figure}
\includegraphics[width=\linewidth]{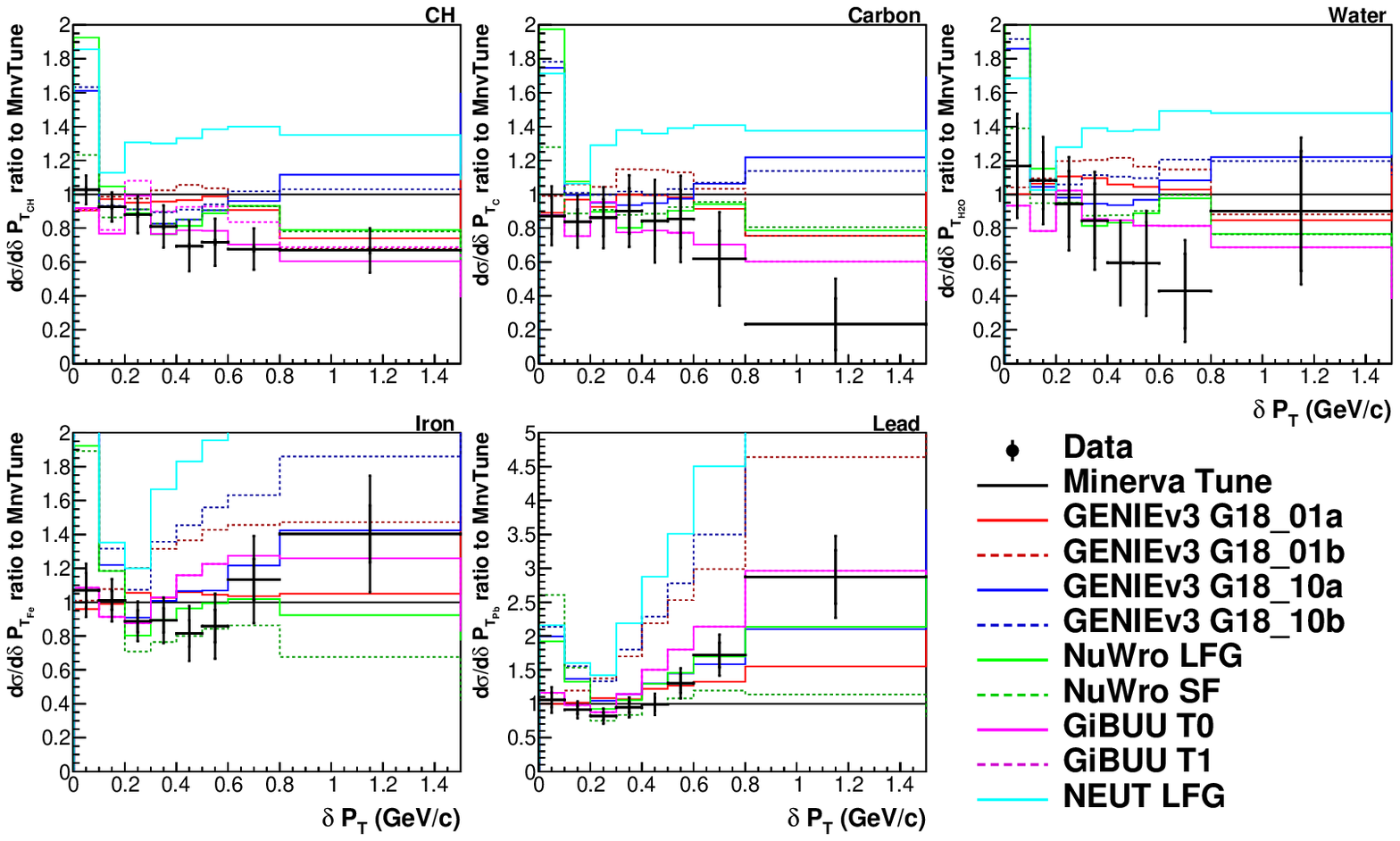}
    \caption[\tkidelta \ cross-section comparison to the MINERvA tune for multiple targets ]{Ratio of absolute cross section measurements to the MINERvA tune for different targets as a function of \tkidelta\ and ratios of different model predictions to the MINERvA tune. 
    }
    \label{fig:models_dpt_rat}
\end{figure}
\begin{figure}
    \centering
\includegraphics[width=\linewidth]{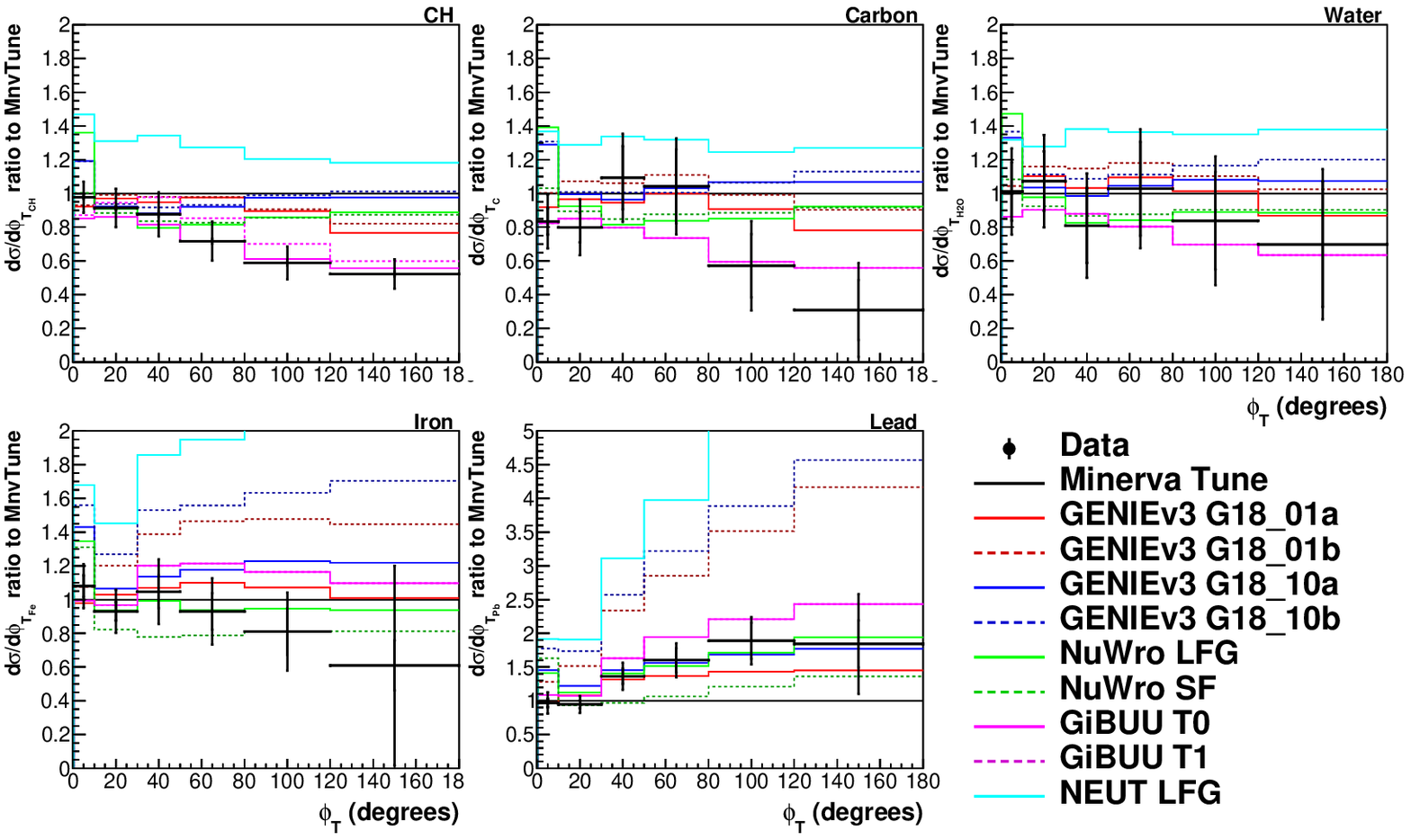}
    \caption[\tkicoplanar \ cross-section ratio to the MINERvA tune for multiple targets ]{Ratio of absolute cross section measurements to the MINERvA tune for different targets as a function of  \tkicoplanar\ and ratios of different model predictions to the MINERvA tune.}
    \label{fig:models_coplan_rat}
\end{figure}
\begin{figure}
    \centering
    \includegraphics[width=\linewidth]{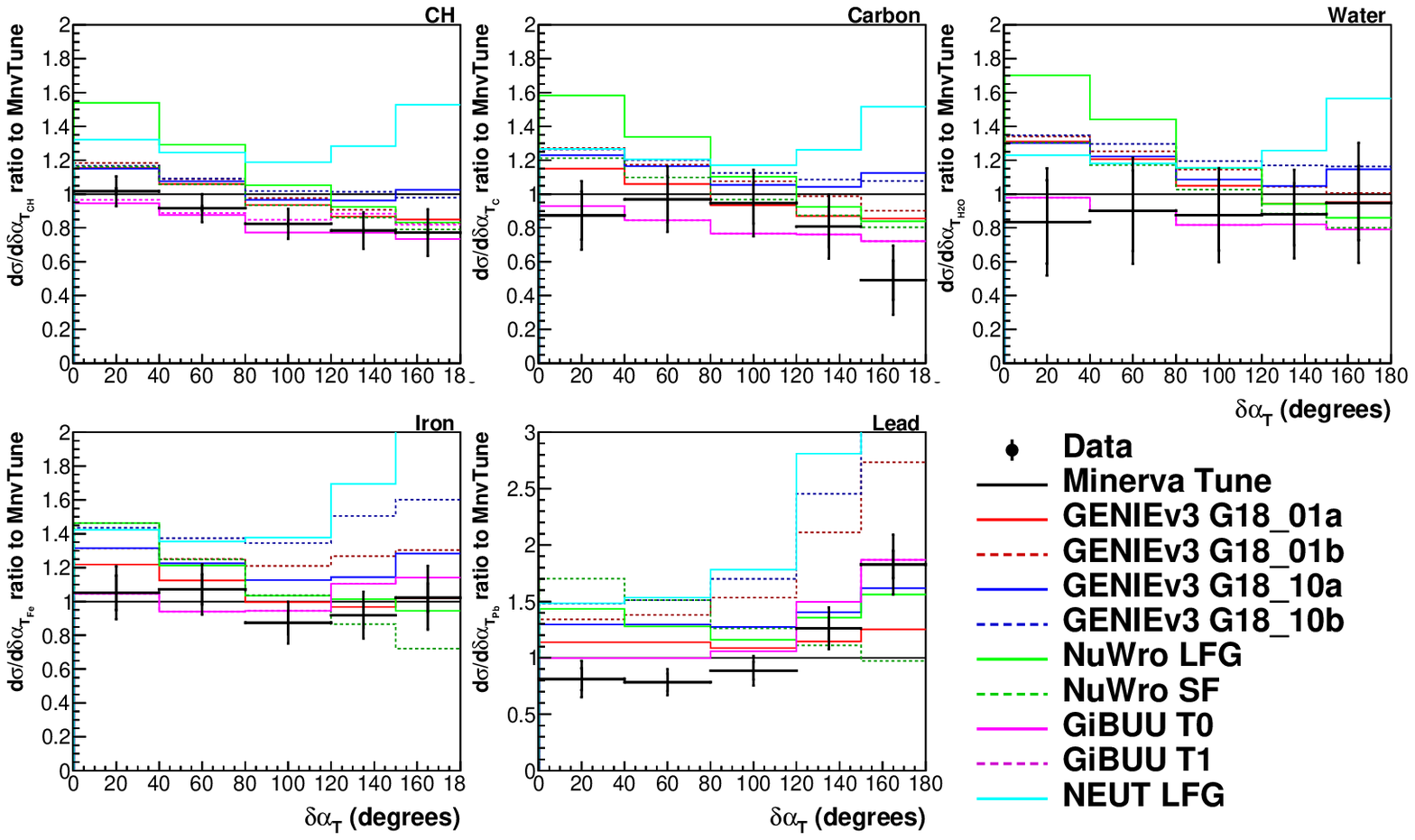}
    \caption[\tkialpha \ cross-section comparison for multiple targets ]{Ratio of absolute cross section measurements to the MINERvA tune for different targets as a function of  \tkialpha\ and ratios of different model predictions to the MINERvA tune.}
    \label{fig:models_alphat_rat}
\end{figure}
\begin{figure}
    \centering
\includegraphics[width=\linewidth]{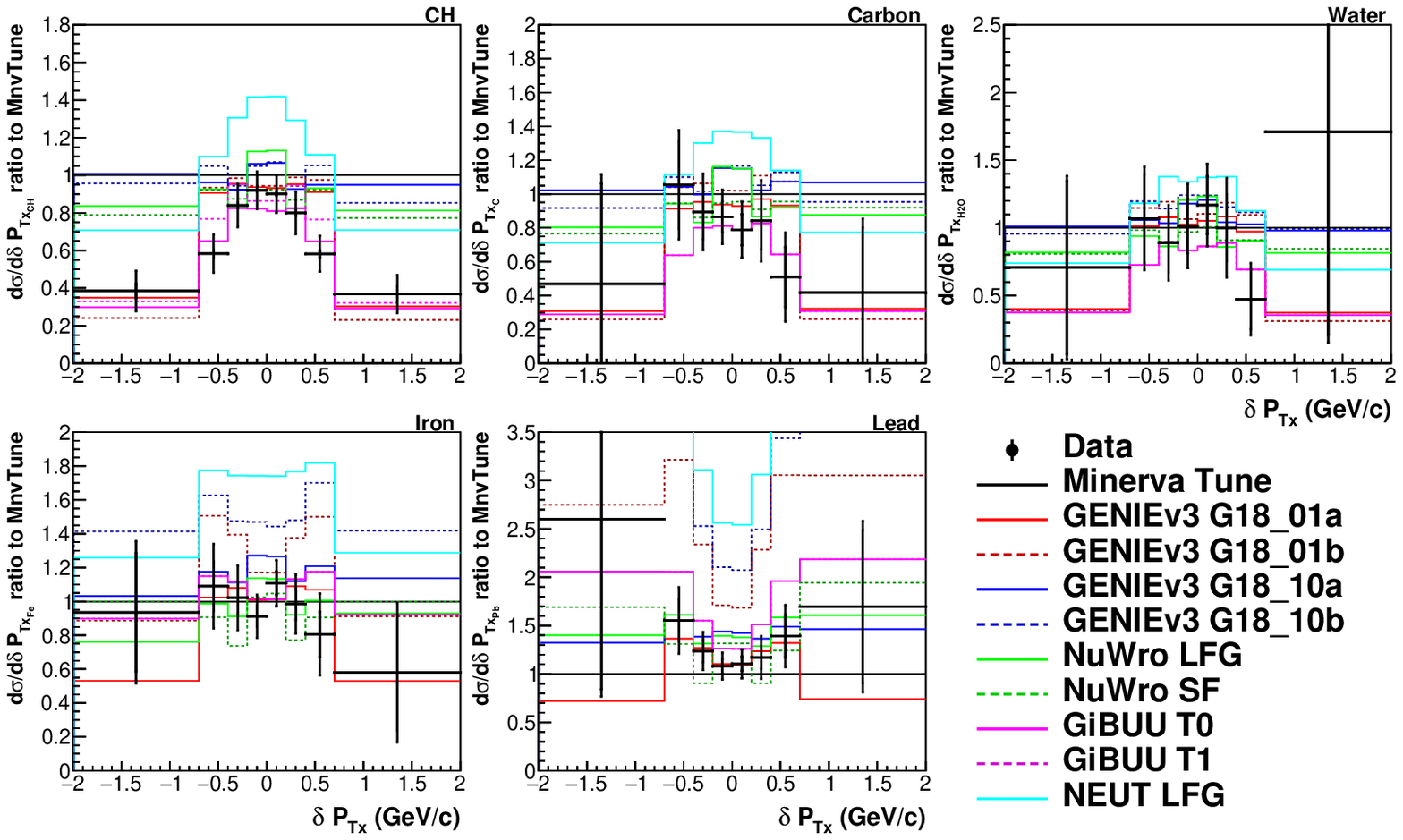}
    \caption{Ratio of absolute cross section measurements to the MINERvA tune for different targets as a function of  \tkidptx\ and ratios of different model predictions to the MINERvA tune.}
    \label{fig:models_dptx_rat}
\end{figure}
\begin{figure}
    \centering
    \includegraphics[width=\linewidth]{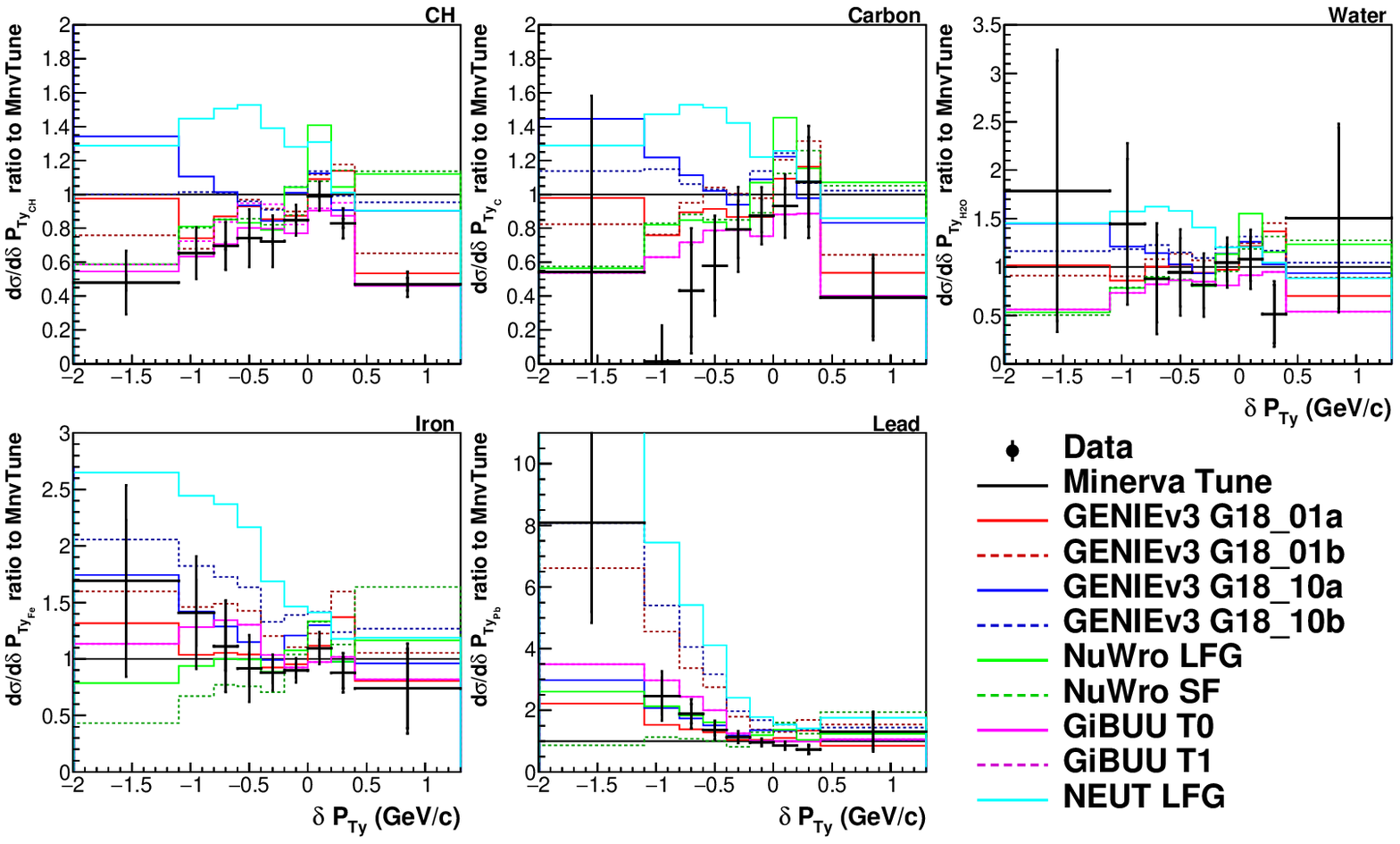}
     \caption{Ratio of absolute cross section measurements to the MINERvA tune for different targets as a function of  \tkidpty\ and ratios of different model predictions to the MINERvA tune. }
    \label{fig:models_dpty_rat}
\end{figure}
\begin{figure}
    \centering
\includegraphics[width=\linewidth]{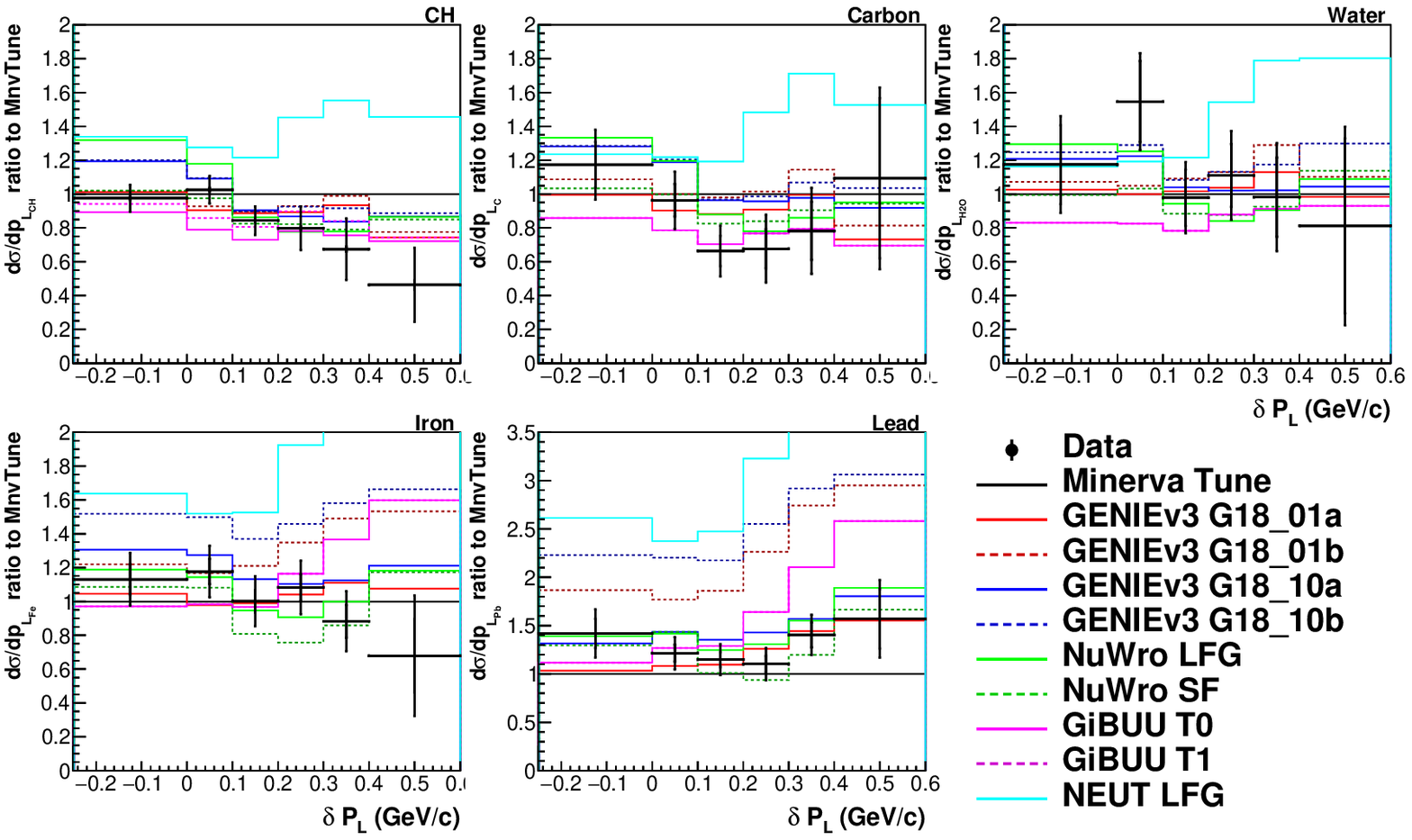}
    \caption{Ratio of absolute cross section measurements to the MINERvA tune for different targets as a function of  \tkipl\ and ratios of different model predictions to the MINERvA tune.}
    \label{fig:models_pl_rat}
\end{figure}
\begin{figure}
    \centering            \includegraphics[width=\linewidth]{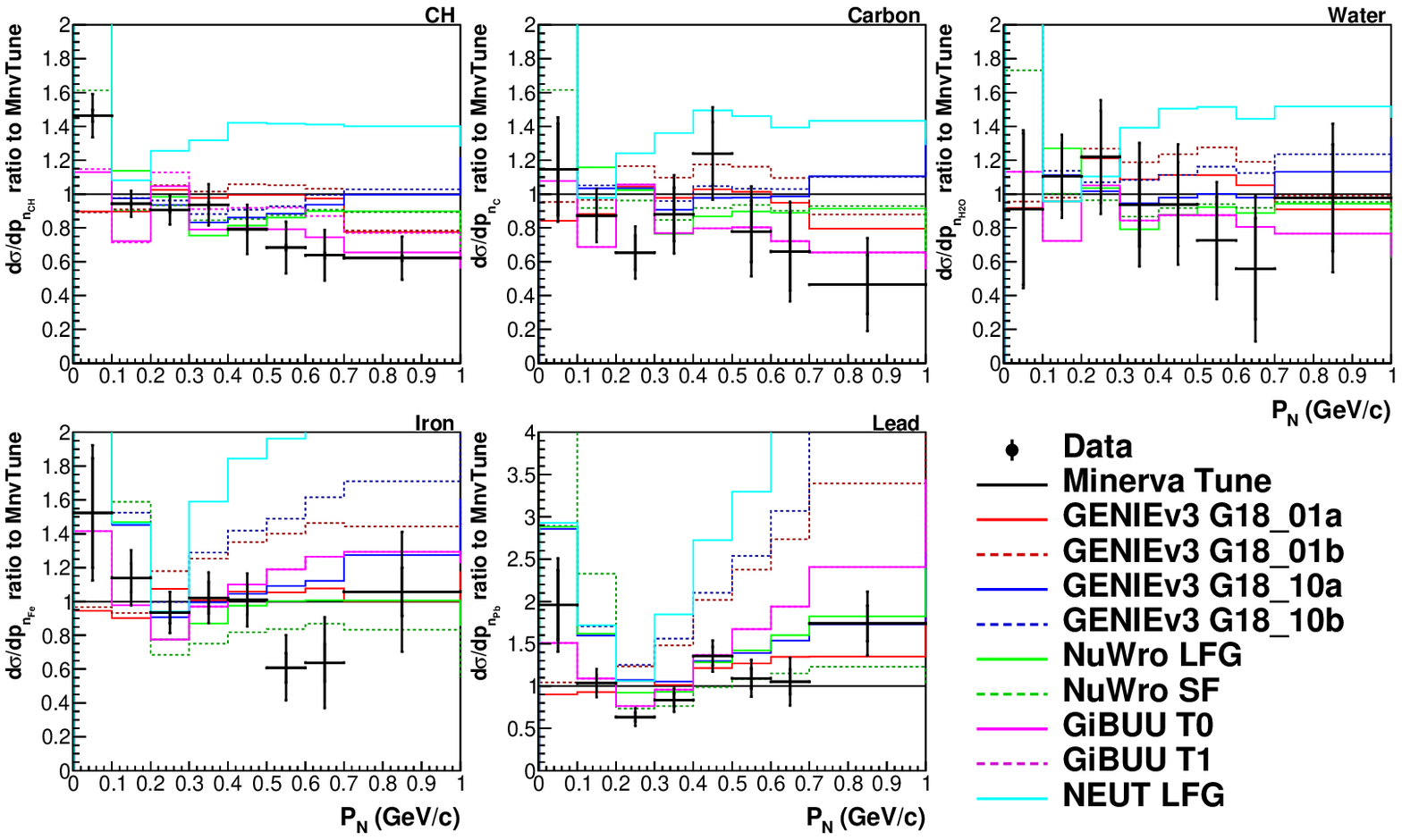}
    \caption{Ratio of absolute cross section measurements to the MINERvA tune for different targets as a function of  \tkipn\ and ratios of different model predictions to the MINERvA tune.}
    \label{fig:models_pn_rat}
\end{figure}
\begin{figure}
    \centering
\includegraphics[width=\linewidth]{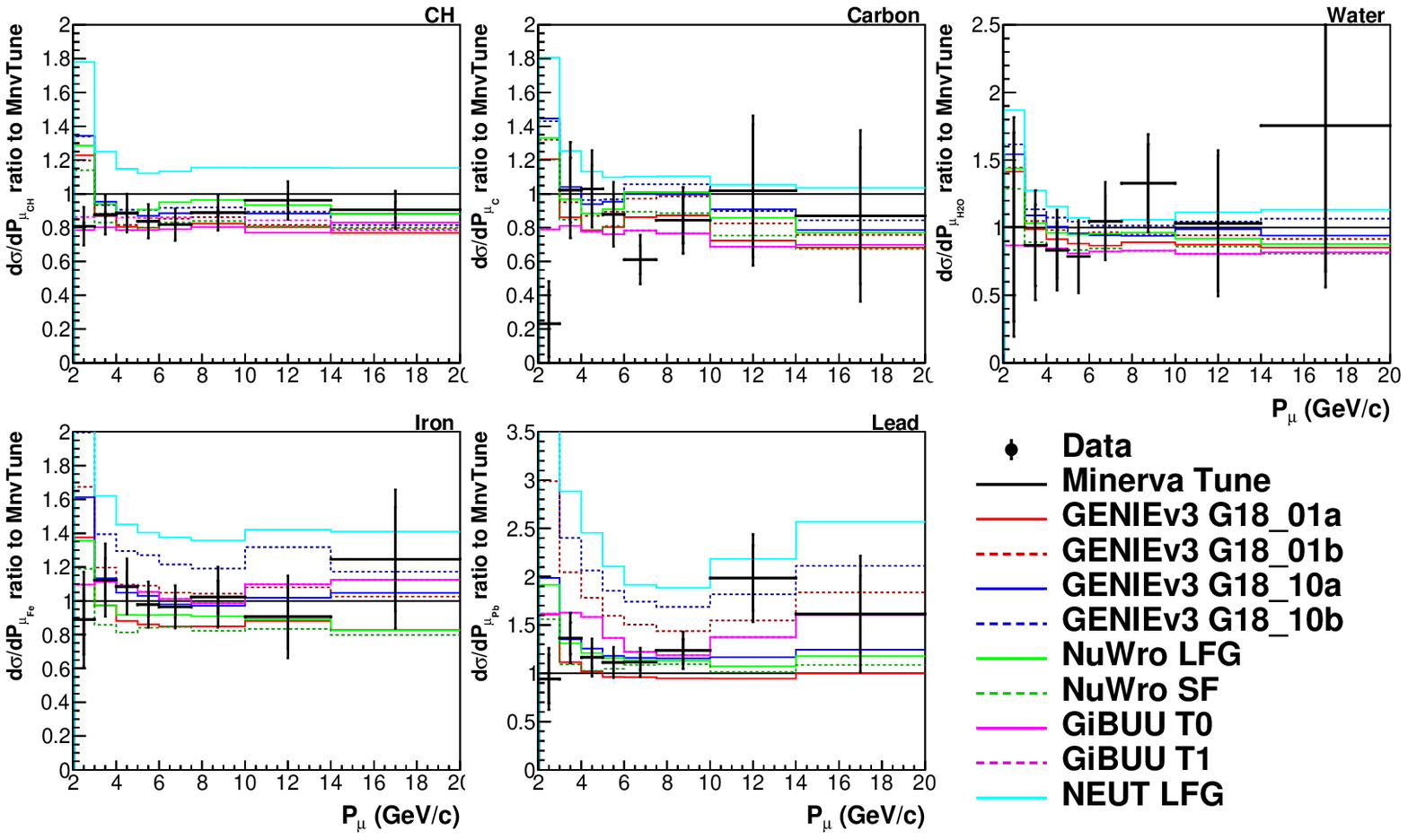}
    \caption{Ratio of absolute cross section measurements to the MINERvA tune for different targets as a function of  P$_\mu$ and ratios of different model predictions to the MINERvA tune.}
    \label{fig:models_muon_p_rat}
\end{figure}
\begin{figure}
    \centering
    \includegraphics[width=\linewidth]{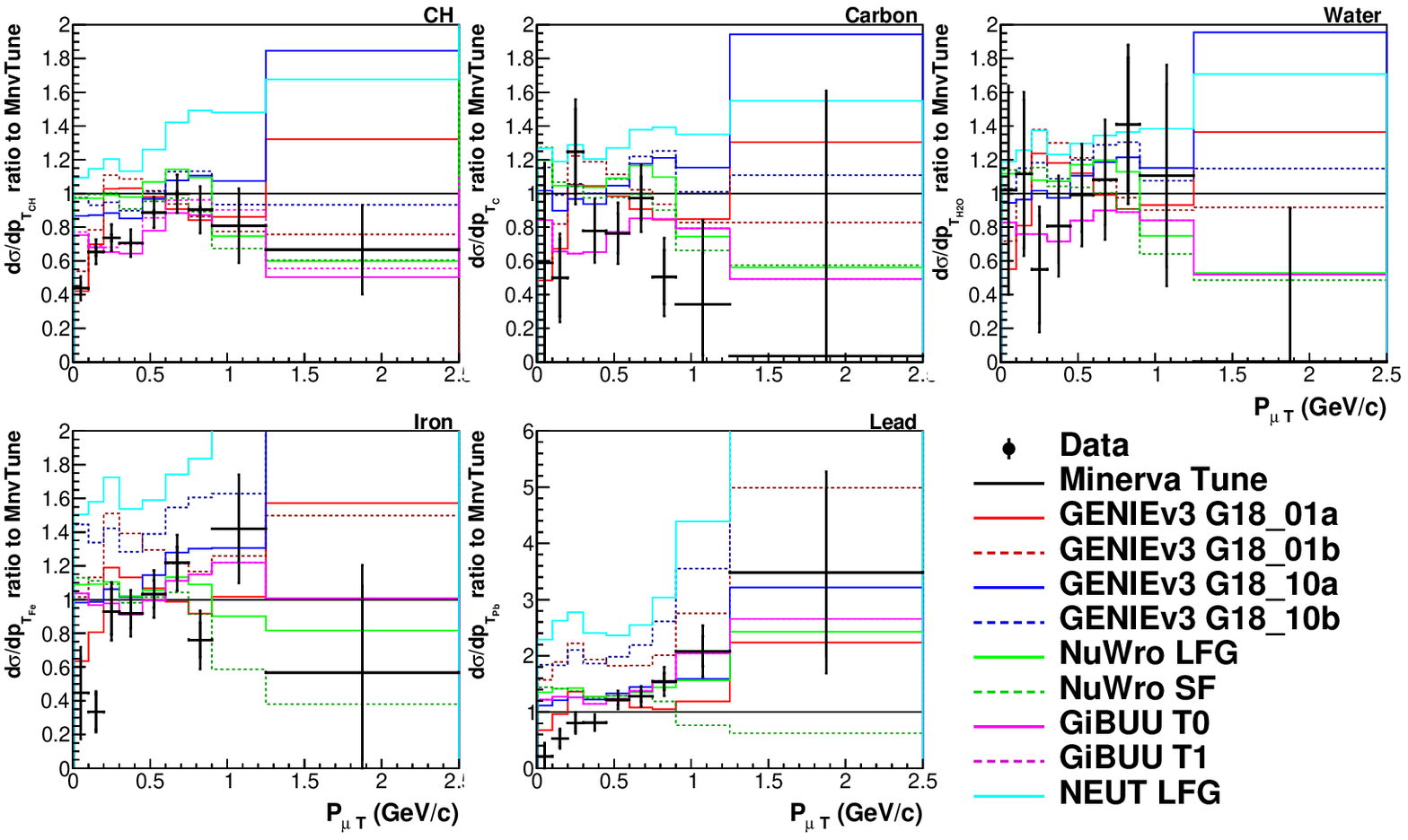}
    \caption[\tkiptmu \ cross-section comparison for multiple targets ]{Ratio of absolute cross section measurements to the MINERvA tune for different targets as a function of  \tkiptmu\ and ratios of different model predictions to the MINERvA tune.}
    \label{Figure:Gencompare_ptmu_rat}
\end{figure}
\begin{figure}
    \centering
\includegraphics[width=\linewidth]{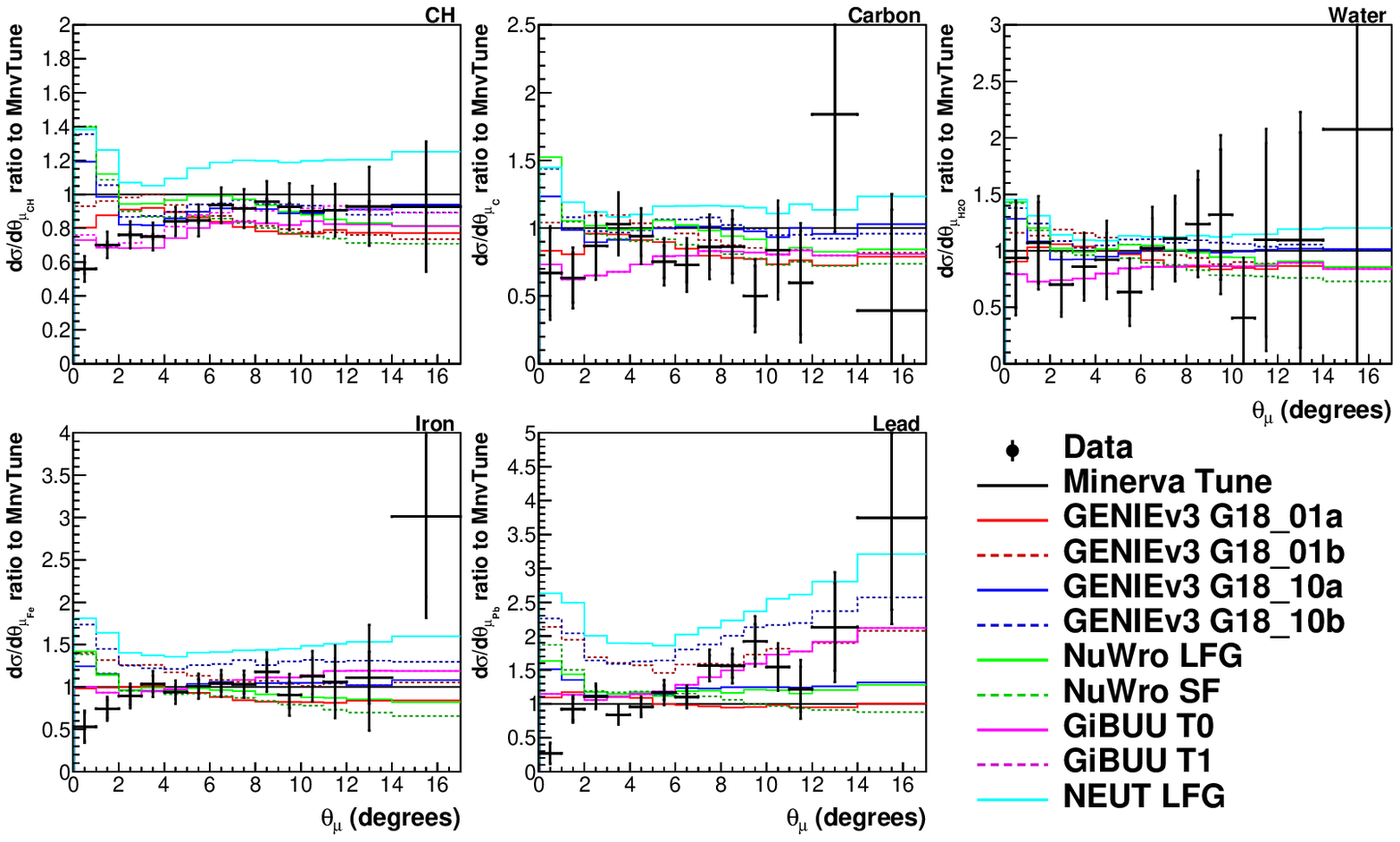}
    \caption{Ratio of absolute cross section measurements to the MINERvA tune for different targets as a function of $\theta_{\mu}$ and ratios of different model predictions to the MINERvA tune.}
    \label{fig:models_muon_theta_rat}
\end{figure}
\begin{figure}
    \centering
 \includegraphics[width=\linewidth]{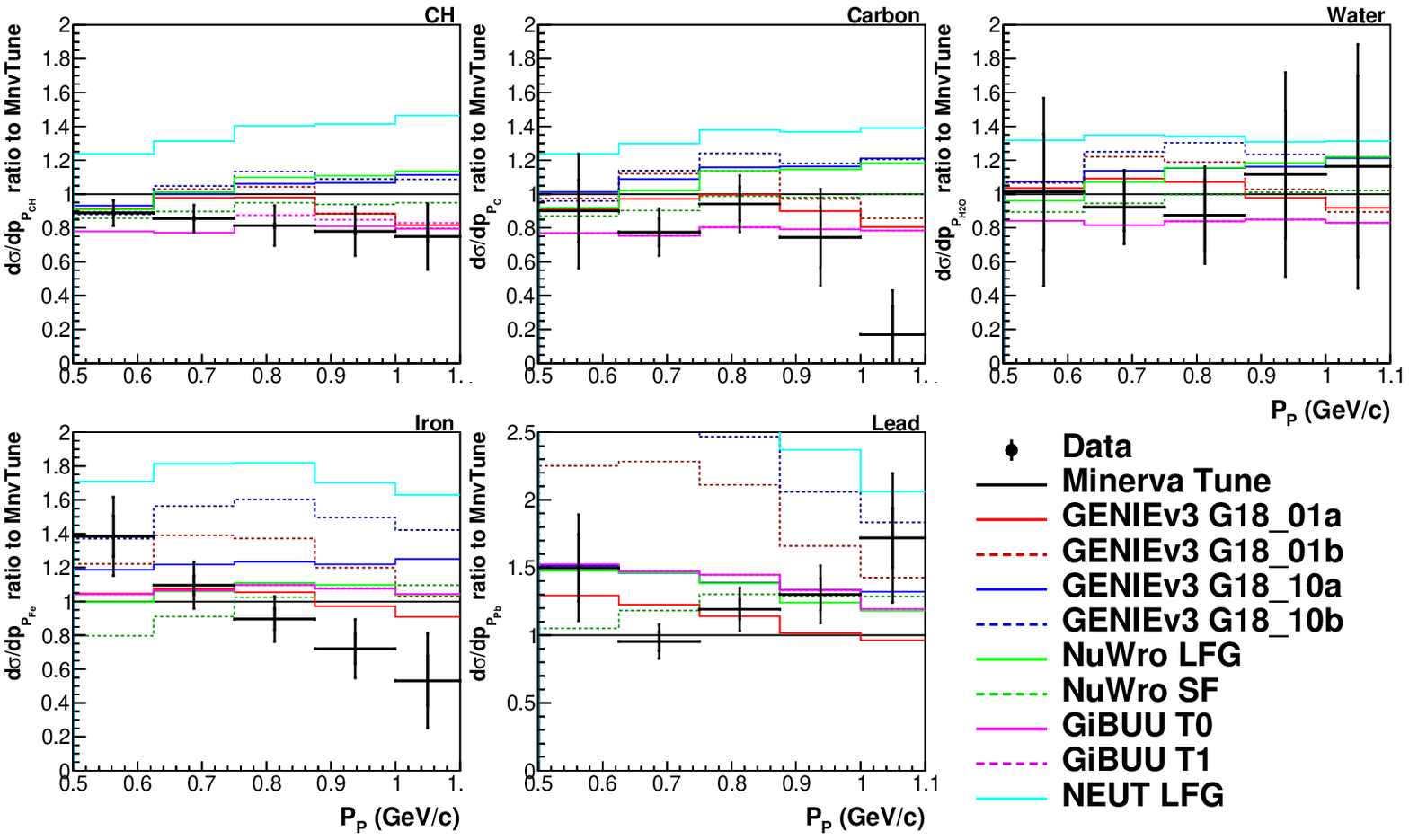}
    \caption{Ratio of absolute cross section measurements to the MINERvA tune for different targets as a function of P$_{p}$ and ratios of different model predictions to the MINERvA tune.}
    \label{fig:models_proton_p_rat}
\end{figure}
\begin{figure}
    \centering
\includegraphics[width=\linewidth]{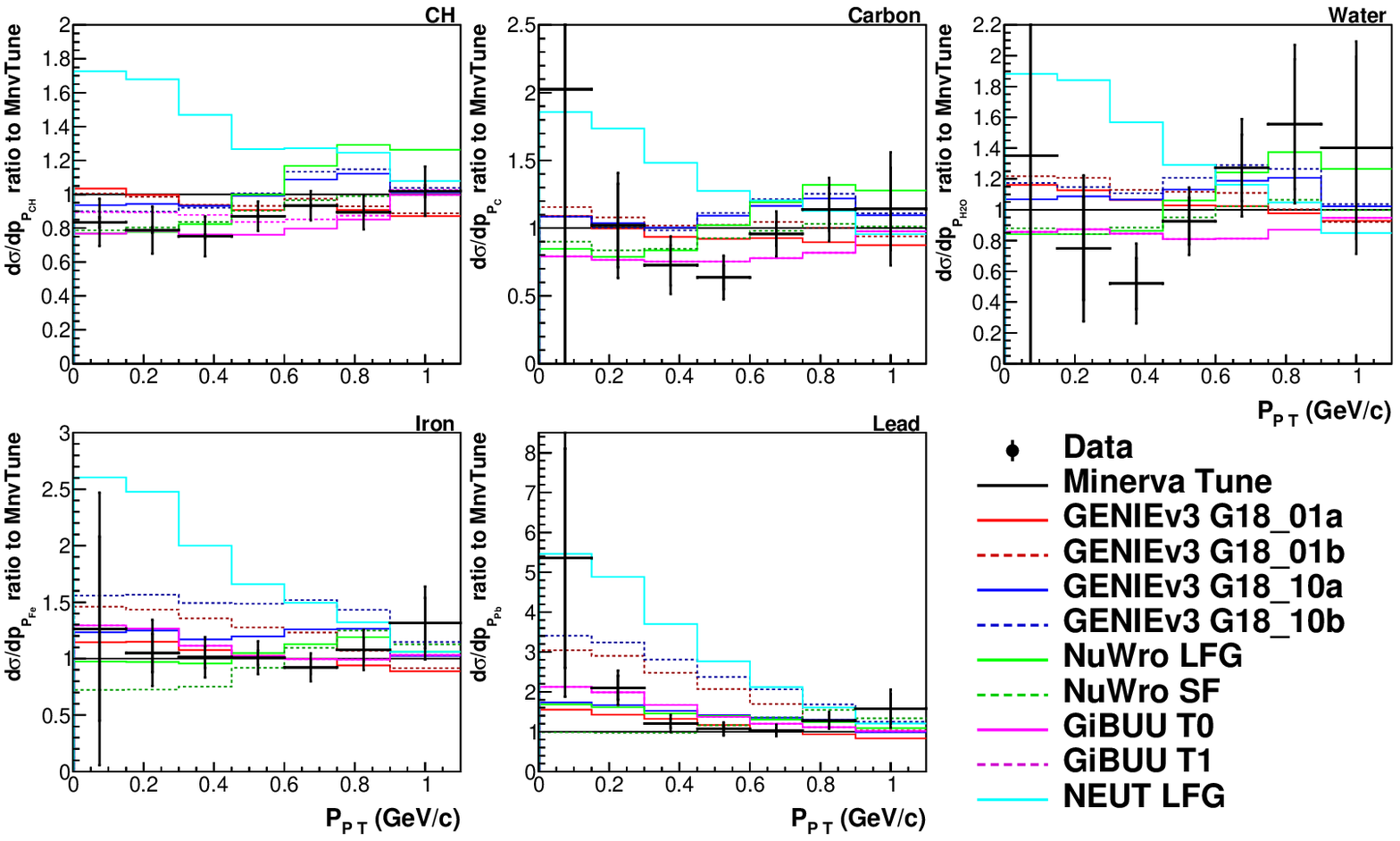}
    \caption{Ratio of absolute cross section measurements to the MINERvA tune for different targets as a function of P$_{p T}$ and ratios of different model predictions to the MINERvA tune.}
    \label{fig:models_proton_pt_rat}
\end{figure}
\begin{figure}
    \centering
    \includegraphics[width=\linewidth]{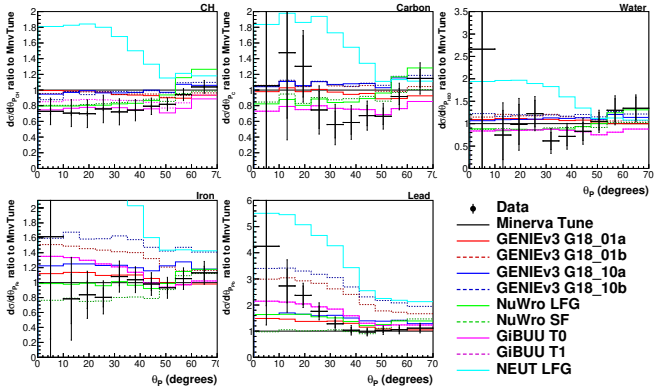}
    \caption{Ratio of absolute cross section measurements to the MINERvA tune for different targets as a function of $\theta_{p}$ and ratios of different model predictions to the MINERvA tune.}
    \label{fig:models_proton_theta_rat}
\end{figure}

\clearpage
\subsection{Supplemental:  Uncertainties in Cross Sections and Ratios} 
\label{sec:appendixB}
The uncertainties on the absolute cross sections and cross-section ratios to scintillator (CH) as a function of all the kinematic variables besides described in the main body of the paper (except \tkidelta\ ) are presented in this Appendix.  The format for each figure is the same:  the top five plots show the uncertainties on the absolute cross sections for CH, carbon, water, iron and lead.  Note that because the statistical uncertainties in carbon and water are large compared to the other targets, those uncertainties have been scaled by 0.5 to be plotted next to the uncertainties on the CH target.  The bottom four plots show the uncertainties in the ratios of the absolute cross sections to CH.  Note that the flux and muon reconstruction uncertainties largely cancel in the ratio, and the statistical uncertainty in the ratio is close to that of the absolute cross sections due to the large mass of the scintillator (CH) target compared to the passive targets.  
\begin{figure}
	\centering
\includegraphics[width=\linewidth]{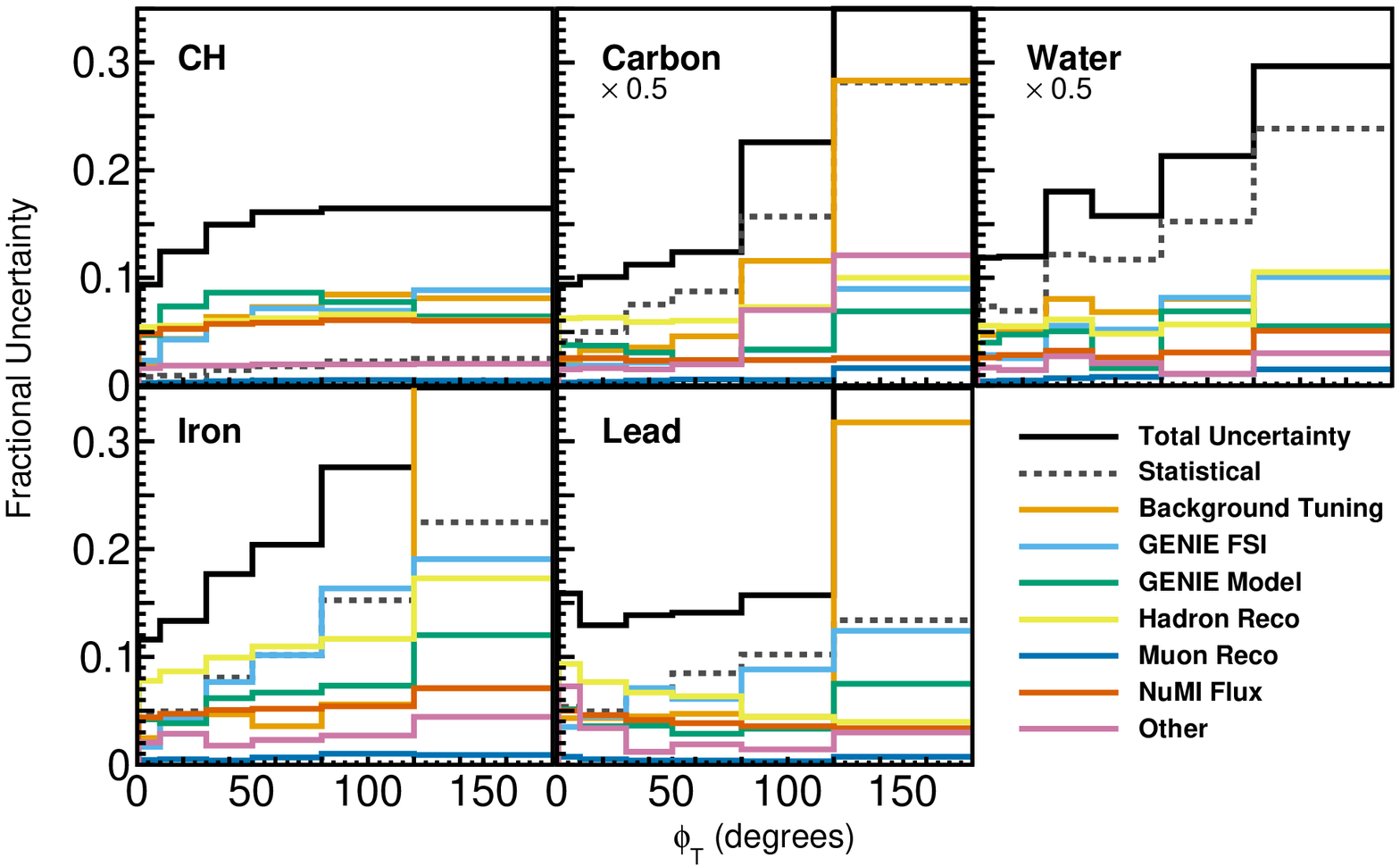}
    \includegraphics[width=\linewidth]{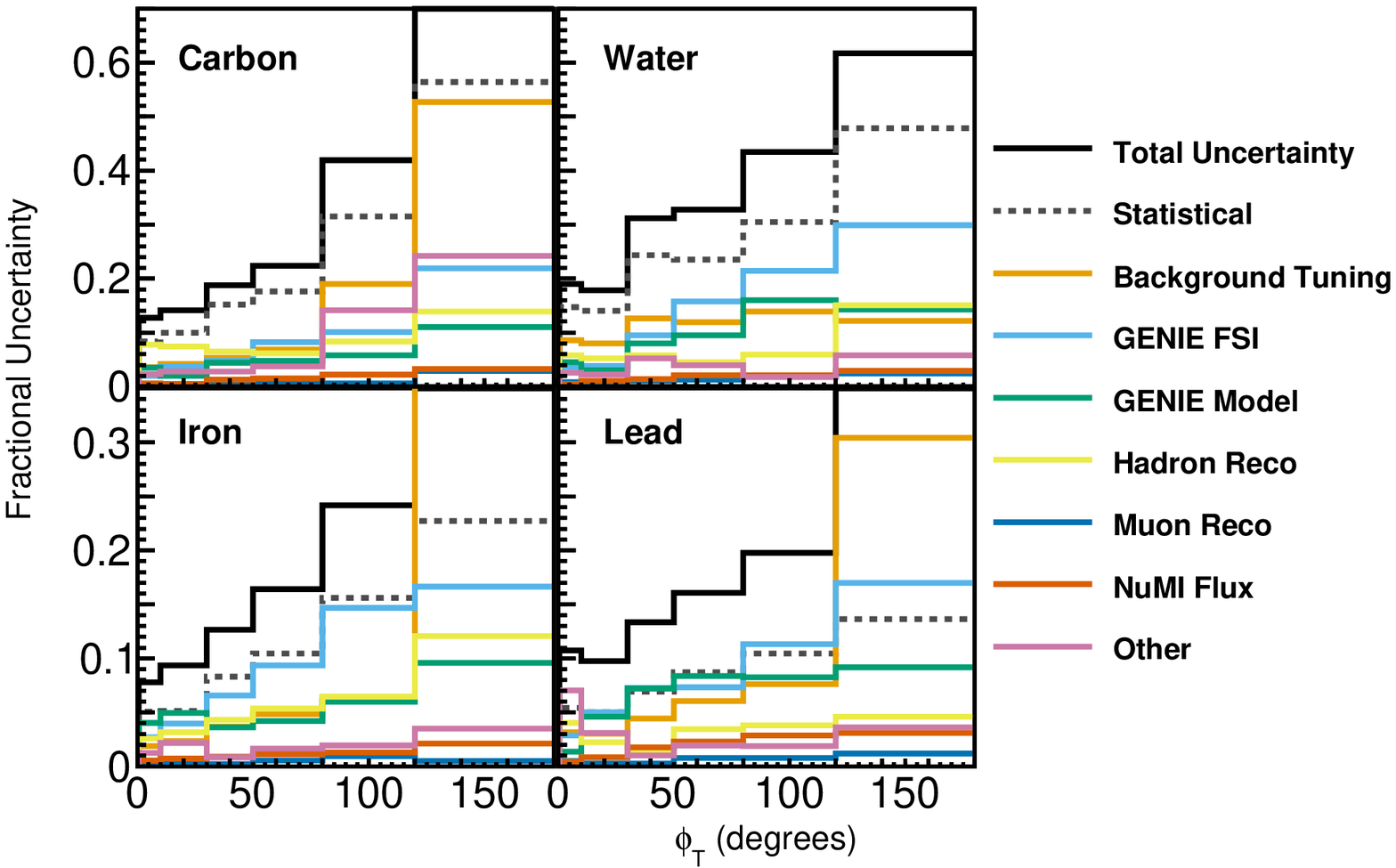}
    \caption[\tkicoplanar cross-section uncertainties for multiple targets]{Uncertainties on the absolute cross section for each target (upper five plots) and on the ratio to CH (lower four plots) for \tkicoplanar.}
    \label{fig:xsec_err_coplan}
\end{figure}
%
%
\begin{figure}
	\centering
\includegraphics[width=\linewidth]{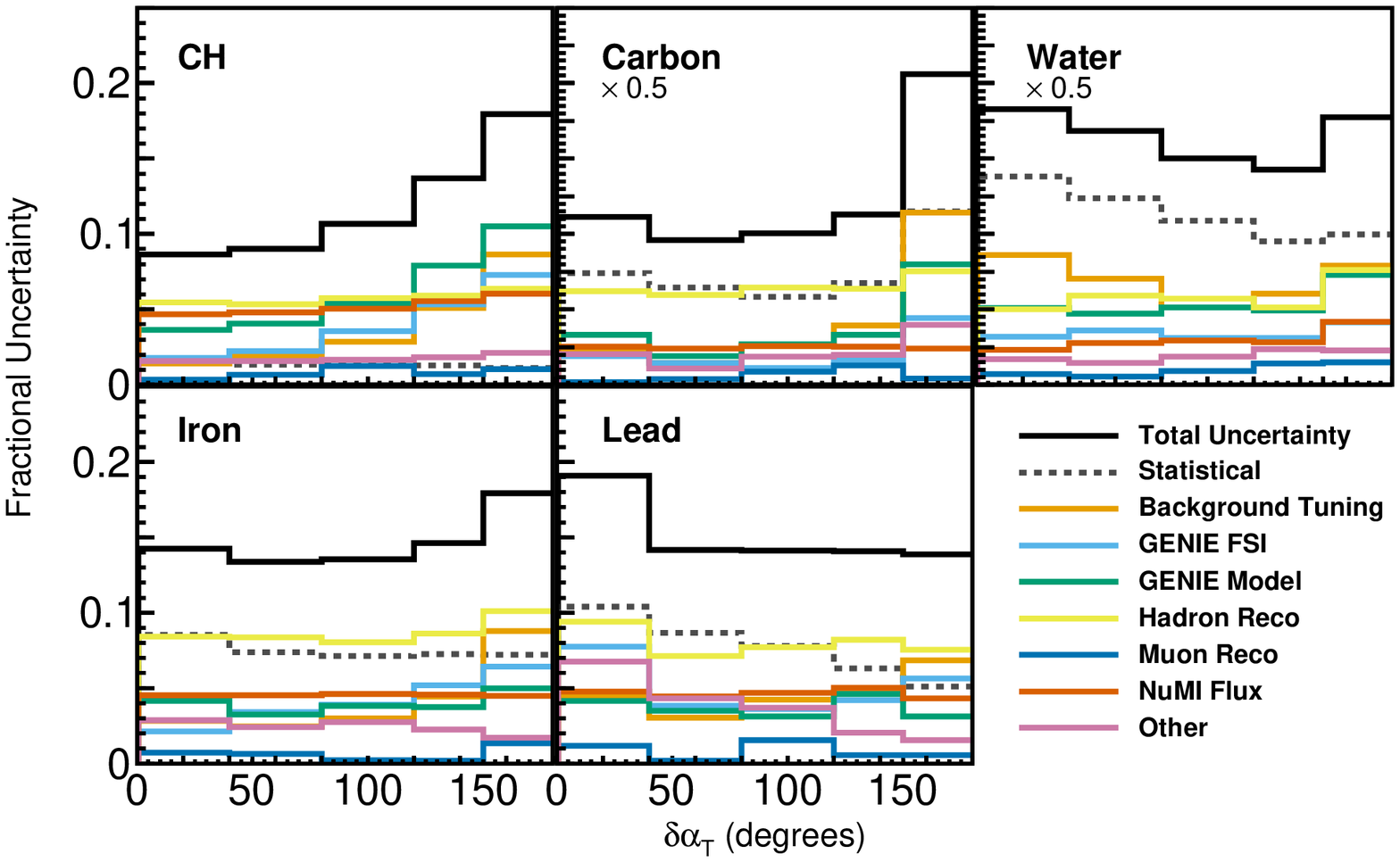}
\includegraphics[width=\linewidth]{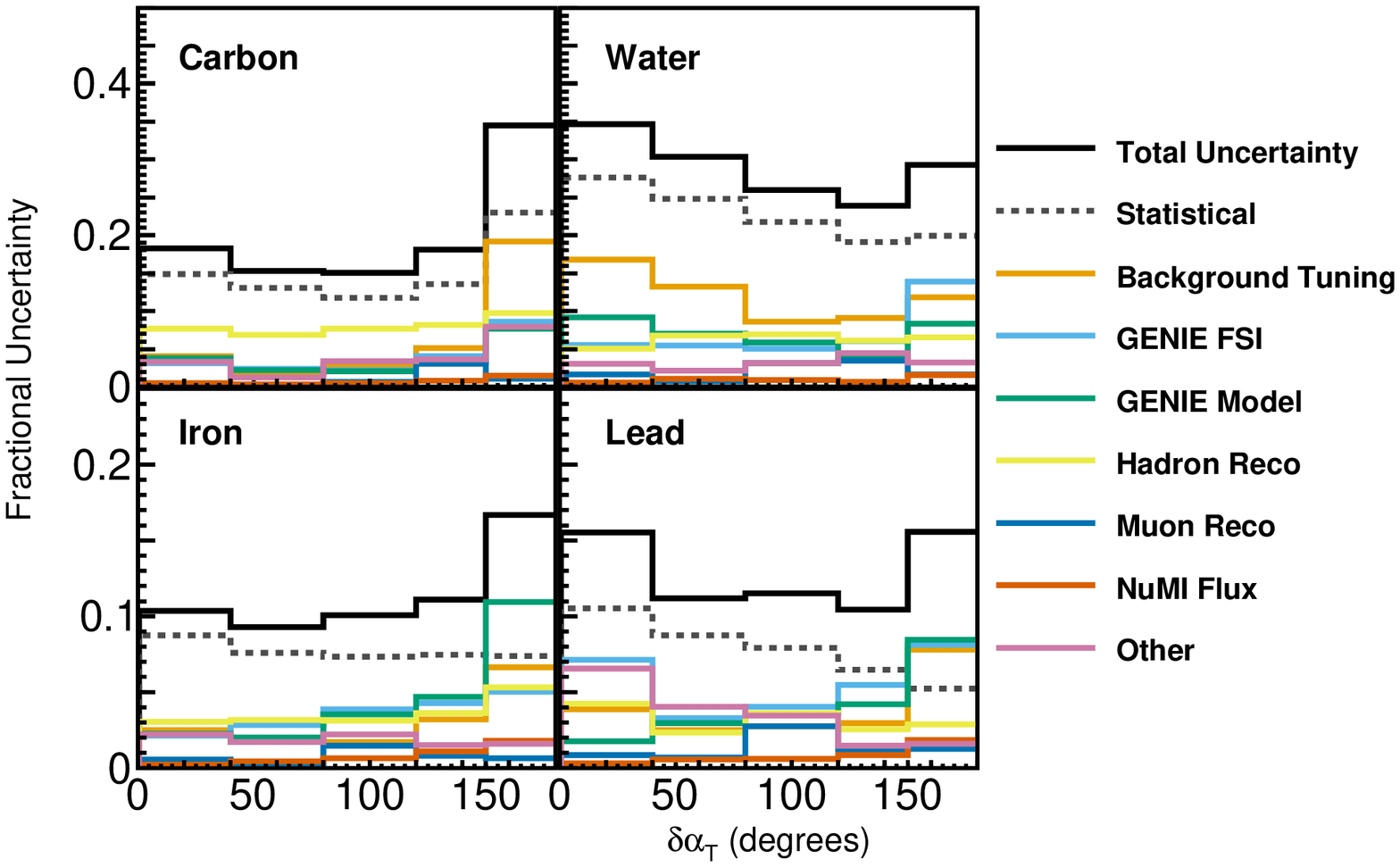}
    \caption[\tkialpha cross-section uncertainties for multiple targets]{Uncertainties on the absolute cross section for each target (upper five plots) and on the ratio to CH (lower four plots) for \tkialpha.}
    \label{fig:xsec_err_alpha}
\end{figure}
\begin{figure}
	\centering
\includegraphics[width=\linewidth]{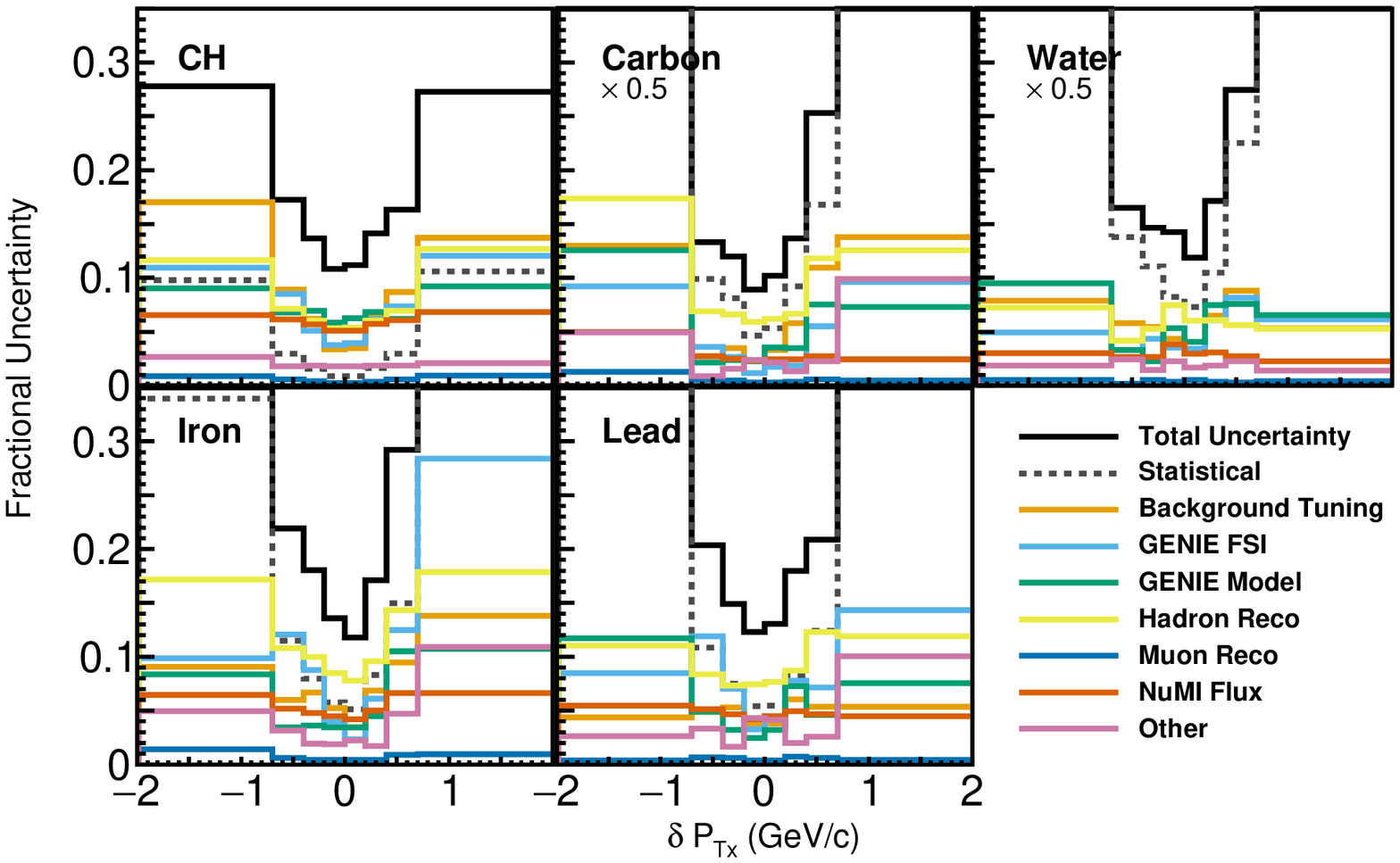}
    \includegraphics[width=\linewidth]{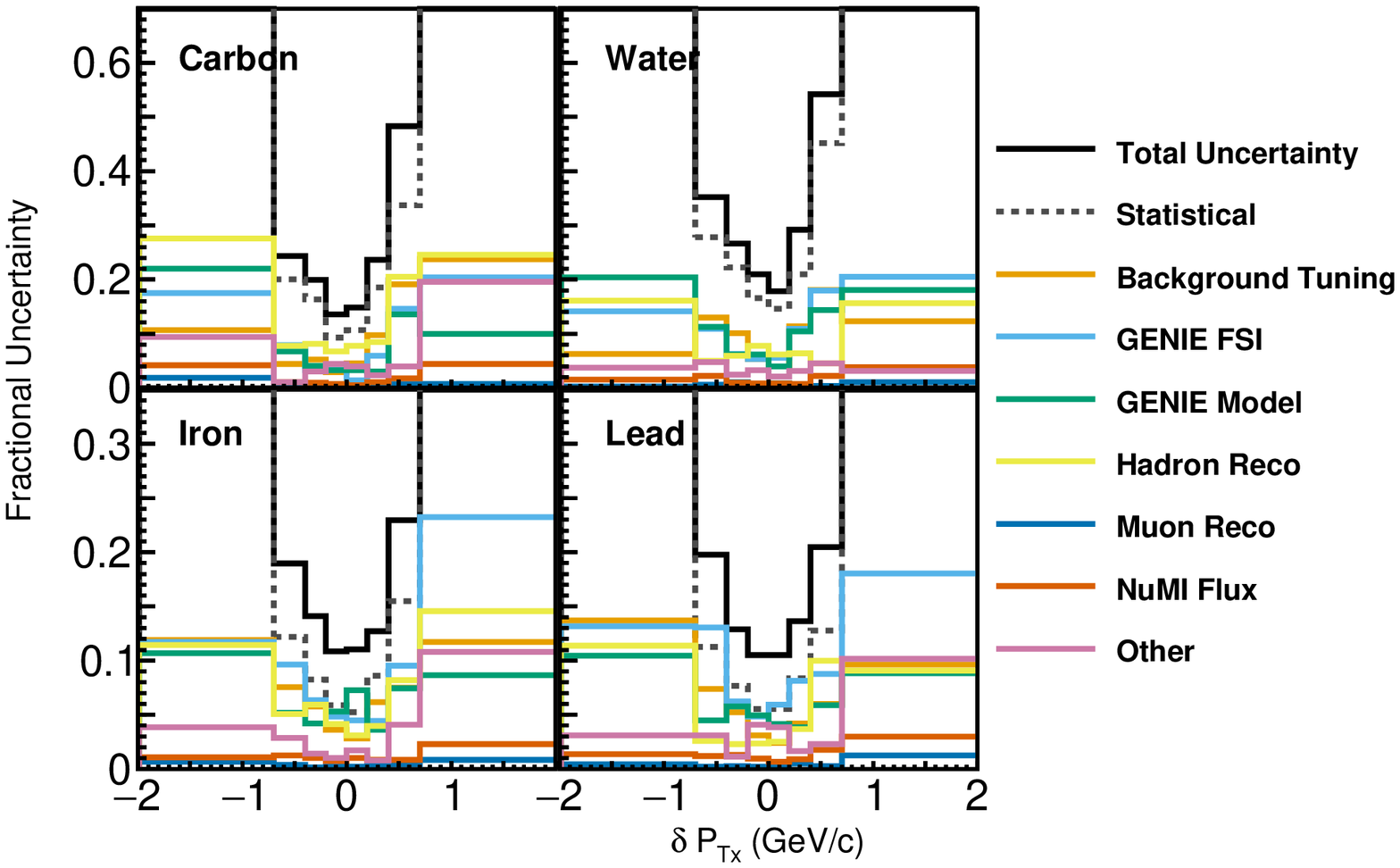}
    \caption[\tkidptx cross-section uncertainties for multiple targets]{Uncertainties on the absolute cross section for each target (upper five plots) and on the ratio to CH (lower four plots) as a function of \tkidptx.}
    \label{fig:xsec_err_dptx}
\end{figure}
\begin{figure}
	\centering
\includegraphics[width=\linewidth]{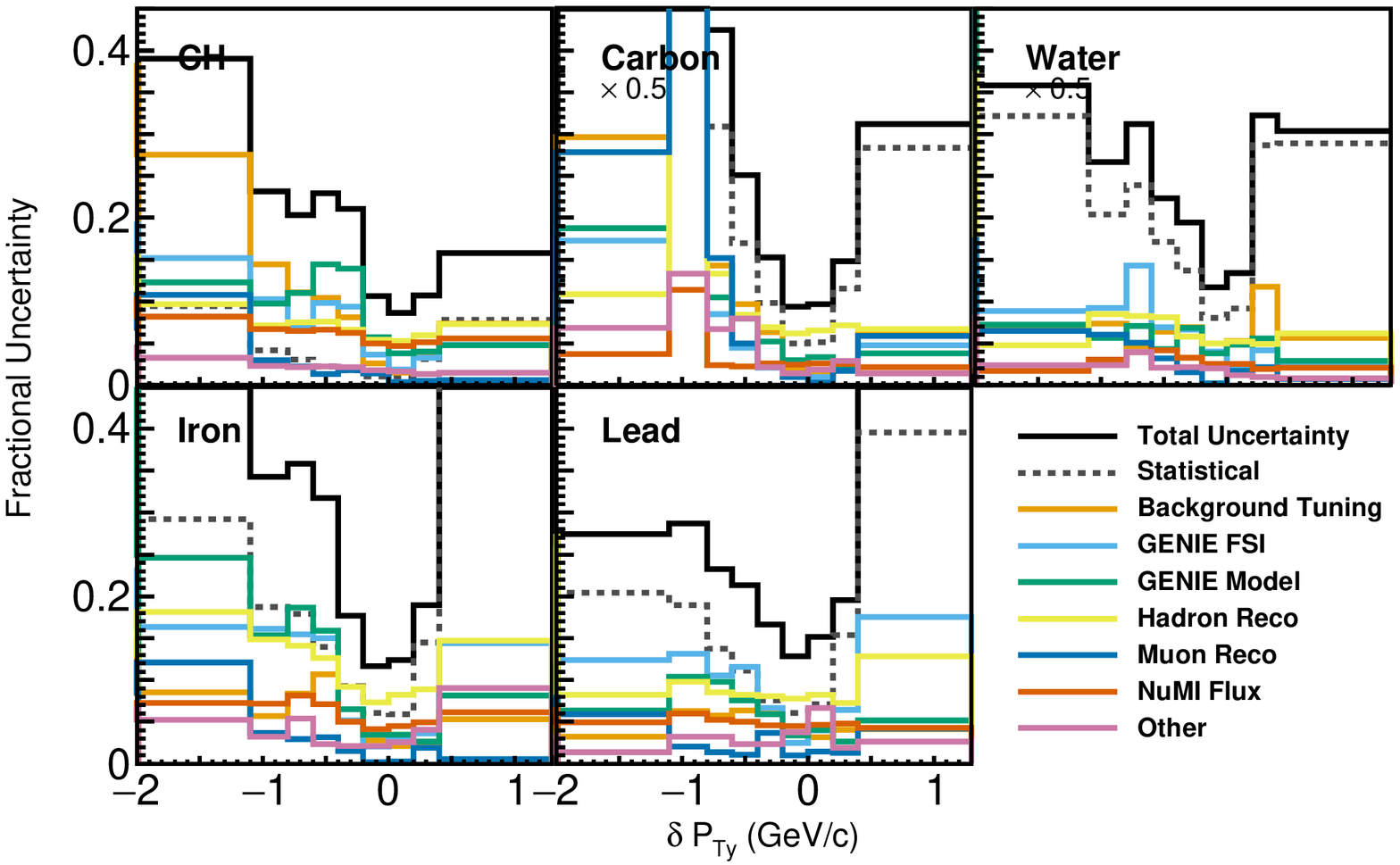}
    \includegraphics[width=\linewidth]{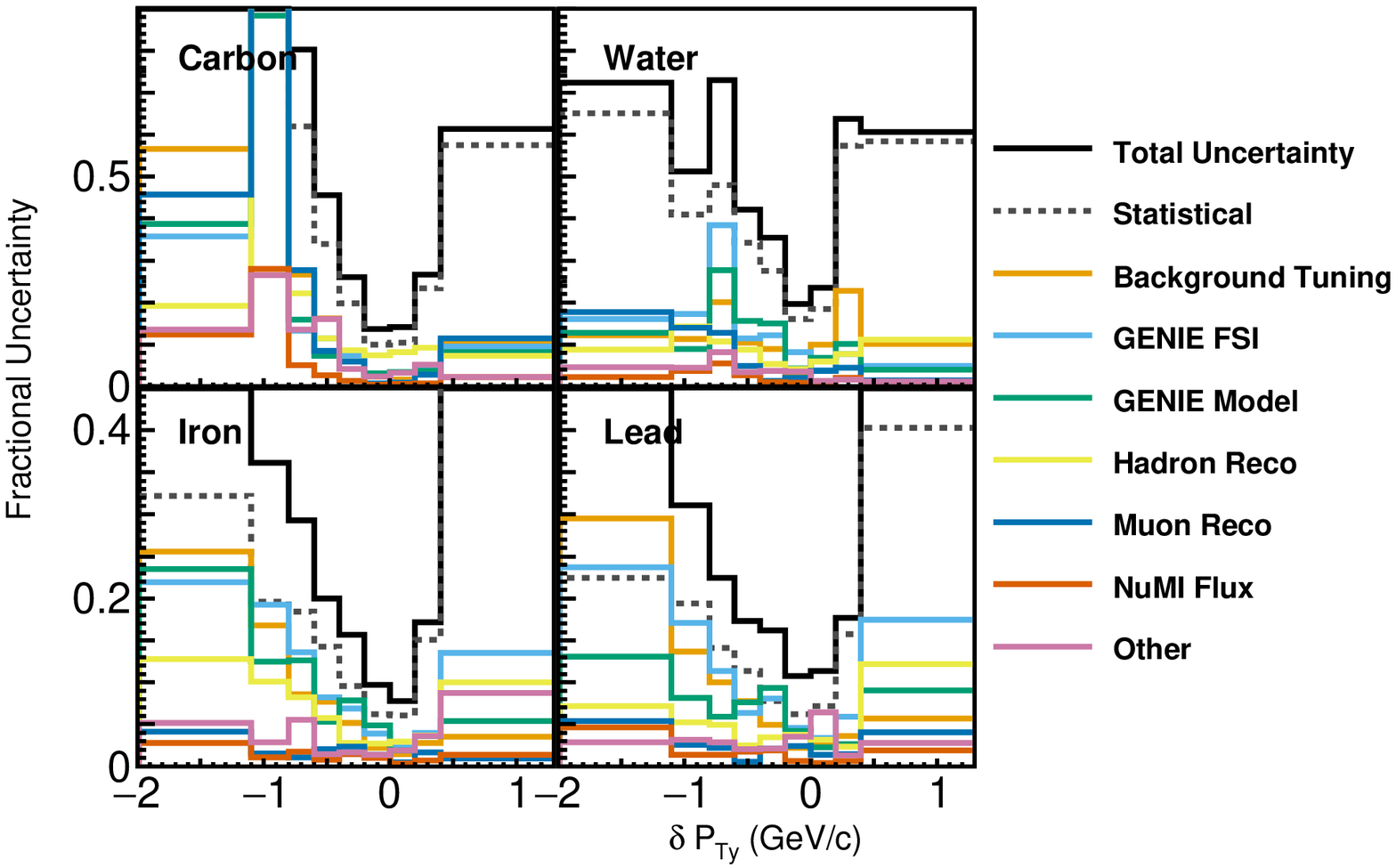}
    \caption[\tkidpty cross-section uncertainties for multiple target ratios]{Uncertainties on the absolute cross section for each target (upper five plots) and on the ratio to CH (lower four plots) as a function of \tkidpty.}
    \label{fig:xsec_err_dpty}
\end{figure}
\begin{figure}
	\centering
\includegraphics[width=\linewidth]{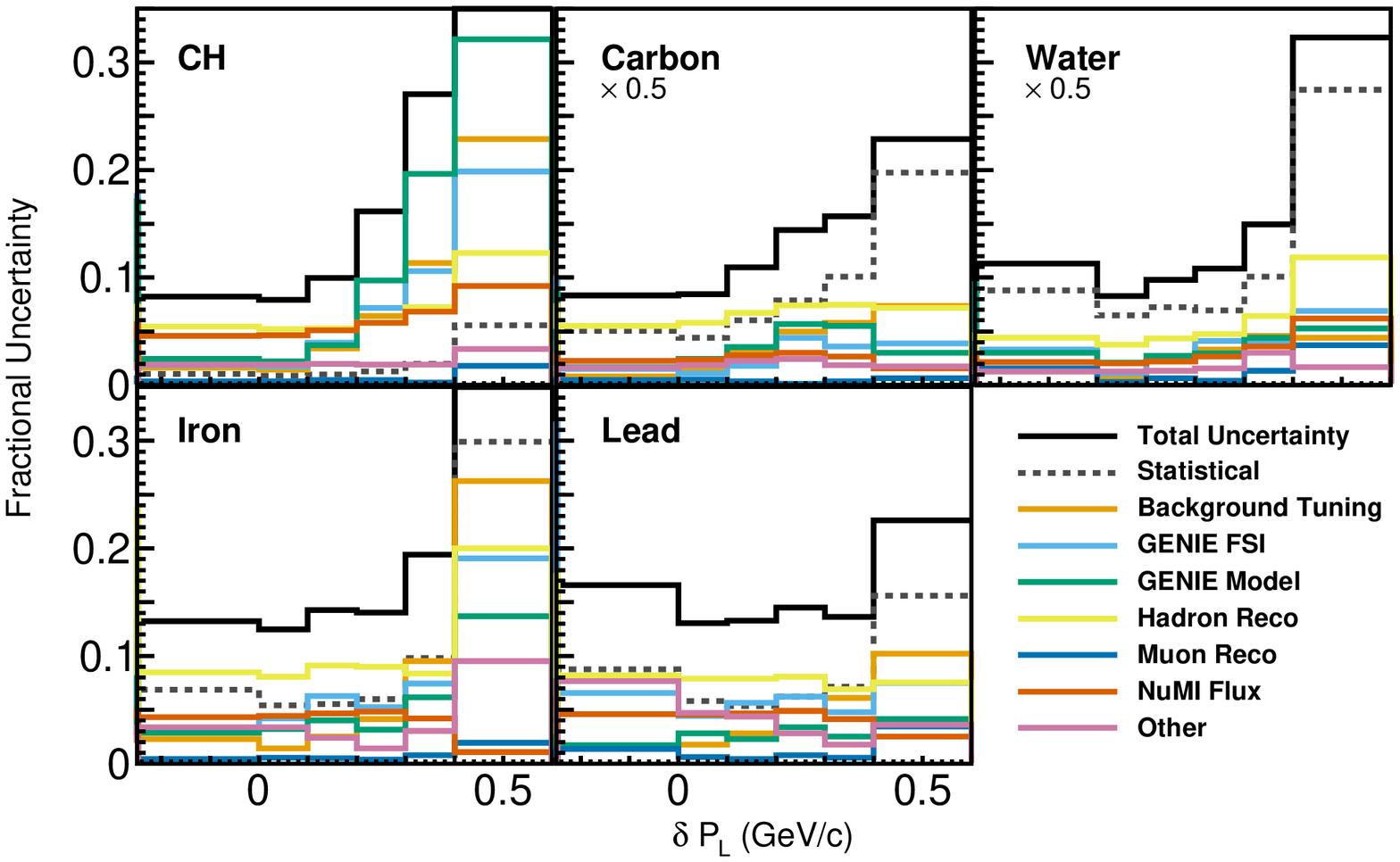}
\includegraphics[width=\linewidth]{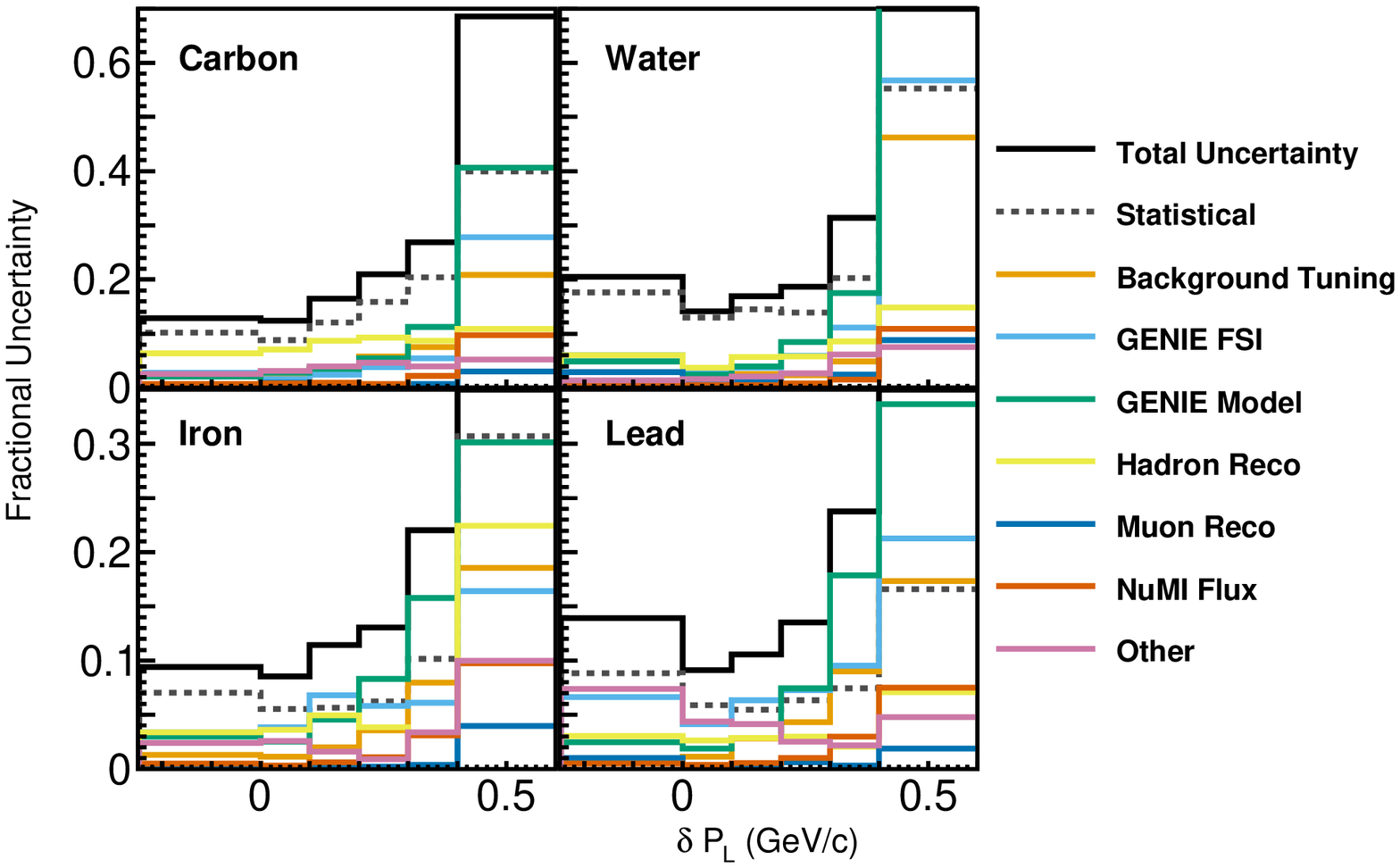}    \caption[\tkipl cross-section uncertainties for multiple targets]{Uncertainties on the absolute cross section for each target (upper five plots) and on the ratio to CH (lower four plots) as a function of \tkipl.}
    \label{fig:xsec_err_pl}
\end{figure}
\begin{figure}
	\centering
    \includegraphics[width=\linewidth]{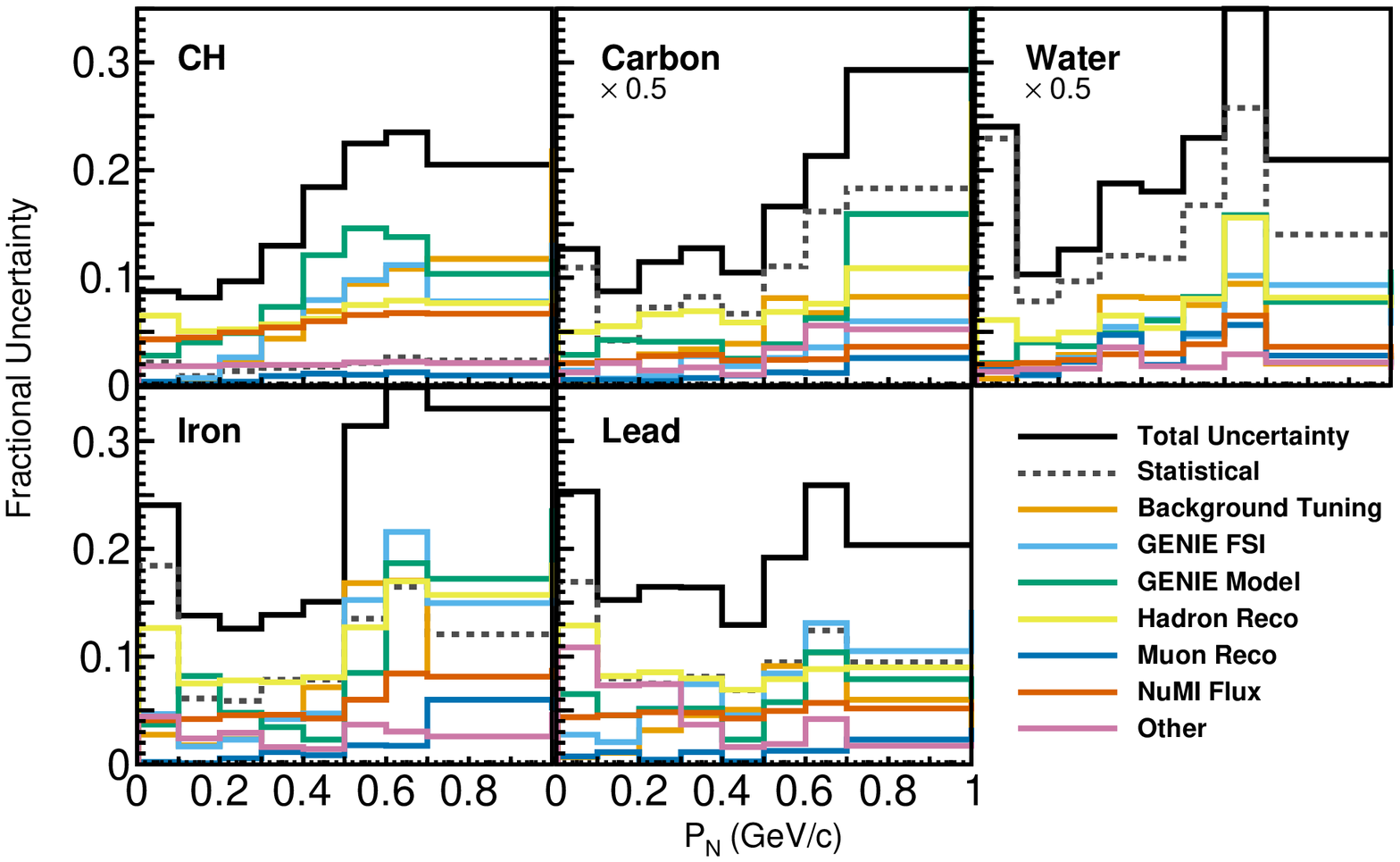}
    \includegraphics[width=\linewidth]{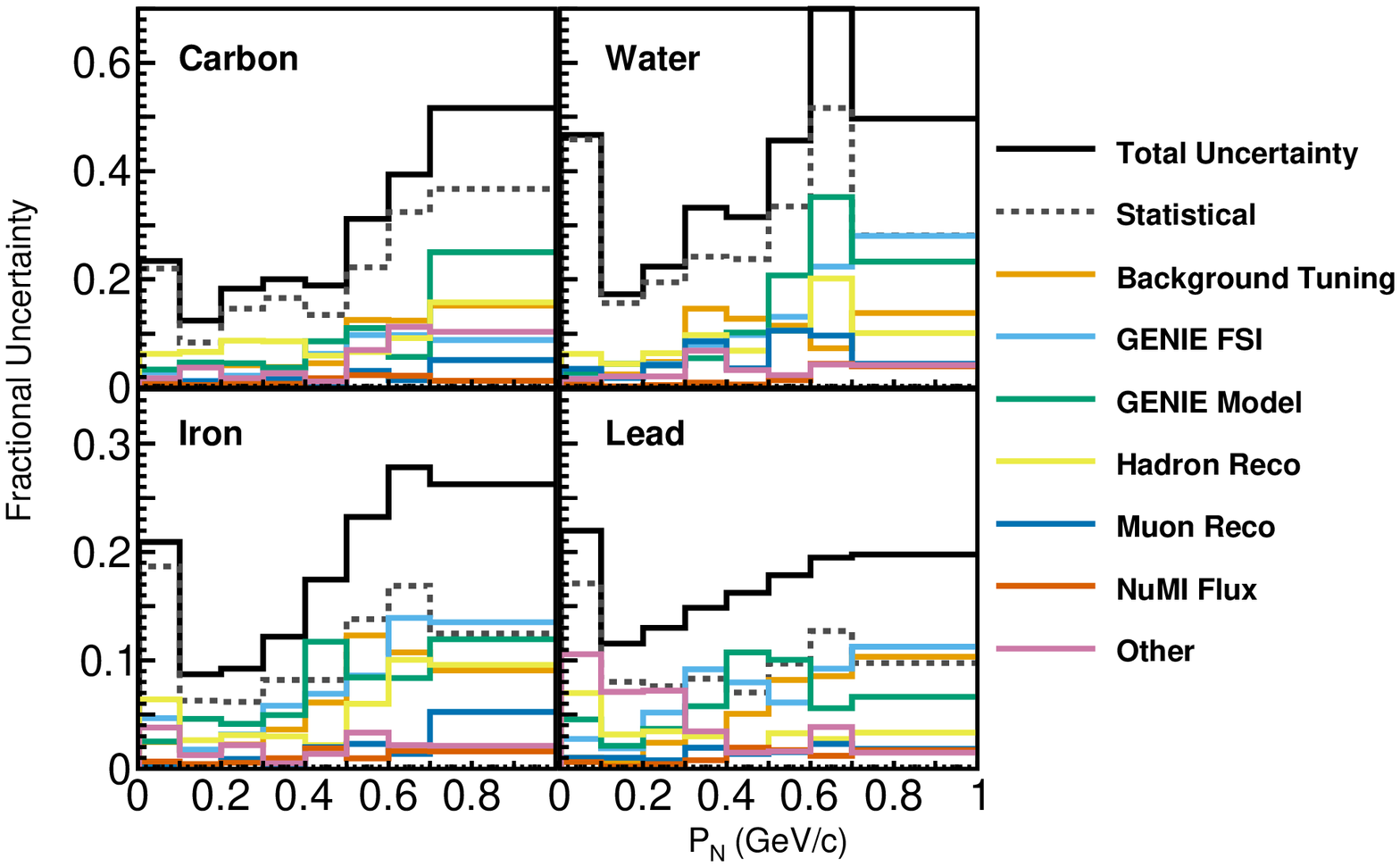}\caption[\tkipn cross-section ratio uncertainties for multiple targets]{Uncertainties on the absolute cross section for each target (upper five plots) and on the ratio to CH (lower four plots) as a function of \tkipn .}
    \label{fig:xsec_err_pn}
\end{figure}
\begin{figure}
	\centering
    \includegraphics[width=\linewidth]{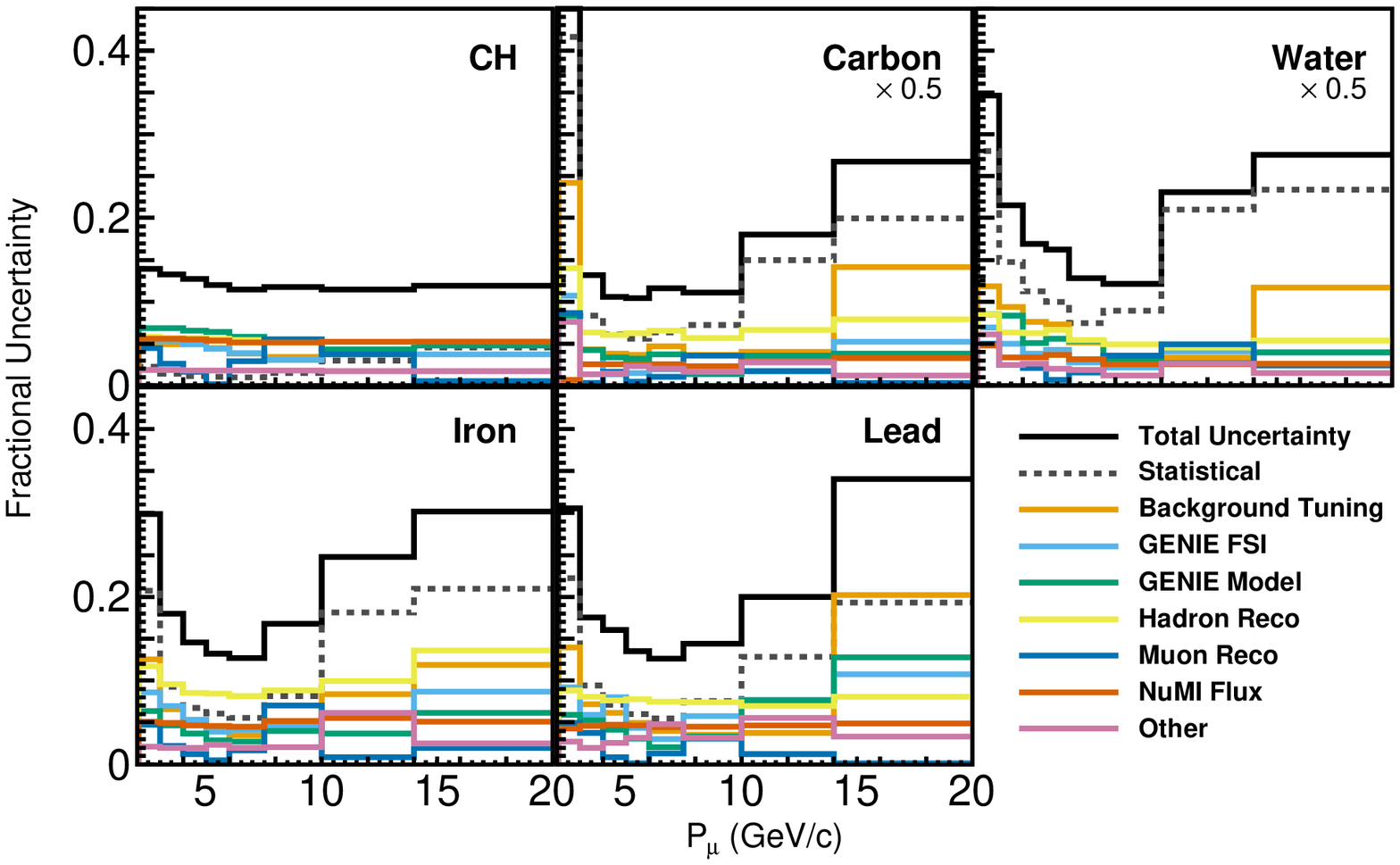}
        \includegraphics[width=\linewidth]{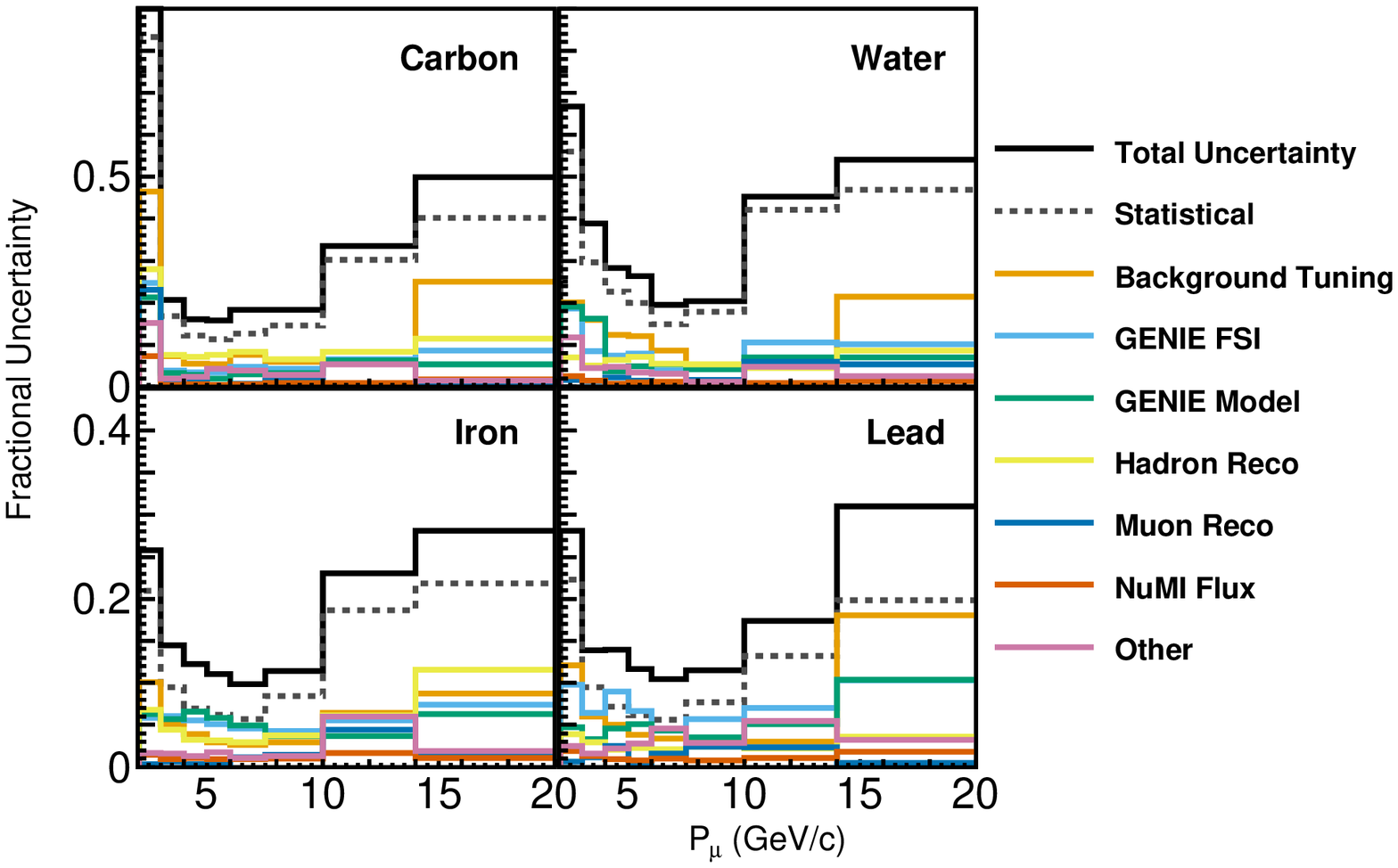}

    \caption[Muon momentum cross-section uncertainties for multiple targets ]{Uncertainties on the absolute cross section for each target (upper five plots) and on the ratio to CH (lower four plots) as a function of muon momentum.}
    \label{fig:xsec_err_muon_p}
\end{figure}
\begin{figure}
	\centering
    \includegraphics[width=\linewidth]{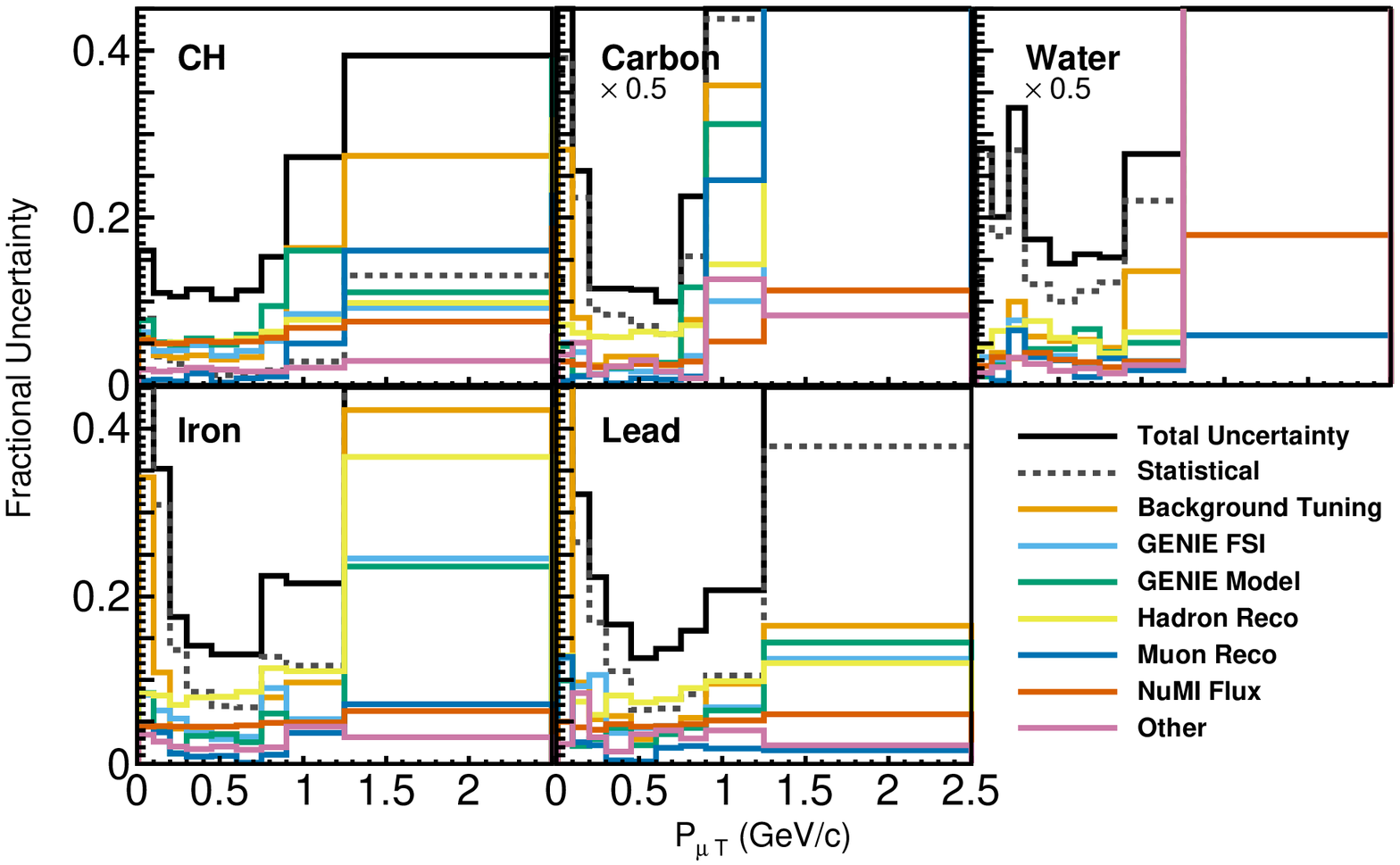}
        \includegraphics[width=\linewidth]{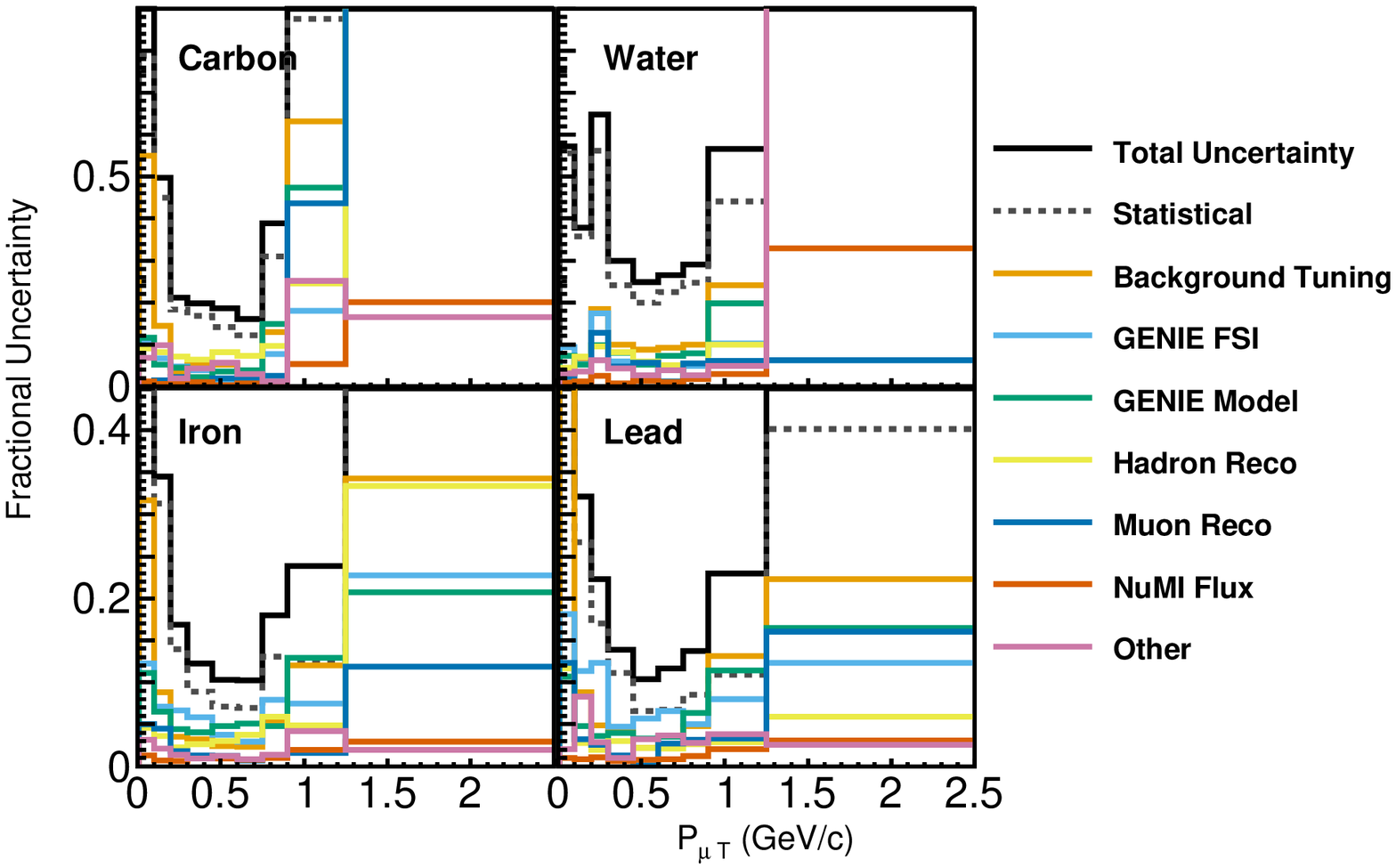}
    \caption[Muon transverse momentum cross-section uncertainties for multiple targets]{Uncertainties on the absolute cross section for each target (upper five plots) and on the ratio to CH (lower four plots) as a function of muon transverse momentum.}
    \label{fig:xsec_err_muon_pt}
\end{figure}
\begin{figure}
	\centering
    \includegraphics[width=\linewidth]{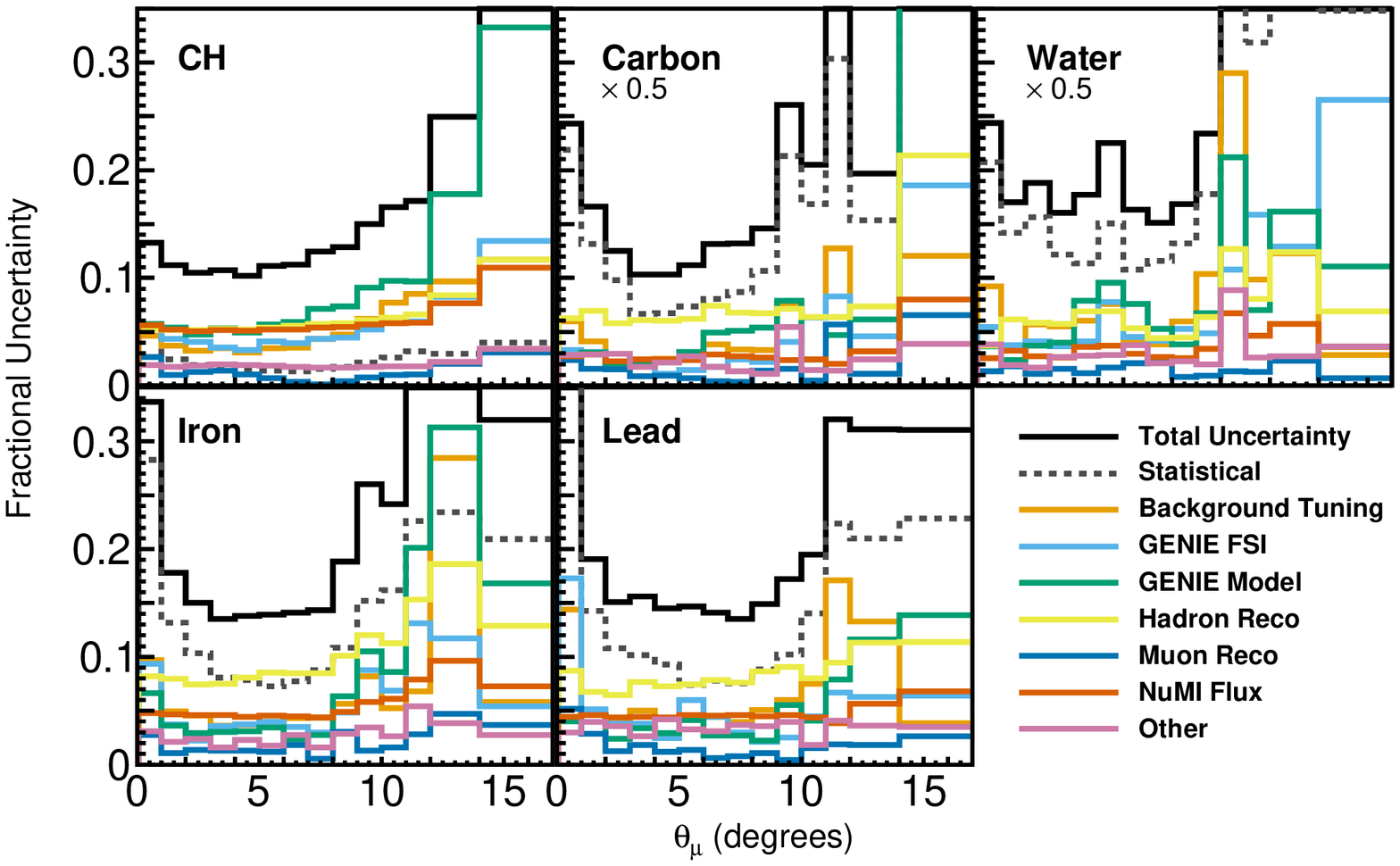}
    \includegraphics[width=\linewidth]{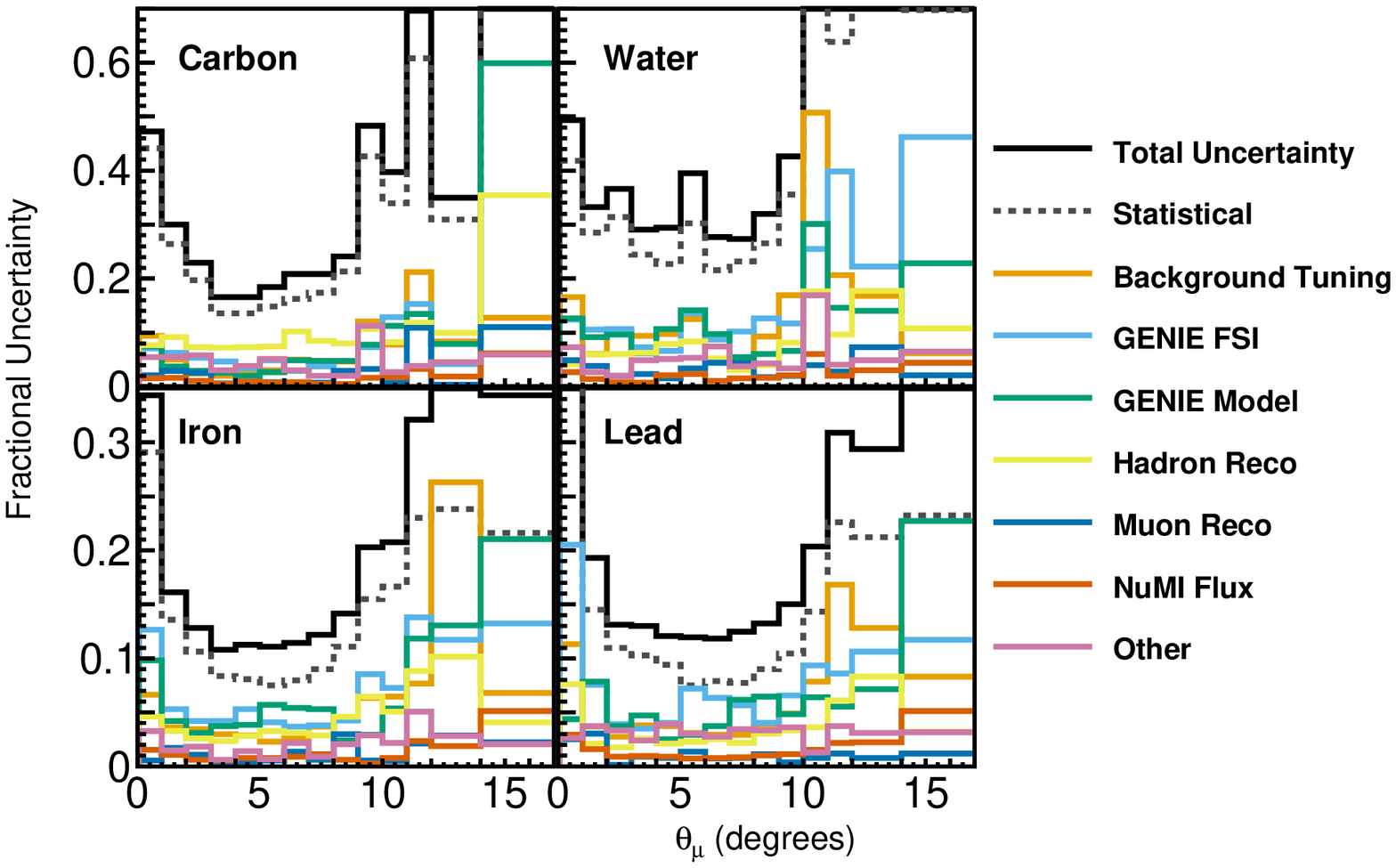}
    \caption[$\theta_\mu$ cross-section uncertainties for multiple targets ]{Uncertainties on the absolute cross section for each target (upper five plots) and on the ratio to CH (lower four plots) as a function of $\theta_\mu$.}
    \label{fig:xsec_err_muon_theta}
\end{figure}
\begin{figure}
	\centering
    \includegraphics[width=\linewidth]{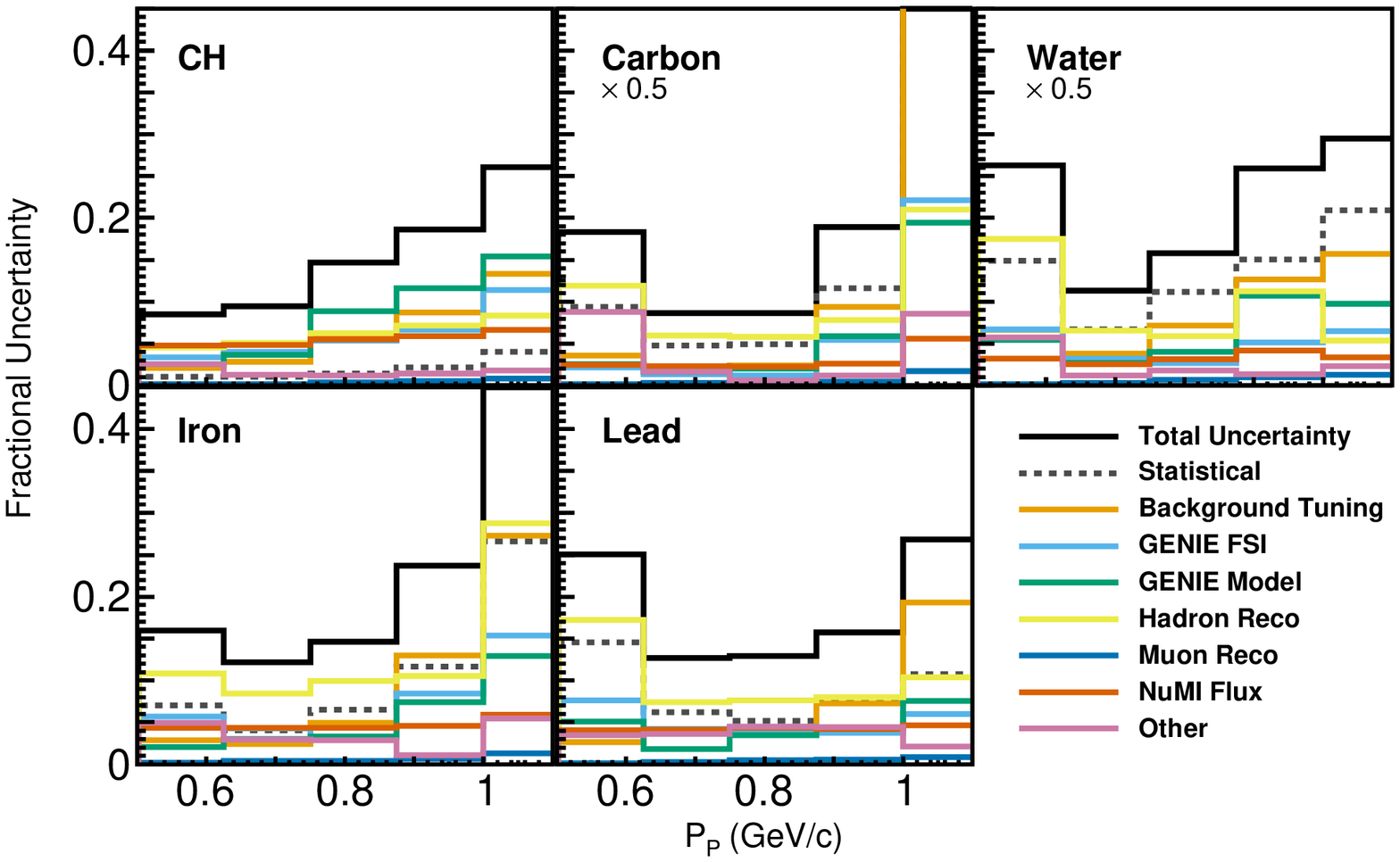}
        \includegraphics[width=\linewidth]{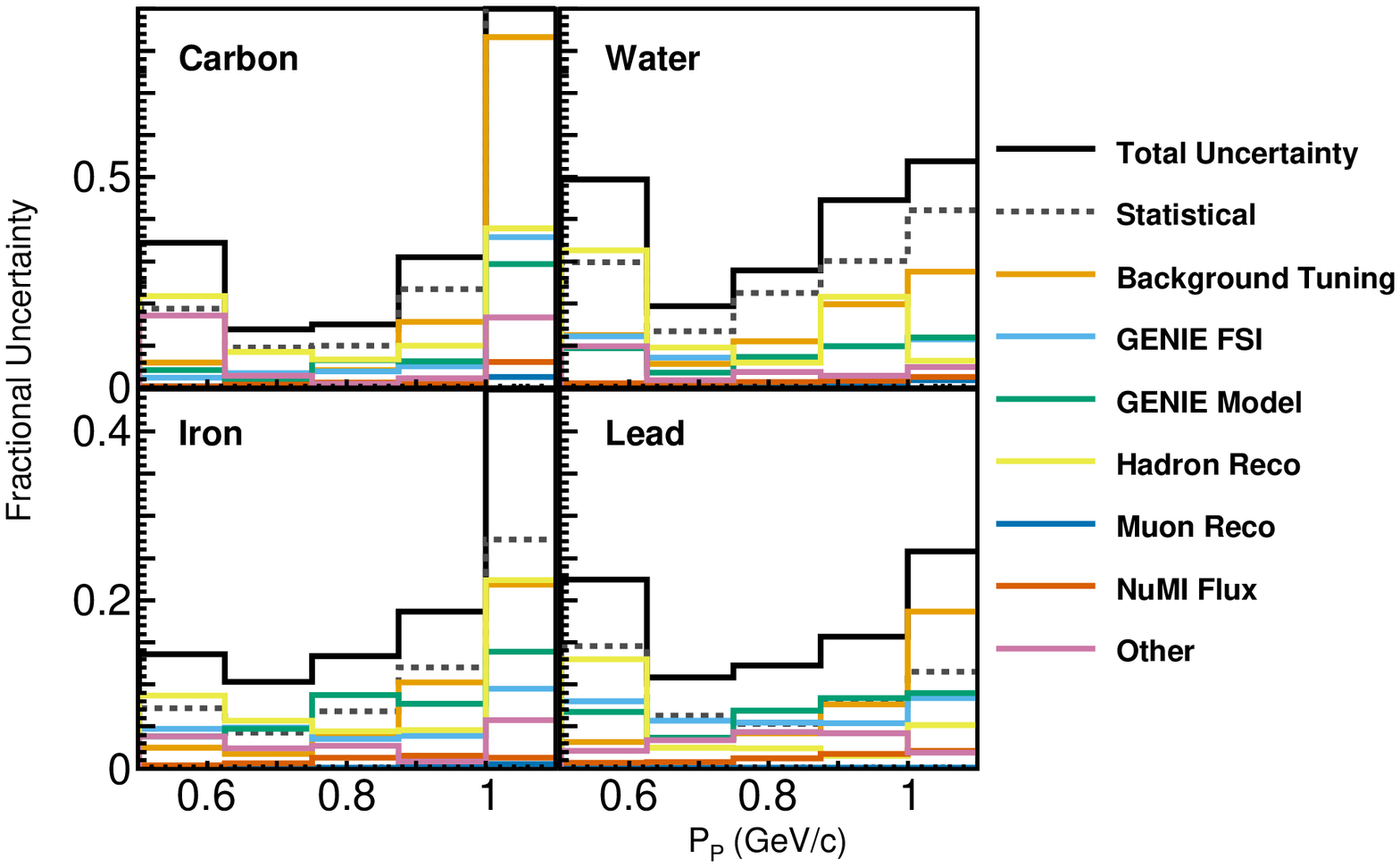}
    \caption[Proton momentum cross-section uncertainties for multiple targets ]{Uncertainties on the absolute cross section for each target (upper five plots) and on the ratio to CH (lower four plots) as a function of proton momentum.}
    \label{fig:xsec_err_proton_p}
\end{figure}
\begin{figure}
	\centering
    \includegraphics[width=\linewidth]{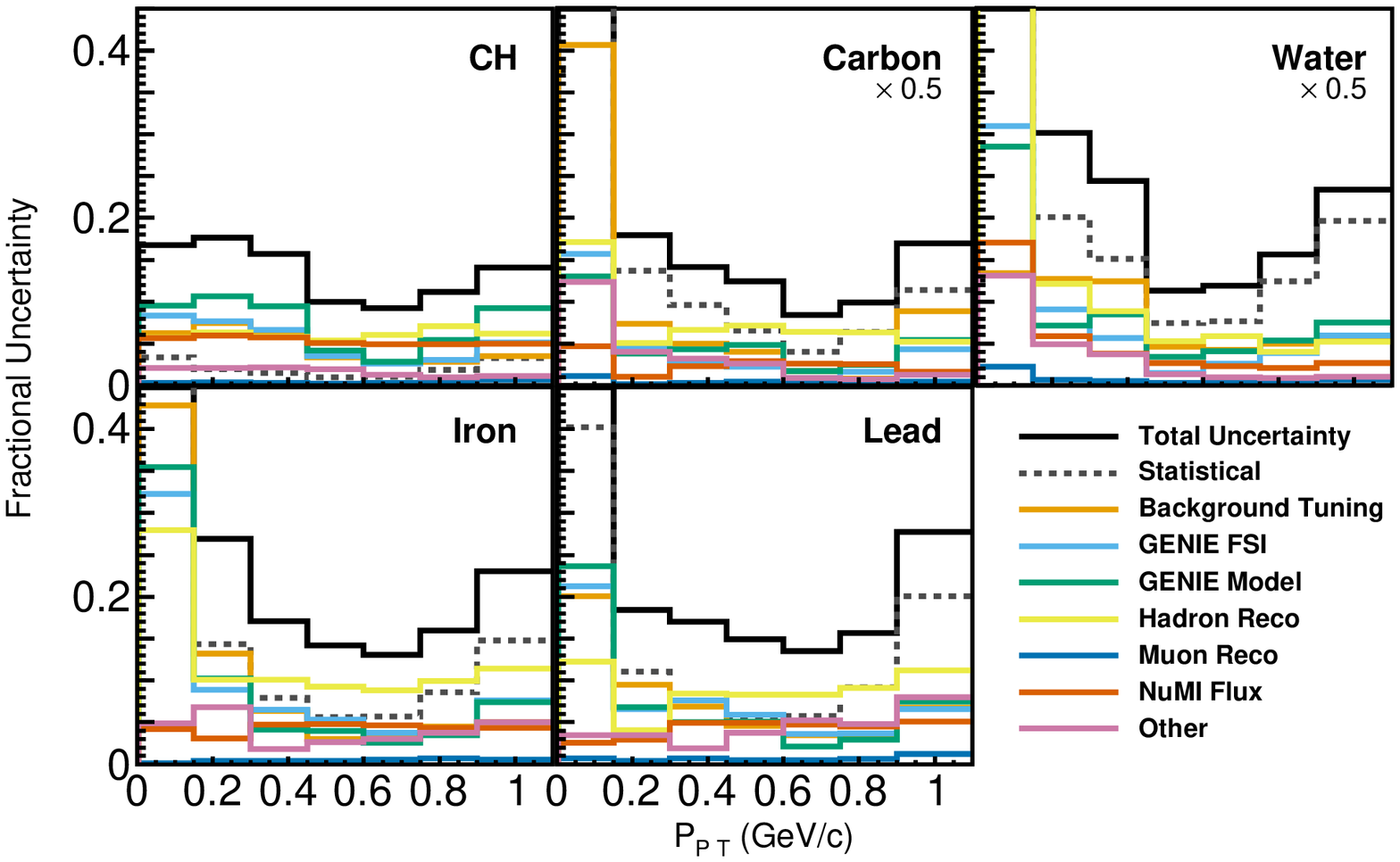}
        \includegraphics[width=\linewidth]{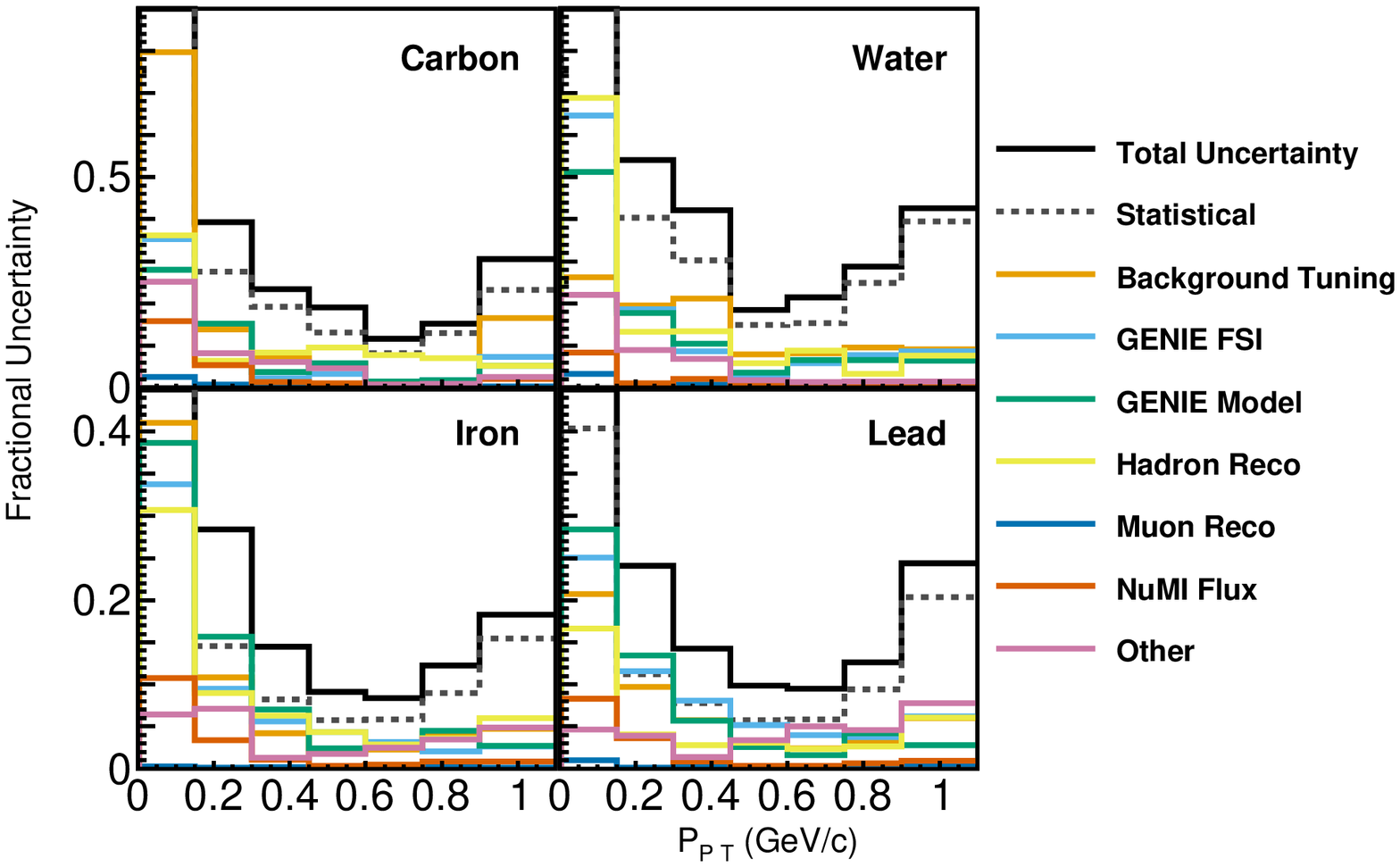}
    \caption[cross-section uncertainties for multiple targets ]{Uncertainties on the absolute cross section for each target (upper five plots) and on the ratio to CH (lower four plots) as a function of proton transverse momentum.}
    \label{fig:xsec_err_proton_pt}
\end{figure}
\begin{figure}
	\centering
    \includegraphics[width=\linewidth]{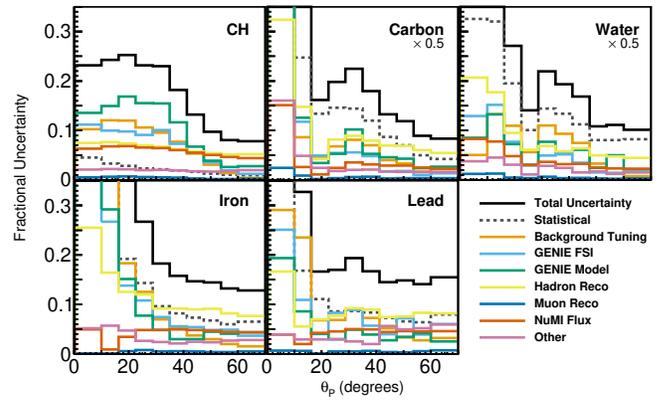}
    \includegraphics[width=\linewidth]{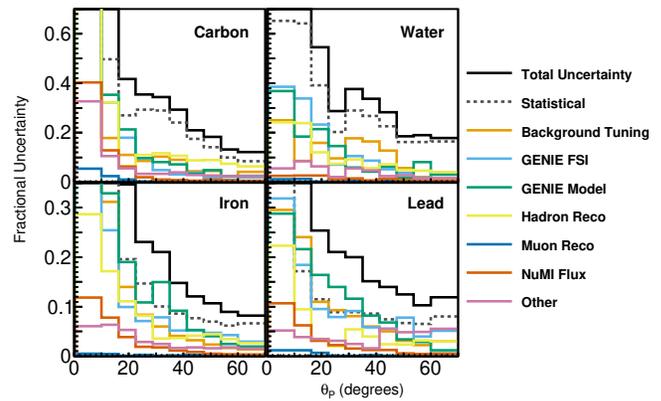}
    \caption[$\theta_p$ cross-section uncertainties for multiple targets ]{Uncertainties on the absolute cross section for each target (upper five plots) and on the ratio to CH (lower four plots) as a function of proton angle with respect to the neutrino beam.}
    \label{fig:xsec_err_proton_theta}
\end{figure}

\clearpage
\section{Comparisons of \chisq\ between models and measurements }
\label{sect:chi2tables}
This section summarizes the comparisons between the measured cross sections and the different predictions described in Table~\ref{tab:generators} by listing the \chisq\ between the measured cross section and the prediction from each generator.  All correlations in the measurement uncertainties are taken into account:  both those between bins in one distribution and those between any passive target and the scintillator cross section.  The number of degrees of freedom, $N{dof}$ is provided in each table's caption.  

\begin{table*}[h]
    \centering
    \begin{tabular}{|Q|P|P|P|P|P|P|P|P|P|}
    \hline
    Model & CH \chisq & C & C/CH & \water & \water/CH & Fe & Fe/CH & Pb & Pb/CH \\ \hline
    Minerva Tune & 26 & 7.8 & 5.4 & 0.51 & 1.1 & 5.0 & 4.5 & 44 & 31  \\ \hline
    GENIEv3 G18 01a & 4.3 & 7.8 & 5.5 & 2.6 & 1.6 & 1.6 & 1.8 & 41 & 21  \\ \hline
    GENIEv3 G18 01b & 5.3 & 7.8 & 6.0 & 2.8 & 1.4 & 9.2 & 5.4 & 29 & 27  \\ \hline
    GENIEv3 G18 10a & 9.2 & 15 & 6.2 & 2.2 & 1.5 & 5.1 & 3.3 & 45 & 35  \\ \hline
    GENIEv3 G18 10b & 8.5 & 12 & 6.3 & 3.1 & 1.7 & 23 & 10 & 53 & 44  \\ \hline
    GiBUU T0 & 6.3 & 5.2 & 5.2 & 0.66 & 0.92 & 7.4 & 8.9 & 6.7 & 13  \\ \hline
    GiBUU T1 & 19 & 5.2 & 6.1 & 0.66 & 1.7 & 7.4 & 3.3 & 6.7 & 10  \\ \hline
    NEUT LFG & 44 & 32 & 5.5 & 4.7 & 0.77 & 60 & 4.1 & 140 & 21  \\ \hline
    NuWro LFG & 71 & 18 & 5.0 & 10 & 1.8 & 9.9 & 4.9 & 49 & 6.7  \\ \hline
    NuWro SF & 5.1 & 6.8 & 5.4 & 3.4 & 2.0 & 17 & 12 & 160 & 88  \\ \hline
    \end{tabular}
    \caption{Comparison of models and data for \tkialpha cross sections  corresponding to Figs. 
    \ref{fig:models_alphat} and \ref{Figure:Gencompare_alpha_ratio}, with $\ndf=5$.}
    \label{tab:dalphat}
\end{table*}

\begin{table*}[h]
    \centering
    \begin{tabular}{|Q|P|P|P|P|P|P|P|P|P|}
    \hline
    Model & CH \chisq & C & C/CH & \water & \water/CH & Fe & Fe/CH & Pb & Pb/CH \\ \hline
    Minerva Tune & 35 & 5.9 & 3.6 & 8.3 & 3.0 & 13 & 8.5 & 35 & 21  \\ \hline
    GENIEv3 G18 01a & 29 & 4.5 & 3.2 & 6.9 & 1.2 & 16 & 7.5 & 21 & 13  \\ \hline
    GENIEv3 G18 01b & 32 & 5.4 & 5.2 & 10 & 1.5 & 30 & 8.9 & 84 & 27  \\ \hline
    GENIEv3 G18 10a & 140 & 24 & 5.1 & 8.4 & 0.88 & 17 & 10 & 39 & 26  \\ \hline
    GENIEv3 G18 10b & 130 & 20 & 5.2 & 11 & 1.4 & 28 & 19 & 100 & 62  \\ \hline
    GiBUU T0 & 21 & 2.2 & 3.0 & 5.1 & 1.8 & 20 & 7.7 & 27 & 21  \\ \hline
    GiBUU T1 & 41 & 2.2 & 5.2 & 5.1 & 1.7 & 20 & 6.1 & 27 & 14  \\ \hline
    NEUT LFG & 120 & 18 & 2.1 & 20 & 3.4 & 49 & 4.8 & 210 & 20  \\ \hline
    NuWro LFG & 250 & 32 & 3.8 & 5.3 & 0.72 & 25 & 7.6 & 42 & 11  \\ \hline
    NuWro SF & 55 & 8.3 & 4.0 & 3.9 & 0.7 & 32 & 39 & 140 & 110  \\ \hline
    \end{tabular}
    \caption{Comparison of models and data for \tkidelta cross sections  corresponding to Figs. \ref{Figure:Gencompare_dpt} and \ref{Figure:Gencompare_dpt_ratio}, with $\ndf=6$.}
    \label{tab:dpt}
\end{table*}

\begin{table*}[h]
    \centering
    \begin{tabular}{|Q|P|P|P|P|P|P|P|P|P|}
    \hline
    Model & CH \chisq & C & C/CH & \water & \water/CH & Fe & Fe/CH & Pb & Pb/CH \\ \hline
    Minerva Tune & 85 & 6.2 & 5.4 & 6.8 & 4.9 & 8.8 & 11 & 6.2 & 19  \\ \hline
    GENIEv3 G18 01a & 170 & 4.8 & 5.1 & 7.5 & 3.5 & 13 & 9.5 & 5.0 & 9.0  \\ \hline
    GENIEv3 G18 01b & 250 & 7.7 & 6.4 & 11 & 3.3 & 20 & 21 & 61 & 50  \\ \hline
    GENIEv3 G18 10a & 110 & 10 & 6.9 & 6.5 & 3.8 & 20 & 14 & 15 & 12  \\ \hline
    GENIEv3 G18 10b & 220 & 11 & 7.8 & 9.5 & 4.0 & 31 & 28 & 84 & 82  \\ \hline
    GiBUU T0 & 23 & 3.3 & 4.9 & 4.7 & 3.9 & 9.8 & 15 & 6.3 & 12  \\ \hline
    GiBUU T1 & 48 & 3.3 & 6.3 & 4.7 & 4.8 & 9.8 & 13 & 6.3 & 7.5  \\ \hline
    NEUT LFG & 120 & 19 & 4.3 & 11 & 5.1 & 56 & 12 & 160 & 45  \\ \hline
    NuWro LFG & 240 & 13 & 5.6 & 5.8 & 3.9 & 18 & 10 & 12 & 9.6  \\ \hline
    NuWro SF & 100 & 5.5 & 5.9 & 5.4 & 3.6 & 19 & 20 & 26 & 34  \\ \hline
    \end{tabular}
    \caption{Comparison of models and data for \tkidptx cross sections  corresponding to Figs. \ref{fig:models_dptx} and \ref{Fig:ratios_dptx}, with $\ndf=8$.}
    \label{tab:dptx}
\end{table*}

\begin{table*}[h]
    \centering
    \begin{tabular}{|Q|P|P|P|P|P|P|P|P|P|}
    \hline
    Model & CH \chisq & C & C/CH & \water & \water/CH & Fe & Fe/CH & Pb & Pb/CH \\ \hline
    Minerva Tune & 140 & 28 & 11 & 4.7 & 5.2 & 8.7 & 4.7 & 32 & 22  \\ \hline
    GENIEv3 G18 01a & 56 & 19 & 11 & 8.9 & 5.3 & 11 & 4.3 & 44 & 17  \\ \hline
    GENIEv3 G18 01b & 45 & 21 & 13 & 9.9 & 5.1 & 20 & 7.2 & 51 & 34  \\ \hline
    GENIEv3 G18 10a & 92 & 47 & 14 & 3.2 & 5.1 & 10 & 4.6 & 37 & 26  \\ \hline
    GENIEv3 G18 10b & 72 & 44 & 14 & 5.2 & 5.3 & 21 & 13 & 55 & 53  \\ \hline
    GiBUU T0 & 18 & 9.9 & 10 & 6.3 & 5.0 & 10 & 8.6 & 16 & 18  \\ \hline
    GiBUU T1 & 28 & 9.9 & 9.5 & 6.3 & 5.2 & 10 & 4.3 & 16 & 12  \\ \hline
    NEUT LFG & 120 & 69 & 11 & 9.7 & 6.0 & 45 & 3.6 & 120 & 26  \\ \hline
    NuWro LFG & 160 & 33 & 11 & 7.7 & 5.2 & 9.2 & 5.3 & 39 & 7.4  \\ \hline
    NuWro SF & 150 & 24 & 11 & 7.9 & 5.5 & 20 & 13 & 120 & 70  \\ \hline
    \end{tabular}
    \caption{Comparison of models and data for \tkidpty cross sections  corresponding to Figs. \ref{fig:models_dpty} and \ref{fig:ratiomodels_dpty}, with $\ndf=12$.}
    \label{tab:dpty}
\end{table*}

\begin{table*}[h]
    \centering
    \begin{tabular}{|Q|P|P|P|P|P|P|P|P|P|}
    \hline
    Model & CH \chisq & C & C/CH & \water & \water/CH & Fe & Fe/CH & Pb & Pb/CH \\ \hline
    Minerva Tune & 57 & 23 & 12 & 4.4 & 4.9 & 4.0 & 5.0 & 9.7 & 16  \\ \hline
    GENIEv3 G18 01a & 350 & 30 & 14 & 5.0 & 5.1 & 13 & 2.8 & 21 & 9.6  \\ \hline
    GENIEv3 G18 01b & 310 & 37 & 17 & 5.0 & 5.3 & 16 & 4.5 & 72 & 33  \\ \hline
    GENIEv3 G18 10a & 400 & 41 & 16 & 5.6 & 5.5 & 15 & 2.4 & 28 & 8.3  \\ \hline
    GENIEv3 G18 10b & 410 & 43 & 17 & 5.7 & 5.3 & 27 & 8.8 & 110 & 43  \\ \hline
    GiBUU T0 & 33 & 16 & 12 & 4.5 & 4.8 & 3.3 & 6.1 & 21 & 21  \\ \hline
    GiBUU T1 & 38 & 16 & 12 & 4.5 & 5.4 & 3.3 & 2.9 & 21 & 15  \\ \hline
    NEUT LFG & 740 & 57 & 12 & 6.9 & 6.2 & 44 & 3.9 & 190 & 32  \\ \hline
    NuWro LFG & 370 & 39 & 15 & 4.7 & 5.7 & 12 & 4.4 & 28 & 11  \\ \hline
    NuWro SF & 280 & 32 & 16 & 4.9 & 5.6 & 12 & 4.0 & 20 & 8.2  \\ \hline
    \end{tabular}
    \caption{Comparison of models and data for P$_\mu$ cross sections  corresponding to Figs. \ref{fig:models_muon_p} and \ref{Figure:Gencompare_pmu_ratio}, with $\ndf=8$.}
    \label{tab:muonmomentum}
\end{table*}

\begin{table*}[h]
    \centering
    \begin{tabular}{|Q|P|P|P|P|P|P|P|P|P|}
    \hline
    Model & CH \chisq & C & C/CH & \water & \water/CH & Fe & Fe/CH & Pb & Pb/CH \\ \hline
    Minerva Tune & 200 & 12 & 11 & 5.0 & 4.8 & 54 & 21 & 45 & 16  \\ \hline
    GENIEv3 G18 01a & 360 & 9.4 & 10 & 12 & 4.1 & 39 & 23 & 64 & 18  \\ \hline
    GENIEv3 G18 01b & 360 & 12 & 12 & 13 & 4.1 & 79 & 38 & 190 & 70  \\ \hline
    GENIEv3 G18 10a & 110 & 16 & 12 & 7.2 & 4.2 & 53 & 21 & 51 & 18  \\ \hline
    GENIEv3 G18 10b & 120 & 16 & 12 & 5.7 & 4.4 & 100 & 38 & 150 & 64  \\ \hline
    GiBUU T0 & 92 & 11 & 9.9 & 3.1 & 4.6 & 50 & 35 & 46 & 38  \\ \hline
    GiBUU T1 & 86 & 11 & 8.9 & 3.1 & 5.4 & 50 & 32 & 46 & 34  \\ \hline
    NEUT LFG & 140 & 23 & 8.4 & 10 & 5.5 & 140 & 30 & 330 & 82  \\ \hline
    NuWro LFG & 130 & 14 & 10 & 4.9 & 3.9 & 67 & 23 & 79 & 18  \\ \hline
    NuWro SF & 160 & 11 & 10 & 7.0 & 4.3 & 79 & 23 & 110 & 22  \\ \hline
    \end{tabular}
    \caption{Comparison of models and data for \tkiptmu cross sections  corresponding to Figs. \ref{Figure:Gencompare_ptmu} and \ref{Figure:Gencompare_ptmu_ratio}, with $\ndf=10$.}
    \label{tab:muonpt}
\end{table*}

\begin{table*}[h]
    \centering
    \begin{tabular}{|Q|P|P|P|P|P|P|P|P|P|}
    \hline
    Model & CH \chisq & C & C/CH & \water & \water/CH & Fe & Fe/CH & Pb & Pb/CH \\ \hline
    Minerva Tune & 130 & 16 & 15 & 8.0 & 6.1 & 21 & 10 & 64 & 28  \\ \hline
    GENIEv3 G18 01a & 170 & 10 & 15 & 10 & 6.3 & 28 & 9.8 & 91 & 35  \\ \hline
    GENIEv3 G18 01b & 240 & 14 & 17 & 13 & 6.6 & 63 & 17 & 260 & 90  \\ \hline
    GENIEv3 G18 10a & 250 & 20 & 17 & 7.2 & 6.1 & 38 & 9.6 & 110 & 39  \\ \hline
    GENIEv3 G18 10b & 360 & 25 & 17 & 9.9 & 7.2 & 82 & 23 & 250 & 87  \\ \hline
    GiBUU T0 & 54 & 13 & 14 & 6.6 & 5.9 & 18 & 16 & 55 & 50  \\ \hline
    GiBUU T1 & 69 & 13 & 14 & 6.6 & 6.2 & 18 & 13 & 55 & 41  \\ \hline
    NEUT LFG & 280 & 29 & 13 & 11 & 6.6 & 100 & 14 & 380 & 89  \\ \hline
    NuWro LFG & 370 & 24 & 15 & 9.9 & 6.3 & 49 & 8.7 & 140 & 28  \\ \hline
    NuWro SF & 440 & 20 & 14 & 9.9 & 6.4 & 54 & 8.8 & 190 & 44  \\ \hline
    \end{tabular}
    \caption{Comparison of models and data for $\theta_{\mu}$ cross sections  corresponding to Figs. \ref{fig:models_muon_theta} and \ref{Figure:Gencompare_mutheta_ratio}, with $\ndf=14$.}
    \label{tab:muontheta}
\end{table*}

\begin{table*}[h]
    \centering
    \begin{tabular}{|Q|P|P|P|P|P|P|P|P|P|}
    \hline
    Model & CH \chisq & C & C/CH & \water & \water/CH & Fe & Fe/CH & Pb & Pb/CH \\ \hline
    Minerva Tune & 120 & 16 & 12 & 4.7 & 6.6 & 17 & 11 & 98 & 53  \\ \hline
    GENIEv3 G18 01a & 140 & 18 & 11 & 5.3 & 5.2 & 27 & 10 & 91 & 46  \\ \hline
    GENIEv3 G18 01b & 140 & 21 & 15 & 7.3 & 5.5 & 39 & 16 & 93 & 61  \\ \hline
    GENIEv3 G18 10a & 420 & 52 & 14 & 21 & 6.3 & 49 & 17 & 59 & 68  \\ \hline
    GENIEv3 G18 10b & 440 & 51 & 14 & 24 & 6.8 & 60 & 30 & 79 & 110  \\ \hline
    GiBUU T0 & 200 & 28 & 11 & 7.3 & 6.5 & 25 & 37 & 29 & 42  \\ \hline
    GiBUU T1 & 240 & 28 & 11 & 7.3 & 6.8 & 25 & 34 & 29 & 32  \\ \hline
    NEUT LFG & 540 & 48 & 9.7 & 24 & 7.9 & 100 & 30 & 170 & 54  \\ \hline
    NuWro LFG & 780 & 70 & 12 & 38 & 6.2 & 64 & 18 & 50 & 35  \\ \hline
    NuWro SF & 70 & 19 & 11 & 9.3 & 6.5 & 83 & 78 & 170 & 170  \\ \hline
    \end{tabular}
    \caption{Comparison of models and data for \tkipn cross sections  corresponding to Figs. \ref{fig:models_pn} and \ref{Figure:Gencompare_pn_ratio}, with $\ndf=10$.}
    \label{tab:neutronmomentum}
\end{table*}

\begin{table*}[h]
    \centering
    \begin{tabular}{|Q|P|P|P|P|P|P|P|P|P|}
    \hline
    Model & CH \chisq & C & C/CH & \water & \water/CH & Fe & Fe/CH & Pb & Pb/CH \\ \hline
    Minerva Tune & 100 & 13 & 7.9 & 1.5 & 2.0 & 5.4 & 5.4 & 23 & 21  \\ \hline
    GENIEv3 G18 01a & 120 & 9.4 & 7.3 & 1.2 & 1.3 & 9.3 & 3.1 & 9.6 & 17  \\ \hline
    GENIEv3 G18 01b & 120 & 12 & 9.5 & 2.6 & 1.5 & 23 & 5.7 & 52 & 36  \\ \hline
    GENIEv3 G18 10a & 210 & 30 & 9.6 & 5.2 & 1.6 & 18 & 4.4 & 13 & 34  \\ \hline
    GENIEv3 G18 10b & 240 & 33 & 11 & 6.5 & 1.7 & 30 & 17 & 78 & 100  \\ \hline
    GiBUU T0 & 14 & 7.0 & 6.7 & 1.6 & 1.6 & 7.0 & 5.0 & 2.5 & 11  \\ \hline
    GiBUU T1 & 17 & 7.0 & 9.3 & 1.6 & 2.1 & 7.0 & 1.5 & 2.5 & 6.0  \\ \hline
    NEUT LFG & 200 & 27 & 6.1 & 7.7 & 2.2 & 53 & 2.2 & 190 & 21  \\ \hline
    NuWro LFG & 250 & 42 & 8.3 & 8.6 & 1.7 & 14 & 5.3 & 12 & 22  \\ \hline
    NuWro SF & 110 & 16 & 8.0 & 2.2 & 1.7 & 22 & 19 & 63 & 49  \\ \hline
    \end{tabular}
    \caption{Comparison of models and data for \tkiphi cross sections  corresponding to Figs. \ref{fig:models_coplan} and \ref{Figure:Gencompare_phi_ratio}, with $\ndf=6$.}
    \label{tab:phi}
\end{table*}

\begin{table*}[h]
    \centering
    \begin{tabular}{|Q|P|P|P|P|P|P|P|P|P|}
    \hline
    Model & CH \chisq & C & C/CH & \water & \water/CH & Fe & Fe/CH & Pb & Pb/CH \\ \hline
    Minerva Tune & 57 & 14 & 7.4 & 6.0 & 8.1 & 5.6 & 5.6 & 11 & 9.7  \\ \hline
    GENIEv3 G18 01a & 630 & 460 & 8.3 & 97 & 6.2 & 230 & 2.5 & 210 & 11  \\ \hline
    GENIEv3 G18 01b & 250 & 190 & 10 & 36 & 5.5 & 170 & 4.1 & 890 & 96  \\ \hline
    GENIEv3 G18 10a & 1300 & 1100 & 9.5 & 240 & 5.9 & 580 & 3.1 & 490 & 18  \\ \hline
    GENIEv3 G18 10b & 360 & 400 & 12 & 82 & 4.2 & 300 & 10 & 1700 & 140  \\ \hline
    GiBUU T0 & 51 & 7.0 & 7.3 & 15 & 10 & 21 & 21 & 33 & 58  \\ \hline
    GiBUU T1 & 53 & 7.0 & 5.6 & 15 & 14 & 21 & 17 & 33 & 34  \\ \hline
    NEUT LFG & 520 & 390 & 9.1 & 110 & 9.8 & 550 & 4.2 & 3300 & 98  \\ \hline
    NuWro LFG & 190 & 92 & 8.0 & 11 & 9.1 & 49 & 8.4 & 180 & 21  \\ \hline
    NuWro SF & 140 & 81 & 7.0 & 10 & 9.9 & 17 & 13 & 28 & 23  \\ \hline
    \end{tabular}
    \caption{Comparison of models and data for \tkipl cross sections  corresponding to Figs. \ref{fig:models_pl} and \ref{Figure:Gencompare_pl_ratio}, with $\ndf=8$.}
    \label{tab:pl}
\end{table*}

\begin{table*}[h]
    \centering
    \begin{tabular}{|Q|P|P|P|P|P|P|P|P|P|}
    \hline
    Model & CH \chisq & C & C/CH & \water & \water/CH & Fe & Fe/CH & Pb & Pb/CH \\ \hline
    Minerva Tune & 4.0 & 13 & 8.4 & 0.62 & 0.81 & 11 & 17 & 11 & 11  \\ \hline
    GENIEv3 G18 01a & 6.6 & 8.5 & 7.7 & 1.8 & 0.59 & 4.0 & 11 & 26 & 14  \\ \hline
    GENIEv3 G18 01b & 33 & 13 & 9.3 & 4.2 & 0.92 & 11 & 7.3 & 210 & 130  \\ \hline
    GENIEv3 G18 10a & 4.2 & 17 & 9.7 & 1.9 & 0.72 & 8.7 & 7.1 & 37 & 25  \\ \hline
    GENIEv3 G18 10b & 28 & 19 & 10 & 4.4 & 1.1 & 25 & 8.0 & 290 & 190  \\ \hline
    GiBUU T0 & 6.2 & 9.6 & 8.2 & 0.54 & 0.62 & 13 & 12 & 40 & 71  \\ \hline
    GiBUU T1 & 2.3 & 9.6 & 8.9 & 0.54 & 0.99 & 13 & 15 & 40 & 47  \\ \hline
    NEUT LFG & 41 & 28 & 7.9 & 5.9 & 1.2 & 40 & 3.5 & 520 & 210  \\ \hline
    NuWro LFG & 8.0 & 16 & 8.6 & 1.3 & 0.55 & 11 & 10 & 44 & 32  \\ \hline
    NuWro SF & 1.5 & 11 & 8.6 & 0.43 & 0.52 & 19 & 24 & 14 & 13  \\ \hline
    \end{tabular}
    \caption{Comparison of models and data for P$_{p}$ cross sections  corresponding to Figs. \ref{fig:models_proton_p} and \ref{Figure:Gencompare_proton_p_ratio}, with $\ndf=5$.}
    \label{tab:protonmomentum}
\end{table*}

\begin{table*}[h]
    \centering
    \begin{tabular}{|Q|P|P|P|P|P|P|P|P|P|}
    \hline
    Model & CH \chisq & C & C/CH & \water & \water/CH & Fe & Fe/CH & Pb & Pb/CH \\ \hline
    Minerva Tune & 36 & 12 & 7.9 & 8.0 & 5.3 & 3.7 & 7.2 & 12 & 14  \\ \hline
    GENIEv3 G18 01a & 41 & 14 & 8.6 & 11 & 5.9 & 5.4 & 5.5 & 16 & 11  \\ \hline
    GENIEv3 G18 01b & 32 & 17 & 10 & 13 & 7.0 & 18 & 16 & 110 & 100  \\ \hline
    GENIEv3 G18 10a & 51 & 15 & 10 & 10 & 6.2 & 13 & 8.2 & 21 & 13  \\ \hline
    GENIEv3 G18 10b & 48 & 15 & 9.9 & 11 & 7.2 & 32 & 24 & 150 & 130  \\ \hline
    GiBUU T0 & 20 & 7.8 & 8.4 & 7.2 & 6.3 & 3.7 & 15 & 18 & 41  \\ \hline
    GiBUU T1 & 22 & 7.8 & 7.4 & 7.2 & 5.5 & 3.7 & 9.5 & 18 & 25  \\ \hline
    NEUT LFG & 140 & 41 & 11 & 31 & 8.5 & 66 & 11 & 310 & 95  \\ \hline
    NuWro LFG & 52 & 8.8 & 8.2 & 3.9 & 5.0 & 7.3 & 6.4 & 18 & 15  \\ \hline
    NuWro SF & 35 & 9.7 & 7.8 & 6.2 & 5.0 & 13 & 15 & 24 & 24  \\ \hline
    \end{tabular}
    \caption{Comparison of models and data for \tkiptproton cross sections  corresponding to Figs. \ref{fig:models_proton_pt} and \ref{Figure:Gencompare_proton_pt_ratio}, with $\ndf=7$.}
    \label{tab:protonpt}
\end{table*}

\begin{table*}[h]
    \centering
    \begin{tabular}{|Q|P|P|P|P|P|P|P|P|P|}
    \hline
    Model & CH \chisq & C & C/CH & \water & \water/CH & Fe & Fe/CH & Pb & Pb/CH \\ \hline
    Minerva Tune & 49 & 15 & 6.0 & 8.6 & 7.9 & 6.5 & 6.5 & 22 & 10  \\ \hline
    GENIEv3 G18 01a & 34 & 12 & 6.4 & 9.9 & 6.2 & 7.4 & 3.0 & 24 & 11  \\ \hline
    GENIEv3 G18 01b & 20 & 13 & 7.7 & 11 & 6.2 & 14 & 9.1 & 92 & 93  \\ \hline
    GENIEv3 G18 10a & 34 & 13 & 7.1 & 8.4 & 6.1 & 11 & 6.8 & 34 & 15  \\ \hline
    GENIEv3 G18 10b & 25 & 12 & 7.7 & 8.6 & 5.7 & 20 & 17 & 140 & 110  \\ \hline
    GiBUU T0 & 29 & 7.9 & 6.5 & 11 & 7.7 & 8.3 & 12 & 23 & 39  \\ \hline
    GiBUU T1 & 23 & 7.9 & 5.6 & 11 & 9.3 & 8.3 & 7.4 & 23 & 21  \\ \hline
    NEUT LFG & 73 & 44 & 7.9 & 30 & 10 & 70 & 8.4 & 230 & 77  \\ \hline
    NuWro LFG & 37 & 6.8 & 6.5 & 4.7 & 6.3 & 5.8 & 7.0 & 22 & 13  \\ \hline
    NuWro SF & 15 & 8.3 & 6.3 & 7.5 & 6.6 & 10 & 9.9 & 43 & 27  \\ \hline
    \end{tabular}
    \caption{Comparison of models and data for $\theta_{p}$ cross sections  corresponding to Figs. \ref{fig:models_proton_theta} and \ref{Figure:Gencompare_proton_theta_ratio}, with $\ndf=10$.}
    \label{tab:protontheta}
\end{table*}

\end{document}